\documentclass[12pt,pdftex]{article}

\usepackage{amsmath,amssymb,epsfig,ulem}
\pdfoutput=1

\allowdisplaybreaks

\usepackage{color}
\usepackage[bookmarks=false]{hyperref}

\def\bm#1{\mbox{\boldmath$#1$\unboldmath}}
\def\sgn{\mbox{sgn}}
\def\Mkk{M_{\rm KK}}

\newcommand{\beq}{\begin{equation}}
\newcommand{\eeq}{\end{equation}}
\newcommand{\ord}{{\cal O}}
\newcommand{\ie}{{\it i.e.}}

\newcommand{\sws}{s_w^2}
\newcommand{\swq}{s_w^4}
\newcommand{\cws}{c_w^2}
\newcommand{\cwq}{c_w^4}
\newcommand{\sw}{s_w}
\newcommand{\cw}{c_w}
\newcommand{\no}{\nonumber}
\newcommand{\epseps}{{\epsilon'_K/\epsilon_K}}

\begin{document}
\normalem

\begin{titlepage}

\begin{flushright}
MZ-TH/09-46
\end{flushright}

\vspace{0.2cm}
\begin{center}
\Large\bf
Flavor Physics in the Randall-Sundrum Model\\[0.2cm]
\large\bf
II.~Tree-Level Weak-Interaction Processes
\end{center}

\vspace{0.2cm}
\begin{center}
  {\sc M. Bauer$^{a}$, S. Casagrande$^{b}$, U. Haisch$^{a}$
    and M. Neubert$^{a}$}\\
  \vspace{10pt} {$^{a}$ \sl Institut f\"ur Physik (WA THEP),
    Johannes Gutenberg-Universit\"at\\
    D-55099 Mainz, Germany \\} \vspace{5pt} {$^{b}$ \sl Excellence
    Cluster Universe,
    Technische Universit\"at M\"unchen \\
    D-85748 Garching, Germany}
\end{center}

\vspace{0.2cm}
\begin{abstract}\noindent
  A comprehensive analysis of tree-level weak interaction processes at
  low energy is presented for the Randall-Sundrum (RS) model with
  $SU(2)_L \times U(1)_Y$ bulk gauge symmetry and brane-localized
  Higgs sector. The complete form of the effective weak Hamiltonian is
  obtained, which results from tree-level exchange of Kaluza-Klein
  (KK) gluons and photons, the $W^\pm$ and $Z^0$ bosons and their KK
  excitations, as well as the Higgs boson. Exact expressions are used
  for the bulk profiles of the various fields, and for the exchange of
  entire towers of KK gauge-boson states. A detailed phenomenological
  analysis is performed for potential new-physics effects in
  neutral-meson mixing and in rare decays of kaons and $B$ mesons,
  including both inclusive and exclusive processes. We find that while
  the predictions for $\Delta F = 2$ observables are rather
  model-independent, $\Delta F = 1$ processes depend sensitively on
  the exact realizations of the electroweak gauge and the fermionic
  sector. In this context, we emphasize that the localization of the
  right-handed top quark in the extra dimension plays a crucial role
  in the case of rare $Z^0$-mediated decays, as it determines the
  relative size of left- to right-handed couplings. We also extend
  earlier studies of quark flavor-changing neutral currents by
  examining observables which up to now attracted little
  attention. These include $D$--$\bar D$ mixing, $B \to \tau
  \nu_\tau$, $B \to X_s (K^\ast) \, l^+ l^-$,
  $\epsilon^\prime_K/\epsilon_K$, $\bar B \to \pi \bar K$, $\bar B^0
  \to \phi K_S$, $\bar B^0 \to \eta^\prime K_S$, and $B^+ \to \pi^+
  \pi^0$.
\end{abstract}
\vfil

\end{titlepage}

\tableofcontents
\newpage

\section{Introduction}
\label{sec:intro}

Models with a warped extra dimension provide a compelling geometrical
explanation of a number of mysteries left unexplained by the Standard
Model (SM), most notably the gauge hierarchy and the flavor
problem. In these scenarios, first proposed by Randall and Sundrum
(RS) \cite{Randall:1999ee}, one studies the SM on a background
consisting of Minkowski space embedded in a slice of five-dimensional
(5D) anti-de-Sitter space (AdS$_5$). The fifth dimension is an
$S^1/Z_2$ orbifold of size $r$ labeled by a coordinate
$\phi\in[-\pi,\pi]$. The metric is given by
\beq
   ds^2 = e^{-2\sigma(\phi)}\,\eta_{\mu\nu}\,dx^\mu dx^\nu 
    - r^2 d\phi^2 \,, \qquad
   \sigma(\phi) = kr|\phi| \,,
\eeq
where $x^\mu$ denote the coordinates on the four-dimensional (4D)
hypersurfaces of constant $\phi$ with metric $\eta_{\mu\nu} =
\mbox{diag}(1, -1, -1, -1)$, and $e^{\sigma(\phi)}$ is called the warp
factor. Three-branes are placed at the orbifold fixed points $\phi=0$
as well as $\phi=\pi$. The brane at $\phi=0$ is called Planck or
ultra-violet (UV) brane, while the brane at $\phi=\pi$ is called TeV
or infra-red (IR) brane. In the original formulation of the RS model
all SM fields were confined to the IR brane, while gravity probed the
bulk of the higher-dimensional space-time. The hierarchy problem is
then resolved through the warping of space along the fifth
dimension. While the RS model only has a single fundamental scale $k$,
the AdS$_5$ curvature, the effective scale of the model exponentially
varies along the extra dimension due to the presence of the warp
factor. For fields near the IR brane the effective Planck scale is
thus redshifted to be of order the weak scale.

In later work, variations on the original setup were considered, where
also the SM fields except for the Higgs boson were allowed to
propagate in the bulk \cite{Davoudiasl:1999tf, Pomarol:1999ad,
  Grossman:1999ra, Chang:1999nh, Gherghetta:2000qt}. It was soon
realized that this framework provides an interesting new approach to
the flavor problem, as now also the hierarchical structures observed
in the masses and mixing of the SM fermions could be explained in
terms of geometrical effects \cite{Grossman:1999ra,
  Gherghetta:2000qt, ArkaniHamed:1999dc, Huber:2000ie,
  Huber:2003tu}. Since the fermion zero modes are exponentially
localized near either one of the two branes, the effective Yukawa
couplings resulting from their wave-function overlap with the Higgs
boson naturally exhibit exponential hierarchies. In this way one
obtains an extra-dimensional realization \cite{Huber:2003tu,
  Casagrande:2008hr, Blanke:2008zb} of the Froggatt-Nielsen mechanism
\cite{Froggatt:1978nt}, in which the flavor structure is accounted
for apart from ${\cal O}(1)$ factors.

Addressing the flavor hierarchies via warping in an extra dimension
makes distinctive predictions for flavor-changing processes as
well. Various new sources of flavor violation arise in RS models as a
consequence of non-trivial overlap factors between fermions and gauge
(or Higgs) bosons, which generically are non-diagonal in the mass
basis. While the new flavor-changing effects generically arise already
at tree level, a dynamical mechanism referred to as RS-GIM mechanism
\cite{Gherghetta:2000qt, Agashe:2004ay, Agashe:2004cp} ensures that
these effects are suppressed, for most observables, to an acceptable
level. During the past years, diverse studies of the flavor structure
of RS models have been performed. Properties of the generalized CKM
matrix, neutral-meson mixing, and CP violation were studied in
\cite{Huber:2003tu}. $Z^0$-mediated flavor-changing neutral currents
(FCNCs) in the kaon system were considered in \cite{Burdman:2002gr},
and effects of Kaluza-Klein (KK) gauge bosons on CP asymmetries in
rare hadronic $B$-meson decays induced by $b\to s$ transitions were
explored in \cite{Burdman:2003nt}. A first general survey of $\Delta
F=2$ and $\Delta F=1$ processes in the RS framework was presented in
\cite{Agashe:2004ay, Agashe:2004cp}. The branching ratios for the
flavor-changing top-quark decays $t\to c Z^0(\gamma,g)$ were examined
in \cite{Agashe:2006wa}. The first complete study of all operators
relevant to $K$--$\bar K$ mixing was presented in
\cite{Csaki:2008zd}.  Comprehensive analyses of $B_{d,s}$--$\bar
B_{d,s}$ mixing \cite{Blanke:2008zb}, rare leptonic $K$- and
$B$-meson decays \cite{Blanke:2008yr, Chang:2007uz} as well as of the
radiative $B \to X_s \gamma$ decay \cite{Agashe:2008uz} have been
performed quite recently. Higgs- \cite{Azatov:2009na} and
radion-mediated \cite{Azatov:2008vm, Davoudiasl:2009xz} FCNCs have
also been investigated. It has been recognized that the only
observables where some fine-tuning of parameters appears to be
unavoidable are CP-violating effects in the neutral kaon system
\cite{Csaki:2008zd, Gedalia:2009ws} and the neutron electric dipole
moment \cite{Agashe:2004ay, Agashe:2004cp}, which for generic choices
of parameters turn out to be too large unless the masses of the
lightest KK gauge bosons lie above (10--20)\,TeV. In view of these
problems, several modifications of the quark flavor sector of warped
extra-dimension models have been proposed. Most of them try to
implement the notion of minimal flavor violation (MFV)
\cite{Buras:2000dm, D'Ambrosio:2002ex} into the RS framework by using
a bulk flavor symmetry \cite{Cacciapaglia:2007fw, Fitzpatrick:2007sa,
  Santiago:2008vq, Csaki:2008eh, Csaki:2009bb, Csaki:2009wc}.  The
problem of too large electric dipole moments has been addressed using
the idea of spontaneous CP violation in the context of warped extra
dimensions \cite{Cheung:2007bu}.

As our benchmark scenario we consider in this work the simplest
implementation of the RS model, featuring an $SU(2)_L \times U(1)_Y$
bulk gauge symmetry and a minimal, brane-localized Higgs sector. Our
analysis is thus in some sense orthogonal to that of
\cite{Blanke:2008zb, Blanke:2008yr}, where $\Delta F = 2$ and $\Delta
F = 1$ processes were investigated in a setup with enlarged gauge
symmetry, namely $SU(2)_L \times SU(2)_R \times U(1)_X \times P_{LR}$,
where $P_{LR}$ exchanges the $SU(2)$ groups. The comparison between
the findings of the latter articles and the results obtained here
allows us to address the important question about the model
(in)dependence of the predictions for neutral-meson mixing and rare
leptonic decays of $K$- and $B$-mesons. We find that while $\Delta F =
2$ observables are relatively model-independent, predictions for
$\Delta F = 1$ processes depend strongly on the exact realization of
both the gauge and fermionic sectors. In this context, we emphasize
that the degree of compositeness of the right-handed top quark,
characterized by its localization in the extra dimension, plays a
crucial role in the case of rare $Z^0$-mediated decays, as it
determines the relative size of left- to right-handed couplings. The
present paper also extends previous studies of quark FCNCs, since it
examines in detail observables which partly or even fully escaped the
attention of the flavor community so far. The list of new observables
includes, among others, $D$--$\bar D$ mixing, $B \to \tau \nu_\tau$,
$B \to X_s (K^\ast) \, l^+ l^-$, $\epsilon^\prime_K/\epsilon_K$, and
$\bar B \to \pi \bar K$. The main question that we will address in the
context of the decays $B \to \tau \nu_\tau$ and $\bar B \to \pi \bar
K$ is whether possible discrepancies, as suggested by experiment, can
be explained in the minimal RS scenario. In the case of $\epseps$, we
will point out that the strong sensitivity of this ratio to the
electroweak penguin sector leads to interesting correlations with the
rare $K\to\pi\nu\bar\nu$ and $K_L\to\pi^0 l^+ l^-$ decays in scenarios
with warped extra dimensions.

This paper is a sequel to our recent work \cite{Casagrande:2008hr},
where we have derived the theoretical foundations forming the basis of
the present analysis and studied in detail the implications of the RS
model for electroweak precision physics. In particular, we have
derived the exact solutions for the bulk profiles of the various gauge
and matter fields, and given closed expressions for sums over KK gauge
bosons in tree-level low-energy processes. We have also presented a
systematic analysis of the flavor-changing couplings of gauge bosons
and of the Higgs boson to the SM fermions. The discussion in the
present paper is self-contained as far as phenomenology is concerned,
but it relies on many equations and results derived in our previous
work, which for the sake of brevity will not be rederived here. The
two papers should therefore be studied together. We will refer to
equations from our paper \cite{Casagrande:2008hr} by adding a roman
``I'' before the equation number. In Section~\ref{sec:prelim} we
briefly recall some important notations, which will be needed
throughout the paper. In Section~\ref{sec:4ferm} we review the
different interactions between fermions and the gauge and Higgs bosons
in the RS model, many of which give rise to new flavor-changing
effects not present in the SM. In Section~\ref{sec:treelevel} we
present a detailed analysis of various weak-interaction processes,
including neutral-meson mixing, as well as rare non-leptonic and
leptonic decays of kaons and $B$ mesons. In each case we present the
relevant effective weak Hamiltonians at tree level in the minimal RS
model and collect the formulas needed for a phenomenological
analysis. A detailed numerical study of all these processes is
performed in Section~\ref{sec:numerics}. We first discuss how we
efficiently perform the scan over the vast parameter space of the RS
model. We then define a few benchmark parameter scenarios, which we
will use to illustrate the type of effects that typically arise in
such models. After a numerical analysis of quark mixing matrices we
then analyse the various decay modes in
detail. Section~\ref{sec:concl} contains a summary of our main results
and some conclusions. In Appendix~\ref{app:RGE} we present formulas
describing the renormalization group (RG) evolution of the Wilson
coefficients of the QCD and electroweak penguin operators, whereas in
Appendix~\ref{app:input} we collect the input values for the SM
parameters used in the course of the article.

\section{Preliminaries}
\label{sec:prelim}

Before going into the details of the structure of flavor-changing
interactions in the RS model, we need to review some important
notations and definitions from \cite{Casagrande:2008hr}, which are
needed for the further discussion. The reader is referred to this
reference for more details.

In order to address the hierarchy between the fundamental Planck and
the electroweak scale, the logarithm of the warp factor
\beq
   L\equiv kr\pi
   \equiv \ln\frac{\Lambda_{\rm UV}}{\Lambda_{\rm IR}}
   \equiv -\ln\epsilon \,, 
\eeq 
has to be chosen such that $L\approx\ln(10^{16})\approx 37$. Below we
will sometimes refer to $L$ as the ``volume'' of the extra
dimension. The warp factor also sets the mass scale for the low-lying
KK excitations of the SM fields to be of order of the ``KK scale''
\beq
   \Mkk\equiv k\epsilon = k\,e^{-kr\pi} 
   = \ord(\mbox{few TeV}) \,.
\eeq
For instance, the masses of the first KK photon and gluon are
approximately equal to $2.45\Mkk$.

Introducing a coordinate $t=\epsilon\,e^{\sigma(\phi)}$ along the
extra dimension \cite{Grossman:1999ra}, which runs from $t=\epsilon$
on the UV brane to $t=1$ on the IR brane, we write the KK
decompositions of the left-handed (right-handed) components of the 5D
$SU(2)_L$ doublet (singlet) quark fields as
\beq\label{KKdecomp}
\begin{split}
  q_L(x,t) &\propto \mbox{diag}\left[ F(c_{Q_i})\,t^{c_{Q_i}} \right]
  \bm{U}_q\,q_L^{(0)}(x) + \ord\bigg( \frac{v^2}{\Mkk^2} \bigg)
  \mbox{}+ \mbox{KK excitations} \,, \\[1mm]
  q_R^c(x,t) &\propto \mbox{diag}\left[ F(c_{q_i})\,t^{c_{q_i}}
  \right] \bm{W}_q\,q_R^{(0)}(x) + \ord\bigg( \frac{v^2}{\Mkk^2}
  \bigg) \mbox{}+ \mbox{KK excitations} \,,
\end{split}
\eeq 
where $v\approx 246\,$GeV denotes the Higgs vacuum expectation value,
and $q=u,d$ stands for up- and down-type quarks, respectively. The
fields are three-component vectors in flavor space. The 5D fields on
the left-hand side refer to interaction eigenstates, while the 4D
fields appearing on the right are mass eigenstates. The superscript
``(0)'' is used to denote the light SM fermions, often called ``zero
modes''. In this article we will not deal with heavy KK fermions.

The zero-mode profile \cite{Grossman:1999ra, Gherghetta:2000qt}
\beq
   F(c) = \mbox{sgn}[\cos(\pi c)]\, 
   \sqrt{\frac{1+2c}{1-\epsilon^{1+2c}}}
\eeq
is exponentially suppressed in the ``volume factor'' $L$ if the bulk
mass parameters $c_{Q_i}=+M_{Q_i}/k$ and $c_{q_i}=-M_{q_i}/k$ are
smaller than the critical value $-1/2$, in which case $F(c)\sim
e^{L(c+\frac12)}$. Here $M_{Q_i}$ and $M_{q_i}$ denote the masses of
the 5D $SU(2)_L$ doublet and singlet fermions. This mechanism explains
in a natural way the large hierarchies observed in the spectrum of the
quark masses \cite{Gherghetta:2000qt, Huber:2000ie}, which follow
from the eigenvalues of the effective Yukawa matrices
\beq\label{Yeff}
   \bm{Y}_q^{\rm eff} = \mbox{diag}\left[ F(c_{Q_i}) \right]
   \bm{Y}_q\,\mbox{diag}\left[ F(c_{q_i}) \right]
   = \bm{U}_q\,\bm{\lambda}_q\,\bm{W}_q^\dagger \,.
\eeq
The 5D Yukawa matrices $\bm{Y}_q$ are assumed to be anarchic, {\em
  i.e.}, non-hierarchical with $\ord(1)$ complex elements, and
$\bm{\lambda}_q$ are diagonal matrices with entries
$(\lambda_q)_{ii}=\sqrt2\,m_{q_i}/v$. The unitary matrices $\bm{U}_q$
and $\bm{W}_q$ in (\ref{KKdecomp}) and (\ref{Yeff}) have a
hierarchical structure given by
\beq\label{eq:UqWq}
   (U_q)_{ij} \sim
   \begin{cases} 
    \frac{\displaystyle F(c_{Q_i})}{\displaystyle F(c_{Q_j})} \,, \!\!
    & \, i\le j \,, \\[3mm]
    \frac{\displaystyle F(c_{Q_j})}{\displaystyle F(c_{Q_i})} \,, \!\!
    & \, i>j \,, 
   \end{cases} \qquad 
   (W_q)_{ij} \sim
   \begin{cases} 
    \frac{\displaystyle F(c_{q_i})}{\displaystyle F(c_{q_j})} \,, \!\! 
    & \, i\le j \,, \\[3mm]
    \frac{\displaystyle F(c_{q_j})}{\displaystyle F(c_{q_i})} \,, \!\! 
    & \, i>j \,. 
   \end{cases}
\eeq

Terms omitted in (\ref{KKdecomp}) are suppressed by at least two
powers of the small ratio $v/\Mkk$ and can be neglected to a good
approximation.  We refer to this as the ``zero-mode approximation''
(ZMA). It allows us to derive compact analytic formulas for the flavor
mixing matrices that are both transparent and accurate. While such an 
approach in general lead to sensible results \cite{Goertz:2008vr}, in
our numerical analysis in Section~\ref{sec:numerics} we nevertheless
employ the exact expressions for the various matrices derived in
\cite{Casagrande:2008hr}.

\section{Induced Four-Fermion Interactions at Low Energies}
\label{sec:4ferm}

Of special interest are the dimension-six operators describing the
effective four-fermion interactions at low energies induced by heavy
gauge-boson exchange. In particular, these will give rise to $\Delta
F=2$ and $\Delta F=1$ FCNC processes, which are loop-suppressed in the
SM. In the RS model, such processes arise already at tree level.
However, as mentioned earlier, they are protected by the RS-GIM
mechanism.

\subsection{Exchange of KK Photons and Gluons}

\begin{figure}
  \begin{center}
    \includegraphics{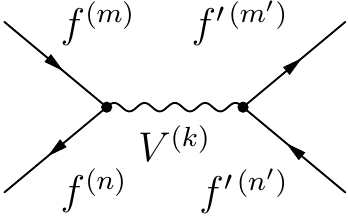}\hspace{2cm}\includegraphics{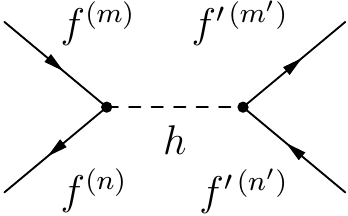}
  \end{center} \vspace{-0.5cm}
  \begin{center}
    \parbox{15.5cm}{\caption{\label{fig:fourfermion} Contributions to the
        effective four-fermion interactions arising from the tree-level
        exchange of the gauge bosons $V=\gamma,g,Z^0,W^\pm$ and their KK
        excitations (left), and of the Higgs boson (right).}}
  \end{center}
\end{figure}

We begin with a discussion of the interactions induced by the exchange
of KK photons and gluons. The graph on the left in
Figure~\ref{fig:fourfermion} shows an example of a diagram giving rise
to such contributions. The relevant sums over KK modes can be
evaluated by means of~(I:34). In the case of KK photon exchange, we
find that the effective Hamiltonian at low energies is given by
\beq\label{photoncouplings}
\begin{split}
  {\cal H}_{\rm eff}^{(\gamma)} &= \frac{2\pi\alpha}{\Mkk^2}
  \sum_{f,f'}\,Q_f\,Q_{f'}\, \bigg\{ \frac{1}{2L} \left( \bar
    f\gamma^\mu f \right) \left( \bar f'\gamma_\mu f' \right) - 2
  \left( \bar f_L\gamma^\mu\bm{\Delta}'_F f_L + \bar
    f_R\gamma^\mu\bm{\Delta}'_f f_R \right) \left( \bar f'\gamma_\mu
    f' \right) \\
  &\quad\mbox{}+ 2L \left( \bar f_L\gamma^\mu\bm{\widetilde\Delta}_F
    f_L + \bar f_R\gamma^\mu\bm{\widetilde\Delta}_f f_R \right)
  \otimes \left( \bar f'_L\gamma_\mu\bm{\widetilde\Delta}_{F'} f'_L +
    \bar f'_R\gamma_\mu\bm{\widetilde\Delta}_{f'} f'_R \right) \bigg\}
  \,.
\end{split}
\eeq
Here the sum over fermions implicitly includes the sum over all KK
modes. The matrices $\bm{\Delta}'_A$ have been defined in
(I:122). These are infinite-dimensional matrices in the space of
flavor and KK modes. In addition, we have defined the new mixing
matrices (with $F=U,D$ and $f=u,d$, and similarly in the lepton
sector) \cite{Bauer:2008xb}
\beq\label{Deltaotimes}
\begin{split}
  \big( \widetilde\Delta_F \big)_{mn}\otimes \big(
  \widetilde\Delta_{f'} \big)_{m'n'} &= \frac{2\pi^2}{L^2\epsilon^2}
  \int_\epsilon^1\!dt \int_\epsilon^1\!dt'\,t_<^2 \\
  &\quad\times \left[ a_m^{(F)\dagger}\,\bm{C}_m^{(Q)}(\phi)\,
    \bm{C}_n^{(Q)}(\phi)\,a_n^{(F)} +
    a_m^{(f)\dagger}\,\bm{S}_m^{(f)}(\phi)\,
    \bm{S}_n^{(f)}(\phi)\,a_n^{(f)} \right] \\
  &\quad\times \left[
    a_{m'}^{(f')\dagger}\,\bm{C}_{m'}^{(f')}(\phi')\,
    \bm{C}_{n'}^{(f')}(\phi')\,a_{n'}^{(f')} +
    a_{m'}^{(F')\dagger}\,\bm{S}_{m'}^{(Q)}(\phi')\,
    \bm{S}_{n'}^{(Q)}(\phi')\,a_{n'}^{(F')} \right] ,
\end{split}
\eeq
{\it etc.} Notice that the matrices $\bm{\widetilde\Delta}_A \otimes
\bm{\widetilde\Delta}_B$ are not defined individually, but only as
tensor products, as indicated by the $\otimes$ symbol. The couplings
to SM fermions are encoded in the upper-left $3\times 3$ blocks of
each $\bm{\widetilde\Delta}_A \otimes \bm{\widetilde\Delta}_B$
matrix. We emphasize that the result (\ref{photoncouplings}) is
exact. In particular, no expansion in powers of $v^2/\Mkk^2$ has been
performed. The effective interactions arising from KK gluon exchange
have a very similar structure, except that we need to restrict the sum
over fermions in (\ref{photoncouplings}) to quarks and replace
$\alpha\,Q_f\,Q_{f'}$ by $\alpha_s\,t^a\otimes t^a$, where the color
matrices $t^a$ must be inserted inside the quark bi-linears.

The four-fermion operators induced by KK gluon exchange give the by
far dominant (leading) contribution to the effective weak Hamiltonians
describing $K$--$\bar K$ ($B_{d,s}$--$\bar B_{d,s}$ and $D$--$\bar D$)
mixing. This observation is important, as it implies that mixing
phenomena mainly probe the extra-dimensional aspects of the strong
interactions, but are to first approximation insensitive to the
precise embedding of the electroweak gauge symmetry in the
higher-dimensional geometry.

\boldmath \subsection{Exchange of the $Z^0$ Boson and its KK
  Excitations} \unboldmath

The induced interactions arising from the exchange of the $Z^0$ boson
and its KK excitations have a richer structure. In this case the
relevant sum over profiles is evaluated by means of
relation~(I:33). We obtain
\begin{align}
   {\cal H}_{\rm eff}^{(Z)} 
   &= \frac{4\pi\alpha}{\sws\cws\,m_Z^2}
    \left[ 1 + \frac{m_Z^2}{2\Mkk^2} \left( 1 - \frac{1}{2L} \right)
    + \ord\left( \frac{m_Z^4}{\Mkk^4} \right) \right] \no \\
   &\quad\times \sum_{f,f'} \left[ \bar f_L\gamma^\mu T_3^f 
    (\bm{1} - \bm{\delta}_F) f_L 
    + \bar f_R\gamma^\mu T_3^f \bm{\delta}_f f_R
    - \sws\,Q_f \bar f\gamma^\mu f \right] \no \\
   &\hspace{11mm}\times \left[ \bar f'_L\gamma_\mu T_3^{f'} ( \bm{1} -
    \bm{\delta}_{F'}) f'_L + \bar f'_R\gamma_\mu T_3^{f'}
    \bm{\delta}_{f'} f'_R
    - \sws\,Q_{f'} \bar f'\gamma_\mu f' \right] \no \\
   &\quad\mbox{}+ \frac{4\pi\alpha L}{\sws\cws\,\Mkk^2} \\
   &\quad\times \sum_{f,f'} \Bigg\{ - \Big[ \bar f_L\gamma^\mu T_3^f
    (\bm{\Delta}_F - \bm{\varepsilon}_F) f_L + \bar f_R\gamma^\mu T_3^f
    \bm{\varepsilon}_f f_R - \sws\,Q_f \left( \bar
    f_L\gamma^\mu\bm{\Delta}_F f_L
    + \bar f_R\gamma^\mu\bm{\Delta}_f f_R \right) \Big] \no \\
   &\hspace{19.5mm}\times \left[ \bar f'_L\gamma_\mu T_3^{f'} ( \bm{1} -
    \bm{\delta}_{F'}) f'_L + \bar f'_R\gamma_\mu T_3^{f'}
    \bm{\delta}_{f'} f'_R
    - \sws\,Q_{f'} \bar f'\gamma_\mu f' \right] \no \\
   &\qquad\mbox{}+ \Big[ \bar f_L\gamma^\mu T_3^f
    (\bm{\widetilde\Delta}_F - \bm{\widetilde\varepsilon}_F) f_L + \bar
    f_R\gamma^\mu T_3^f \bm{\widetilde\varepsilon}_f f_R - \sws\,Q_f
    \left( \bar f_L\gamma^\mu\bm{\widetilde\Delta}_F f_L
    + \bar f_R\gamma^\mu\bm{\widetilde\Delta}_f f_R \right) \Big] \no \\
   &\qquad\otimes \Big[ \bar f'_L\gamma_\mu T_3^{f'}
    (\bm{\widetilde\Delta}_{F'} - \bm{\widetilde\varepsilon}_{F'}) f'_L
    + \bar f'_R\gamma_\mu T_3^{f'} \bm{\widetilde\varepsilon}_{f'} f'_R
    - \sws\,Q_{f'} \left( \bar f'_L\gamma_\mu\bm{\widetilde\Delta}_{F'}
    f'_L + \bar f'_R\gamma_\mu\bm{\widetilde\Delta}_{f'} f'_R \right)
    \Big] \Bigg\} \,. \no
\end{align}
Here we have introduced the shorthand notation $\sw\equiv\sin\theta_w$
and $\cw\equiv\cos\theta_w$ for the sine and cosine of the weak mixing
angle renormalized at a scale $\mu_{\rm KK}=\ord(\Mkk)$. Note that
this result is exact except for the expansion performed in the overall
normalization of the first term, as indicated by the
$\ord(m_Z^4/\Mkk^4)$ symbol in brackets. Apart from this, no expansion
in powers of $v^2/\Mkk^2$ has been performed. However, the rather
complicated expression obtained above can be simplified by noting that
the matrices $\bm{\delta}_A$ and $\bm{\varepsilon}^{(\prime)}_A$ are
of $\ord(v^2/\Mkk^2)$. It follows that the latter ones can be
discarded to very good approximation, whereas the contributions from
the $\bm{\delta}_A$ matrices must only be kept to first order in
$v^2/\Mkk^2$. We then obtain the simpler expression
\begin{align}\label{HZsimple} 
   {\cal H}_{\rm eff}^{(Z)} &= \frac{4\pi\alpha}{\sws \cws\,m_Z^2}
    \left[ 1 + \frac{m_Z^2}{2\Mkk^2} \left( 1 - \frac{1}{2L} \right)
    \right] J_Z^\mu\,J_{Z\mu} \nonumber \\
   &\mbox{} \phantom{xx} - \frac{8\pi\alpha}{\sws\cws\,m_Z^2} \sum_f
    \left[ \bar f_L\gamma^\mu T_3^f \bm{\delta}_F f_L - \bar
    f_R\gamma^\mu T_3^f \bm{\delta}_f f_R \right]
    J_{Z\mu} \nonumber \\
   &\mbox{} \phantom{xx} - \frac{4\pi\alpha L}{\sws\cws\,\Mkk^2} \sum_f
    \Big[ \left( T_3^f - \sws\,Q_f \right) \bar f_L\gamma^\mu
    \bm{\Delta}_F f_L - \sws\,Q_f\,\bar f_R\gamma^\mu\bm{\Delta}_f f_R
    \Big]\,J_{Z\mu} \nonumber \\
   &\mbox{} \phantom{xx} + \frac{4\pi\alpha L}{\sws\cws\,\Mkk^2}
    \sum_{f,f'} \Bigg\{ \Big[ \left( T_3^f - \sws\,Q_f \right) \bar
    f_L\gamma^\mu \bm{\widetilde\Delta}_F f_L - \sws\,Q_f\,
    \bar f_R\gamma^\mu\bm{\widetilde\Delta}_f f_R \Big] \nonumber \\
   &\hspace{30mm}\phantom{xx} \otimes \Big[ \left( T_3^{f'} -
    \sws\,Q_{f'} \right) \bar f'_L\gamma_\mu
   \bm{\widetilde\Delta}_{F'} f'_L - \sws\,Q_{f'}\, \bar
   f'_R\gamma_\mu\bm{\widetilde\Delta}_{f'} f'_R \Big] \Bigg\} \,,
\end{align}
which is valid up to corrections of $\ord(v^4/\Mkk^4)$. Here
\beq
   J_Z^\mu\equiv \sum_f \left[ \left( T_3^f - \sws\,Q_f \right) 
   \bar f_L\gamma^\mu f_L - \sws\,Q_f\bar f_R\gamma^\mu f_R \right]
\eeq
is the familiar SM expression for the neutral current.

\boldmath \subsection{Exchange of $W^\pm$ bosons and their KK
  excitations} \unboldmath

Using (I:33) it is also straightforward to derive the effective
four-fermion interactions induced by the exchange of the charged weak
gauge bosons $W^\pm$ and their KK excitations. In this case, of
course, flavor-changing effects are unsuppressed already in the
SM. Restricting ourselves to the phenomenologically most relevant case
with leptons in the final state, and including corrections up to
${\cal O} (m_W^2/\Mkk^2)$, we find
\beq\label{eq:HeffW}
   {\cal H}_{\rm eff}^{(W)} = \frac{2\pi\alpha}{\sws\,m_W^2}\,
   \sum_l\,\Big\{ \big[ \bar{u}_L\gamma^\mu \bm{\cal V}_L d_L 
   + \bar{u}_R\gamma^\mu \bm{\cal V}_R d_R \big ] 
   (\bar l_L\gamma_\mu \nu_{l\hspace{0.05mm} L}) + {\rm h.c.} 
   \Big\} \,,
\eeq 
where
\beq\label{eq:calVLR}
\begin{split} 
   ({\cal V}_L)_{mn} 
   &= \frac{2 \pi}{L \epsilon} \int_\epsilon^1\!dt
    \left[ 1 - \frac{m_W^2}{2 \Mkk^2} \left ( L\,t^2 - 1
    + \frac{1}{2 L} \right) \right] a_m^{(U) \dagger}\,
    \bm{C}_m^{(Q)}(\phi) \bm{C}_n^{(Q)}(\phi)\,a_n^{(D)} \,, \\[1mm]
   ({\cal V}_R)_{mn} 
   &= \frac{2 \pi}{L \epsilon} \int_\epsilon^1 \! dt
    \left [ 1 - \frac{m_W^2}{2 \Mkk^2} \left ( L\,t^2 - 1
    + \frac{1}{2 L} \right ) \right ] a_m^{(U) \dagger} \,
    \bm{S}_m^{(Q)}(\phi) \bm{S}_n^{(Q)}(\phi) \, a_n^{(D)} \,.
\end{split}
\eeq 
Notice that $({\cal V}_L)_{mn}$ and $({\cal V}_R)_{mn}$ differ from
the left- and right-handed charged-current mixing matrix $(V_L)_{mn}$
and $(V_R)_{mn}$ introduced in (I:132), as they contain besides the
contribution due to the ground-state $W^\pm$ bosons also corrections
from the tower of KK excitations. Like in the case of muon decay
\cite{Casagrande:2008hr}, the latter contributions are suppressed by
the logarithm of the warp factor with respect to the former ones, so
that in leading $L$ the elements $({\cal V}_{L,R})_{mn}$ reduce to
$(V_{L,R})_{mn}$. Due to the suppression proportional to two powers of
light quark masses, which is evident from (I:134), the contribution
from the right-handed operator is strongly suppressed compared to the
one of its left-handed counterpart. Notice finally that in order to
arrive at the effective Hamiltonian (\ref{eq:HeffW}), we have made the
simplifying assumption that the left- and right-handed 5D leptonic
fields all have the same bulk mass parameter and that they are
localized sufficiently close to the the UV brane so as not to violate
the constraints imposed by electroweak precision tests. By
construction the interactions of the SM leptons with the $W^\pm$ boson
and its KK excitations are therefore flavor universal and numerically
insignificant.

\subsection{Exchange of the Higgs Boson}

The Higgs boson is a brane-localized field in the simplest RS model,
so it does not have KK partners. Nevertheless, due to the mixing of
fermion zero-modes with their KK excitations, the interactions of
fermions with the Higgs boson are no longer flavor diagonal in the
mass basis of the fermions \cite{Agashe:2006wa}. Unlike in the SM,
the Higgs boson of the RS model thus has FCNC couplings at tree level.

In unitary gauge, the effective four-fermion interactions resulting
from Higgs-boson exchange follow directly from (I:134). The graph on
the right in Figure~\ref{fig:fourfermion} shows an example of a
diagram giving rise to such a contribution. We obtain
\beq \label{eq:Hh}
\begin{split}
  {\cal H}_{\rm eff}^{(h)} &= \frac{1}{m_h^2} \sum_{f,f'} \left[ \bar
    f_L\,\frac{\bm{m}_f}{v}\,f_R - \bar f_L \left(
      \frac{\bm{m}_f}{v}\,\bm{\delta}_f +
      \bm{\delta}_F\,\frac{\bm{m}_f}{v} + \bm{\Delta
        \tilde{g}}_h^f \right) f_R
    + \mbox{h.c.} \right] \\
  &\hspace{1.45cm}\times \left[ \bar f'_L\,\frac{\bm{m}_{f'}}{v}\,f'_R
    - \bar f'_L \left( \frac{\bm{m}_{f'}}{v}\,\bm{\delta}_{f'} +
      \bm{\delta}_{F'}\,\frac{\bm{m}_{f'}}{v} + \bm{\Delta
        \tilde{g}}_h^{f'} \right) f'_R + \mbox{h.c.} \right] ,
\end{split}
\eeq
where $\bm{m}_f\equiv\mbox{diag}(m_{f_1},m_{f_2},m_{f_3})$ represents
a diagonal matrix containing the masses of the zero-mode
fermions. This exact expression can be simplified by noting that the
matrices $\bm{\delta}_A$ and $\bm{\Delta \tilde{g}}_h^f$ are of
$\ord(v^2/\Mkk^2)$, and terms involving two such matrices can be
neglected to very good approximation. We then obtain the simpler
expression
\beq\label{Hhsimple}
{\cal H}_{\rm eff}^{(h)} = \frac{1}{m_h^2\,v^2} \sum_{f,f'}\,\Big[
\left( \bar f\,\bm{m_f} f - 2 \left[ \bar f_L \left(
      \bm{m}_f\,\bm{\delta}_f + \bm{\delta}_F\,\bm{m}_f + v \,
      \bm{\Delta \tilde{g}}_h^f \right) f_R + \mbox{h.c.} \right]
\right) \left( \bar f'\,\bm{m_{f'}} f' \right) \Big] \,,
\eeq
which is valid up to corrections of $\ord(v^4/\Mkk^4)$. For the sake
of completeness we will include all ${\cal O} (v^2/M_{\rm KK}^2)$
terms due to Higgs-boson exchange in the analytic formulas of the
effective weak Hamiltonians presented in Section~\ref{sec:treelevel}.

At this point a couple of comments are in order. Analytic expressions
for the relevant interactions have been first presented within the RS
model with $SU(2)_L \times U(1)_Y$ bulk symmetry and minimal,
brane-localized Higgs sector in \cite{Casagrande:2008hr}. Our
previous work did however not include the Yukawa couplings that
involve $Z_2$-odd fermion profiles. Although these terms are not
needed to generate the SM fermion masses and can therefore technically
be set to zero they are not excluded by any symmetry and appear
naturally in the brane limit of a bulk Higgs scenario. This omission
has been noticed in \cite{Azatov:2009na}, where it has been pointed
out that the latter terms embody the dominant tree-level corrections
to the flavor-changing Higgs-boson couplings in the RS framework. We
have carefully revisited the issue of flavor-misalignment of fermion
masses and Yukawa couplings and confirm the results of the latter
work. In (\ref{eq:Hh}) and (\ref{Hhsimple}) the terms induced by the
$Z_2$-odd Yukawa couplings are encoded in the elements of the matrices
$\bm{\Delta \tilde{g}}_h^{f}$. They take the form
\begin{equation} \label{eq:Deltagtilde}
  (\Delta \tilde{g}_h^f )_{mn} = - \sqrt{2} \,
  \frac{2\pi}{L\epsilon}\int_\epsilon^1\!dt \, \delta(t-1) \, 
  a_m^{\hspace{0.25mm} f\,\dagger}\, \bm{S}_m^{\hspace{0.25mm} f} (t) \,
  \bm{Y}_{f}^{\dagger} \, \bm{S}_n^F(t)\,  a_n^F \,.
\end{equation}
In contrast to the $\bm{\delta}_{A}$ terms, the latter contributions
do not receive an additional chiral suppression and thus represent the
main source of flavor violation in the Higgs sector (if light quark
transitions are considered).

Numerically, we find that the effect of the Higgs exchange in $\Delta
F = 1$ transitions is negligible compared to the one of the $Z^0$
boson, due to the chiral suppression at the flavor-preserving
vertex. In the case of $\Delta F = 2$ processes the importance of
Higgs FCNCs turns out to be limited too. The most pronounced effects
occur in the case of the CP-violating parameter $\epsilon_K$, but even
here they are typically smaller than the corrections due to KK gluon
exchange \cite{Duling:2009pj}. This result can be readily understood
by noticing that the only Higgs-boson contributions in $\Delta F = 2$
transitions that are not chirally suppressed are proportional to
$(\Delta\tilde{g}_h^f)^2_{mn}$, making them smaller by a factor of
$v^2/\Mkk^2$ than the tensor products defined in
(\ref{Deltaotimes}). Since possible Higgs-boson effects are
numerically insignificant in all $\Delta F=2$ and $\Delta F=1$
processes discussed in this work, we will simply neglect them in our
numerical analysis.

Let us finally mention, that a model-independent analysis of the
flavor misalignment of the SM fermion masses and the Yukawa couplings
has been presented in \cite{Agashe:2009di}. There it has been shown
that at the level of dimension six, chirally unsuppressed
contributions to flavor-changing Higgs-boson vertices generically will
arise from composite operators like $\bar q_L^{\hspace{0.5mm} i}
\hspace{0.25mm} H \hspace{0.25mm} q_R^j \hspace{0.5mm} (H^\dagger H)$
in models where the Higgs is a bound state of a new
strongly-interacting theory. If present, the latter terms will
dominate over the chirally suppressed contributions originating from
operators of the form $\bar q_L^{\hspace{0.5mm} i} \hspace{0.25mm} D
\!\!\!\!/\, \hspace{0.25mm} q_L^j \hspace{0.5mm} (H^\dagger H)$,
because the couplings $y_{q \ast}$ of the composite Higgs to the other
strong interacting states can be large, resulting in $y_{q \ast}^2/(16
\pi^2) \gg m_q/v$. Notice that in our concrete model, considering all
relevant dimension-six operators in lowest-order of the mass insertion
approximation, allows to recover the results for the ${\cal O}
(v^2/M_{\rm KK}^2)$ terms of the matrices $\bm{\delta}_A$ and
$\bm{\Delta \tilde{g}}_h^{f}$ (see \cite{Azatov:2009na} for a
illuminating discussion). In another publication \cite{Casagrande:2010si}
we have shown in detail how all new-physics effects induced by the mass
insertions, corresponding to both the non-derivative and derivative
dimension-six operators, can be resummed to all orders in $v^2/\Mkk^2$
at tree level.

\subsection{Mixing Matrices in the ZMA}

It is trivial to read off from (\ref{photoncouplings}),
(\ref{HZsimple}), and (\ref{Hhsimple}) the resulting flavor-violating
transitions involving quark and lepton fields. Those will be discussed
in the following section. We complete the present discussion by
collecting the expressions for the various mixing matrices valid in
the ZMA. We restrict ourselves to the $3\times 3$ submatrices
governing the couplings of the SM fermion fields. The elements of the
$\bm{\Delta}_A^{(\prime)}$ matrices that arise from the non-trivial
overlap of gauge and fermion profiles are given explicitly in (I:122).
In the ZMA the corresponding expressions simplify considerably (see
also \cite{Burdman:2002gr, Agashe:2003zs}). We find
\beq \label{eq:DeltaZMA}
\begin{split}
   \bm{\Delta}_F &\to \bm{U}_f^\dagger\,\,\mbox{diag} \left[
    \frac{F^2(c_{F_i})}{3+2c_{F_i}} \right] \bm{U}_f \,, \\
   \bm{\Delta}_f &\to \bm{W}_f^\dagger\,\,\mbox{diag} \left[
    \frac{F^2(c_{f_i})}{3+2c_{f_i}} \right] \bm{W}_f \,, \\
   \bm{\Delta}'_F &\to \bm{U}_f^\dagger\,\,\mbox{diag} \left[
    \frac{5+2c_{F_i}}{2(3+2c_{F_i})^2}\,F^2(c_{F_i}) \right]
    \bm{U}_f \,, \\
   \bm{\Delta}'_f &\to \bm{W}_f^\dagger\,\,\mbox{diag} \left[
    \frac{5+2c_{f_i}}{2(3+2c_{f_i})^2}\,F^2(c_{f_i}) \right]
    \bm{W}_f \,, \\
\end{split}
\eeq 
where the diagonal matrices contain the elements shown in
brackets. Note further that, to a good approximation, we have
$\bm{\Delta}'_A\approx\bm{\Delta}_A$ for $A=F,f$, since all $c_i$
parameters are near $-1/2$. Explicit expressions for the unitary
rotations $\bm{U}_f$ and $\bm{W}_f$ can be found in (I:97) to
(I:100). The matrices $\bm{\varepsilon}^{(\prime)}_A$ vanish at
leading order in the ZMA, meaning that they are suppressed by an extra
factor of $v^2/M_{\rm KK}^2$. Thus we do not report them here.

The matrices $\bm{\delta}_A$ and $\bm{\Delta \tilde{g}}_h^f$ arise
from the fact that the fermion profiles are not orthonormal on each
other. Their elements take the form (I:123) and
(\ref{eq:Deltagtilde}). The corresponding ZMA results are given by
\beq\label{ZMA2}
\begin{split}
  \bm{\delta}_F &\to \bm{x}_f\,\bm{W}_f^\dagger\,\, \mbox{diag}\left[
    \frac{1}{1-2c_{f_i}} \left( \frac{1}{F^2(c_{f_i})} - 1 +
      \frac{F^2(c_{f_i})}{3+2c_{f_i}} \right) \right]
  \bm{W}_f\,\bm{x}_f \,, \\
  \bm{\delta}_f &\to \bm{x}_f\,\bm{U}_f^\dagger\,\, \mbox{diag}\left[
    \frac{1}{1-2c_{F_i}} \left( \frac{1}{F^2(c_{F_i})} - 1 +
      \frac{F^2(c_{F_i})}{3+2c_{F_i}} \right) \right]
  \bm{U}_f\,\bm{x}_f \,, \\
  {\bm{\Delta \tilde{g}}}_h^f & \, \to \frac
    {\sqrt{2} \, v^2 }{3 M_{\rm KK}^2} \; {\bm U}_f^\dagger \; {\rm
      diag} \left[ F(c_{F_i}) \right] \hspace{0.25mm} \bm{Y}_f
    \hspace{0.25mm} \bm{Y}_f^\dagger \hspace{0.25mm} \bm{Y}_f \; {\rm
      diag } \left[ F(c_{f_i}) \right] {\bm W}_f\,,
\end{split}
\eeq
where $\bm{x}_f \equiv\mbox{diag}(m_{f_1},m_{f_2},m_{f_3})/M_{\rm KK}$
is a diagonal matrix containing the masses of the SM quarks in units
of $M_{\rm KK}$.

As mentioned above, the expressions for the new flavor matrices
defined in (\ref{Deltaotimes}) are tensor products, which do not
factorize in the form of simple matrix products. In component
notation, we obtain \cite{Bauer:2008xb}
\beq\label{eq:DDcomponent} 
   \big( \widetilde\Delta_F \big)_{m n}\!\otimes \big(
   \widetilde\Delta_{f'} \big)_{m' n'} 
   \to \big( U_f^\dagger \big)_{mi}\,\big( U_f \big)_{in}\,  
   (\widetilde{\Delta}_{Ff})_{ij} \, 
   \big( W_f^\dagger \big)_{m' j} \, \big( W_f \big)_{j n'} \,,
\eeq
where
\beq
   (\widetilde{\Delta}_{Ff})_{ij} 
   = \frac{F^2(c_{F_i})}{3+2c_{F_i}}\,
   \frac{3+c_{F_i}+c_{f_j}}{2(2+c_{F_i}+c_{f_j})}\,
   \frac{F^2(c_{f_j})}{3+2c_{f_j}} \,,
\eeq
and a summation over the indices $i,j$ is understood. Analogous
expressions hold for the remaining combinations of indices $F$ and
$f$. Using the fact that all $c_i$ parameters except $c_{u_3}$ are
very close to $-1/2$, it is a reasonable approximation to replace
$(3+c_{F_i}+c_{f_j})/(2+c_{F_i}+c_{f_j})$ by 2, in which case we
obtain the approximate result
\beq
   \bm{\widetilde\Delta}_A\otimes \bm{\widetilde\Delta}_B
   \to \bm{\Delta}_A\,\bm{\Delta}_B \,.
\eeq
In the same approximation, there is no need to distinguish between the
$\bm{\Delta}'_A$ and $\bm{\Delta}_A$ matrices.

\section{Tree-Level Weak Decay Processes}
\label{sec:treelevel}

We now discuss a few prominent applications of the general results
obtained in the previous section, focusing on FCNC processes in the
quark sector. The generalization to the lepton sector is trivial
(apart from the issue of Majorana mass terms). We begin with a
discussion of $\Delta F=2$ processes, giving rise to the mixing of
neutral mesons with their antiparticles. We then study a plethora of
inclusive and exclusive $\Delta F=1$ transitions in the kaon and
$B$-meson sectors. The purpose of this section is to compile in a
concise way the theoretical ingredients needed for the
phenomenological analysis of the various observables considered in
this work. The reader primarily interested in phenomenological results
can skip this section in a first reading.

\subsection{Neutral-Meson Mixing}

We will discuss the new contributions to neutral-meson mixing arising
in the RS model on the basis of $K$--$\bar K$ mixing, where the
resulting constraints have been shown to be particularly severe
\cite{Csaki:2008zd, Santiago:2008vq, Blanke:2008zb, Agashe:2008uz,
  Bauer:2008xb}, confirming model-independent considerations presented
in \cite{Bona:2007vi, Davidson:2007si}. With a simple substitution of
indices, the expressions we obtain can be applied to the mixing of
$B_{d,s}$ and $D$ mesons.

\boldmath \subsubsection{Effective $\Delta S=2$ Hamiltonian}
\unboldmath
\label{sec:effectiveDeltaS2}

We adopt the following general parametrization of new-physics effects
in $K$--$\bar K$ mixing \cite{Bagger:1997gg, Ciuchini:1998ix,
  Buras:2000if}
\beq\label{eq:HeffDelta2}
   {\cal H}_{\rm eff}^{\Delta S=2} 
   = \sum_{i=1}^5 C_i\,Q_i^{sd} 
    + \sum_{i=1}^3 \tilde C_i\,\tilde Q_i^{sd} \,,
\eeq
where 
\beq\label{eq:Qisd}
\begin{split}
   Q_1^{sd} &= (\bar d_L \gamma^\mu s_L )\,(\bar d_L \gamma_\mu s_L )
    \,, \qquad \tilde Q_1^{sd}
    = (\bar d_R \gamma^\mu s_R )\,(\bar d_R \gamma_\mu s_R) \,, \\
   Q_2^{sd} &= (\bar d_R s_L )\,(\bar d_R s_L) \,, \hspace{1.65cm}
    \tilde Q_2^{sd}
    = (\bar d_L s_R)\,(\bar d_L s_R) \,, \\
   Q_3^{sd} &= (\bar d_R^\alpha s_L^\beta)\,(\bar d_R^\beta s_L^\alpha)
    \,, \hspace{1.65cm} \tilde Q_3^{sd}
    = (\bar d_L^\alpha s_R^\beta)\,(\bar d_L^\beta s_R^\alpha) \,, \\
   Q_4^{sd}
   &= (\bar d_R s_L)\,(\bar d_L s_R) \,, \\
   Q_5^{sd} &= (\bar d_R^\alpha s_L^\beta)\,(\bar d_L^\beta s_R^\alpha)
    \,.
\end{split}
\eeq 
A summation over color indices $\alpha,\beta$ is understood. We write
the Wilson coefficients as a sum of a SM and a new-physics
contribution, $C_i \equiv C_i^{\rm SM} + C_i^{\rm RS}$, where in the
SM only $C_1^{\rm SM}$ is non-zero. Using the general results from the
previous section along with standard Fierz identities, we obtain for
the contributions arising in the RS model
\beq\label{eq:Cmix}
\begin{split}
   C_1^{\rm RS} &= \frac{4\pi L}{\Mkk^2}\, \big( \widetilde\Delta_D
    \big)_{12}\otimes \big( \widetilde\Delta_D \big)_{12} \left[
    \frac{\alpha_s}{2} \left( 1 - \frac{1}{N_c} \right) +
    Q_d^2\,\alpha + \frac{(T_3^d-\sws\,Q_d)^2\,\alpha}%
    {\sws\cws} \right] , \\
   \tilde C_1^{\rm RS} &= \frac{4\pi L}{\Mkk^2}\, \big(
    \widetilde\Delta_d \big)_{12}\otimes \big( \widetilde\Delta_d
    \big)_{12} \left[ \frac{\alpha_s}{2} \left( 1 - \frac{1}{N_c}
    \right) + Q_d^2\,\alpha + \frac{(\sws\,Q_d)^2\,\alpha}%
    {\sws\cws} \right] , \\
   C_4^{\rm RS} &= \frac{4\pi L}{\Mkk^2}\, \big( \widetilde\Delta_D
    \big)_{12}\otimes \big( \widetilde\Delta_d \big)_{12} \left[
    - 2\alpha_s \right] , \\
   C_5^{\rm RS} &= \frac{4\pi L}{\Mkk^2}\,
    \big( \widetilde\Delta_D \big)_{12}\otimes 
    \big( \widetilde\Delta_d \big)_{12} \left[
    \frac{2\alpha_s}{N_c} 
    - 4 Q_d^2\,\alpha
    + \frac{4\sws\,Q_d\,(T_3^d-\sws\,Q_d)\,\alpha}{\sws\cws} \right] ,
\end{split}
\eeq 
where $Q_d=-1/3$, $T_3^d=-1/2$, and $N_c=3$. The expressions in
brackets refer, in an obvious way, to the contributions from KK
gluons, KK photons, and from the $Z^0$ boson and its KK
excitations. As can bee seen from (\ref{eq:Hh}), $\Delta F = 2$
contributions from flavor-changing Higgs-boson exchange are of
$\ord(v^4/\Mkk^4)$ and we thus refrain from giving them explicitly.
The Wilson coefficients $C_{2,3}$ and $\tilde C_{2,3}$ do not receive
tree-level contributions in the RS model, as they can only arise from
scalar or tensor exchange, but not from gauge interactions. For the
cases of $B_d$--$\bar B_d$ and $B_s$--$\bar B_s$ mixing the same
expressions hold, but one takes the 13 and 23 entries of the mixing
matrices, respectively. For the case of $D$--$\bar D$ mixing one
replaces the mixing matrices $\widetilde\Delta_{D,d}$ with
$\widetilde\Delta_{U,u}$ and takes the 12 entries. Also, in this case
$Q_u=2/3$ and $T_3^u=1/2$ are the appropriate SM charges. Notice that
the QCD contributions are in all cases larger by more than a factor of
3 than the combined QED and electroweak effects, and that the
numerically dominant contribution proportional to $C_4^{\rm RS}$ is a
pure QCD effect. KK gluon exchange thus dominates all mixing
amplitudes in the original RS scenario with SM bulk gauge symmetry. We
will see later that this is {\em not\/} true for $\Delta F=1$
processes. In the present case, although QED and electroweak
corrections are subleading, we will include them in our numerical
analysis.

In passing, we mention that in warped extra-dimension models with
extended $SU(2)_R$ symmetry and custodial protection of the $Z^0 b_L
\bar b_L$ vertex, the corrections arising from new heavy neutral
electroweak gauge bosons can compete with the KK gluon corrections to
$B_{d,s}$--$\bar B_{d,s}$ mixing \cite{Blanke:2008zb}. This implies
that in this class of models the effects in the observables $\Delta
m_{d,s}$, $S_{\psi K_S}$, $S_{\psi \phi}$, $\Delta
\Gamma_{d,s}/\Gamma_{d,s}$, and $A_{\rm SL}^{d,s}$ will typically be
more pronounced than in RS formulations without $SU(2)_R$ symmetry. We
emphasize that the same pattern of enhancement of electroweak relative
to QCD effects is also present in the case of $D$--$\bar D$
mixing. Further comments on the impact of the exact realization of the
electroweak sector on both $B_{d,s}$--$\bar B_{d,s}$ and $D$--$\bar D$
mixing are postponed to Sections~\ref{sec:num_BBmix} and
\ref{sec:num_DDmix}.

The tree-level expressions for the Wilson coefficients given above
refer to a renormalization scale $\mu_{\rm KK}=\ord(\Mkk)$. They must
be evolved down to a scale $\mu\approx 2$\,GeV, where the hadronic
matrix elements of the four-quark operators can be evaluated using
lattice QCD. The evolution is accomplished with the help of formulas
compiled in the literature. For the case of $K$--$\bar K$ mixing these
can be found in Eq.~(2.6) of \cite{Ciuchini:1998ix}, for
$B_{d,s}$--$\bar B_{d,s}$ mixing they are given in Eq.~(9) of
\cite{Becirevic:2001jj}, and for $D$--$\bar D$ mixing they are listed
in Eq.~(8) of \cite{Bona:2007vi}. The hadronic matrix elements of the
various operators are customarily expressed in terms of parameters
$B_i$. For the operators relevant to our analysis, one
has\footnote{Here the meson states are normalized to $\langle
  K^0|K^0\rangle=\langle\bar K^0|\bar K^0\rangle=1$}
\beq
\begin{split}
   \langle K^0|\,Q_1(\mu)\,|\bar K^0\rangle 
   &= \langle K^0|\,\tilde Q_1(\mu)\,|\bar K^0\rangle 
    = \left( 1 + \frac{1}{N_c} \right) \frac{m_K f_K^2}{4}\,B_1(\mu) 
    \,, \\
   \langle K^0|\,Q_4(\mu)\,|\bar K^0\rangle 
   &= \left[ \frac{m_K}{m_s(\mu)+m_d(\mu)} \right]^2
    \frac{m_K f_K^2}{4}\,B_4(\mu) \,, \\
   \langle K^0|\,Q_5(\mu)\,|\bar K^0\rangle 
   &= \frac{1}{N_c} \left[ \frac{m_K}{m_s(\mu)+m_d(\mu)} \right]^2 
    \frac{m_K f_K^2}{4}\,B_5(\mu) \,.
\end{split}
\eeq
These definitions are such that in the vacuum-insertion approximation
(VIA)
\beq
\begin{split}
   \left[ B_1(\mu) \right]_{\rm VIA} 
   &= 1 \,, \\
   \left[ B_4(\mu) \right]_{\rm VIA} 
   &= 1 + \frac{1}{2N_c} \left[ \frac{m_s(\mu)+m_d(\mu)}{m_K} 
    \right]^2 , \\
   \left[ B_5(\mu) \right]_{\rm VIA} 
   &= 1 + \frac{N_c}{2} \left[ \frac{m_s(\mu)+m_d(\mu)}{m_K} 
    \right]^2 .
\end{split}
\eeq
Analogous definitions are used for the other mesons. In our numerical
analysis we employ the $B_i$ parameters from
\cite{Lubicz:2008am}. Furthermore, we use $m_K=497.6$\,MeV and
$f_K=156.1$\,MeV for the kaon mass and decay constant
\cite{Amsler:2008zz}. The hadronic parameters corresponding to
$K$--$\bar K$, $B_{d,s}$--$\bar B_{d,s}$, and $D$--$\bar D$ mixing can
be found in Table~I and Eq.~(22), in Table~II and Eqs.~(24) and (25),
and in Eq.~(28) of \cite{Lubicz:2008am}, respectively. For the sake
of completeness we collect the entire set of $B_i$ parameters entering
our calculations in Appendix~\ref{app:input}.

\subsubsection{Important Formulas for Neutral-Meson Mixing}

In the following we compile the formulas that we will employ in our
numerical analysis of the mixing of neutral mesons with their
antiparticles. While physical observables are phase-convention
independent, we stress that some of the formulas given below hold only
if the standard phase convention for the CKM matrix
\cite{Amsler:2008zz} is adopted.

The $K_L$--$K_S$ mass difference $\Delta m_K$ and the CP-violating
quantity $\epsilon_K$ are given by\footnote{Here and below, we include
  in the full effective Hamiltonian a universal correction to the
  Fermi constant $G_F$ of order $m_W^2/\Mkk^2$
  \cite{Casagrande:2008hr}. Numerically, these contributions are
  insignificant in all the FCNC processes we consider.}
\beq
   \Delta m_K = 2\,\mbox{Re}\,
   \langle K^0|\,{\cal H}_{\rm eff,full}^{\Delta S=2}\, 
   |\bar K^0\rangle \,, 
    \qquad 
   \epsilon_K = \frac{\kappa_\epsilon\,e^{i\varphi_\epsilon}}%
                     {\sqrt 2\,(\Delta m_K)_{\rm exp}}\, 
   \mbox{Im}\,\langle K^0|\,{\cal H}_{\rm eff,full}^{\Delta S=2}\, 
   |\bar K^0\rangle \,,
\eeq
where $\varphi_\epsilon=(43.51\pm 0.05)^\circ$ and
$\kappa_\epsilon=0.92\pm 0.02$ \cite{Buras:2008nn}. The suppression
factor $\kappa_\epsilon$ parametrizes the effects due to the imaginary
part of the isospin-zero amplitude in $K\to\pi\pi$ decays
\cite{Anikeev:2001rk, Andriyash:2003ym, Andriyash:2005ax}.

Taking into account all indirect constraints from the global
unitarity-triangle fit \cite{Charles:2004jd}, we obtain the SM
prediction
\beq \label{epsKSM}
   |\epsilon_K|_{\rm SM} = (1.9\pm 0.4)\cdot 10^{-3} \,,
\eeq
where the quoted error corresponds to a frequentist 68\% confidence
level (CL). The dominant theoretical uncertainties arise from the
lattice QCD prediction $B_1(2\,{\rm GeV})=0.527\pm 0.022$
\cite{Aubin:2009jh} and the error on $\kappa_\epsilon$, which have
been scanned in their uncertainty ranges to obtain the best fit
value. Within errors the SM prediction agrees with the experimental
value $|\epsilon_K|_{\rm exp}=(2.229\pm 0.010)\cdot 10^{-3}$. We do
not attempt a prediction for $\Delta m_K$, which is plagued by very
large hadronic uncertainties.

In the $B_{d,s}$ system one has
\beq
   \Delta m_{B_{q}} 
   = 2\left|\langle B_{q}|\,{\cal H}_{\rm eff,full}^{\Delta B=2}\,
    |\bar B_{q}\rangle\right| , \qquad
   2\varphi_{q} 
   = \mbox{arg} \left( \langle B_{q}|\,
    {\cal H}_{\rm eff,full}^{\Delta B=2}\,|\bar B_{q}\rangle \right) ,
\eeq
for $q=d,s$. In all cases the effective Hamiltonian contains the SM
contribution plus possible contributions from new physics, as
indicated by the subscript ``full''. It is also common to normalize
the matrix elements of the full effective Hamiltonian to those of only
the SM contribution. For instance, one can define \cite{Bona:2005eu}
\beq\label{eq:BBmixpara}
   C_{B_{q}}\,e^{2i\phi_{B_{q}}} 
   = \frac{\langle B_{q}|\,{\cal H}_{\rm eff,full}^{\Delta B=2}\,
           |\bar B_{q}\rangle}%
          {\langle B_{q}|\,{\cal H}_{\rm eff,SM}^{\Delta B=2}\,
           |\bar B_{q}\rangle} \,.
\eeq
The coefficient $C_{B_{q}}$ measures the magnitude of the mass
difference $\Delta m_{q}$ relative to the one in the SM, namely
\beq
   C_{B_{q}} = \frac{\Delta m_{q}}{(\Delta m_{q})_{\rm SM}} \,,
\eeq
while the phases $\phi_{B_{d,s}}$ affect the coefficients of
$\sin(\Delta M_{d,s} t)$ in the time-dependent asymmetries of
$B_d\to\psi K_s$ and $B_s\to\psi\phi$. One obtains
\cite{Blanke:2006ig}
\beq\label{eq:SpsiKsphi}
   S_{\psi K_s} = \sin(2\beta + 2\phi_{B_d}) \,, \qquad
   S_{\psi\phi} = \sin(2|\beta_s| - 2\phi_{B_s}) \,,
\eeq
where 
\beq
   V_{td} = |V_{td}|\,e^{-i\beta} \,, \qquad 
   V_{ts} = -|V_{ts}|\,e^{-i\beta_s} \,,
\eeq
and $\beta\approx 22^\circ$, $\beta_s\approx -1^\circ$. In the
presence of non-vanishing phases $\phi_{B_{d,s}}$, the two
time-dependent asymmetries thus measure $\varphi_d=\beta+\phi_{B_d}$
and $\varphi_s=|\beta_s|-\phi_{B_s}$, and not $\beta$ and
$\beta_s$. We recall that the formulas (\ref{eq:SpsiKsphi}) are
correct only if no weak phase is present in the $B_d\to\psi K_s$ and
$B_s\to\psi\phi$ decay amplitudes, as it happens to be the case, to
excellent approximation, in the SM. Strictly speaking this assumption
does not hold in the case at hand, because in the RS model the
charged-current interactions differ from the ones present in the
SM. The new contributions are, however, suppressed by
$m_W^2/\Mkk^2$. Numerically, it turns out that the value of $\arg
(V_{cb}^\ast V_{cs})$ stays below the level of $0.001^\circ$, and that
the elements $(V_R)_{22}$ and $(V_R)_{23}$ appearing in the
right-handed charged-current interactions do not exceed the level of
$10^{-5}$ in magnitude \cite{Casagrande:2008hr}. These tiny
corrections can be safely neglected for all practical purposes.

The width differences $\Delta \Gamma_q$ and the semileptonic CP
asymmetries $A_{\rm SL}^q$ take the form
\begin{align}
   \frac{\Delta\Gamma_{q}}{\Gamma_{q}} 
   &= - \left( \frac{\Delta m_{q}}{\Gamma_{q}} \right )_{\rm exp} 
    \left [ {\rm Re} \left(
    \frac{\Gamma_{12}^{q}}{M_{12}^{q}} \right)_{\rm SM} 
    \frac{\cos 2\phi_{B_{q}}}{C_{B_{q}}} 
    - {\rm Im} \left( \frac{\Gamma_{12}^{q}}{M_{12}^{q}} 
    \right)_{\rm SM}
    \frac{\sin 2\phi_{B_{q}}}{C_{B_{q}}} \right] , \nonumber \\[2mm]
   A_{\rm SL}^{q} 
   &= {\rm Im} \left(
    \frac{\Gamma_{12}^{q}}{M_{12}^{q}} \right)_{\rm SM} 
    \frac{\cos 2\phi_{B_{q}}}{C_{B_{q}}} 
    - {\rm Re} \left(
    \frac{\Gamma_{12}^{q}}{M_{12}^{q}} \right)_{\rm SM} 
    \frac{\sin 2\phi_{B_{q}}}{C_{B_{q}}} \,,
\end{align} 
with \cite{Lenz:2006hd} 
\beq\label{eq:ALUN}
\begin{aligned}
   {\rm Re} \left( \frac{\Gamma_{12}^{d}}{M_{12}^{d}} 
    \right)_{\rm SM} 
   &= (-5.26^{+1.15}_{-1.28})\cdot 10^{-3} \,, 
   &\qquad 
   {\rm Im} \left( \frac{\Gamma_{12}^{d}}{M_{12}^{d}} 
    \right)_{\rm SM} 
   &= (-4.8^{+1.0}_{-1.2})\cdot 10^{-4} \,,\\[2mm]
   {\rm Re} \left( \frac{\Gamma_{12}^{s}}{M_{12}^{s}}
    \right)_{\rm SM} 
   &= (-4.97\pm 0.94)\cdot 10^{-3} \,, 
   &\qquad 
   {\rm Im} \left( \frac{\Gamma_{12}^{s}}{M_{12}^{s}} 
    \right)_{\rm SM} 
   &= (2.06\pm 0.57)\cdot 10^{-5} \,.
\end{aligned} 
\eeq
Note that the quoted central values for ${\rm
  Re}\,(\Gamma_{12}^{q}/M_{12}^{q})_{\rm SM}$ and ${\rm
  Im}\,(\Gamma_{12}^{q}/M_{12}^{q})_{\rm SM}$ differ in some cases
notably from older determinations. This is mainly due to a different
choice of the operator basis used in \cite{Lenz:2006hd}, and it is
related to unknown $\ord(\alpha_s^2)$ and $\ord(\alpha_s\Lambda_{\rm
  QCD}/m_b)$ corrections. Although the basis chosen in the latter
article leads to smaller theoretical uncertainties, the shifts
observed in the central values may signal that the effects of
higher-order corrections are larger than previously estimated. While
such a possibility cannot be fully excluded, it would not change the
conclusions drawn in our work.

The theoretical parameters describing $D$--$\bar D$ mixing can be
defined in complete analogy to those for $B_{d,s}$--$\bar B_{d,s}$
mixing. Our sign convention for the phase of the dispersive part of
the $D$--$\bar D$ mixing amplitude is
\beq\label{eq:M12D}
   M_{12}^D = \langle D^0| {\cal H}_{\rm eff,full}^{\Delta C=2}
    |\bar D^0\rangle 
   \equiv |M_{12}^D|\,e^{-2i\varphi_D} .  
\eeq 
In contrast to $M_{12}^D$, the absorptive part of the $D$--$\bar D$
mixing amplitude, $\Gamma_{12}^D$, is dominated by SM tree-level
interactions and therefore essentially unaffected by the presence of
new physics. The small corrections to the charged-current interactions
arising in the RS model do not change this conclusion. Motivated by
the measurements of meson oscillations in the $D$ system by BaBar
\cite{Aubert:2007wf}, Belle \cite{Staric:2007dt,Abe:2007rd}, and CDF
\cite{Aaltonen:2007uc}, we investigate both the magnitude and phase
of the new-physics contributions to (\ref{eq:M12D}) in the RS
model. We find that the available experimental data already have a
non-trivial impact on the allowed model parameters, even if the
theoretical uncertainties entering the SM prediction for $D$--$\bar D$
mixing are treated conservatively.

Studies of CP asymmetries in $D$ decays offer another powerful probe
of new physics. Experimentally the most striking signature would be
the observation of a time-dependent CP asymmetry for a Cabibbo-allowed
final state like $D\to\phi K_S$, in analogy to the case of $B_d\to\psi
K_S$. The coefficient $S_{\phi K_S}^D$ in the time-dependent CP
asymmetry of $D\to\phi K_S$ can be written as
\begin{equation}
  S_{\phi K_S}^D = \hspace{0.25mm} y_D  \, \big ( \big | q/p \big |_D -  
  \big | p/q \big |_D  \big ) \hspace{0.25mm} \cos 2 \tilde \varphi -
  x_D \, \big ( \big | q/p \big |_D +  
  \big | p/q \big |_D  \big ) \hspace{0.25mm} \sin 2 \tilde \varphi \,,
\end{equation}
where 
\beq
   x_D = \frac{\Delta M_D}{\Gamma_D} \,, \qquad 
   y_D = \frac{\Delta\Gamma_D}{2 \Gamma_D} \,,
\eeq
and 
\begin{gather}
  (q/p)_D = \sqrt{\frac{M_{12}^{D \, \ast} - \frac{i}{2} \,
      \Gamma_{12}^{D \, \ast}}{M_{12}^{D} - \frac{i}{2} \,
      \Gamma_{12}^{D}}} \,, \qquad
  \tilde \varphi = \frac{1}{2} \, {\rm arg} \, (q/p)_D \,, \nonumber \\[1mm]
  \Delta M_D = 2\,{\rm Re} \left[ \left| M_{12}^D \right|^2 - \frac14
    \left| \Gamma_{12}^D \right|^2 - i\,{\rm Re} \left( \Gamma_{12}^D
      M_{12}^{D\,\ast} \right)
  \right]^{1/2} , \nonumber \\[1mm]
  \Delta\Gamma_D = -4\,{\rm Im} \left[ \left| M_{12}^D \right|^2 -
    \frac14 \left| \Gamma_{12}^D \right|^2 - i\,{\rm Re} \left(
      \Gamma_{12}^D M_{12}^{D\,\ast} \right) \right]^{1/2} , 
\end{gather}
while $\Gamma_D=1/\tau_D$ denotes the mean width. Under the
assumption of negligible direct CP violation, one can then show that
the CP asymmetry in semileptonic $D$ decays
\beq\label{eq:ASLDdef} 
A_{\rm SL}^D = \frac{\Gamma (D(t) \to l^- \bar \nu K^{+ (\ast)}) -
  \Gamma(D(t) \to l^+ \bar \nu K^{- (\ast)})}{\Gamma (D(t) \to l^-
  \bar \nu K^{+ (\ast)}) + \Gamma(D(t) \to l^+ \bar \nu K^{- (\ast)})}
= \frac{ \big | q/p \big |_D^{\, 4} - 1}{\big | q/p \big |_D^{\, 4} +
  1} \,,
\eeq
can be expressed in the limit $\big | |q/p |_D - 1 \big | \ll 1$ and
$x_D \approx y_D$ through the CP-violating parameter $S_{\phi K_S}^D$
as \cite{Bigi:2009df, Grossman:2009mn, Kagan:2009gb}
\beq\label{eq:ASLD} 
A_{\rm SL}^D =  \frac{y_D}{x_D^2+y_D^2}\,S_{\phi K_S}^D \,.
\eeq
Analyzing in detail the predictions for $A_{\rm SL}^D$ and $S_{\phi
 K_S}^D$ and their correlation in the RS model, we find CP-violating
effects in $D$--$\bar D$ mixing that exceed by far the SM
expectation. The observation of such large mixing-induced CP
asymmetries would be a clear signal of new physics. We will also
comment briefly on a number of other charm decays besides $D\to\phi
K_S$, which could offer further insight into the dynamics of CP
violation in $D$ decays.

\boldmath \subsection{Rare Non-Leptonic Decays of Kaons and $B$
  Mesons} \unboldmath

We now study $\Delta F=1$ processes in the quark sector, focusing on
the important FCNC decays mediated by four-quark operators with flavor
structure $b\to s q\bar q$. It is straightforward to generalize the
discussion to other cases. Processes with leptons will be considered
in Section~\ref{subsec:rare_leptons}.

\boldmath \subsubsection{Effective $\Delta B=1$ Hamiltonian}
\unboldmath

We write the new-physics contributions to these decays using the usual
operator basis $Q_{1-10}$ augmented with chirality-flipped operators
$\tilde Q_{1-10}$ that are obtained from the original ones by
exchanging left- by right-handed fields everywhere. Explicitly, we use
\beq\label{eq:HeffDeltaF1}
\begin{split} 
   {\cal H}_{\rm eff}^{\Delta B=1} 
   &= \frac{G_F}{\sqrt2} \sum_{q=u,c} \lambda_q^{(sb)} 
    \, \bigg( C_1\,Q_1^q + C_2\,Q_2^q + \sum_{i=3}^{10} C_i\,Q_i
    + C_{7}^{\gamma}\,Q_{7}^{\gamma} + C_{8}^{g}\,Q_{8}^{g} \bigg) \\
   &\quad\mbox{}+ \sum_{i=3}^{10} \left( C_i^{\rm RS}\,Q_i + \tilde
    C_i^{\rm RS}\,\tilde Q_i \right) ,
\end{split}
\eeq
where $\lambda^{(pr)}_q \equiv V_{qp}^\ast V_{qr}$ and $Q_{1,2}^q$ are
the left-handed current-current operators arising from $W^\pm$-boson
exchange, $Q_{3-6}$ and $Q_{7-10}$ are QCD and electroweak penguin
operators, and $Q_{7}^{\gamma}$ and $Q_{8}^{g}$ are the
electromagnetic and chromomagnetic dipole operators. The operators
relevant to our discussion are defined as
\beq
\begin{split}
   Q_3 &= 4\,(\bar s_L\gamma^\mu b_L) \sum{}_{\!q}\, (\bar
    q_L\gamma_\mu q_L) \,, \hspace{1.4cm} \quad Q_4 = 4\,(\bar
    s_L^\alpha\gamma^\mu b_L^\beta) \sum{}_{\!q}\,
    (\bar q_L^\beta\gamma_\mu q_L^\alpha) \,, \\
   Q_5 &= 4\,(\bar s_L\gamma^\mu b_L) \sum{}_{\!q}\, (\bar
    q_R\gamma_\mu q_R) \,, \hspace{1.4cm} \quad Q_6 = 4\,(\bar
    s_L^\alpha\gamma^\mu b_L^\beta) \sum{}_{\!q}\,
    (\bar q_R^\beta\gamma_\mu q_R^\alpha) \,, \\
   Q_7 &= 6\,(\bar s_L\gamma^\mu b_L) \sum{}_{\!q}\, Q_q\,(\bar
    q_R\gamma_\mu q_R) \,, \qquad \quad Q_8 = 6\,(\bar
    s_L^\alpha\gamma^\mu b_L^\beta) \sum{}_{\!q}\,
    Q_q\,(\bar q_R^\beta\gamma_\mu q_R^\alpha) \,, \\
   Q_9 &= 6\,(\bar s_L\gamma^\mu b_L) \sum{}_{\!q}\, Q_q\,(\bar
    q_L\gamma_\mu q_L) \,, \hspace{0.7cm} \quad Q_{10} = 6\,(\bar
    s_L^\alpha\gamma^\mu b_L^\beta) \sum{}_{\!q}\, Q_q\,(\bar
    q_L^\beta\gamma_\mu q_L^\alpha) \,.
\end{split}
\eeq
A summation over color indices $\alpha,\beta$ and flavors
$q=u,d,s,c,b$ is implied. To leading order in small parameters, we
obtain
\begin{eqnarray}\label{eq:Cpenguin}
\begin{aligned}
   C_3^{\rm RS} &=
    \frac{\pi\alpha_s}{\Mkk^2}\,\frac{(\Delta_D')_{23}}{2N_c} -
    \frac{\pi\alpha}{6\sws\cws\,\Mkk^2} (\Sigma_D)_{23} \,, & \quad
    \tilde C_3^{\rm RS}
   &= \frac{\pi\alpha_s}{\Mkk^2}\,\frac{(\Delta_d')_{23}}{2N_c} \,, \\
   C_4^{\rm RS} &= C_6^{\rm RS} = -
    \frac{\pi\alpha_s}{2\Mkk^2}\,(\Delta_D')_{23} \,, & \quad \tilde
   C_4^{\rm RS} &= \tilde C_6^{\rm RS}
    = - \frac{\pi\alpha_s}{2\Mkk^2}\,(\Delta_d')_{23} \,, \\
   C_5^{\rm RS} &=
    \frac{\pi\alpha_s}{\Mkk^2}\,\frac{(\Delta_D')_{23}}{2N_c} \,, &
    \quad 
   \tilde C_5^{\rm RS} &=
    \frac{\pi\alpha_s}{\Mkk^2}\,\frac{(\Delta_d')_{23}}{2N_c}
    + \frac{\pi\alpha}{6\sws\cws\,\Mkk^2} (\Sigma_d')_{23} \,, \\
    C_7^{\rm RS} &= \frac{2\pi\alpha}{9\Mkk^2}\,(\Delta_D')_{23} -
     \frac{2\pi\alpha}{3\cws\,\Mkk^2} (\Sigma_D)_{23} \,, & \quad
     \tilde C_7^{\rm RS} &= \frac{2\pi\alpha}{9\Mkk^2}\,(\Delta_d')_{23}
      - \frac{2\pi\alpha}{3\sws\,\Mkk^2} (\Sigma_d')_{23} \,, \\
    C_8^{\rm RS} &= C_{10}^{\rm RS} = 0 \,, & \quad
     \tilde C_8^{\rm RS} &= \tilde C_{10}^{\rm RS} = 0 \,, \\
    C_9^{\rm RS} &= \frac{2\pi\alpha}{9\Mkk^2}\,(\Delta_D')_{23} +
     \frac{2\pi\alpha}{3\sws\,\Mkk^2} (\Sigma_D)_{23} \,, & \quad
     \tilde C_9^{\rm RS} &= \frac{2\pi\alpha}{9\Mkk^2}\,(\Delta_d')_{23}
     + \frac{2\pi\alpha}{3\cws\,\Mkk^2} (\Sigma_d')_{23} \,,
     \hspace{1cm}
\end{aligned}
\end{eqnarray}
where 
\beq \label{eq:Sigmas}
   \bm{\Sigma}_A \equiv L \left( \frac12 - \frac{\sws}{3} \right) {}
   \bm{\Delta}_A + \frac{\Mkk^2}{m_Z^2}\,\bm{\delta}_A \,, \qquad
   \bm{\Sigma}_A' \equiv L\,\frac{\sws}{3}\,\bm{\Delta}_A +
   \frac{\Mkk^2}{m_Z^2}\,\bm{\delta}_A \,.
\eeq
It is an interesting fact that in this case the QCD contributions from
KK gluon exchange no longer give the dominant contributions to the
Wilson coefficients. The effects due to $Z^0$-boson exchange are
enhanced by an extra factor of $L$, which compensates for the smaller
gauge couplings. Numerically one finds that relative to $C_{4,6}^{\rm
  RS}$ the Wilson coefficients $C_3^{\rm RS}$, $C_5^{\rm RS}$,
$C_7^{\rm RS}$, and $C_9^{\rm RS}$ are typically a factor of about
$2$, $1/3$, $3$, and $9$ bigger/smaller in magnitude. The
chirality-flipped coefficients $\tilde{C}_{3-10}$ are generically very
small in the model at hand, since they involve right-handed fermion
profiles that are naturally more UV-localized than their left-handed
counterparts. We will come back to this point in
Section~\ref{sec:numkaons}.

\boldmath \subsection{Important Formulas for $\epseps$} \unboldmath

The Wilson coefficients of the various penguin operators in
(\ref{eq:HeffDeltaF1}) enter the prediction for $\epseps$ which
measures the ratio of the direct and indirect CP-violating
contributions to $K \to \pi \pi$. Unfortunately, unlike the rare $K$
decays with leptons in the final state, the observable $\epseps$ is
affected by large hadronic uncertainties. In spite of the large theory
errors, a study of $\epseps$ can provide useful information on the
flavor structure of the underlying theory, since this ratio depends
very sensitively on the relative size of QCD and electroweak penguin
contributions, which cancel to a large extent in the SM. The strong
sensitivity of $\epseps$ to the electroweak penguin sector can
furthermore lead to interesting correlations with the rare
$K\to\pi\nu\bar\nu$ and $K_L\to\pi^0 l^+ l^-$ decays. One goal of our
work is to investigate to which extent such correlations exist in
the RS scenario.

In the presence of new-physics contributions to the Wilson
coefficients of the operators $Q_{3-10}$ and their chirality-flipped
partners $\tilde Q_{3-10}$, the ratio $\epseps$ can be approximated
by\footnote{The fact that, accidentally,
  $\arg(\epsilon_K)\approx\arg(\epsilon'_K)$, implies that ${\rm
    Re}\,(\epsilon'_K/\epsilon_K)\approx\epseps$.}
\beq\label{eq:epsepsRS}
   \frac{\epsilon'_K}{\epsilon_K} 
   = -{\rm Im}\,\Big( \lambda_t^{(ds)}\,F_{\rm SM} + F_{\rm RS} \Big)\,
   \frac{|\epsilon_K|_{\rm exp}}{|\epsilon_K|} \,,
\eeq
where 
\begin{equation}\label{eq:feps}
\begin{aligned}
   F_{\rm SM} &= - 1.4 + 13.6\,R_6 - 6.4\,R_8 \,, \\
   F_{\rm RS} &= 27.1\,K_3 - 56.1\,K_4 + 8.7\,K_5 + 36.0\,K_6
    - 544.4\,K_7 - 1663.5\,K_8 \\ 
   &\quad\mbox{}+ 141.0\,K_9 + 56.1\,K_{10}
    - \big( 11.4\,K_3 + 61.5\,K_4 - 177.1\,K_5 - 479.1\,K_6 \\ 
   &\quad\mbox{}+ 6.0\,K_7 + 27.5\,K_8 - 18.7\,K_9 + 16.4\,K_{10} 
    \big)\,R_6 - \big( 17.6\,K_3 - 45.0\,K_4 \\ 
   &\quad\mbox{}+ 90.8\,K_5 + 218.6\,K_6 - 8976.4\,K_7
    - 28190.5\,K_8 + 102.4\,K_9 + 23.8\,K_{10} \big)\,R_8 \,,
\end{aligned}
\end{equation}
and $|\epsilon_K|_{\rm exp} = (2.229 \pm 0.010) \cdot 10^{-3}$
\cite{Amsler:2008zz}, while $|\epsilon_K|$ denotes the value obtained
in the RS model. This formula has been derived following the approach
outlined in detail in \cite{Buras:1993dy, Bosch:1999wr}. The function
$F_{\rm SM}$ comprises the information on the SM contributions to the
Wilson coefficients of the $\Delta S = 1$ weak effective Hamiltonian
at the next-to-leading order \cite{Buras:1991jm, Buras:1992tc,
  Buras:1992zv, Ciuchini:1992tj, Ciuchini:1993vr, Gambino:2000fz}. The
given expression corresponds to a charm-quark renormalization scale of
$\mu_c = 1.3 \ {\rm GeV}$ and the central values of the SM parameters
$\alpha_s (m_Z) = 0.118 \pm 0.003$ \cite{Amsler:2008zz} and
$m_{t,{\rm pole}} = (172.6 \pm 1.4) \, {\rm GeV}$
\cite{Group:2008nq}. We have employed two-loop formulas to calculate
$\alpha_s (\mu)$ from $\alpha_s (m_Z)$ and to convert the top-quark
mass from the pole into the $\overline{\rm MS}$ scheme.

For the coefficients $K_i$ entering the new-physics contribution
$F_{\rm RS}$ we obtain
\beq\label{eq:Ki}
   K_i = \frac{2 \hspace{0.25mm} \sws \cws m_Z^2}{\pi \alpha} \left (
   C^{\rm RS}_i - \tilde C^{\rm RS}_i \right ) , \qquad 
   i= 3, \dots, 10 \,,
\eeq
where the minus sign is a consequence of $\langle \pi \pi | \tilde Q_i
| K \rangle = - \langle \pi \pi | Q_i | K \rangle$. The
electromagnetic coupling constant $\alpha$ as well as the Wilson
coefficients $C^{\rm RS}_i$ and $\tilde C^{\rm RS}_i$ are taken to be
renormalized at the scale $m_W$. In our numerical analysis we
determine $C^{\rm RS}_i$ and $\tilde C^{\rm RS}_i$ from the initial
conditions in (\ref{eq:Cpenguin}) using the leading-order RG equations
for the evolution from $\Mkk$ down to $m_W$. Notice that the inclusion
of running effects is necessary to obtain correct results in this
case, because a non-zero value of $K_8$, having the largest numerical
coefficient in (\ref{eq:feps}), is only generated through operator
mixing. Details on the latter issue can be found in
Appendix~\ref{app:RGE}.  We also stress that the coefficients
$K_{3-10}$ (\ref{eq:Ki}), which in the case of $\epseps$ are
understood to contain the 12 elements of the mixing matrices, have to
be calculated in the standard phase convention for the CKM matrix
\cite{Amsler:2008zz} in order to obtain correct results. The same
applies to all Wilson coefficients appearing hereafter.

The non-perturbative parameters $R_6$ and $R_8$ are given in terms of
the hadronic parameters $B_{6}^{(1/2)} \equiv B_{6}^{(1/2)} (m_c)$ and
$B_{8}^{(3/2)} \equiv B_{8}^{(3/2)} (m_c)$ and the strange- and
down-quark masses as
\beq \label{eq:Ri}
   R_i \equiv B_i^{(j)} \left [ 
   \frac{121\hspace{0.5mm}{\rm MeV}}{m_s(m_c) + m_d(m_c)} \right ]^2 .
\eeq
The hadronic parameters present the dominant source of theoretical
uncertainty in the prediction of the ratio $\epseps$. The current
status of the calculation of the $\langle \pi\pi | Q_{6,8} | K
\rangle$ matrix elements is reviewed in \cite{Buras:2003zz}. While
$B_8^{(3/2)} = 1.0 \pm 0.2$ is obtained in various approaches, the
situation with $B_6^{(1/2)}$ is far less clear. For instance, in the
framework of the $1/N$ expansion, values for $B_6^{(1/2)}$ notably
above unity are obtained \cite{Bijnens:2000uq, Bijnens:2001ps,
  Knecht:2001bc, Hambye:2003cy}. For example, the articles
\cite{Bijnens:2000uq, Bijnens:2001ps} find $B_6^{(1/2)} = 2.5 \pm
0.4$ and $B_8^{(3/2)} = 1.1 \pm 0.3$. On the other hand, while the
lattice values of $B_8^{(3/2)}$ are compatible with unity
\cite{Blum:2001xb, Boucaud:2004aa}, they are lower than 1 for
$B_6^{(1/2)}$ \cite{Bhattacharya:2001uw, Bhattacharya:2004qu}. 
First unquenched determinations of the parameters $B_6^{(1/2)}$
and $B_8^{(3/2)}$ exist too \cite{Li:2008kc}. Yet the latter results
are limited by a combination of statistical and systematic uncertainties
and we do not include them since they, if correct, would require a
sizable new-physics contribution in $\epseps$ to bring the theory
prediction in agreement with experiment.

In view of the rather uncertain value of $B_6^{(1/2)}$, we adopt a
conservative point of view and scan the hadronic parameters in our SM
analysis of $\epseps$ over the ranges
\beq \label{eq:bagranges}
   B_6^{(1/2)} = [0.8, 2.0] \,, \qquad B_8^{(3/2)} = [0.8, 1.2] \,.
\eeq
requiring in addition $B_6^{(1/2)} > B_8^{(3/2)}$, as suggested by
studies of the $\langle \pi\pi | Q_{6,8} | K \rangle$ matrix elements
in the framework of the $1/N$ expansion \cite{Bijnens:2000uq,
  Bijnens:2001ps, Knecht:2001bc, Hambye:2003cy}. We furthermore employ
the values
\beq \label{eq:massranges}
   m_s(m_c) = (115 \pm 20) \, {\rm MeV} \,, \qquad 
   m_d(m_c) = (6 \pm 2)\, {\rm MeV}
\eeq
for the running quark masses. The quoted central value of the
strange-quark mass corresponds to $m_s (2 \, {\rm GeV}) = 100 \, {\rm
  GeV}$, which is in the ballpark of a number of recent determinations
\cite{Amsler:2008zz}.

Our SM prediction reads 
\beq \label{eq:epsepsSM}
   \left( \frac{\epsilon'_K}{\epsilon_K} \right)_{\rm SM} 
   = \Big( 8.8^{+41.2}_{-~4.8} \Big) \cdot 10^{-4} \,,
\eeq
where the quoted central value has been obtained from
(\ref{eq:epsepsRS}) by setting $F_{\rm RS} = 0$ and $R_{6,8} = 1$. In
particular, it includes the normalization factor $|\epsilon_K|_{\rm
  exp}/|\epsilon_K|_{\rm SM}$ with $|\epsilon_K|_{\rm SM}$ taken from
(\ref{epsKSM}). Rather than corresponding to a $68\%$ CL, the given
uncertainties represent the ranges in which we believe that the true
value of $(\epseps)_{\rm SM}$ lies within a high probability. Note
that within errors the SM prediction is in agreement with the
experimental value $(\epseps)_{\rm exp} = (16.5 \pm 2.6) \cdot
10^{-4}$ \cite{Amsler:2008zz} as well as other SM predictions
\cite{Bosch:1999wr, Buras:2003zz, Pallante:2001he}. In our numerical
analysis of $\epseps$ in the RS scenario, we scan independently over
the ranges given in (\ref{eq:bagranges}) and (\ref{eq:massranges}),
requiring $B_6^{(1/2)} > B_8^{(3/2)}$, and check whether it is
possible to achieve agreement with the measured value. We will see
later on that even with such a conservative treatment of errors the
constraint from $\epseps$ has a non-negligible effect on the possible
new-physics effects in rare $K$ decays within the RS model.

\boldmath \subsection{Important Formulas for Non-Leptonic $B$
  Decays} \unboldmath

In the following we will present formulas for a few of the most
interesting observables in the vast array of measurements in the field
of exclusive hadronic decays of $B$ mesons. We will consider two-body
decays of the type $\bar B\to PP$ and $\bar B\to PV$, where $P$ and
$V$ stand for light pseudoscalar and longitudinally polarized vector
mesons, respectively. Parity invariance of the strong interactions
allows us to relate the hadronic matrix elements of the
opposite-chirality operators $\tilde Q_i$ to those of the
corresponding SM operators $Q_i$. It follows that for $\bar B\to PP$
decay modes the new-physics contributions enter via the differences
$(C_i^{\rm RS}-\tilde C_i^{\rm RS})$ of Wilson coefficients, while for
$\bar B\to PV$ decay modes they enter via the sums $(C_i^{\rm
  RS}+\tilde C_i^{\rm RS})$. The structure of the effective
Hamiltonian (\ref{eq:HeffDeltaF1}) then implies that the RS
contributions can be included by the replacement rule
\begin{equation}\label{replacement_rule}
\begin{split}
  \frac{G_F}{\sqrt2} \left( \lambda_u^{(sb)} + \lambda_c^{(sb)}
  \right) C_i &\,\to \,\frac{G_F}{\sqrt2} \left( \lambda_u^{(sb)} +
    \lambda_c^{(sb)} \right) C_i
  + \big( C_i^{\rm RS}\mp\tilde C_i^{\rm RS} \big) \\
  &~= \frac{G_F}{\sqrt2} \left[ \left( \lambda_u^{(sb)} +
        \lambda_c^{(sb)} \right) C_i \, +
      \stackrel{\scalebox{0.4}{(}-\scalebox{0.4}{)}}{K_i} \right] ,
  \qquad i=3,\dots,10 \,,
\end{split}
\end{equation}
where the upper (lower) sign refers to $PP$ ($PV$) modes, and the
coefficients $K_i$ have been defined in (\ref{eq:Ki}). For $PV$ modes,
the $K_i$ are replaced by coefficients $\bar K_i$ defined analogously,
but with a relative plus sign between $C_i^{\rm RS}$ and $\tilde
C_i^{\rm RS}$. The remaining Wilson coefficients $C_{1,2}$,
$C_{7}^\gamma$, and $C_{8}^g$ are unchanged at tree level.

Most of the complications in the analysis of exclusive hadronic decays
originate from complicated strong-interaction physics encoded in the
hadronic matrix elements of the local four-quark operators in the
effective weak Hamiltonian evaluated between meson states. However,
the basis of operators in the RS model is the same as in the SM, apart
from the addition of the chirality-flipped operators, whose matrix
elements are linked to the ones of the SM by parity invariance as
already noted above. The strong-interaction physics is therefore the
same as in the SM. For the purposes of our analysis we will adopt the
QCD factorization approach \cite{Beneke:1999br, Beneke:2000ry,
  Beneke:2001ev}, which allows for a systematic analysis of the
relevant hadronic matrix elements at leading and partially at
subleading order in the heavy-quark expansion. Our phenomenological
analysis will closely follow corresponding analyses in
\cite{Beneke:2001ev} for $\bar B\to\pi\bar K$ decays, and
\cite{Beneke:2002jn, Beneke:2003zv} for the analysis of other decays
into $PP$ or $PV$ final states. We will also resort to $SU(3)$
symmetry and Fierz relations first employed in
\cite{Neubert:1998pt}. A general analysis of new-physics effects in
the decays $\bar B \to \pi \bar K$ in a wide class of models giving
rise to SM four-quark operators and their chirality-flipped
counterparts was presented in \cite{Grossman:1999av}. The RS model
falls into this class of models, and we will apply some of the
strategies proposed in the latter reference in our context.

In the QCD factorization approach to $\bar B\to M_1 M_2$ decays, the
decay amplitudes are decomposed into a basis a flavor operators
multiplied by coefficients $\alpha_i^q(M_1 M_2)$ and $\beta_i^q(M_1
M_2)$ \cite{Beneke:2003zv}, the latter of which account for the
effects of weak annihilation. The index $q=u,c$ indicates that these
coefficients are associated with a CKM factor $\lambda_q^{(sb)}$. The
$\alpha_i^q(M_1 M_2)$ and $\beta_i^q(M_1 M_2)$ parameters are
themselves given in terms of explicit expressions involving the Wilson
coefficient functions $C_i$ of the SM.\interfootnotelinepenalty=10000
\footnote{{\nopagebreak[4]See Eqs.~(9), (18),
  and (19) of \cite{Beneke:2003zv} for the precise definitions of
  these coefficients, and Eqs.~(31) to (55) for explicit
  expressions in terms of convolutions of hard-scattering kernels with
  light-cone distribution amplitudes.}}
\interfootnotelinepenalty=100
In order to include the RS
contributions, all that is required now is to apply the replacement
rule (\ref{replacement_rule}), if not stated otherwise.

We emphasize that our study will be exploratory and focus only on a
small number of relevant observables, which are particularly
interesting with regard to searches for new physics. We will not be
concerned here with a comprehensive analysis of strong-interaction
uncertainties. For the same reason, we will not consider alternative
theoretical schemes such as the perturbative QCD approach
\cite{Keum:2000ph,Keum:2000wi}, the phenomenological $SU(3)$ approach
\cite{Gronau:1994rj,Gronau:1995hn}, or an approach based on
soft-collinear effective theory \cite{Bauer:2004tj,Bauer:2005kd},
which is similar in spirit to QCD factorization but differs in the
implementation \cite{Beneke:2004bn}. Our main conclusions will not
depend on the scheme used to evaluate the hadronic matrix elements.

We will illustrate the potential impact of the RS contributions by
investigating a couple of interesting observables, which probe
different aspects of the decay amplitudes such as tree, QCD, and
electroweak penguin contributions. The color-allowed SM tree
amplitudes are so large that no viable model of new physics can give
rise to effects that could compete with them, as this would require
new-physics contributions to the Wilson coefficients of ${\cal
  O}(1)$. Moreover, in the SM the QCD penguin amplitudes are larger
than the electroweak penguin amplitudes by about an order of
magnitude. The non-vanishing RS contributions to the electroweak
penguin coefficients are instead at least as big as the coefficients
of the QCD penguins, so that we expect to find larger new-physics
effects in the electroweak penguin sector.

In the case of $\bar B \to \pi \bar K$ decays, the electroweak penguin
coefficients are denoted by $\alpha_{3,{\rm EW}}^q (\pi \bar K)$ and
$\alpha_{4,{\rm EW}}^q (\pi \bar K)$ in \cite{Beneke:2003zv}. They
receive contributions from the Wilson coefficients $C_{7-10}$. In the
RS model the largest contribution typically stems from the coefficient
$C_9$. The coefficients $C_{8,10}$ vanish at the matching scale and
remain moderately small when the RG evolution from the KK scale down
to the weak scale is included. This is shown explicitly in
Appendix~\ref{app:RGE}. Similarly, the $C_7$ terms are numerically
subdominant because they are suppressed by a factor of roughly
$-\sws/\cws\approx-1/3$ with respect to $C_9$. The dominance of $C_9$
implies that, like in the SM, $SU(3)$ symmetry and Fierz relations can
be used to obtain a model-independent prediction for the leading
electroweak penguin effects in terms of a single parameter $q$, which
is free of hadronic uncertainties \cite{Neubert:1998pt}. In the RS
model this parameter is accompanied by a CP-odd phase, which we shall
call $\phi$. Neglecting small effects arising from the QCD penguins
$C_{3-6}$, we obtain the approximate formula
\begin{equation} \label{eq:qnaive} 
    q = |q|\,e^{i\phi} \approx q_{\rm SM}\,\Big[ \hspace{0.25mm} 
    1 - \left ( 138 - 45 \, i \right ) K_7 - \left (8084 + 236 \, i 
    \right ) K_8 - \left ( 3430 + 45 \, i \right ) (K_9 +
    K_{10}) \hspace{0.25mm} \Big] \,, 
\end{equation}
where $q_{\rm SM}=(0.69\pm 0.18)\,e^{i\hspace{0.25mm}(0.2\pm
  2.1)^\circ}$ is calculable in the SM in terms of the top-quark mass
and electroweak parameters. The quoted number corresponds to scenario
S4 of hadronic parameters introduced in \cite{Beneke:2003zv}, for
which $\gamma=(70\pm 20)^\circ$. In addition we fix both parameters
$\rho_{H}$ and $\rho_A$, which are related to the endpoint
singularities appearing in the hard spectator scattering and weak
annihilation kernels, to 1. The same set of input will be used for all
other SM predictions given below. A complete list of all the relevant
theoretical input parameters can be found in
Appendix~\ref{app:input}. The coefficients $K_{7-10}$ have been
defined in (\ref{eq:Ki}) and are understood to be evaluated at the
electroweak scale. While (\ref{eq:qnaive}) allows for a quantitative
understanding of the importance of the different contributions, we
will use the complete expressions for $\alpha_{3,{\rm EW}}^q (\pi \bar
K)$ and $\alpha_{4,{\rm EW}}^q (\pi\bar K)$ of \cite{Beneke:2003zv}
supplemented by the shift (\ref{replacement_rule}) to determine $q$ in
our numerical analysis. Corrections due to $SU(3)$ breaking, QCD
penguins, and electromagnetic effects in the RG evolution influence
the obtained results only in a minor way.

New-physics contributions to the electroweak penguin coefficients can
also have a visible impact on the results for the two ratios
\begin{equation} \label{eq:RastR00}
\begin{split}
  R_\ast &= \frac{\Gamma(B^-\to\pi^-\bar K^0) + \Gamma(B^+\to\pi^+
    K^0)} {2\big[ \Gamma(B^-\to\pi^0 K^-)
    + \Gamma(B^+\to\pi^0 K^+)\big]} \,, \\[1mm]
  R_{00} &= \frac{2\big[ \Gamma(\bar B^0\to\pi^0\bar K^0) +
    \Gamma(B^0\to\pi^0 K^0) \big]} {\Gamma(B^-\to\pi^-\bar K^0) +
    \Gamma(B^+\to\pi^+ K^0)} \,,
\end{split}
\end{equation}
of CP-averaged rates. The current experimental values of these ratios
are $(R_\ast)_{\rm exp}=0.79\pm 0.08$ and $(R_{00})_{\rm exp}=0.86\pm
0.09$ \cite{Barberio:2008fa}. Adding individual errors in quadrature,
the theoretical SM predictions obtained using QCD factorization read
$(R_\ast)_{\rm SM}=0.88\pm 0.14$ and $(R_{00})_{\rm SM}=0.86\pm 0.08$.

As another interesting example, we consider the ``puzzle'' of the
observed difference $\Delta A_{\rm CP}$ in the direct CP asymmetries
$A_{\rm CP} (B^-\to\pi^0 K^-)$ and $A_{\rm CP} (\bar B^0\to\pi^+
K^-)$. Hereafter we define the asymmetries as
\begin{equation}
  A_{\rm CP}(\bar B \to f) \equiv \frac{\Gamma (\bar B \to \bar f) 
    - \Gamma (B \to f)}{\Gamma (\bar B \to \bar f) + \Gamma (B \to f)} \,.
\end{equation}
The present world average of the considered quantity is $(\Delta
A_{\rm CP})_{\rm exp} = (14.7 \pm 2.7) \%$ \cite{Barberio:2008fa}. In
the SM, theoretical expectations for this difference are very small,
typically no more than a few percent. For our reference parameter
scenario, we find numerically $(\Delta A_{\rm CP})_{\rm SM}=(0.7\pm
2.9) \%$ which is about $3.5\sigma$ below the experimental
determination.

To quantify the impact of new physics on $R_\ast$, $R_{00}$, and
$\Delta A_{\rm CP}$ one has to determine the dependence of the
relevant $\bar B \to \pi \bar K$ decay amplitudes on the coefficients
$K_{3-10}$. Following the QCD factorization approach of
\cite{Beneke:2003zv}, we find the expressions
\begin{align} \label{eq:amplitudes} {\cal A} (B^- \to \pi^- \bar K^0 )
  \, = & \, \Big [ \, (123.0 - 10.1 \, i) \hspace{0.25mm}
  \lambda_u^{(sb)} + (112.9 - 18.6 \, i) \hspace{0.25mm}
  \lambda_c^{(sb)} - (270.2 - 102.5 \, i) \hspace{0.25mm} K_3
  \nonumber \\[0mm] & \, - (368.0 + 3.1 \, i) \hspace{0.25mm} K_4 -
  (1079.9 - 384.4 \,i) \hspace{0.25mm} K_5 - (2951.5 - 1163.3 \, i)
  \hspace{0.25mm} K_6 \nonumber \\[0mm] & \, + (268.1 + 343.1 \, i)
  \hspace{0.25mm} K_7 + (963.3 + 1038.2 \, i) \hspace{0.25mm} K_8 +
  (19.5 + 18.4 \, i) \hspace{0.25mm} K_9 \nonumber \\[0mm] & \, +
  (512.3 - 16.8 \, i) \hspace{0.25mm} K_{10} \Big ]
  \hspace{0.25mm} \cdot 10^{-8} \, {\rm GeV} \,, \nonumber \\[1mm]
  \sqrt{2} \hspace{0.25mm} {\cal A} (B^- \to \pi^0 K^- ) \, = & \,
  \Big [ - (1120.9 - 45.1 \, i) \hspace{0.25mm} \lambda_u^{(sb)} +
  (127.7 - 19.5 \, i) \hspace{0.25mm} \lambda_c^{(sb)} - (275.8 -
  102.8 \, i) \hspace{0.25mm} K_3 \nonumber \\[0mm] & \, - (383.2 +
  2.3 \, i) \hspace{0.25mm} K_4 - (1074.6 - 383.9 \,i) \hspace{0.25mm}
  K_5 - (2926.6 - 1162.1 \, i) \hspace{0.25mm} K_6 \nonumber \\[0mm] &
  \, + (159.1 + 379.8 \, i) \hspace{0.25mm} K_7 - (3551.7 - 1249.6 \,
  i) \hspace{0.25mm} K_8 - (1794.7 - 128.7 \, i) \hspace{0.25mm} K_9
  \nonumber \\[0mm] & \, - (1354.4 - 68.3 \, i) \hspace{0.25mm} K_{10}
  \Big ] \hspace{0.25mm} \cdot 10^{-8} \, {\rm GeV} \,, \nonumber \\[1mm]
  \sqrt{2} \hspace{0.25mm} {\cal A} (\bar B^0 \to \pi^0 \bar K^0 ) \,
  = & \, \Big [ - (527.9 - 71.9 \, i) \hspace{0.25mm} \lambda_u^{(sb)}
  - (101.4 - 18.6 \, i) \hspace{0.25mm} \lambda_c^{(sb)} + (264.5 -
  103.0 \, i) \hspace{0.25mm} K_3 \nonumber \\[0mm] & \, + (364.4 +
  5.6 \, i) \hspace{0.25mm} K_4 + (1076.7 - 388.3 \,i) \hspace{0.25mm}
  K_5 + (2956.0 - 1172.9 \, i) \hspace{0.25mm} K_6 \nonumber \\[0mm] &
  \, + (721.2 + 212.9 \, i) \hspace{0.25mm} K_7 - (1799.6 - 763.0 \,
  i) \hspace{0.25mm} K_8 - (1448.3 - 7.0 \, i) \hspace{0.25mm} K_9
  \nonumber \\[0mm] & \, - (1013.7 - 119.2 \, i) \hspace{0.25mm}
  K_{10} \Big ] \hspace{0.25mm} \cdot 10^{-8} \, {\rm GeV} \,,
  \nonumber \\[1mm]
  {\cal A} (\bar B^0 \to \pi^+ K^- ) \, = & \, \Big [- (716.0 + 16.7
  \, i) \hspace{0.25mm} \lambda_u^{(sb)} + (116.2 - 19.5 \, i)
  \hspace{0.25mm} \lambda_c^{(sb)} - (270.0 - 103.3 \, i)
  \hspace{0.25mm} K_3 \nonumber \\[0mm] & \, - (379.6 + 4.8 \, i)
  \hspace{0.25mm} K_4 - (1071.4 - 387.8 \,i) \hspace{0.25mm} K_5 -
  (2931.1 - 1171.7 \, i) \hspace{0.25mm} K_6 \nonumber \\[0mm] & \, -
  (830.2 + 176.2 \, i) \hspace{0.25mm} K_7 - (2715.4 + 551.6 \, i)
  \hspace{0.25mm} K_8 - (365.9 - 103.2 \, i) \hspace{0.25mm} K_9
  \hspace{6mm} \nonumber \\[0mm] & \, - (853.0 + 34.1 \, i)
  \hspace{0.25mm} K_{10} \Big ] \hspace{0.25mm} \cdot 10^{-8} \, {\rm
    GeV} \,.
\end{align}
Analogous formulas hold in the case of the CP-conjugated amplitudes
with the replacements $\lambda_q^{(sb)} \to \lambda_q^{(sb) \ast}$ and
$K_i \to K_i^\ast$. Notice that the coefficients multiplying the QCD
penguin contributions $K_{3-6}$ in the different amplitudes of
(\ref{eq:amplitudes}) are, up to possible overall signs, of a similar
magnitude. As a result, the quantities $R_\ast$, $R_{00}$, and $\Delta
A_{\rm CP}$ are to first approximation independent of new physics
affecting the QCD penguin sector. For a similar reason also the
contribution from $K_7$ ($K_8$) cancels to a large extent in the
ratios $R_\ast$ ($R_{00}$), irrespectively of the precise nature of
new physics. In the RS framework, the observables $R_\ast$, $R_{00}$,
and $\Delta A_{\rm CP}$ receive the largest correction from $K_9$,
while the remaining electroweak penguin coefficients lead to
subleading but non-negligible effects.

We will use the penguin-dominated decay mode $B^-\to\pi^-\bar K^0$ to
probe both the magnitude and CP-odd phase of the leading QCD penguin
amplitude in the $\bar B \to \pi \bar K$ system. In the SM the decay
$B^-\to\pi^-\bar K^0$ is to a very good approximation dominated by the
QCD penguin amplitude $\hat\alpha_4^c (\pi \bar K) \equiv\alpha_4^c
(\pi \bar K) +\beta_3^c ( \pi \bar K)$, and as a result the direct CP
asymmetry $A_{\rm CP}(B^- \to \pi^-\bar K^0)$ is expected to be very
small, not exceeding the level of a few percent
\cite{Beneke:2003zv}. In our reference parameter scenario, we find
$A_{\rm CP}(B^- \to \pi^-\bar K^0)_{\rm SM} = (0.4 \pm 0.7) \%$. A
significant new-physics contribution to both the QCD and electroweak
penguin coefficients with a new CP-odd phase could change this
conclusion. The present experimental value $A_{\rm CP}(B^- \to
\pi^-\bar K^0)_{\rm exp} = (0.9 \pm 2.5) \%$ \cite{Barberio:2008fa}
still allows for larger effects, so it is interesting to see whether
it is possible to saturate the experimental limits in the RS
framework.

As a second measure of new-physics effects in the QCD penguin
amplitudes we consider
\begin{equation}
  R_{P/T} = \left| \frac{V_{ub}}{V_{cb}} \right| \frac{f_\pi}{f_K} \left[
    \frac{\Gamma(B^-\to\pi^-\bar K^0) + \Gamma(B^+\to\pi^+ K^0)}%
    {2\left[ \Gamma(B^-\to\pi^-\pi^0)
        + \Gamma(B^+\to\pi^+\pi^0) \right]} 
  \right]^{1/2} ,
\end{equation}
which in the SM determines to very good accuracy the absolute value of
the penguin-to-tree ratio $\hat\alpha_4^c(\pi\bar K)/(\alpha_1(\pi\pi)
+ \alpha_2(\pi\pi))$. We find $(R_{P/T})_{\rm SM} = 0.10 \pm 0.03$,
which should be compared to the current experimental value
$(R_{P/T})_{\rm exp} = 0.15 \pm 0.01$ \cite{Barberio:2008fa}.

To calculate $R_{P/T}$ in the RS model one needs in addition to
(\ref{eq:amplitudes}) the decay amplitude for $B^- \to \pi^- \pi^0$
and its CP-conjugated counterpart. In terms of the coefficients
$K_{3-10}$, we obtain in QCD factorization
\begin{eqnarray} \label{eq:amplitude}
\begin{aligned}  
  \sqrt{2} \hspace{0.25mm} {\cal A} (B^- \to \pi^- \pi^0 ) \, = & \,
  \Big [ - (1004.5 - 47.4 \, i) \hspace{0.25mm} \lambda_u^{(db)} +
  (12.4 - 0.5 \, i) \hspace{0.25mm} \lambda_c^{(db)} - (4.6 - 0.2 \,
  i) \hspace{0.25mm} K_3 \\[0mm] & \, - (12.5 - 0.7 \, i)
  \hspace{0.25mm} K_4 + (4.5 - 0.4 \,i) \hspace{0.25mm} K_5 + (20.5 -
  1.0 \, i) \hspace{0.25mm} K_6 \\[0mm] & \, - (107.3 - 29.7 \, i)
  \hspace{0.25mm} K_7 - (3711.2 - 170.6 \, i) \hspace{0.25mm} K_8 -
  (1508.6 - 71.9 \, i) \hspace{0.25mm} K_9 \hspace{8mm} \\
  & \, - (1519.5 - 72.6 \, i) \hspace{0.25mm} K_{10} \Big ]
  \hspace{0.25mm} \cdot 10^{-8} \, {\rm GeV} \,,
\end{aligned}
\end{eqnarray}
and $\sqrt{2} \hspace{0.25mm} {\cal A} (B^+ \to \pi^+ \pi^0 ) $ is
obtained from the above expression by replacing $\lambda_q^{(db)} \to
\lambda_q^{(db) \ast}$ and $K_i \to K_i^\ast$. In the case of $A_{\rm
  CP}(B^- \to \pi^-\bar K^0)$, the initial conditions of $C_{3,6,7,9}$
can all contribute in a similar fashion to the prediction, although
the corrections arising from the electroweak penguin sector are
usually the more important ones. For $R_{P/T}$, on the other hand, one
can easily convince oneself that the coefficient $C_9$ always gives
the dominant correction, irrespectively of the exact values of the new
CP-odd phases appearing in the remaining coefficients.

New-physics contributions to either the QCD or the electroweak penguin
amplitudes can affect the determination of $\sin2\beta$ from the
time-dependent CP asymmetries in neutral $B$-meson decays into CP
eigenstates $f$. In cases where the decay amplitude ${\cal A} (\bar
B^0\to f)$ does not contain a CP-odd phase, the coefficient $S_f$ in
the general relation
\begin{equation}
  \frac{\Gamma(\bar B^0(t)\to f)-\Gamma(B^0(t)\to f)}
  {\Gamma(\bar B^0(t)\to f)+\Gamma(B^0(t)\to f)}
  = S_f\hspace{0.25mm} \sin(\Delta m_B t) - 
  C_f\hspace{0.25mm}\cos(\Delta m_B t)
\end{equation}
measures $\sin2\beta$. However, if the decay amplitude receives a
small correction with a different weak phase, then this relation is
modified. Parametrizing the admixture as $(1+e^{-i\gamma}
d_f\,e^{i\phi_f})$, one finds \cite{Beneke:2003zv}
\begin{equation}
  S_f = \frac{\sin 2 \beta + 2 d_f \cos \phi_f \sin (2 \beta +
    \gamma) + d_f^2 \sin (2 \beta + 2 \gamma)}{1 + 2 d_f \cos \phi_f
    \cos \gamma + d_f^2} \,.
\end{equation}

Unlike the Wilson coefficients of the SM, that are real functions,
the coefficients parametrizing the RS contributions contain complex,
CP-violating phases. In order to still use the above formula, it is
necessary to decompose these parameters into two terms proportional to
$\lambda_u^{(sb)}$ and $\lambda_c^{(sb)}$, which is always possible
since $\mbox{Im}\big(\lambda_u^{(sb)}\,\lambda_c^{(sb) \ast} \big)\ne
0$. To this end, we define the coefficients
\begin{equation}
    \Delta \! \stackrel{\scalebox{0.4}{(}-\scalebox{0.4}{)}}{C_i^u} 
    \; \equiv \frac{\sqrt2}{G_F}\,
    \frac{\mbox{Im}\big [(C_i^{\rm RS}\mp\tilde C_i^{\rm RS})\, 
      \lambda_c^{(sb) \ast} \big ]}
    {\mbox{Im}\big ( \lambda_u^{(sb)}\,\lambda_c^{(sb) \ast} \big)} \,,
    \qquad
    \Delta \! \stackrel{\scalebox{0.4}{(}-\scalebox{0.4}{)}}{C_i^c} \; 
    \equiv \frac{\sqrt2}{G_F}\,
    \frac{\mbox{Im}\big [(C_i^{\rm RS}\mp\tilde C_i^{\rm RS})\, 
      \lambda_u^{(sb)\ast} \big]}
    {\mbox{Im} \big (\lambda_c^{(sb)}\,\lambda_u^{(sb)\ast} \big)} \,, 
\end{equation}
which are invariant under phase redefinitions of the quark fields.
We then obtain the replacement rule
\begin{equation}\label{Cieffdef}
    \frac{G_F}{\sqrt2} \left( \lambda_u^{(sb)} + \lambda_c^{(sb)} \right)
    C_i \, \to \, \frac{G_F}{\sqrt2}
    \left( \lambda_u^{(sb)} \,
      \stackrel{\scalebox{0.4}{(}-\scalebox{0.4}{)}}{C_i^u}\,  
      + \lambda_c^{(sb)} \, 
      \stackrel{\scalebox{0.4}{(}-\scalebox{0.4}{)}}{C_i^c} \, \right) ,  
    \qquad
    \stackrel{\scalebox{0.4}{(}-\scalebox{0.4}{)}}{C_i^q} \; 
    \equiv C_i + 
    \Delta \! \stackrel{\scalebox{0.4}{(}-\scalebox{0.4}{)}}{C_i^q} \, \,,
\end{equation}
instead of (\ref{replacement_rule}). The RS contributions are now
encoded in the {\em real\/} quantities $\Delta \!
\stackrel{\scalebox{0.4}{(}-\scalebox{0.4}{)}}{C_i^q}$, while the
CP-odd phases are carried by the SM parameters $\lambda_u^{(sb)}$ and
$\lambda_c^{(sb)}$.

In the SM the corresponding shifts $\Delta S_f \equiv S_f - \sin 2
\beta$ for some of the most interesting decay modes are found to be
positive and small \cite{Beneke:2003zv, Beneke:2005pu}. Employing
$\sin 2 \beta = 0.672 \pm 0.023$ \cite{Barberio:2008fa}, we find in
the parameter scenario S4 the results
\begin{equation}
\begin{split}
  (\Delta S_{\phi K_S})_{\rm SM}  &= 0.022 \pm 0.013 \pm 0.010 \,, \\
  (\Delta S_{\eta' K_S})_{\rm SM} &= 0.003 \pm 0.010 \pm 0.010 \,, \\
  (\Delta S_{\pi^0 K_S})_{\rm SM} &= 0.138 \pm 0.054 \pm 0.010 \,,
\end{split}
\end{equation} 
where the second error estimates the theory uncertainty in $\sin 2
\beta$ obtained from $B_d \to \psi K_S$. The current world averages of
the shifts read $(\Delta S_{\phi K_S})_{\rm exp} = (-0.23 \pm 0.18)$,
$(\Delta S_{\eta^\prime K_S})_{\rm exp} = (-0.08 \pm 0.07)$, and
$(\Delta S_{\pi^0 K_S})_{\rm exp} = (-0.10 \pm 0.17)$
\cite{Barberio:2008fa}.  Although the result for each individual mode
does not differ significantly from the SM expectation, the central
values tend to lie below the SM predictions. While much more data is
needed to firmly establish the presence of a new CP-violating phase
beyond the SM for each of these modes, we will investigate whether the
RS model is able to reproduce the observed deviations.

For the necessary decay amplitudes, we find within QCD factorization
\begin{eqnarray} \label{eq:sin2b}
\begin{split}
  {\cal A} (\bar B^0 \to \phi K_S ) \, = & \, \Big [ - (79.7 + 3.4 \,
  i) \hspace{0.25mm} \lambda_u^{(sb)} - (83.7 - 18.2 \, i)
  \hspace{0.25mm} \lambda_c^{(sb)} + (2006.5 - 112.4 \, i)
  \hspace{0.25mm} \bar K_3  \\[0mm]
  & \, + (1570.7 - 108.9 \, i) \hspace{0.25mm} \bar K_4 + (2123.0 -
  325.6 \,i) \hspace{0.25mm} \bar K_5 + (1667.4 - 634.2 \, i)
  \hspace{0.25mm} \bar K_6 \hspace{6mm}  \\[0mm] & \, -
  (1070.5 - 151.0 \, i) \hspace{0.25mm} \bar K_7 - (1152.3 - 295.2 \,
  i) \hspace{0.25mm} \bar K_8 - (953.6 - 43.1 \, i) \hspace{0.25mm}
  \bar K_9  \\[0mm] & \, - (1019.0 - 58.1 \, i)
  \hspace{0.25mm} \bar K_{10}
  \Big ] \hspace{0.25mm} \cdot 10^{-8} \, {\rm GeV} \,,  \\[1mm]
  {\cal A} (\bar B^0 \to \eta^\prime K_S ) \, = & \Big [ \, (40.6 +
  38.5 \, i) \hspace{0.25mm} \lambda_u^{(sb)} + (220.2 - 36.0 \, i)
  \hspace{0.25mm} \lambda_c^{(sb)} - (2046.2 - 87.5 \, i)
  \hspace{0.25mm} K_3  \\[0mm] & \, - (985.3 - 109.3 \, i)
  \hspace{0.25mm} K_4 - (39.7 - 644.2 \,i) \hspace{0.25mm} K_5 -
  (4327.2 - 2052.9 \, i) \hspace{0.25mm} K_6 \hspace{6mm} 
  \\[0mm] & \, + (652.0 - 256.3 \, i) \hspace{0.25mm} K_7 + (2965.7 -
  843.5 \, i) \hspace{0.25mm} K_8 + (311.9 - 25.1 \, i)
  \hspace{0.25mm} K_9  \\[0mm] & \, + (841.0 - 34.6 \, i)
  \hspace{0.25mm} K_{10} \Big ]
  \hspace{0.25mm} \cdot 10^{-8} \, {\rm GeV} \,,  \\[1mm]
  {\cal A} (\bar B^0 \to \pi^0 K_S ) \, = & \, - (373.3 - 50.8 \, i)
  \hspace{0.25mm} \lambda_u^{(sb)} - (71.7 - 13.2 \, i)
  \hspace{0.25mm} \lambda_c^{(sb)} + (187.0 - 72.8 \, i)
  \hspace{0.25mm} K_3  \\[0mm] & \, + (257.6 + 4.0 \, i)
  \hspace{0.25mm} K_4 + (761.3 - 274.5 \,i) \hspace{0.25mm} K_5 +
  (2090.2 - 829.4 \, i) \hspace{0.25mm} K_6  \\[0mm] & \, +
  (510.0 + 150.5 \, i) \hspace{0.25mm} K_7 - (1272.5 - 539.5 \, i)
  \hspace{0.25mm} K_8 - (1024.1 - 5.0 \, i) \hspace{0.25mm} K_9
  \\[0mm] & \, - (716.8 - 84.3 \, i) \hspace{0.25mm} K_{10}
  \Big ] \hspace{0.25mm} \cdot 10^{-8} \, {\rm GeV} \,.
\end{split}
\end{eqnarray}
Here the coefficients $\bar K_i$ entering the $\bar B^0 \to \phi K_S$
amplitude are defined in analogy to (\ref{eq:Ki}), but owing to the
vector-like nature of the $\phi$ meson with the minus replaced by a
plus sign. Numerically we find that in the RS model the shifts $\Delta
S_f$ can receive contributions of comparable size from the matching
corrections to $C_{3,6,7,9}$. The corrections arising from the
electroweak penguin sector are typically larger than those due to the
QCD penguins.

\boldmath \subsection{Rare Leptonic Decays of Kaons and $B$ Mesons}
\unboldmath
\label{subsec:rare_leptons}

In the following, we consider $\Delta F=1$ processes in the quark
sector that feature leptons in the final state. Both inclusive and
exclusive transitions induced by neutral- as well as charged-current
interactions are discussed. Before examining the specific decay modes
in the RS model, we introduce the effective Hamiltonians describing
the $s \to d \nu \bar \nu$, $b \to s l^+ l^-$, and $b \to u l \bar
\nu$ transitions.

\boldmath \subsubsection{Effective Hamiltonian for $s \to d \nu \bar
  \nu$ } \unboldmath

The effective Hamiltonian describing $s \to d \nu\bar\nu$ transitions
reads
\beq
   {\cal H}_{\rm eff}^{s\to d\nu\bar\nu} 
   = C_{\nu}\,(\bar d_L\gamma^\mu s_L) \sum_l\,
    (\bar\nu_{l\hspace{0.05mm} L}\gamma_\mu\nu_{l\hspace{0.05mm} L}) 
    + \tilde C_{\nu}\,(\bar d_R\gamma^\mu s_R) \sum_l\,
    (\bar\nu_{l\hspace{0.05mm} L}\gamma_\mu\nu_{l\hspace{0.05mm} L}) \,,
\eeq
where the new-physics contributions arising in the RS model are given by
\beq\label{eq:Cnus}
   C_{\nu}^{{\rm RS}} 
   = \frac{2\pi\alpha}{\sws\cws\,\Mkk^2} (\Sigma_D)_{12} \,,
    \qquad
   \tilde C_{\nu}^{{\rm RS}} 
   = - \frac{2\pi\alpha}{\sws\cws\,\Mkk^2} (\Sigma_d')_{12} \,.
\eeq
Higgs exchange gives no contribution here due to the tininess of
neutrino masses.

\boldmath \subsubsection{Effective Hamiltonian for $b \to s l^+ l^-$ }
\unboldmath

The effective Hamiltonian for $b\to s l^+ l^-$ transitions contains
the following operators in addition to those entering ${\cal H}_{\rm
  eff}^{\Delta B=1}$ as given in the previous section:
\beq
\begin{split}
   {\cal H}_{\rm eff}^{b\to s l^+ l^-} 
   &= C_{l1}\,(\bar s_L\gamma^\mu b_L) \sum_l\,
    (\bar l_L\gamma_\mu l_L)
    + C_{l2}\,(\bar s_L\gamma^\mu b_L) \sum_l\,
    (\bar l_R\gamma_\mu l_R) \\
   &\quad\mbox{}+ \tilde C_{l1}\,(\bar s_R\gamma^\mu b_R) \sum_l\,
    (\bar l_R\gamma_\mu l_R)
    + \tilde C_{l2}\,(\bar s_R\gamma^\mu b_R) \sum_l\,
    (\bar l_L\gamma_\mu l_L) \\
   &\quad\mbox{}+ C_{l3}\,(\bar s_L b_R) \sum_l\,(\bar l l)
    + \tilde C_{l3}\,(\bar s_R b_L) \sum_l\,(\bar l l) \,, 
\end{split}
\eeq
where in the RS model
\begin{align}
  C_{l1}^{{\rm RS}} &= - \frac{4\pi\alpha}{3\Mkk^2}\,(\Delta_D')_{23}
  - \frac{2\pi\alpha\,(1-2\sws)}{\sws\cws\,\Mkk^2}\,(\Sigma_D)_{23}
  \,, \nonumber \\
  C_{l2}^{{\rm RS}} &= - \frac{4\pi\alpha}{3\Mkk^2}\,(\Delta_D')_{23}
  + \frac{4\pi\alpha}{\cws\,\Mkk^2}\,(\Sigma_D)_{23} \,, \nonumber \\
  \tilde C_{l1}^{{\rm RS}} &= - \frac{4\pi\alpha}{3\Mkk^2}\,
  (\Delta_d')_{23} - \frac{4\pi\alpha}{\cws\,\Mkk^2}\,
  (\Sigma_d')_{23} \,,   \label{eq:Cl} \nonumber \\
  \tilde C_{l2}^{{\rm RS}} &= - \frac{4\pi\alpha}{3\Mkk^2}\,
  (\Delta_d')_{23} + \frac{2\pi\alpha\,(1-2\sws)}{\sws\cws\,\Mkk^2}\,
  (\Sigma_d')_{23} \,, \nonumber \\
  C_{l3}^{{\rm RS}} &= - \frac{2m_l}{m_h^2 v} \left[
    \frac{m_s}{v}\,(\delta_{d})_{23} +
    \frac{m_b}{v}\,(\delta_{D})_{23} + (\Delta \tilde g^d_h)_{23}
  \right] , \nonumber \\
  \tilde C_{l3}^{{\rm RS}} & = - \frac{2m_l}{m_h^2 v} \left[
    \frac{m_b}{v}\,(\delta_{d})_{23} +
    \frac{m_s}{v}\,(\delta_{D})_{23} + (\Delta \tilde g^d_h)_{23}
  \right] , 
\end{align}
and $m_h$ denotes the mass of the Higgs boson. Even for a
Higgs-boson mass as low as $115 \, {\rm GeV}$ the corrections to $b\to
s l^+ l^-$ arising from $C_{l3}^{\rm RS}$ and $\tilde C_{l3}^{\rm RS}$
turn out to be negligibly small since they are either of order
$\ord(m_l m_b/\Mkk^2)$ or $\ord(v^4/\Mkk^4)$. Consequently we will
ignore them in our numerical analysis.

Notice that electromagnetic dipole operators enter the effective
Hamiltonian for $b \to s l^+ l^-$ first at the one-loop level and thus
are formally subleading with respect to the contributions from
semileptonic operators. Whether this formal suppression translates
into a numerical one can only be seen by calculating the complete
one-loop matching corrections to the Wilson coefficients of the
electromagnetic dipole operators in the RS model. Such a computation
seems worthwhile but is beyond the scope of this work. The possibility
that the inclusion of such one-loop effects could have a
non-negligible impact on our results for the various $b \to s l^+ l^-$
observables should however be kept in mind.

\boldmath \subsubsection{Effective Hamiltonian for $b\to u l\bar\nu$}
\unboldmath

The effective Hamiltonian inducing $b \to u l \bar \nu$ transitions
reads
\beq
   {\cal H}_{\rm eff}^{b\to u l\bar\nu} 
   = C_l\,(\bar u_L\gamma^\mu b_L) \sum_l\,
    (\bar \nu_{l\hspace{0.05mm} L}\gamma_\mu l_L) 
    + \tilde{C}_l\,(\bar u_R \gamma^\mu b_R) \sum_l\, 
    (\bar \nu_{l\hspace{0.05mm} L} \gamma_\mu l_L) 
    + {\rm h.c.} \,,
\eeq
where the contributions arising in the RS model are given by
\beq \label{eq:CClRS}
   C_l^{\rm RS} = \frac{2\pi\alpha}{\sws\cws\,m_Z^2} 
    ({\cal V}_L)_{13} \,, \qquad 
   \tilde C_l^{\rm RS} = \frac{2\pi\alpha}{\sws\cws\,m_Z^2}
    ({\cal V}_R)_{13} \,.
\eeq
For simplicity, we have dropped contributions stemming from
Higgs-boson exchange, which are strongly chirally suppressed by the
masses of the leptons.

\boldmath \subsection{Important Formulas for $K \to \pi \nu \bar
  \nu$, $K_L \to \mu^+ \mu^-$, and $K_L \to \pi^0 l^+ l^-$}
\unboldmath

Below we gather the formulas that we will use in our numerical
analysis of rare decays involving kaons. We begin with the ``golden
modes'' $K \to \pi \nu \bar \nu$ and then discuss the theoretically
less clean $K_L \to \mu^+ \mu^-$ and $K_L \to \pi^0 l^+ l^-$ channels,
emphasizing that the latter modes can add useful information on the
chiral nature of the flavor structure of possible non-standard
interactions.

After summation over the three neutrino flavors, the branching ratios
for the $K \to \pi \nu \bar \nu$ modes can be written as
\beq \label{eq:BRKpivv}
\begin{split}
  {\cal B} (K_L \to \pi^0 \nu \bar \nu) & = \kappa_L \hspace{0.5mm}
  \big ( {\rm Im} \hspace{0.5mm} X  \big )^2 \,, \\[2mm]
  {\cal B} (K^+ \to \pi^+ \nu \bar \nu (\gamma)) & = \kappa_+
  \hspace{0.5mm} (1 + \Delta_{\rm EM} ) \hspace{0.5mm} \big | X \big
  |^2 \,,
\end{split}
\eeq
where $\kappa_L = (2.231 \pm 0.013) \cdot 10^{-10} \hspace{0.5mm}
(\lambda/0.225)^8$ and $\kappa_+ = (0.5173 \pm 0.0025) \cdot 10^{-10}
\hspace{0.5mm} (\lambda/0.225)^8 $ capture isospin-breaking
corrections in relating $K \to \pi \nu \bar \nu$ to $K \to \pi e \nu$,
while the factor $\Delta_{\rm EM} = -0.003$ encodes long-distance QED
corrections affecting the charged mode \cite{Mescia:2007kn}.

The SM and RS contributions entering the coefficient $X \equiv X_{\rm
  SM} + X_{\rm RS}$ take the form
\beq \label{eq:XSMRS}
X_{\rm SM} = \frac{\lambda_t^{(ds)}}{\lambda^5} \hspace{0.5mm} X_t +
\frac{{\rm Re} \hspace{0.5mm} \lambda_c^{(ds)}}{\lambda}
\hspace{0.5mm} P_{c,u} \,, \qquad X_{\rm RS} =\ \frac{\swq \cws
  m_Z^2}{\alpha^2 \lambda^5} \left (C_\nu^{\rm RS} + \tilde C_\nu^{\rm
    RS} \right) ,
\eeq
with $\lambda \equiv |V_{us}|$. The top-quark contribution
\cite{Misiak:1999yg, Buchalla:1998ba} is $X_t = 1.464 \pm 0.041$, and
the parameter $P_{c,u} = (0.41 \pm 0.04) \hspace{0.5mm}
(0.225/\lambda)^4$ includes dimension-six and -eight charm-quark
effects and genuine long-distance contributions due to up-quark loops
\cite{Isidori:2005xm,Buras:2005gr,Buras:2006gb,Brod:2008ss}.\footnote{The
  parameter $P_{c,u}$ is affected by new-physics contributions of
  order $m_{Z,W}^2/\Mkk^2$ to the neutral- and charged-current
  interactions present in the RS scenario. Compared to the SM
  contribution these corrections are negligible. The same applies to
  the charm-quark effects appearing in $K_L \to \mu^+ \mu^-$.}

Adding individual errors in quadrature, we find the following SM
predictions for the two $K \to \pi \nu \bar \nu$ branching fractions:
\beq \label{eq:KvvSM}
\begin{split} 
  {\cal B} (K_L \to \pi^0 \nu \bar \nu)_{\rm SM} & = (2.7 \pm 0.4)
  \cdot 10^{-11} \,, \\[2mm]
  {\cal B} (K^+ \to \pi^+ \nu \bar \nu (\gamma))_{\rm SM} & = (8.3 \pm
  0.9) \cdot 10^{-11} \,.
\end{split}
\eeq
The quoted errors are dominated by the uncertainties due to the CKM
input. In view of the expected improvement in the extraction of the
mixing angles, precise measurements of the $K \to \pi \nu \bar \nu$
branching ratios will provide a unique test of the flavor sector of a
variety of models of new physics, in particular of those where the
strong Cabibbo suppression of the SM amplitude is not present.

The branching ratio of the $K_L \to \mu^+ \mu^-$ decay can be
expressed as \cite{Mescia:2006jd}
\beq
{\cal B} (K_L \to \mu^+ \mu^-) = \left (6.7 + \big [ 1.1
  \hspace{0.5mm} {\rm Re} \hspace{0.5mm} Y_A' + y_c \pm
  y_{\gamma\gamma} \big ]^2 + \big [0.08 \, {\rm Im} \hspace{0.5mm}
  Y_S' \big ]^2 \right ) \cdot 10^{-9} \,,
\eeq
where $y_c = (-0.20 \pm 0.03)$ and $y_{\gamma\gamma} = 0.4 \pm 0.5$
encode the charm-quark contribution \cite{Gorbahn:2006bm} and
two-photon correction \cite{Isidori:2003ts}, respectively. The sign
of the latter contribution depends on the sign of the $K_L \to
\gamma\gamma$ amplitude, which itself depends on the sign of an
unknown low-energy constant. Theoretical arguments suggest that the
sign of the $K_L \to \gamma\gamma$ amplitude is positive
\cite{Gerard:2005yk}. Better measurements of $K_S \to \pi^0 \gamma
\gamma$ and $K^+ \to \pi^+ \gamma \gamma$ could settle this issue. The
error on $y_{\gamma\gamma}$ reflects only the uncertainty on the
dispersive part of the two-photon amplitude, which at present is the
dominant individual source of error.

The coefficients $Y_{A,S}'$ are given by 
\beq
   Y_A' = y_A + \frac{\sws\cws m_Z^2}{2\pi\alpha^2\lambda_t^{(ds)}}
    \left( C_{l1}^{\rm RS} - C_{l2}^{\rm RS} 
    + \tilde C_{l1}^{\rm RS} - \tilde C_{l2}^{\rm RS} \right) ,
   \qquad 
  Y_S' = \frac{\swq\cwq m_Z^4}{\alpha^2 m_l m_s}
   \left( C_{l3}^{\rm RS} - \tilde C_{l3}^{\rm RS} \right) ,
\eeq
where $y_A=(-0.68\pm 0.03)$ is the SM contribution to the Wilson
coefficient of the semileptonic axial-vector operator
\cite{Buras:1994qa}, and the coefficients $C_{l1-3}^{\rm RS}$ and
$\tilde C_{l1-3}^{\rm RS}$ are understood to contain the 12 entries of
the relevant mixing matrices. The coefficient $Y_S'$ describes the
correction due to tree-level Higgs-boson exchange. This correction is
scale dependent but numerically insignificant, so that in practice one
can neglect its RG evolution. In our numerical analysis we will
set the corrections arising from Higgs exchange to zero.

In the SM one finds for the $K_L \to \mu^+ \mu^-$ branching ratio 
\beq \label{eq:KmmSM}
{\cal B} (K_L \to \mu^+ \mu^-)_{\rm SM} = \left \{ 7.0 \pm 0.6 , 8.5
  \pm 1.4 \right \} \cdot 10^{-9} \,,
\eeq
in the case of positive (negative) sign of the two-photon
amplitude. The shown uncertainties have been obtained by adding the
individual errors in quadrature. We will see later that, being
measured precisely, the $K_L \to \mu^+ \mu^-$ decay can lead to
interesting constraints in the RS model.

The branching ratios of the decays $K_L \to \pi^0 l^+ l^-$ are
obtained from
\beq \label{eq:KLpi0ll}
{\cal B} (K_L \to \pi^0 l^+ l^-) = \left ( C_{\rm dir}^l \pm C_{\rm
    int}^l \hspace{0.5mm} |a_S|+ C_{\rm mix}^l \hspace{0.5mm} |a_S|^2
  + C_{\gamma\gamma}^l + C_S^l \right ) \cdot 10^{-12} \,,
\eeq
where the chiral-perturbation-theory counterterm $|a_S| = 1.20 \pm
0.20$ has been extracted from the measurements of the $K_S \to \pi^0
l^+ l^-$ branching fraction \cite{Batley:2003mu,Batley:2004wg}. The
coefficients $C_{{\rm dir}, {\rm int}, {\rm mix}, \gamma\gamma, S}^l$
read \cite{Mescia:2006jd}
\beq 
\begin{split}
  C_{\rm dir}^e & = (4.62 \pm 0.24) \left[ ({\rm Im} \hspace{0.5mm}
    Y_A)^2 + ({\rm Im} \hspace{0.5mm} Y_V)^2 \right] ,  \\
  C_{\rm int}^e & = (11.3 \pm 0.3) \, {\rm Im} \hspace{0.5mm} Y_V \,,
   \\
  C_{\rm mix}^e & = 14.5 \pm 0.5 \,,  \\
  C_{\gamma\gamma}^e & \approx C_S^e \approx 0 \,,  \\
  C_{\rm dir}^\mu & = (1.09 \pm 0.05) \left[ 2.32 \hspace{0.5mm} ({\rm
      Im} \hspace{0.5mm} Y_A)^2 + ({\rm Im} \hspace{0.5mm} Y_V)^2
  \right] ,  \\
  C_{\rm int}^\mu & = (2.63 \pm 0.06) \, {\rm Im} \hspace{0.5mm} Y_V
  \,,  \\
  C_{\rm mix}^\mu & = 3.36 \pm 0.20 \,,  \\
  C_{\gamma\gamma}^\mu & = 5.2 \pm 1.6 \,,  \\
  C_S^\mu & = (0.04 \pm 0.01) \hspace{0.75mm} {\rm Re} \hspace{0.5mm}
  Y_S + 0.0041 \hspace{0.5mm} ({\rm Re} \hspace{0.5mm} Y_S)^2 \,.
\end{split}
\eeq
They describe the short-distance direct CP-violating contribution
$C_{\rm dir}^l$, the long-distance indirect CP-violating term $C_{\rm
  mix}^l$ that can be determined from the experimental data on $K_S
\to \pi^0 l^+ l^-$, and a long-distance CP-conserving correction
$C_{\gamma\gamma}^l$ that can be extracted from a measurement of $K_L
\to \pi^0 \gamma \gamma$. The direct and indirect CP-violating
amplitudes interfere, leading to the term $C_{\rm int}^l$. The latest
theoretical analyses \cite{Buchalla:2003sj, Friot:2004yr} point
towards a constructive interference, corresponding to the plus sign in
(\ref{eq:KLpi0ll}). The correction $C_S^l$ encodes additional
contributions due to scalar operators which we have neglected in
our numerical analysis. Both $C_{\gamma\gamma}^l$ and $C_S^l$ are
helicity suppressed and thus have in general a negligible effect on
the $K_L \to \pi^0 e^+ e^-$ branching ratio.

The coefficients $Y_{A,V,S}$ take the form 
\beq
\begin{split} \label{YVAdef}
   Y_A &= y_A + \frac{\sws\cws m_Z^2}{2\pi\alpha^2\lambda_t^{(ds)}} 
    \left( C_{l1}^{\rm RS} - C_{l2}^{\rm RS} - \tilde C_{l1}^{\rm RS} 
    + \tilde C_{l2}^{\rm RS} \right) , \\
   Y_V &= y_V - \frac{\sws\cws m_Z^2}{2\pi\alpha^2\lambda_t^{(ds)}} 
    \left( C_{l1}^{\rm RS} + C_{l2}^{\rm RS} + \tilde C_{l1}^{\rm RS} 
    + \tilde C_{l2}^{\rm RS} \right) , \\
   Y_S &= \frac{\swq\cwq m_Z^4}{\alpha^2 m_l m_s} 
    \left( C_{l3}^{\rm RS} + \tilde C_{l3}^{\rm RS} \right) ,
\end{split}
\eeq
where $y_V=0.73\pm 0.04$ represents the SM contribution to the Wilson
coefficient of the semileptonic vector operator \cite{Buras:1994qa},
and $C_{l1-3}^{\rm RS}$ and $\tilde C_{l1-3}^{\rm RS}$ involve the 12
entries of the appropriate mixing matrices. The coefficients
$C_{l1-3}^{\rm RS}$ and $\tilde C_{l1-3}^{\rm RS}$ entering $Y_{V}$
are understood to be evaluated at a low-energy scale $\mu \approx 1 \,
{\rm GeV}$. In our numerical analysis we use leading-order RG running
to determine these coefficients from the initial conditions of
$C_{3-6,l1-3}^{\rm RS}$ and $\tilde C_{3-6,l1-3}^{\rm RS}$ given in
(\ref{eq:Cpenguin}) and (\ref{eq:Cl}). The effect of the new-physics
contributions is mainly felt in $Y_A$, as the corresponding
contributions in $Y_V$ cancel each other to a large extent and $Y_S$
is highly suppressed. In consequence, RG effects influence the
obtained results only in a minor way.

The SM predictions for the branching ratios of $K_L \to \pi^0 l^+ l^-$
are
\beq \label{eq:BRKllSM}
\begin{split}
  {\cal B} (K_L \to \pi^0 e^+ e^-)_{\rm SM} & = \left \{ 3.5 \pm 0.9,
    1.6 \pm 0.6 \right \} \cdot 10^{-11} \,, \\[2mm]
  {\cal B} (K_L \to \pi^0 \mu^+ \mu^-)_{\rm SM} & = \left \{ 1.4 \pm
    0.3, 0.9 \pm 0.2 \right \} \cdot 10^{-11} \,,
\end{split}
\eeq 
for constructive (destructive) interference. Better measurements of
the $K_S \to \pi^0 l^+ l^-$ decay rate would allow to improve the
quoted errors, which are currently dominated by the uncertainty due to
the chiral-perturbation-theory parameter $|a_S|$.

The integrated forward-backward CP asymmetry for $K_L \to \pi^0 \mu^+
\mu^-$ is given by \cite{Mescia:2006jd}
\beq \label{eq:AFBKll}
A_{\rm FB} (K_L \to \pi^0 \mu^+ \mu^-) = \frac{(1.3 \pm 0.1)
  \hspace{0.5mm} {\rm Im} \hspace{0.5mm} Y_V\mp 0.057 \hspace{0.5mm}
  |a_S| \hspace{0.5mm} {\rm Re} \hspace{0.5mm} Y_S \pm (1.7 \pm 0.2)
  \hspace{0.5mm} |a_S| }{{\cal B} (K_L \to \pi^0 \mu^+ \mu^-) } \cdot
10^{-12} \,, 
\eeq
and the corresponding SM predictions read
\beq \label{eq:AFBmmSM}
A_{\rm FB} (K_L \to \pi^0 \mu^+ \mu^-)_{\rm SM} = \left \{ 21 \pm 4 ,
  -12 \pm 4 \right \} \%
\eeq
in the case of constructive (destructive) interference. Despite the
fact that the predictions of the decays $K_L \to \pi^0 l^+ l^-$ are
not fully under theoretical control, the latter transitions represent
promising channels to study new physics. In particular, the different
impact of helicity-suppressed contributions to the muon and electron
modes makes the comparison of $K_L \to \pi^0 \mu^+ \mu^-$ and $K_L \to
\pi^0 e^+ e^-$ a powerful tool for analyzing the chiral properties of
non-standard flavor interactions.

\boldmath \subsection{Important Formulas for $B \to X_q \nu \bar
  \nu$, $B_q \to \mu^+ \mu^-$, and $B \to X_s l^+ l^-$} \unboldmath

In the following we assemble the formulas that we will employ in the
numerical analysis of rare semileptonic and purely leptonic $B$
decays. We thereby focus on the theoretically cleanest observables
arising from the quark-level transitions $b \to q \nu \bar \nu$ and $b
\to q l^+ l^-$.

Summing over the three neutrino flavors and neglecting effects
proportional to small quark masses, the branching ratio of $B \to X_q
\nu \bar \nu$ can be written as
\beq
   {\cal B}(B\to X_q\nu\bar\nu) 
   = {\cal B}(B\to X_c l\bar\nu)_{\rm exp}\,
    \frac{3\alpha^2}{4\pi^2\swq C}\,
    \frac{\big|\lambda_t^{(qb)}\big|^2}{\lambda_c^{(bb)}} 
    \left( |X_L|^2 + |X_R|^2 \right) ,
\eeq
where ${\cal B} (B \to X_c l \bar \nu)_{\rm exp} = 0.1064 \pm 0.0011$
denotes the experimental value of the semileptonic branching ratio
\cite{Barberio:2008fa}, and the appropriate value of the
electromagnetic coupling is $\alpha(m_Z) = 1/127.9$. In order to
minimize the theoretical uncertainties, we have normalized the $B \to
X_q \nu \bar \nu$ decay rate to the charmless semileptonic rate
$\Gamma (B \to X_u l \bar \nu)$ following \cite{Haisch:2007ia}. Like
in the case of the $B \to X_s \gamma$ decay \cite{Gambino:2001ew},
this modification is offset by the phase-space ratio $C$ defined by
\beq C = \frac{\lambda_u^{(bb)}}{\lambda_c^{(bb)}} \, \frac{\Gamma (B
  \to X_c l \bar \nu)}{\Gamma (B \to X_u l \bar \nu)} \,.  \eeq In our
numerical analyses we employ $C = 0.546^{+0.023}_{-0.033}$, which has
recently been extracted from a global fit to the moments of the
inclusive $B \to X_c l \bar \nu$ and $B \to X_s \gamma$ decay
distributions \cite{Gambino:2008fj}.

The coefficients $X_{L,R}$ take the form 
\beq
   X_L = X_t + \frac{\swq\cws m_Z^2}{\alpha^2\lambda_t^{(qb)}}\,
    C_\nu^{\rm RS} \,, \qquad 
   X_R = \frac{\swq\cws m_Z^2}{\alpha^2\lambda_t^{(qb)}}\,
    \tilde C_\nu^{\rm RS} \,,
\eeq
and the coefficients $C_\nu^{\rm RS}$ and $\tilde C_\nu^{\rm RS}$
contain the 13 or 23 elements of the relevant mixing matrices.

In the SM one finds for the branching ratios of the $B \to X_q \nu
\bar \nu$ decay modes
\beq \label{eq:BXqvvSM}
\begin{split}
  {\cal B} (B \to X_d \nu \bar \nu)_{\rm SM} & = (1.6 \pm 0.1) \cdot
  10^{-6} \,, \\[2mm]
  {\cal B} (B \to X_s \nu \bar \nu)_{\rm SM} & = (3.5 \pm 0.2) \cdot
  10^{-5} \,,
\end{split} 
\eeq
where the error on ${\cal B}(B \to X_d \nu \bar \nu)$ is in equal
shares due to the uncertainty on $C$ and the CKM elements, while the
uncertainty of ${\cal B}(B \to X_s \nu \bar \nu)$ arises almost solely
from the one entering the semileptonic phase-space ratio. In view of
the small theoretical errors, measurements of $B \to X_d \nu \bar \nu$
would constitute unique probes of the flavor sector of the underlying
theory. Besides the inclusive $b \to q \nu \bar \nu$ transitions, also
the $B \to K^{(\ast)} \nu \bar \nu$ channels provide opportunities to
search for new physics. Since the pattern of deviations observed in
these exclusive modes follows the one in $B \to X_s \nu \bar \nu$
decays, our discussion of non-standard effects in $B \to X_q \nu \bar
\nu$ essentially also applies to the modes $B \to K^{(\ast)} \nu \bar
\nu$.

The branching ratios for the $B_q \to \mu^+ \mu^-$ decays can be
expressed as
\beq \label{eq:BRBqmm}
\begin{split}
  {\cal B} (B_q \to \mu^+ \mu^-) & = \frac{G_F^2 \hspace{0.5mm}
    \alpha^2 \hspace{0.5mm} m_{B_q}^3 f_{B_q}^2 \tau_{B_q}}{64 \pi^3
    \swq} \, \big |\lambda_t^{(qb)} \big |^2 \, \sqrt{1 - \frac{4
      m_\mu^2}{m_{B_q}^2}} \\ & \phantom{xx} \times \left ( \frac{4
      m_\mu^2}{m_{B_q}^2} \left | C_A - C_A' \right |^2 + m_{B_q}^2
    \left [ 1 - \frac{4 m_\mu^2}{m_{B_q}^2} \right ] \left | \frac{m_b
        \hspace{0.5mm} C_S - m_q \hspace{0.5mm} C_S'}{m_b + m_q}
    \right |^2 \right) ,
\end{split}
\eeq
where $m_{B_q}$, $f_{B_q}$, and $\tau_{B_q}$ are the mass, decay
constant, and lifetime of the $B_q$ meson. The electromagnetic
coupling $\alpha$ entering the branching ratios should be evaluated at
$m_Z$.

The expressions for the coefficients $C_{A,S}$ and $C_{A,S}'$ read 
\beq\label{eq:CACAp}
\begin{aligned}
   C_A &= c_A - \frac{\swq\cws m_Z^2}{\alpha^2\lambda_t^{(qb)}} 
    \Big( C_{l1}^{\rm RS} - C_{l2}^{\rm RS} \Big) \,, 
   &\qquad 
   C_A' &= \frac{\swq\cws m_Z^2}{\alpha^2\lambda_t^{(qb)}}
    \Big( \tilde C_{l1}^{\rm RS} - \tilde C_{l2}^{\rm RS} \Big) \,, \\
   C_S &= \frac{2\swq\cws m_Z^2}{\alpha^2 m_b\lambda_t^{(qb)}}\,
    C_{l3}^{\rm RS} \,, 
   &\qquad  
   C_S' &= \frac{2\swq\cws m_Z^2}{\alpha^2 m_q\lambda_t^{(qb)}}\, 
    \tilde C_{l3}^{\rm RS} \,,
\end{aligned}
\eeq
where $c_A = 0.96 \pm 0.02$ denotes the SM contribution to the Wilson
coefficient of the axial-vector current
\cite{Misiak:1999yg,Buchalla:1998ba}, and the coefficients
$C_{l1-3}^{\rm RS}$ and $\tilde C_{l1-3}^{\rm RS}$ contain the 13 or
23 elements of the mixing matrices in the case of $B_d \to \mu^+
\mu^-$ and $B_s \to \mu^+ \mu^-$, respectively. While $C_A$ and $C_A'$
are scale independent, the coefficients $C_S$ and $C_S'$ have a
non-trivial RG evolution. Since the scalar interactions arising from
tree-level Higgs-boson exchange are numerically insignificant, we have
neglected them in our numerical analysis.

The SM branching ratios of the $B_q \to \mu^+ \mu^-$ decay channels
evaluate to 
\beq \label{eq:BqmmSM}
\begin{split}
  {\cal B} (B_d \to \mu^+ \mu^-)_{\rm SM} & = (1.2 \pm 0.1) \cdot
  10^{-10} \,, \\[2mm]
  {\cal B} (B_s \to \mu^+ \mu^-)_{\rm SM} & = (3.8 \pm 0.4) \cdot
  10^{-9} \,.
\end{split} 
\eeq 
These predictions are obtained by normalizing the decay rates to the
well-measured meson mass differences $(\Delta m_q)_{\rm exp}$. This
eliminates the dependence on CKM parameters and the bulk of the
hadronic uncertainties by trading the decay constants for less
uncertain hadronic parameters \cite{Buras:2003td}. The dominant
source of error is nevertheless still provided by the hadronic
input. Leptonic $B$-meson decays belong to the channels that can be
studied by three out of the four major CERN Large Hadron Collider
(LHC) experiments, namely ATLAS, CMS, and LHCb. These experiments will
probe the branching fraction of $B_s \to \mu^+ \mu^-$ down to its SM
value and might reveal a signal of new physics well ahead of the
direct searches.

Normalizing the $B \to X_s l^+ l^-$ decay width to $\Gamma (B \to X_u
l \bar \nu)$ and neglecting effects proportional to the strange-quark
mass, the branching ratio of the rare semileptonic decay integrated
over the range $s \in [s_0, s_1]$ of the rescaled invariant dilepton
mass $s \equiv q^2/m_{b, {\rm pole}}^2 \in [0,1]$ can be written as
\begin{eqnarray}
{\cal B} (B \to X_s l^+ l^-)^{s \in [s_0, s_1]} = \frac{{\cal B} (B
  \to X_c e \bar \nu)_{\rm exp} }{C} \,
\frac{\lambda_u^{(bb)}}{\lambda_c^{(bb)}} \int_{s_0}^{s_1} \! ds\,
\frac{1}{\Gamma (B \to X_u e \bar \nu)}\, \frac{d\Gamma (B \to X_s l^+
  l^-)}{ds} \,, \hspace{6mm}
\end{eqnarray}
where   
\begin{eqnarray} 
\begin{aligned}
  \frac{d\Gamma (B \to X_s l^+ l^-)}{ds} & = \frac{G_F^2
    \hspace{0.5mm} m_{b,{\rm pole}}^5 \hspace{0.5mm} \big
    |\lambda_t^{(sb)} \big |^2}{48 \pi^3} \left ( \frac{\alpha}{4 \pi}
  \right )^2 \, (1 - s)^2 \Bigg [ \left ( 4 + \frac{8}{s} \right )
  \left | C_{7, {\rm BR}}^{\gamma \, \rm eff} (s) \right |^2 \\
  & \phantom{xx} + (1 + 2 s) \left ( \left | C_{9, {\rm BR}}^{l \, \rm
        eff} (s) \right |^2 + \left | C_{10, {\rm BR}}^{l \, \rm eff}
      (s) \right |^2 + \left | \tilde C_{9, {\rm BR}}^{l \, \rm eff}
      (s) \right |^2 + \left | \tilde C_{10, {\rm BR}}^{l \, \rm eff}
      (s) \right |^2\right ) \hspace{6mm} \\
  & \phantom{xx} + 12 \, {\rm Re} \hspace{0.5mm} \Big ( C_{7, {\rm
      BR}}^{\gamma \, \rm eff} (s) \hspace{0.5mm} C_{9, {\rm BR}}^{l
    \, {\rm eff}} (s)^\ast \Big ) + \frac{d\Gamma^{\rm brems} (B \to
    X_s l^+ l^-)}{ds} \, \Bigg ] \,.
\end{aligned}  
\end{eqnarray}
Here $\alpha$ is taken to be renormalized at the scale $m_b$, and
$C_{7,{\rm BR}}^{\gamma\,\rm eff}(s)$, $C_{9,10,{\rm BR}}^{l\,\rm
  eff}(s)$, and $\tilde C_{9,10,{\rm BR}}^{l\,\rm eff}(s)$ denote the
low-energy Wilson coefficients of the electromagnetic dipole, vector,
and axial-vector semileptonic operators and their counterparts
obtained by exchanging left- with right-handed quark fields. The
effective Wilson coefficients $C_{9,10,{\rm BR}}^{l\,\rm eff}(s)$
include all known real and virtual QCD and electroweak SM corrections
\cite{Bobeth:1999mk, Asatrian:2001de, Asatryan:2001zw,
  Gambino:2003zm, Bobeth:2003at, Gorbahn:2004my, Huber:2005ig}, while
the last term encodes the finite bremsstrahlung corrections calculated
in \cite{Asatryan:2002iy}. At leading order the coefficients
$C_{7,{\rm BR}}^{\gamma\,\rm eff}(s)$ and $C_{9,10,{\rm BR}}^{l\,\rm
  eff}(s)$ are identical to $C_7^\gamma$ and $C_{9,10}^l$. Beyond
leading order the definitions of the effective Wilson coefficients are
affected by the presence of both the QCD and the electroweak penguin
operators $Q_{3-6}$ and $Q_{7-10}$ as well as their chirality-flipped
analogs. In our analyses of the $b\to s l^+ l^-$ observables we
include all new-physics corrections to $C_{9,{\rm BR}}^{l\,\rm
  eff}(s)$ and $\tilde C_{9,{\rm BR}}^{l\,\rm eff}(s)$ stemming from the
one-loop matrix elements of the operators $Q_{3-10}$ and $\tilde
Q_{3-10}$. However, the inclusion of these effects has only a minor
impact on the results.

At the matching scale $\mu_{\rm KK} = \ord (\Mkk)$ the new
contributions to the Wilson coefficients $C_{9,10}^l$ and
$\tilde C_{9,10}^l$ arising in the RS scenario are
\beq\label{eq:CCt910l}
\begin{split}
   C_{9}^{l\,{\rm RS}} 
   &= \frac{\alpha_s\sws\cws m_Z^2}{4\pi\alpha^2\lambda_t^{(sb)}}\,
    \Big( C_{l1}^{\rm RS} + C_{l2}^{\rm RS} \Big) \,, \qquad 
   C_{10}^{l\,{\rm RS}} 
    = - \frac{\alpha_s\sws\cws m_Z^2}{4\pi\alpha^2\lambda_t^{(sb)}}\,
    \Big( C_{l1}^{\rm RS} - C_{l2}^{\rm RS} \Big) \,, \\
   \tilde C_{9}^{l\,{\rm RS}} 
   &= \frac{\alpha_s\sws\cws m_Z^2}{4\pi\alpha^2\lambda_t^{(sb)}}\,
    \Big( \tilde C_{l1}^{\rm RS} + \tilde C_{l2}^{\rm RS} \Big) \,,
    \qquad 
   \tilde C_{10}^{l\,{\rm RS}} 
    = \frac{\alpha_s\sws\cws m_Z^2}{4\pi\alpha^2\lambda_t^{(sb)}}\,
    \Big( \tilde C_{l1}^{\rm RS} - \tilde C_{l2}^{\rm RS} \Big) \,,
\end{split}
\eeq
where the coefficients $C_{l1-3}^{\rm RS}$ and $\tilde C_{l1-3}^{\rm
  RS}$ depend on the 23 entries of the appropriate mixing
matrices. Notice that $C_7^\gamma$ does not receive corrections at
tree level. The Wilson coefficients $C_{9}^l$ and $\tilde C_{9}^l$
depend of the renormalization scale, while $C_{10}^l$ and $\tilde
C_{10}^l$ are scale independent. In our analysis we include the
running of $C_{9}^l$ and $\tilde C_{9}^l$ at leading order in the
strong coupling constant. Although RG effects change the values of
$C_{9}^l$ and $\tilde C_{9}^l$ drastically, the inclusion of the
running has only a slight impact on the $b \to s l^+ l^-$ observables,
because the contributions due to the semileptonic vector operators are
generically much smaller than the corresponding axial-vector
corrections.

In order to allow for an understanding of the importance of the
different new-physics contributions to $B \to X_s l^+ l^-$ in the RS
model, we derive an approximate formula for the branching ratio
integrated over the low-$q^2$ region with $q^2 \in [1, 6] \, {\rm
  GeV}^2$. Neglecting new-physics effects arising from the one-loop
matrix elements of $Q_{3-10}$ and $\tilde Q_{3-10}$, we find in terms
of the Wilson coefficients evaluated at the low-energy scale $\mu =
m_b$ the expression
\begin{equation}
\begin{aligned}
  {\cal B} (B \to X_s l^+ l^-)^{q^2 \in [1, 6] \, {\rm GeV}^2} & =
  1.68 \, \bigg [ \, 1 + 0.17 \, {\rm Re} \hspace{0.5mm} C_9^{l \,
    {\rm RS}} - 0.29 \, {\rm Re} \hspace{0.5mm}
  C_{10}^{l \, {\rm RS}} \\
  & \hspace{14.75mm} + 0.03 \left ( \big | C_9^{l \, {\rm RS}} \big
    |^2 + \big | C_{10}^{l \, {\rm RS}} \big |^2 + \big | \tilde
    C_9^{l \, {\rm RS}} \big |^2 + \big | \tilde C_{10}^{l \, {\rm
        RS}} \big |^2\right ) \bigg ] \cdot 10^{-6} \,.
\end{aligned} 
\end{equation}
The corresponding SM prediction amounts to 
\beq 
\begin{split} \label{eq:BRBXsllSM}
  {\cal B} (B \to X_s l^+ l^-)_{\rm SM}^{q^2 \in [1, 6] \, {\rm
      GeV}^2} = (1.7 \pm 0.2 ) \cdot 10^{-6} \,,
\end{split} 
\eeq
when individual uncertainties are added in quadrature. At present the
dominant source of error is associated to non-perturbative effects
related to the cut on the invariant hadronic mass $M_{X_s}$
\cite{Lee:2005pk, Lee:2005pwa, Lee:2008xc} and to unknown enhanced
non-local power corrections of $\ord (\alpha_s \Lambda_{\rm QCD}/m_b)$
\cite{Lee:2006wn}. We assume that in combination these
non-perturbative effects introduce an uncertainty of $5\%$.  The
sensitivity of the SM prediction on the Wilson coefficients
$C_{9,10}^l$ and $\tilde C_{9,10}^l$ in the high-$q^2$ region, $q^2 >
14 \, {\rm GeV}^2$, is essentially the same as the one for low $q^2$,
while at high $q^2$ theory uncertainties are considerably larger than
in the low-$q^2$ range. We therefore restrict our attention to the
low-$q^2$ region as far as $B \to X_s l^+ l^-$ is concerned. We
emphasize, however, that all of our findings also apply in a similar
fashion to the high-$q^2$ range.

We also consider the so-called normalized $B \to X_s l^+ l^-$
forward-backward asymmetry defined as
\beq
\bar{A}_{\rm FB} (B \to X_s l^+ l^-) = \left ( \frac{d \Gamma(B \to
    X_s l^+ l^-)}{d s} \right )^{-1} \int_{-1}^{1} \! dz \, \sgn (z)
\, \frac{d^2 \Gamma(B \to X_s l^+ l^-)}{d s \hspace{0.5mm} d z} \,,
\eeq
where for a vanishing strange-quark mass one has 
\beq 
\begin{split}
  \int_{-1}^{1} \! dz \, \sgn (z) \, \frac{d^2 \Gamma(B \to X_s l^+
    l^-)}{d s \hspace{0.5mm} d z} & = \frac{G_F^2 \hspace{0.5mm}
    m_{b,{\rm pole}}^5 \hspace{0.5mm} \big |\lambda_t^{(sb)} \big
    |^2}{48 \pi^3} \left ( \frac{\alpha}{4 \pi} \right )^2
  \, (1 - s)^2 \\
  & \hspace{-3cm} \times \Bigg [ -3 \hspace{0.5mm} s \, {\rm Re}
  \hspace{0.5mm} \Big ( C_{9, {\rm FB}}^{l \, \rm eff} (s)
  \hspace{0.5mm} C_{10, {\rm FB}}^{l \, \rm eff} (s)^\ast - \tilde
  C_{9, {\rm FB}}^{l \, \rm eff} (s) \hspace{0.5mm}
  \tilde C_{10, {\rm FB}}^{l \, \rm eff} (s)^\ast \Big ) \\
  & \hspace{-2.6cm} \quad -6 \, {\rm Re} \hspace{0.5mm} \Big ( C_{7,
    {\rm FB}}^{\gamma \, \rm eff} (s) \hspace{0.5mm} C_{10, {\rm
      FB}}^{l \, \rm eff} (s)^\ast \Big ) + A_{\rm FB}^{\rm brems} (B
  \to X_s l^+ l^-) \, \Bigg ] \,.
\end{split}
\eeq 
The variable $z \equiv \cos \theta$ denotes the cosine of the angle
between the directions of the momenta of the decaying $B$ meson and
the positively-charged lepton measured in the dilepton center-of-mass
frame. We include all known effects \cite{Bobeth:1999mk,
  Asatryan:2001zw, Gambino:2003zm, Bobeth:2003at, Gorbahn:2004my,
  Huber:2005ig, Ghinculov:2002pe, Asatrian:2002va, Asatrian:2003yk}
associated with $C_{9, {\rm FB}}^{l \, \rm eff} (s)$, $\tilde C_{9,
  {\rm FB}}^{l \, \rm eff} (s)$, and $A_{\rm FB}^{\rm brems} (B \to
X_s l^+ l^-)$ in our numerical analysis. The matching corrections to
the original Wilson coefficients $C_{9,10}^l$ and $\tilde C_{9,10}^l$
have been given in (\ref{eq:CCt910l}).

The position of the zero of the forward-backward asymmetry, $q_0^2$,
is known to be especially sensitive to non-standard contributions,
because in the SM it can be predicted without significant perturbative
and hadronic uncertainties. Assuming that $q_0^2$ does not differ
much from its SM prediction, and suppressing subleading non-standard
effects associated to the one-loop matrix elements of the operators
$Q_{3-10}$ and $\tilde Q_{3-10}$, we obtain the approximate relation
\begin{eqnarray}
  q_0^2 = 3.56 \! \left [ \frac{1 - 0.09 \, {\rm Re} \hspace{0.5mm} 
      C_9^{l \, {\rm RS}} - 0.22 \, {\rm Re} \hspace{0.5mm} 
      C_{10}^{l \, {\rm RS}} 
      + 0.02 \, {\rm Re} \left ( C_9^{l \, {\rm RS}} 
        C_{10}^{l \, {\rm RS} \ast} 
        - \tilde C_9^{l \, {\rm RS}} \tilde C_{10}^{l \, {\rm RS} \ast}
      \right )}{1 + 0.16 \, {\rm Re} \hspace{0.5mm} C_9^{l \, {\rm RS}} 
      - 0.22 \, {\rm Re} \hspace{0.5mm} C_{10}^{l \, {\rm RS}} - 0.03 
      \, {\rm Re} \left ( C_9^{l \, {\rm RS}} C_{10}^{l \, {\rm RS} \ast}
        - \tilde C_9^{l \, {\rm RS}} \tilde C_{10}^{l \, {\rm RS} \ast} 
      \right )} \right ] \! {\rm GeV}^2 , \hspace{5mm}
\end{eqnarray}
where the Wilson coefficients are understood to be evaluated at the
scale $m_b$. The SM prediction for the latter quantity reads
\beq \label{eq:q02SM}
   q_{0, {\rm SM}}^2 = (3.6 \pm 0.3) \, {\rm GeV}^2 \,,
\eeq
when individual uncertainties are added in quadrature. The bulk of the
quoted error stems from non-perturbative effects related to the
$M_{X_s}$ cut \cite{Lee:2005pk,Lee:2005pwa,Lee:2008xc} and unknown
$\ord(\alpha_s \Lambda_{\rm QCD}/m_b)$ power corrections
\cite{Lee:2006wn}. We assume again that these non-perturbative
effects combined lead to an error of $5\%$. Precision measurements of
the decay distributions of $B \to X_s l^+ l^-$, possible at a super
flavor factory, will allow to probe the nature of the weak
interactions encoded in those Wilson coefficients of the semileptonic
operators which are not accessible to $B \to X_s\gamma$.

\boldmath \subsection{Important Formulas for $B\to K^{\ast} l^+
  l^-$} \unboldmath

We focus on the decay $B \to K^\ast l^+ l^-$ as one of the
phenomenologically most relevant $b \to s l^+ l^-$ transition and
discuss the normalized forward-backward asymmetry $\bar A_{\rm
  FB}(B\to K^\ast l^+ l^-)$ and the longitudinal $K^\ast$ polarization
$F_{\rm L}(B \to K^\ast l^+ l^-)$, which are already accessible
experimentally. It is straightforward to extend the discussion to
other decay channels \cite{Bobeth:2007dw}, angular distributions
\cite{Kruger:2005ep,Lunghi:2006hc,Egede:2008uy,Altmannshofer:2008dz},
and CP asymmetries \cite{Altmannshofer:2008dz,Bobeth:2008ij}.

In the kinematic region where the dilepton invariant mass is
sufficiently above the real photon pole and below the charm threshold,
$q^2 \in [1, 6] \, {\rm GeV}^2$, the relevant transversity amplitudes
can be written within QCD factorization as
\cite{Bobeth:2008ij,Beneke:2001at, Beneke:2004dp}
\beq \label{eq:Ailow}
\begin{split}
  A_\perp^{L,R} & = \frac{\sqrt{2} N \hspace{0.25mm} (m_B^2 -
    q^2)}{m_B} \, \bigg \{ \! \left [ \left ( C_9^l + \tilde C_9^l
    \right ) \mp \left ( C_{10}^l + \tilde C_{10}^l \right ) \right ]
  \xi_\perp + \frac{2 \hspace{0.25mm} m_b \hspace{0.25mm} m_B}{q^2}
  \left ( {\cal T}_\perp +
    \tilde {\cal T}_\perp \right ) \! \bigg \} \, , \\[2mm]
  A_\parallel^{L,R} & = - \frac{\sqrt{2} N \hspace{0.25mm} (m_B^2 -
    q^2)}{m_B} \, \bigg \{ \! \left [ \left ( C_9^l - \tilde C_9^l
    \right ) \mp \left ( C_{10}^l - \tilde C_{10}^l \right ) \right ]
  \xi_\perp + \frac{2 \hspace{0.25mm} m_b \hspace{0.25mm} m_B}{q^2}
  \left ( {\cal T}_\perp - \tilde {\cal T}_\perp \right )
  \! \bigg \} \, , \\[2mm]
  A_0^{L,R} & = -\frac{N \hspace{0.25mm} (m_B^2 - q^2)^2}{2
    \hspace{0.25mm} m_B \hspace{0.25mm} m_{K^\ast} \sqrt{q^2}} \,
  \bigg \{ \! \left [ \left ( C_9^l - \tilde C_9^l \right ) \mp \left
      ( C_{10}^l - \tilde C_{10}^l \right ) \right ] \xi_\parallel -
  \frac{2 \hspace{0.25mm} m_b}{m_B} \left ( {\cal T}_\parallel -
    \tilde {\cal T}_\parallel \right ) \! \bigg \} \,,
\end{split} 
\eeq
with 
\beq \label{eq:N}
N \equiv \Bigg [ \, \frac{G_F^2 \hspace{0.5mm} \alpha^2}{3027 \, \pi^5
  \, m_B^3} \, \big |\lambda_t^{(sb)} \big |^2 \; q^2 \, \sqrt{q^4 +
  m_B^4 + m_{K^\ast}^4 - 2 \left ( q^2 \hspace{0.1mm} m_B^2 + q^2
    \hspace{0.1mm} m_{K^\ast}^2 + m_B^2 \hspace{0.1mm} m_{K^\ast}^2
  \right )} \, \Bigg ]^{1/2} .
\eeq
Kinematical terms suppressed by $m_{K^\ast}^2/m_B^2$ have been
neglected here and the mass of the charged leptons has been set to
zero. The form factors $\xi_{\perp, \parallel}$ are discussed in
Appendix~\ref{app:input}. The functions ${\cal T}_{\perp, \parallel}$
include besides the usual one-loop matrix elements of the four-quark
operators $Q_{3-6}$ also the two-loop matrix elements of the
current-current operators $Q_{1,2}$ and hard spectator-scattering
effects, which give sizable contributions to the imaginary
parts. Explicit expressions for ${\cal T}_{\perp, \parallel}$ can be
found in \cite{Beneke:2001at,Beneke:2004dp}. The functions $\tilde
{\cal T}_{\perp, \parallel}$ are defined in analogy to ${\cal
  T}_{\perp, \parallel}$, and include the one-loop matrix elements of
the chirality-flipped operators $\tilde Q_{3-6}$. The electromagnetic
coupling constant and the Wilson coefficients entering
(\ref{eq:Ailow}) and (\ref{eq:N}) are evaluated at the scale $m_b$.

In terms of the transversity amplitudes, the dilepton spectrum is
given by
\beq
  \frac{d\Gamma (B \to K^\ast l^+ l^-)}{dq^2} = \big |A_\perp \big |^2 +
  \big |A_\parallel \big |^2 + \big |A_0 \big |^2 \,,
\eeq
where $A_i A_j^\ast \equiv A_i^L A_j^{L \ast} + A_i^R A_j^{R \ast}$
for $i,j = \, \perp, \parallel, 0$. Expressed through $A_i^{L,R}$ and
$d\Gamma/dq^2$, the normalized forward-backward asymmetry and the
longitudinal $K^\ast$ polarization of $B \to K^\ast l^+ l^-$ take the
simple form 
\beq \label{eq:AFBFLlow}
\begin{split}
  \bar A_{\rm FB} (B \to K^\ast l^+ l^-) & = \frac{3}{2} \left (
    \frac{d\Gamma (B \to K^\ast l^+ l^-)}{dq^2} \right )^{-1} {\rm Re}
  \, \Big ( A_\perp^R A_\parallel^{R \ast}
  -  A_\perp^L A_\parallel^{L \ast} \Big ) \,, \\[2mm]
  F_{\rm L} (B \to K^\ast l^+ l^-) &= \left (
    \frac{d\Gamma (B \to K^\ast l^+ l^-)}{dq^2} \right )^{-1} \, \big
  | A_0 \big |^2 \,.
\end{split}
\eeq

In order to elucidate the importance of the different new-physics
contributions to the exclusive $b \to s l^+ l^-$ observables in the RS
model, we present approximate formulas for the forward-backward
asymmetry and the longitudinal $K^\ast$ polarization integrated over
$q^2 \in [1, 6] \, {\rm GeV}^2$. Ignoring small effects due to
new-physics contributions stemming from the one-loop matrix elements
of $Q_{3-10}$ and $\tilde Q_{3-10}$, and neglecting terms suppressed
by $m_{K^\ast}^2/m_B^2$, we obtain for the quantities appearing in
(\ref{eq:AFBFLlow}) the approximations\footnote{Terms with
  coefficients smaller than 0.01 have been discarded here. The same
  applies to (\ref{eq:approxhigh}).}
\begin{align} \label{eq:approxlow}
  & \int^{6 \, {\rm GeV}^2}_{1 \, {\rm GeV}^2} dq^2 \, \frac{d\Gamma
    (B \to K^\ast l^+ l^-)}{dq^2} = 1.59 \, \Bigg [ \, 1 + 0.16 \,
  {\rm Re} \hspace{0.5mm} C_9^{l \, {\rm RS}} - 0.28 \, {\rm Re}
  \hspace{0.5mm} C_{10}^{l \, {\rm RS}} - 0.01 \, {\rm Im}
  \hspace{0.5mm} C_9^{l \, {\rm RS}} \nonumber \\ & \hspace{4cm} - 0.17 \, {\rm
    Re} \hspace{0.5mm} \tilde C_9^{l \, {\rm RS}} + 0.20 \, {\rm Re}
  \hspace{0.5mm} \tilde C_{10}^{l \, {\rm RS}} - 0.05 \, {\rm Re}
  \left ( C_9^{l \, {\rm RS}} \tilde C_9^{l \, {\rm RS} \ast} +
    C_{10}^{l \, {\rm RS}} \tilde C_{10}^{l \, {\rm RS} \ast} \right )
  \nonumber \\ & \hspace{4cm} + 0.03 \left ( \big | C_9^{l \, {\rm RS}} \big |^2
    + \big | C_{10}^{l \, {\rm RS}} \big |^2 + \big | \tilde C_9^{l \,
      {\rm RS}} \big |^2 + \big | \tilde C_{10}^{l \, {\rm RS}}
    \big |^2 \right ) \Bigg ] \cdot 10^{-7} \, {\rm ps}^{-1} \,, \nonumber \\[2mm]
  & \int^{6 \, {\rm GeV}^2}_{1 \, {\rm GeV}^2} dq^2 \; \frac{3}{2} \,
  {\rm Re} \, \Big ( A_\perp^R A_\parallel^{R \ast} - A_\perp^L
  A_\parallel^{L \ast} \Big ) = -0.08 \, \Bigg [ \, 1 - 1.38 \, {\rm
    Re} \hspace{0.5mm} C_9^{l \, {\rm RS}} - 0.23 \, {\rm Re}
  \hspace{0.5mm} C_{10}^{l \, {\rm RS}} \nonumber \\ & \hspace{4cm} - 0.15 \,
  {\rm Im} \hspace{0.5mm} C_{10}^{l \, {\rm RS}} + 0.32 \, {\rm Re}
  \left ( C_9^{l \, {\rm RS}} C_{10}^{l \, {\rm RS} \ast} - \tilde
    C_9^{l \, {\rm RS}} \tilde C_{10}^{l \, {\rm RS} \ast} \right )
  \Bigg ] \cdot 10^{-7} \, {\rm ps}^{-1} \,, \hspace{8mm} \nonumber \\[2mm]
  & \int^{6 \, {\rm GeV}^2}_{1 \, {\rm GeV}^2} dq^2 \; \big | A_0 \big
  |^2 = 1.17 \, \Bigg [ \, 1 + 0.23 \, {\rm Re} \, \Big ( C_9^{l \,
    {\rm RS}} - \tilde C_9^{l \, {\rm RS}} \Big ) - 0.27 \, {\rm Re}
  \, \Big ( C_{10}^{l \, {\rm RS}} - \tilde C_{10}^{l \, {\rm RS}}
  \Big ) \nonumber \\ & \hspace{4cm} - 0.06 \, {\rm Re} \, \Big ( C_9^{l \, {\rm
      RS}} \tilde C_9^{l \, {\rm RS} \ast} + C_{10}^{l \, {\rm RS}}
  \tilde C_{10}^{l \, {\rm RS} \ast} \Big ) \nonumber \\ & \hspace{4cm} + 0.03
  \left ( \big | C_9^{l \, {\rm RS}} \big |^2 + \big | C_{10}^{l \,
      {\rm RS}} \big |^2 + \big | \tilde C_9^{l \, {\rm RS}} \big |^2
    + \big | \tilde C_{10}^{l \, {\rm RS}} \big |^2 \right ) \Bigg ]
  \cdot 10^{-7} \, {\rm ps}^{-1} \,,
\end{align}
where $C_{9,10}^{l \, {\rm RS}}$ and $\tilde C_{9,10}^{l \, {\rm RS}}$
denote the low-energy Wilson coefficients. The corresponding initial
conditions are given in (\ref{eq:CCt910l}).

The SM predictions for the normalized forward-backward asymmetry and
the longitudinal $K^\ast$ polarization read
\beq \label{eq:bsllexcllowSM}
\begin{split} 
  \bar A_{\rm FB} (B \to K^\ast l^+ l^-)^{q^2 \in [1, 6] \, {\rm
      GeV}^2}_{\rm SM}
  & = (-0.05 \pm 0.03)\,, \\[2mm]
  F_{\rm L} (B \to K^\ast l^+ l^-)^{q^2 \in [1, 6] \, {\rm
      GeV}^2}_{\rm SM} & = 0.74 \pm 0.12 \,.
\end{split}
\eeq 
They are plagued by sizable uncertainties due to the form factors
$\xi_{\perp, \parallel}$ and unknown power corrections of $\ord
(\Lambda_{\rm QCD}/m_b)$. Our estimates include a $10\%$ error due to
the latter terms. The given numbers can already be compared with
existing measurements performed at BaBar
\cite{Aubert:2006vb,Aubert:2008ju} and Belle
\cite{Ishikawa:2006fh, Wei:2009zv}. The large increase of statistics
expected at the LHCb experiment will allow for precision measurements
of these observables. Full angular analyses in $B \to K l^+ l^-$
\cite{Bobeth:2007dw} and $B \to K^\ast l^+ l^-$ decays
\cite{Egede:2008uy, Altmannshofer:2008dz} as well as experimental
studies of CP-violating observables \cite{Altmannshofer:2008dz,
  Bobeth:2008ij} also seem promising, and offer a unique window on
physics at and above the electroweak scale.

In the kinematic range above the charm resonances, $q^2 > 14 \, {\rm
  GeV}^2$, the relevant transversity amplitudes can be calculated in a
rigorous way by performing an expansion in $\Lambda_{\rm QCD}/m_b$ and
$m_c^2/m_b^2$. Following \cite{Grinstein:2004vb} one finds the
expressions
\begin{eqnarray} \label{eq:Aihigh}
\begin{aligned}
  A_\perp^{L,R} & = - \sqrt{2} N \hspace{0.25mm} (m_B^2 - q^2) \,
  \bigg \{ \! \left [ \left ( C_{9,A}^{l \, {\rm eff}} (q^2) + \tilde
      C_{9,A}^{l \, {\rm eff}} (q^2) \right ) \mp \left ( C_{10}^l +
      \tilde C_{10}^l \right ) \right ] g (q^2) \\ & \hspace{2.5cm} -
  \frac{2 m_b}{q^2} \, g_+ (q^2) \, C_{7,A}^{\gamma \,{\rm eff}} (q^2)
  \bigg \} \,, \\[2mm]
  A_\parallel^{L,R} & = -\sqrt{2} N \, \bigg \{ \! \left [ \left (
      C_{9,A}^{l \, {\rm eff}} (q^2) - \tilde C_{9,A}^{l \, {\rm eff}}
      (q^2) \right ) \mp \left ( C_{10}^l - \tilde C_{10}^l \right )
  \right ] f(q^2) \\
  & \hspace{2.5cm} + \frac{2 m_b}{q^2} \, \Big [ (m_B^2 -
  m_{K^\ast}^2) \, g_+ (q^2) + q^2 \, g_- (q^2) \Big ] \,
  C_{7,A}^{\gamma \,{\rm eff}}
  (q^2) \bigg \} \,, \\[2mm]
  A_0^{L,R} & = -\frac{N \hspace{0.25mm} (m_B^2 - q^2)}{2
    \hspace{0.25mm} m_{K^\ast} \sqrt{q^2}} \, \bigg \{ \!  \left [
    \left ( C_{9,A}^{l \, {\rm eff}} (q^2) - \tilde C_{9,A}^{l \, {\rm
          eff}} (q^2) \right ) \mp \left ( C_{10}^l - \tilde C_{10}^l
    \right ) \right ] \! \!
  \Big [ f(q^2) + (m_B^2 - q^2) \, a_+(q^2) \Big ] \\
  & \hspace{2.5cm} + \frac{2 m_b}{m_B^2} \, \Big [ (q^2 -
  m_{K^\ast}^2) \, g_+ (q^2) + q^2 \left ( \hspace{0.25mm} g_- (q^2) +
    (m_B^2 - q^2) \, h(q^2) \hspace{0.25mm} \right ) \Big ]
  \hspace{0.25mm} C_{7,A}^{\gamma \,{\rm eff}} (q^2) \bigg \} \,,
  \hspace{0mm}
\end{aligned} 
\end{eqnarray} 
with the form factors $f(q^2)$, $g(q^2)$, $a_+ (q^2)$, $g_\pm (q^2)$,
and $h(q^2)$ given in Appendix~\ref{app:input}. The explicit
expressions for the effective Wilson coefficients $C_{9,A}^{l \, {\rm
    eff}} (q^2)$ and $C_{7,A}^{\gamma \, {\rm eff}} (q^2)$ can be
found in \cite{Grinstein:2004vb}. The coefficient $\tilde C_{9,A}^{l
  \, {\rm eff}} (q^2)$ is defined in analogy to $C_{9,A}^{l \, {\rm
    eff}} (q^2)$ with the one-loop matrix elements $Q_{3-6}$ replaced
by those of $\tilde Q_{3-6}$. All Wilson coefficients are evaluated at
the scale $m_b$.

Omitting again small new-physics contributions from one-loop matrix
elements of $Q_{3-10}$ and $\tilde Q_{3-10}$, we obtain for the
relevant quantities integrated from above the charm resonances to the
endpoint of the spectrum, $q^2 \in [14 \, {\rm GeV}^2, (m_B -
m_{K^\ast})^2] = [14, 19.2] \, {\rm GeV}^2$, the approximate
expressions
\beq \label{eq:approxhigh} 
\begin{split} 
  & \int^{19.2 \, {\rm GeV}^2}_{14 \, {\rm GeV}^2} dq^2 \,
  \frac{d\Gamma (B \to K^\ast l^+ l^-)}{dq^2} = 1.86 \, \Bigg [ \, 1 +
  0.24 \, {\rm Re} \hspace{0.5mm} C_9^{l \, {\rm RS}} - 0.28 \, {\rm
    Re} \hspace{0.5mm} C_{10}^{l \, {\rm RS}}  \\ &
  \hspace{4cm} - 0.16 \, {\rm Re} \hspace{0.5mm} \tilde C_9^{l \, {\rm
      RS}} + 0.19 \, {\rm Re} \hspace{0.5mm} \tilde C_{10}^{l \, {\rm
      RS}} + 0.02 \, {\rm Im} \hspace{0.5mm} C_9^{l \, {\rm RS}} -
  0.01 \, {\rm Im} \hspace{0.5mm} \tilde C_9^{l \, {\rm RS}} 
  \\[2mm] & \hspace{4cm} - 0.04 \, {\rm Re} \left ( C_9^{l \, {\rm
        RS}} \tilde C_9^{l \, {\rm RS} \ast} + C_{10}^{l \, {\rm RS}}
    \tilde C_{10}^{l \, {\rm RS} \ast} \right )  \\ &
  \hspace{4cm} + 0.03 \left ( \big | C_9^{l \, {\rm RS}} \big |^2 +
    \big | C_{10}^{l \, {\rm RS}} \big |^2 + \big | \tilde C_9^{l \,
      {\rm RS}} \big |^2 + \big | \tilde C_{10}^{l \, {\rm RS}} \big
    |^2 \right ) \Bigg
  ] \cdot 10^{-7} \, {\rm ps}^{-1} \,,  \\[2mm]
  & \int^{19.2 \, {\rm GeV}^2}_{14 \, {\rm GeV}^2} dq^2 \; \frac{3}{2}
  \, {\rm Re} \, \Big ( A_\perp^R A_\parallel^{R \ast} - A_\perp^L
  A_\parallel^{L \ast} \Big ) = 0.75 \, \Bigg [ \, 1 + 0.30 \, {\rm
    Re} \hspace{0.5mm} C_9^{l \, {\rm RS}} - 0.24 \, {\rm Re}
  \hspace{0.5mm} C_{10}^{l \, {\rm RS}}  \\ & \hspace{4cm} -
  0.02 \, {\rm Im} \hspace{0.5mm} C_{10}^{l \, {\rm RS}} - 0.07 \,
  {\rm Re} \left ( C_9^{l \, {\rm RS}} C_{10}^{l \, {\rm RS} \ast} -
    \tilde C_9^{l \, {\rm RS}} \tilde C_{10}^{l \, {\rm RS} \ast}
  \right ) \Bigg ] \cdot 10^{-7} \, {\rm ps}^{-1} \,, \hspace{8mm} 
    \\[2mm]
  & \int^{19.2 \, {\rm GeV}^2}_{14 \, {\rm GeV}^2} dq^2 \; \big | A_0
  \big |^2 = 0.76 \, \Bigg [ \, 1 + 0.24 \, {\rm Re} \, \Big ( C_9^{l
    \, {\rm RS}} - \tilde C_9^{l \, {\rm RS}} \Big ) - 0.27 \, {\rm
    Re} \, \Big ( C_{10}^{l \, {\rm RS}} - \tilde C_{10}^{l \, {\rm
      RS}} \Big )  \\ & \hspace{4cm} \phantom{xx} + 0.02 \,
  {\rm Im} \, \Big ( C_9^{l \, {\rm RS}} - \tilde C_9^{l \, {\rm RS}}
  \Big ) - 0.07 \, {\rm Re} \, \Big ( C_9^{l \, {\rm RS}} \tilde
  C_9^{l \, {\rm RS} \ast} + C_{10}^{l \, {\rm RS}} \tilde C_{10}^{l
    \, {\rm RS} \ast} \Big )  \\ & \hspace{4cm} \phantom{xx}
  + 0.03 \left ( \big | C_9^{l \, {\rm RS}} \big |^2 + \big |
    C_{10}^{l \, {\rm RS}} \big |^2 + \big | \tilde C_9^{l \, {\rm
        RS}} \big |^2 + \big | \tilde C_{10}^{l \, {\rm RS}} \big |^2
  \right ) \Bigg ] \cdot 10^{-7} \, {\rm ps}^{-1} \,. \hspace{4mm}
\end{split}
\eeq

In the high-$q^2$ region the SM predictions are given by
\beq \label{eq:bsllexclhighSM}
\begin{split} 
  \bar A_{\rm FB} (B \to K^\ast l^+ l^-)^{q^2 \in [14 , 19.2] \, {\rm
      GeV}^2}_{\rm SM}
  & = 0.40 \pm 0.11 \,, \\[2mm]
  F_{\rm L} (B \to K^\ast l^+ l^-)^{q^2 \in [14, 19.2] \, {\rm
      GeV}^2}_{\rm SM} & = 0.36 \pm 0.07 \,.
\end{split}
\eeq 
Like in the low-$q^2$ region, the quoted uncertainties arise mainly
from the form factors and unknown power corrections. The additional
non-perturbative error due to $\ord (\Lambda_{\rm QCD}/m_b)$ terms is
estimated to be about $10\%$ and included in the total
uncertainties. The different sensitivity to short-distance dynamics in
the low- and high-$q^2$ ranges makes combinations of measurements of
$\bar A_{\rm FB} (B \to K^\ast l^+ l^-)$ and $F_{\rm L} (B \to K^\ast
l^+ l^-)$ in different kinematic regions powerful probes of new
physics.

\boldmath \subsection{Important Formulas for $B \to \tau \nu_\tau$}
\unboldmath

Among the purely leptonic and semileptonic decays of kaons and $B$
mesons induced by the charged-current interactions, we focus on the $B
\to \tau \nu_\tau$ decay, which in the SM provides a direct
measurement of the product of the CKM matrix element $|V_{ub}|$ and
the $B$-meson decay constant. The plethora of available $K \to (\pi) l
\nu$ data \cite{Antonelli:2008jg} will be used in Section
\ref{sec:ckmnumerics} to set bounds on the violation of the unitarity
of the first row of the generalized CKM matrix. Another interesting
test in the field of semileptonic kaon decays consists in the
comparison of the value of $|V_{us}|$ determined in the
helicity-suppressed $K \to l \nu$ decays with the one extracted from
the helicity-allowed $K \to \pi l \nu$ modes
\cite{Antonelli:2008jg}. While the presence of non-vanishing
right-handed currents can lead to different extractions of $|V_{us}|$,
the possible shifts in the $Wu_Rs_R$ coupling predicted in the RS
framework are, due to the strong chiral suppression, too small to lead
to any observable effect.

In the RS model, the branching ratio of the $B \to \tau \nu_\tau$
decay can be written in terms of the coefficients (\ref{eq:CClRS}) as
\beq
  {\cal B} (B \to \tau \nu_\tau) = \frac{\tau_B f_B^2 \, m_B
    \hspace{0.25mm} m_\tau^2}{16 \pi} \left ( 1 - \frac{m_\tau^2}{m_B^2}
  \right )^2 \, \left | C_l^{\rm RS} - \tilde C_l^{\rm RS} \right |^2 .
\eeq
The dependence on the lepton mass arises from helicity conservation,
which suppresses the muon and electron channels.

From the global fit of the unitarity triangle, we obtain the following
SM prediction
\beq \label{eq:BtaunuSM} 
  {\cal B} (B \to \tau \nu_\tau)_{\rm SM} = (0.90 \pm 0.19) \cdot
  10^{-4} \,.
\eeq 
The major part of the total error stems from the uncertainty due to
the $B$-meson decay constant $f_B \equiv f_{B_d} = (0.200 \pm 0.020)
\, {\rm GeV}$ \cite{Lubicz:2008am}. Notice that the indirect fit
prediction (\ref{eq:BtaunuSM}) is $2.1 \sigma$ below the current world
average ${\cal B} (B \to \tau \nu_\tau)_{\rm exp} = (1.73 \pm 0.35)
\cdot 10^{-4}$ \cite{Ikado:2006un, Aubert:2007xj, Adachi:2008ch,
  Aubert:2008gx}. While at the moment systematic errors in the
lattice determinations of $f_B$ in conjunction with limited
experimental statistics do not allow to draw a definite conclusion
about the presence of new physics in $B \to \tau \nu_\tau$, we will
investigate whether the RS model can in principle reproduce the
observed deviation.

\section{Numerical Analysis}
\label{sec:numerics}

This part of our article is devoted to a thorough numerical analysis
of new-physics effects in the quark flavor sector. We begin our
discussion with a concise description of the algorithm used to scan
the parameter space of the RS model. We then introduce four benchmark
scenarios, three of them specifically devised to suppress potentially
dangerous contributions to processes such as $Z^0 \to b \bar b$ and
$K$--$\bar K$ mixing. We will investigate in detail to which extent
the different scenarios allow us to relax the strong constraints
arising from electroweak precision tests and quark flavor
physics. Next we analyze the potential size of new-physics effects in
the charged-current sector, paying special attention to the
non-unitarity of the CKM matrix and the extraction of the elements
$|V_{ub}|$, $|V_{cb}|$, and $|V_{tb}|$. Then follows a comprehensive
study of $\Delta F = 2$ and $\Delta F = 1$ flavor-changing
transitions. We find that rare decays of $K$ and $B_s$ mesons are
particularly interesting in the minimal RS framework, as their
branching ratios can be enhanced considerably with respect to the SM
even after taking into account the constraints imposed by $Z^0 \to b
\bar b$ and $\epsilon_K$. We emphasize that whereas many observables
within the $K$, $B_{d,s}$, and $D$ systems are strongly correlated in
the RS model, there are in general no correlations between observables
belonging to different sectors. While at first sight the observed
correlations appear to be very promising to distinguishing the RS
scenario from other extensions of the SM, we will show that they
indeed arise in a wide class of models of new physics and hence do not
constitute unmistakable signals of warped extra dimensions. In each
case we will discuss the origin of the observed pattern. Comments on
how our results would change in a RS model with custodial protection
round off our phenomenological survey.

\subsection{Parameter Scan}

In a first step we determine sets of Yukawa matrices $\bm{Y}_{u,d}$
that allow one to reproduce the observed values of the Wolfenstein
parameters $\bar \rho$ and $\bar \eta$. Since the latter two
quantities are to leading order in hierarchies independent of the
zero-mode profiles $F(c_{A_i})$ \cite{Casagrande:2008hr}, it turns
out to be computationally expensive to find suitable pairs of Yukawa
matrices $\bm{Y}_{u,d}$ by a simple random sampling. To generate
proper sets we proceed in the following way. We randomly pick one
element of the Yukawa matrices keeping its phase $\phi$ and modulus
$y$ arbitrary. Using uniform initial distribution the remaining
elements are then generated in the ranges ${\rm arg} \left
  ((Y_{u,d})_{ij} \right ) \in [0, 2 \pi[$ and $|(Y_{u,d})_{ij}| \in
[1/10, Y_{\rm max}]$, where the lower limit enforces Yukawa entries of
natural size, while the upper limit $Y_{\rm max}$ will vary between
the different benchmark scenarios. Next we calculate the Wolfenstein
parameters $\bar \rho$ and $\bar \eta$ by means of (I:102) and
minimize the function
\beq 
\chi^2 (x) = \sum_n \left ( \frac{x_{\rm exp} (n) - x_{\rm theo} (n)
  }{\sigma_{\rm exp} (n)} \right )^2 \,, \qquad x = \{ \bar \rho, \bar
\eta \} \,,
\eeq 
with respect to $\phi$ and $y$, requiring $1/10 \leq y \leq
  Y_{\rm max}$. Here
$x_{\rm exp} (n)$ and $x_{\rm theo} (n)$ denote the experimental and
theoretical value of the $n^{\rm th}$ observable and $\sigma_{\rm exp}
(n)$ is the standard deviation of the corresponding measurement.

After the elements of the Yukawa matrices $\bm{Y}_{u,d}$ have been
fixed, we choose a random value for the bulk mass parameter $c_{u_3}
\in\, ]\!-1/2, c_{u_3}^{\rm max}]$ with $c_{u_3}^{\rm max} = 2$ or
$5/2$ depending on the scenario, and calculate the whole set of
observables $x = \{m_u, m_d, m_s, m_c, m_b, m_t, A, \lambda, \bar
\rho, \bar \eta \}$ in terms of the zero-mode profiles of the
remaining $c_{A_i}$ using the leading-order Froggatt-Nielsen relations
(I:96) and (I:102).\footnote{The choice of $c_{u_3}$ as a prior is
  motivated by the fact that this bulk mass parameter is special in
  the sense that it determines the degree of compositeness of the top
  quark. While the lower limit of the allowed range of $c_{u_3}$ is
  motivated by the fact that the right-handed top quark should be
  localized near the IR brane, the upper limit is chosen in a somewhat
  {\it ad hoc} way. Allowing for $c_{u_3}^{\rm max} \gg 1$ would
  however take away an attractive feature of the RS model, namely that
  it explains the quark hierarchies in terms of fundamental parameters
  of ${\cal O} (1)$.} The values of the zero-mode profiles
$F(c_{A_i})$ are then determined from the best fit to the $\chi^2 (x)$
function supplemented by the constraints $F(c_{A_i}) \leq
F(c_{u_3}^{\rm max})$. Points with $\chi^2 (x)/{\rm dof} > 11.5/10$,
corresponding to 68\% CL, are rejected, while for points that pass the
test, we recompute $\chi^2 (x)$ using exact formulas for the quark
masses and Wolfenstein parameters and remove all points that show a
deviation of more than $3 \sigma$ in at least one observable.

In order to assure that our algorithm populates the whole parameter
space without introducing spurious correlations, we have inspected the
final distributions of parameters. While the magnitudes and phases of
the elements of the Yukawa matrices are all nearly flatly distributed,
the shapes of the distributions of the quark masses and Wolfenstein
parameters are almost Gaussian with a width of at most twice the
corresponding experimental uncertainty. We consider this as a strong
indication that we achieve full coverage of the parameter space in an
unbiased way. These features are basic prerequisites guaranteeing that
correlations between different observables have indeed a physical
origin and are not artifacts of an imperfect Monte Carlo sampling.

\subsection{Benchmark Scenarios}
\label{sec:benchmarks}

In order to assess the robustness of predictions, it is important to
investigate how sensitively they depend on the values of the most
relevant parameters of the model. We therefore study four different
benchmark scenarios that differ by the allowed maximal magnitude of
the Yukawa couplings, the structure of the bulk masses matrices, and
the value of the logarithm of the warp factor. Three out of the four
benchmark scenarios are designed specifically to suppress harmful
contributions to electroweak precision and quark flavor observables
and therefore present particular cases of viable models of warped
extra dimensions with improved prospects for discovery at the LHC. Our
benchmark scenarios are defined as follows:

\begin{itemize}

\item Scenario 1 (S1): ``standard''

  The magnitudes of the entries $|(Y_{u, d})_{ij}|$ of the Yukawa
  matrices have to be bounded from above in order for the Yukawa
  couplings to be perturbative. In our first scenario, we employ
  $Y_{\rm max} = 3$, which coincides with the upper limit estimated by
  means of naive dimensional analysis \cite{Csaki:2008zd}. This is
  the standard choice for $Y_{\rm max}$ employed in several articles
  on flavor effects in warped extra-dimension models
  \cite{Csaki:2008zd, Santiago:2008vq, Csaki:2008eh, Blanke:2008zb,
    Agashe:2008uz, Bauer:2008xb, Blanke:2008yr}. No further
  restrictions on the bulk mass parameters are imposed and the
  logarithm of the warp factor is set to $L = \ln (10^{16}) \approx
  37$, as needed to explain the hierarchy between the Planck and the
  electroweak scales.

\item Scenario 2 (S2): ``aligned''

  In this scenario we consider only the subclass of models
  characterized by common bulk masses $c_{d_i}$ in the sector of
  right-handed down-type quarks. This is a viable solution, since by
  virtue of the moderate hierarchy, the mass splittings in the
  down-type quark sector can be naturally accommodated by ${\cal O}
  (1)$ variations of the Yukawa couplings. The equality of bulk mass
  parameters can be achieved by imposing an $U(3)$ flavor symmetry,
  under which the fields that give rise to the right-handed down-type
  quark zero modes transform as triplets and all other fields
  transform as singlets \cite{Santiago:2008vq}. Besides the natural
  suppression of vast corrections to $\epsilon_K$
  \cite{Santiago:2008vq}, this scenario has the appealing feature
  that it leads to a unique pattern of deviations in the $K \to \pi
  \nu \bar \nu$ sector.
  
\item Scenario 3 (S3): ``little''
  
  From a purely phenomenological point of view, it is possible to
  lower the UV cutoff from the Planck scale to a value only few orders
  of magnitude above the TeV scale, even though in this case a true
  solution to the hierarchy problem is postponed to higher
  energy. Since many amplitudes in the RS model are enhanced by $L$,
  it is worthwhile to address the question to what extent certain
  experimental constraints can be avoided by such a choice
  \cite{Davoudiasl:2008hx}. This will be done in our last scenario,
  where we consider a ``volume-truncated'' variant of our first
  scenario characterized by $L = \ln (10^3) \approx 7$. Unfortunately,
  in this ``little'' RS scenario \cite{Davoudiasl:2008hx} no
  improvement concerning $\epsilon_K$ can be achieved compared to the
  standard benchmark scenario \cite{Bauer:2008xb}.

\item Scenario 4 (S4): ``large''

  This scenario differs from the first one by the choice of $Y_{\rm
    max}$. Since it is {\it a priori} not clear if the theory should
  be weakly coupled in the Yukawa or in any other sector, and up to
  what cutoff scale weak coupling should hold, the bound $Y_{\rm max}
  = 3$ might be regarded as being too restrictive. In fact, in order
  to avoid fine-tuning in the Higgs sector the UV cutoff has to be
  very close to the KK scale, and requiring weak coupling up to $\Mkk$
  allows for Yukawa couplings that are larger by a factor of 4 than
  the latter limit \cite{Casagrande:2008hr}. In our second benchmark
  scenario we entertain this possibility and consequently allow for
  $Y_{\rm max} = 12$. This choice has the attractive property that it
  helps to reduce some particularly dangerous flavor-changing
  couplings in $\epsilon_K$ \cite{Csaki:2008zd}.

\end{itemize}

\begin{table}
  \begin{center}
    \begin{tabular}[h]{|c|c|c|c|c|}
      \hline
      ~ & S1 & S2 & S3 & S4 \\ \hline
      \rule{0pt}{12pt} $c_{Q_1}$ & $-0.63\pm0.03$ &
      $-0.66\pm0.02$ & $-1.34\pm0.16$ & $-0.67\pm0.03$ \\
      \rule{0pt}{12pt} $c_{Q_2}$ & $-0.57\pm0.05$ &
      $-0.59\pm0.03$ & $-1.04\pm0.18$ & $-0.61\pm0.03$ \\
      \rule{0pt}{12pt} $c_{Q_3}$ & $-0.34 \pm 0.32$ &
      $-0.24\pm0.43$ & $-0.49\pm0.34$ & $-0.52\pm0.09$ \\
      \rule{0pt}{12pt} $c_{u_1}$ & $-0.68\pm0.04$ &
      $-0.65\pm0.03$ & $-1.58\pm0.18$ & $-0.68\pm0.04$ \\
      \rule{0pt}{12pt} $c_{u_2}$ & $-0.51\pm0.12$ &
      $-0.50\pm0.12$ & $-0.79\pm0.26$ & $-0.51\pm0.12$ \\
      \rule{0pt}{12pt} $c_{u_3}$ & $]-1/2, 2]$ &
      $]-1/2, 2]$ & $]-1/2, 5/2]$ & $]-1/2, 2]$ \\
      \rule{0pt}{12pt} $c_{d_1}$ & $-0.65\pm0.03$ &
      $-0.60\pm0.02$ & $-1.44\pm0.17$ & $-0.66\pm0.03$ \\
      \rule{0pt}{12pt} $c_{d_2}$ & $-0.62\pm0.03$ &
      $-0.60\pm0.02$ & $-1.28\pm0.17$ & $-0.62\pm0.03$ \\
      \rule{0pt}{12pt} $c_{d_3}$ & $-0.58\pm0.03$ &
      $-0.60\pm0.02$ & $-1.05\pm0.13$ & $-0.58\pm0.03$ \\[2pt] 
      \hline
    \end{tabular}
  \end{center}
  \begin{center}
    \parbox{15.5cm}{\caption{\label{tab:bulkmasses} Central values and
        statistical uncertainties of the bulk mass parameters in the
        different benchmark scenarios. The shown errors correspond to $1
        \sigma$ ranges when fitting the distributions to a Gaussian
        function. See text for details.}}
  \end{center}
\end{table}
  
The statistical approach outlined in the last section provides us with
distributions of bulk mass parameters rather than with their precise
values. The obtained results are different in the individual benchmark
scenarios, and in order to make our article self-contained, we
summarize the central values and statistical uncertainties of the
parameters $c_{A_i}$ for the four cases in
Table~\ref{tab:bulkmasses}. All distributions are nearly Gaussian,
apart from those of $c_{Q_3}$, which feature a small tail to higher
values, and those of $c_{u_3}$, which are chosen as flat priors in the
range $]-1/2, 2]$ for all scenarios, except scenario S3, for which we
allow for $]-1/2, 5/2]$. Notice that while the central values of the
bulk mass parameters in the benchmark scenarios S1, S2 and S4 are very
similar, the splitting between the individual $c_{A_i}$ parameters is
on average much bigger in scenario S3 due to the smaller ``volume
factor'' $L$. We have verified that the precise form of the
distributions is essentially independent of the details of the
algorithm used to scan the parameter space. Besides inspecting the
coverage of our Monte Carlo sampling, we also asses the numerical and
perturbative stability of each single parameter point. For this
purpose we use the quantitative measurement for fine-tuning
$\Delta_{\rm{BG}}(\mathcal{O}) = \max_{p \in \mathcal{P}}
\big| \partial_{\log(p)} \log(\mathcal{O}) \big|$ introduced by
Barbieri and Giudice \cite{Barbieri:1987fn}.  In our case the set of
model parameters $\mathcal{P}$ includes the Yukawa couplings, their
complex conjugates, and the bulk masses.

We identify two different situations, where
$\Delta_{\rm{BG}}(\mathcal{O})$ can become large, rendering the
results unstable under small variations of the input parameters.  The
first case corresponds to large values of the observable
$\mathcal{O}$, where the variation is (almost) entirely dominated by a
single contribution. In this case it is sufficient to apply the
fine-tuning measure to the matrices $\bm{\Delta}_A^{(')}$ and
$\bm{\delta}_A$, which constitute the new-physics contributions to all
the relevant couplings (KK photons and gluons, $Z^0$ boson and its KK
excitations) in our analysis. In practice, we use the analytic form
(I:A2) and (I:A3) of those matrices, obtained from a Froggatt-Nielsen
analysis to leading order in $\lambda$.  The contributions to
$\Delta_{\rm{BG}}(\bm{\Delta}_A^{(')}, \, \bm{\delta}_A)$ stemming
from variations of the bulk mass parameters $c_{A_i}$ scale either as
$L$ or $1$ depending on whether $c_{A_i} <-1/2$ or $c_{A_i}
>-1/2$. Similarly, variations with respect to the Yukawa entries
typically lead to contributions of $\mathcal{O}(1)$, but can also
become larger if individual elements $(Y_{u,d})_{ij}$ appearing in the
denominators of the expressions for $\bm{\Delta}_A^{(')}$ and
$\bm{\delta}_A$ cancel each other. However, even in such a case the
resulting contribution to $\Delta_{\rm{BG}}(\bm{\Delta}_A^{(')}, \,
\bm{\delta}_A)$ never exceeds 100 for points consistent with the
observed quark masses and mixing, implying no severe fine-tuning.

The situation is different for untypically small values of
$\mathcal{O}$. As we will show explicitly below, this situation occurs
in the case of $|\epsilon_K|$, where the experimental constraint
forces the new-physics contribution to be small when compared to the
typical values arising in the model at hand. In this case we calculate
$\Delta_{\rm{BG}}(|\epsilon_K|)$ taking into account the full
expression for $|\epsilon_K|$ including all interferences and RG
effects, and reject parameter points that show an excessive
fine-tuning of $1000$ or larger. We have verified that (reasonable)
variations of the chosen cut on $\Delta_{\rm{BG}}(|\epsilon_K|)$ do
neither change the shape of the distribution of allowed scatter points
nor affect any conclusions reached below. In contrast, restricting the
amount of allowed fine-tuning to values of around 10 would result in
distributions that show a sharp cutoff at the lower end of the
experimental allowed region of $|\epsilon_K|$, below which no solution
can be found.

As it turns out, the majority of flavor observables is largely
insensitive to the choice of the benchmark scenario. For all these
observables we will only present plots showing the results for a
parameter scan in our standard scenario S1 and describe possible
differences occurring in the remaining benchmark scenarios in words.

\subsection{Quark Mixing Matrices}
\label{sec:ckmnumerics}

In this section we study non-standard effects in the flavor-mixing
matrices appearing in the charged-current sector, focusing on the
generalized CKM matrix. In all cases we adjust the phases of the SM
quark fields according to the standard CKM phase convention
\cite{Amsler:2008zz}, which is defined by the requirements that the
matrix elements $V_{ud}$, $V_{us}$, $V_{cb}$, $V_{tb}$ are real, and
that
\beq \label{eq:standardckm}
   \mbox{Im}\,V_{cs} = \frac{V_{us} V_{cb}}{V_{ud}^2 + V_{us}^2}\,
   \mbox{Im}\,V_{ub} \,.
\eeq
These five conditions fix the phase differences between the six quark
fields uniquely.

The physically most meaningful definition of the CKM matrix is based
on the effective four-fermion interactions induced by the exchange of
the entire tower of $W^\pm$ bosons and their KK excitations. Unlike
the CKM matrix in the SM, the left-handed quark mixing matrix ${\bm{
    \cal V}}_L$ appearing in (\ref{eq:HeffW}) is not a unitary matrix
\cite{Huber:2003tu, Casagrande:2008hr, Cheung:2007bu,
  Buras:2009ka}. Following our previous work
\cite{Casagrande:2008hr}, we consider as two measures of this effect
the deviation from unity of the sum of the squares of the matrix
elements in the first row and the lack of closure of the unitarity
triangle \cite{Bjorken:1988ni}. To quantify the deviations in these
two observables, we define

\beq \label{eq:unitarityviolation} 
\Delta_1^{\rm non} \equiv 1 - \left( |V_{ud}|^2 + |V_{us}|^2 +
  |V_{ub}|^2 \right) \,, \qquad \Delta_2^{\rm non} \equiv 1 +
\frac{V_{ud} V_{ub}^\ast}{V_{cd} V_{cb}^\ast} + \frac{V_{td}
  V_{tb}^\ast}{V_{cd} V_{cb}^\ast} \,,
\eeq 
with 
\beq \label{eq:defVij}
V_{ij} \equiv \left[ 1 + \frac{m_W^2}{2\Mkk^2} \left( 1 - \frac{1}{2L}
  \right) \right]^{-1} ({\cal V}_L)_{ij} \,, \quad i,j = 1, 2, 3 \,.
\eeq
The prefactor in the last relation arises because an extraction of a
CKM element generally involves normalization of the semileptonic
amplitude relative to the Fermi constant $G_F$. The fact that in the
RS model the value of $G_F$ determined from muon decay differs from
the one in the SM by the finite correction (I:141) will play an
important role in what follows. Tests of the unitarity of the CKM
matrix involving the first column or second row and column suffer from
larger experimental uncertainties \cite{Amsler:2008zz} and therefore
cannot compete with the constraints imposed by
(\ref{eq:unitarityviolation}) at present.

\begin{figure}[!t]
  \begin{center}
    \includegraphics[height=6.5cm]{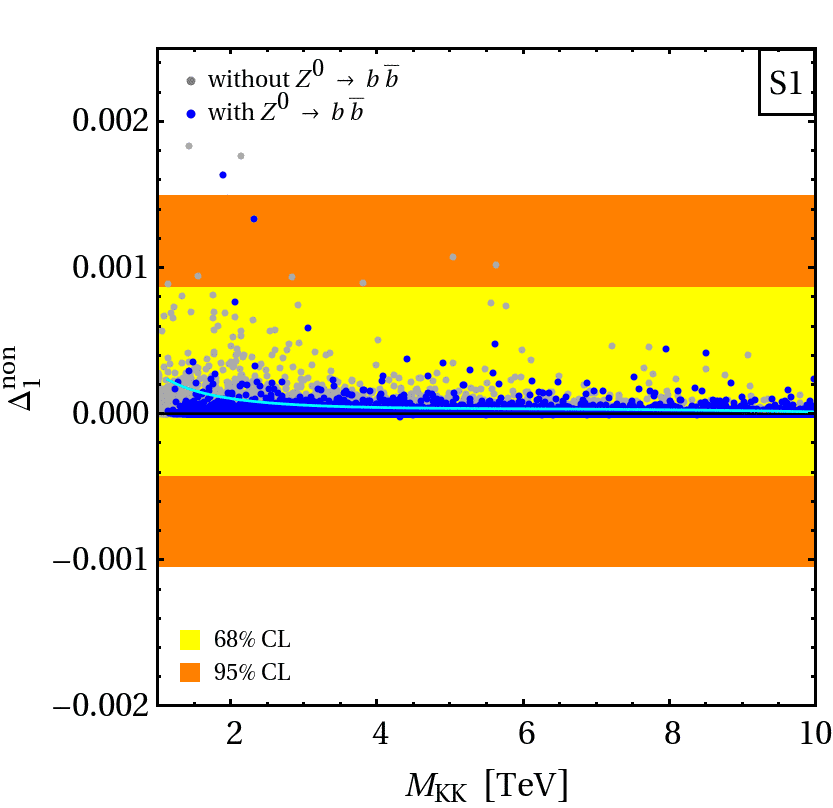}
    \hspace{0.5cm}
    \includegraphics[height=6.35cm]{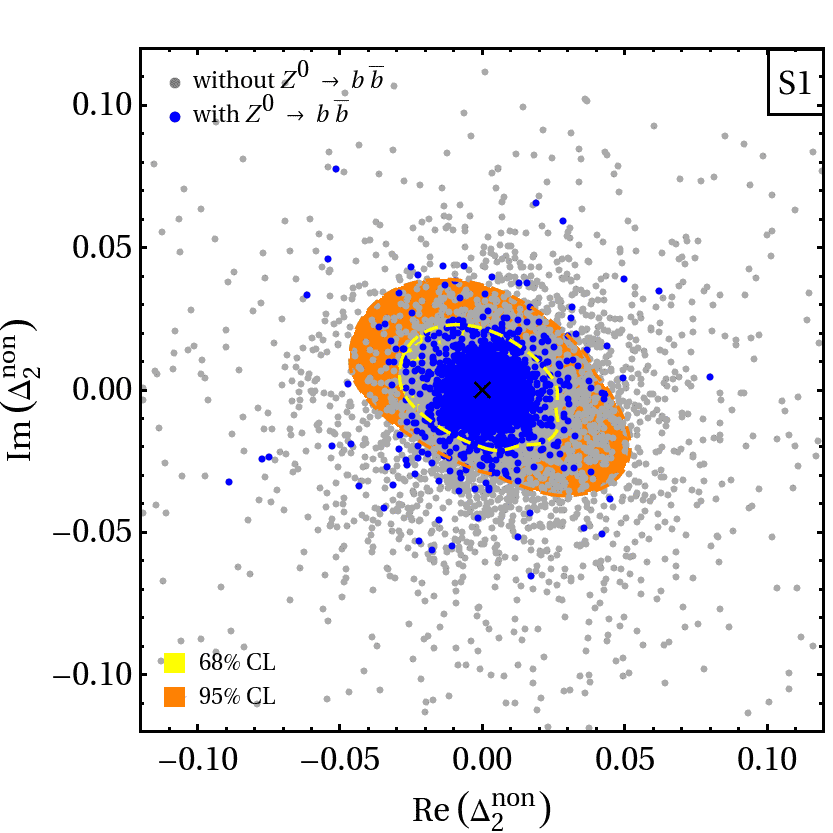}
  \end{center}
  \vspace{-8mm}
  \begin{center}
    \parbox{15.5cm}{\caption{\label{fig:nonunitarity} Predictions for
        $\Delta_1^{\rm non}$ and $\Delta_2^{\rm non}$ in the benchmark
        scenario S1. The blue (light gray) points are consistent
        (inconsistent) with the measured $Z^0 \to b \bar b$ couplings at
        the 99\% CL. The solid black line (left plot) and the black
        cross (right plot) indicate the SM expectations. The cyan
        line illustrates the decoupling behavior with $\Mkk$ obtained
        from a fit to the $99\%$ quantile of the points that are
        consistent with $Z^0 \to b \bar b$. For comparison the
        regions of 68\% (yellow) and 95\% (orange) probability following
        from a combined analysis of $K \to (\pi) l \nu$ branching ratios
        (left) and a global CKM fit (right) are also displayed. See text
        for details.}}
  \end{center}
\end{figure} 

The predictions for $\Delta_1^{\rm non}$ and $\Delta_2^{\rm non}$
obtained in our standard parameter scenario are shown in
Figure~\ref{fig:nonunitarity}, where they are compared with the
current experimental situation. The regions of 68\% and 95\%
probability following from a combined analysis of $K \to (\pi) l \nu$
branching ratios and a global fit to the CKM matrix are indicated by
the shaded bands and the egg-shaped regions. The corresponding central
values and 68\% CL ranges read
\beq \label{eq:expunitarity}
  (\Delta_1^{\rm non})_{\rm exp} = 0.00022 \pm 0.00065 \,, \qquad
  (\Delta_2^{\rm non})_{\rm exp} = (0.001 \pm 0.024) + (0.001 \pm
  0.015) \, i \,.
\eeq
The quoted total uncertainty of $(\Delta_1^{\rm non})_{\rm exp}$ has
been obtained by combining the individual errors $0.00051$ and
$0.00041$ associated to $|V_{ud}|$ and $|V_{us}|$
\cite{Antonelli:2008jg} in quadrature. The prediction for
$(\Delta_2^{\rm non})_{\rm exp}$ has been derived using a customized
version of the CKMfitter package \cite{Charles:2004jd}. The input
parameters entering our global CKM analysis are collected in
Appendix~\ref{app:input}.

The distribution of scatter points in the left plot of
Figure~\ref{fig:nonunitarity} illustrates that unitarity violations in
the first row of the generalized CKM matrix are typically too small to
be observable with present data. The smallness of the corrections to
$\Delta_1^{\rm non}$ has two sources. First, flavor-dependent effects
arising from the $t$-dependent terms in (\ref{eq:calVLR}) are highly
suppressed by products of the zero-mode profiles $F(c_{Q_i})$ and as a
result are negligibly small. Second, the $t$-independent contributions
to (\ref{eq:calVLR}) cancel almost exactly against the multiplicative
factor in (\ref{eq:defVij}) arising from the normalization of the
semileptonic rates to $G_F$. Notice that flavor-independent
corrections also cancel in any ratio of purely leptonic rates
involving light quarks. In consequence, the RS predictions for
$|V_{us}/V_{ud}|$ obtained from the ratio $\Gamma(K \to \mu
\nu)/\Gamma(\pi \to \mu \nu)$ turn out to be basically
indistinguishable from the SM expectation.

A detection of a unitarity violation in the charged-current
interactions by means of $\Delta_2^{\rm non}$ offers much better
prospects. This is illustrated in the right plot of
Figure~\ref{fig:nonunitarity}. We see that the lack of closure of the
unitarity triangle predicted in the RS model, while typically at the
same level as the uncertainty in (\ref{eq:expunitarity}), can also
exceed it. The actual value of $\Delta_2^{\rm non}$ is determined by a
complicated interplay of the contributions arising almost entirely
from the CKM elements $V_{ub}, V_{cb}, V_{td}$, and $V_{tb}$. The
observable $\Delta_2^{\rm non}$ therefore depends very sensitively on
the exact localization of the $SU(2)_L$ quark doublet $(t_L, b_L)$. In
view of the typical size of the corrections to $\Delta_2^{\rm non}$,
an improvement of the determination of the unitarity triangle,
expected from the LHC and later from a super flavor factory, might
allow to detect the non-unitarity of the generalized CKM matrix
induced in the RS framework.

\begin{figure}[!t]
  \begin{center}
    \includegraphics[height=6.5cm]{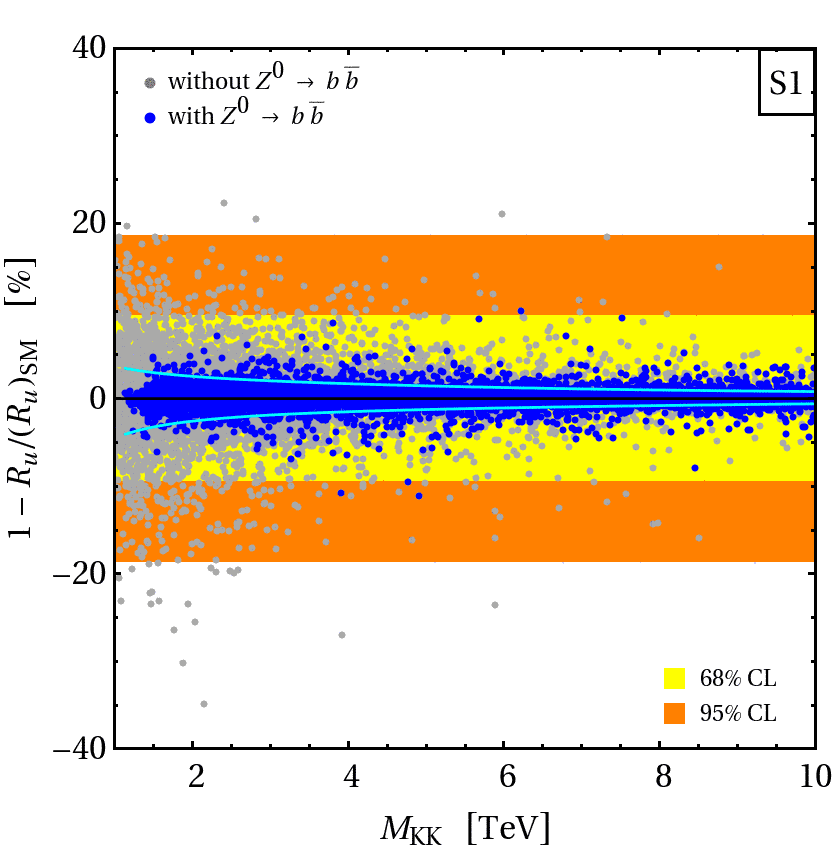}
    \hspace{0.5cm}
    \includegraphics[height=6.5cm]{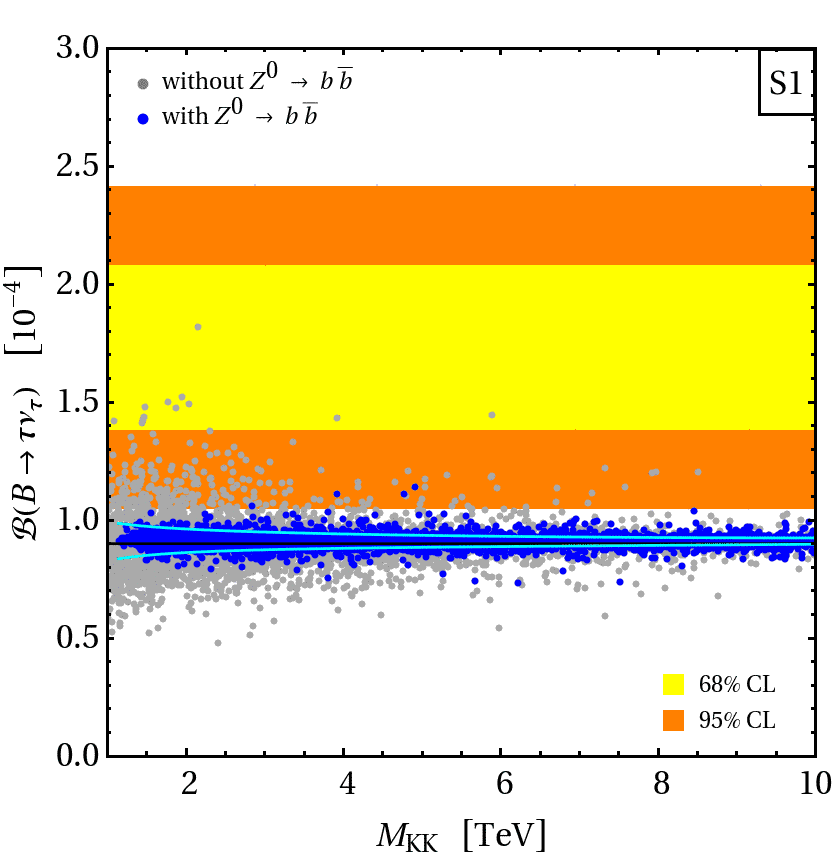}
  \end{center}
  \vspace{-8mm}
  \begin{center}
    \parbox{15.5cm}{\caption{\label{fig:Vij} Relative shift in $R_u$ and
        the branching ratio for $B \to \tau \nu_\tau$ as a function of
        $\Mkk$ in the S1 parameter scenario. The blue (light gray)
        points are consistent (inconsistent) with the measured $Z^0 \to
        b \bar b$ couplings at the 99\% CL. The solid black lines
        indicate the SM predictions. The cyan lines illustrate
        the decoupling behavior with $\Mkk$ obtained from a fit to the
        $1\%$ (lower lines) and $99\%$ (upper lines) quantile of the
        points that are consistent with $Z^0 \to b \bar b$. The
        regions of 68\% (yellow) and 95\% (orange) probability following
        from a combination of semileptonic $B$ decay data (left) and the
        BaBar and Belle measurements of the $B \to \tau \nu_\tau$
        branching ratio (right) are also displayed. See text for
        details.}}
  \end{center}
\end{figure}

We now turn our attention to the direct extraction of the CKM elements
$|V_{ub}|$, $|V_{cb}|$, and $|V_{tb}|$. The value of $|V_{ub}/V_{cb}|$
can be determined from the relative ratio of charmless over charmed
semileptonic $B$ decays. It also enters the expression for the
quantity
\beq 
  R_u \equiv \left | \frac{V_{ud} V_{ub}^\ast}{V_{cd} V_{cb}^\ast}
  \right | = \left ( 1 - \frac{\lambda^2}{2} \right ) \frac{1}{\lambda}
  \left | \frac{V_{ub}}{V_{cb}} \right | ,
\eeq
which measures the length of the side of the rescaled unitarity
triangle opposite to the angle $\beta$. Notice that in order to arrive
at the final expression we have used that the ratio $V_{ud}/V_{cd}$ in
the RS model is to an excellent approximation given by its SM value.

A combination of the present experimental information coming from
inclusive and exclusive measurements of semileptonic $B$ decays gives
\cite{Amsler:2008zz} \beq (R_{u})_{\rm exp} = 0.42 \pm 0.04 \,.  \eeq
The quoted total uncertainty should be compared with the relative
corrections to $R_u$ predicted in the RS framework. The results of a
parameter scan in the standard scenario are displayed on the left in
Figure~\ref{fig:Vij}. It is evident from the plot that the relative
corrections in the ratio $|V_{ub}/V_{cb}|$ are typically smaller than
the current combined experimental and theoretical accuracy of about
10\%. Similar statements apply to the direct extraction of the
elements $|V_{ub}|$ and $|V_{cb}|$ from semileptonic and purely
leptonic $B$ decays. In the case of $|V_{ub}|$ this is illustrated by
the right plot in Figure~\ref{fig:Vij}, which shows the standard
benchmark scenario predictions for the $B \to \tau \nu_\tau$ branching
ratio. We see that in the RS framework it is not possible to explain
the $2.1 \sigma$ deviation between the value (\ref{eq:BtaunuSM}),
following from a global fit to the unitarity triangle, and the current
experimental world average indicated by the shaded bands. The typical
relative corrections in $|V_{cb}|$ do not exceed the 7\% (8\%) level
for scatter points that satisfy (violate) the $Z^0 \to b \bar b$
constraint, and the observed pattern of deviations follows the one
seen in $R_u$ and the branching ratio of $B \to \tau \nu_\tau$.

A direct determination of $|V_{tb}|$ without assuming unitarity is
possible from measurements of the single top-quark production cross
section. {\color{black} The measured cross sections $\sigma (p \bar p
  \to tb + X, tqb + X) = ( 2.3^{+0.6}_{-0.5} ) \, {\rm pb}$
  \cite{Aaltonen:2009jj} and $\sigma (p \bar p \to tb + X, tqb + X) =
  (3.94 \pm 0.88) \, {\rm pb}$ \cite{Abazov:2009ii} translate into
  the limits $|V_{tb}| > 0.78$ and $|V_{tb}| > 0.71$ at 95\% CL by CDF
  and {D\O}, respectively.} We find that in the RS model the
prediction for $|V_{tb}|$ is strictly below the corresponding value in
the SM.  The relative corrections are however safely within the
current experimental limit, as they do not surpass the level of 1\%
(15\%) for parameter points that satisfy (violate) the constraint
imposed by the precision measurement of the $Z^0 \to b \bar b$
couplings. The value of $|V_{tb}|$ will be further constrained as soon
as the LHC will start to take data. Simulation studies by ATLAS
\cite{ATLAS:1999fr} and CMS \cite{Ball:2007zza} suggest that the
cross section of the most promising single-top-production channel,
namely $p p \to tqb + X$, is measurable with a total error of 10\%,
which implies that $|V_{tb}|$ can be determined with 5\% accuracy. At
this level of precision one becomes sensitive to non-standard effects
affecting the $Wt_Lb_L$ coupling in the RS framework.

For the sake of completeness, we mention that the non-unitarity of the
CKM matrix has been analyzed previously in \cite{Huber:2003tu,
  Cheung:2007bu}. However, a thorough discussion of all relevant
effects is missing in these articles. On the other hand, a detailed
discussion of the breakdown of the unitarity of the quark mixing
matrix in the framework of a RS model with custodial protection has
been presented very recently in \cite{Buras:2009ka}. Unfortunately,
in the latter paper the CKM matrix is defined via the $W u_{i
  \hspace{0.25mm} L} d_{j \hspace{0.25mm} L}$ vertex and not the
effective four-fermion interactions induced by the exchange of the
entire tower of the $W^\pm$ boson and its KK excitations. As explained
in detail above, defining $V_{ij}$ as part of the charged-current
coupling is unphysical, since in practice it is impossible to
determine the CKM elements without reference to $G_F$.  The exact
meaning of the results obtained in \cite{Buras:2009ka} thus remains
cloudy, preventing us from a straightforward comparison with our
findings.

Before proceeding, we would like to add two further comments. First,
in Figures \ref{fig:nonunitarity}~and~\ref{fig:Vij} we have only shown
predictions corresponding to our standard benchmark scenario. Plots
for the remaining parameter scenarios have been omitted since they
look essentially the same. Second, we note that the correlations
between the $W u_{i \hspace{0.25mm} L} b_L$ and $W t_L d_{i
  \hspace{0.25mm} L}$ vertices and the $Z^0 b_L \bar b_L$ coupling are
in general different in warped extra-dimension models with an extended
electroweak sector, because the custodial symmetry cannot
simultaneously protect all of these vertices \cite{Agashe:2006at}. We
therefore expect that, depending on the exact realization of the
model, the corrections to the $W u_{i \hspace{0.25mm} L} b_L$ and $W
t_L d_{i \hspace{0.25mm} L}$ couplings could be less correlated with
the ones appearing in the $Z^0 b_L \bar b_L$ vertex. To illustrate
this possibility, we have shown in the plots of Figures
\ref{fig:nonunitarity} and \ref{fig:Vij} scatter points that fulfill
(blue) and violate (light gray) the constraints following from the
measurements of the $Z^0 \to b \bar b$ couplings. It is apparent that
removing the stringent $Z^0 \to b \bar b$ constraint leads to much
brighter experimental prospects for a detection of a unitarity
violation of the generalized CKM matrix or deviations in the branching
ratios of semileptonic and purely leptonic $B$ decays. Detailed
analyses of the effective charged-current interactions in extended RS
models would however be required to make definite statements about how
much these prospects would improve.

\subsection{Neutral-Meson Mixing}

The observation of neutral-meson mixing has played a central role in
unraveling the flavor structure of the SM. While these glory days have
passed, even today measurements of the magnitudes and phases of
$\Delta F = 2$ amplitudes provide some of the most stringent
constraints on models of physics beyond the SM. Due to enhancements in
the RG evolution and in the matrix elements, these constraints turn
out to be particularly severe for scenarios that generate transitions
between quarks of different chiralities. As a result, in such models
the scale of new physics is typically pushed to values beyond the
reach of direct searches at the LHC. The questions that we want to
address in this section is how severe the constraints from $\Delta F =
2$ observables are in the RS framework, and if it is possible to
ameliorate them so as to build viable models with a warped extra
dimension that are accessible at the LHC.

\boldmath \subsubsection{Numerical Analysis of $K$--$\bar K$ Mixing}
\unboldmath \label{sec:numericsKKmixing}

Let us begin this section by giving explicit results for the various
neutral-meson mixing amplitudes in terms of the Wilson coefficients in
(\ref{eq:Cmix}). These formulas will enable us to understand which
Wilson coefficient generically furnishes the dominant contribution to
a given $\Delta F = 2$ observable. Making the dependence on the
matching scale $\mu_{\rm KK}$ explicit, we find the following
approximate formula for the $K$--$\bar K$ mixing amplitude,
\beq \label{eq:HeffKKapp}
  \left \langle K^0 \left | {\cal H}_{\rm eff, RS}^{\Delta S = 2} \right
    | \bar K^0 \right \rangle \, \propto \, C_1^{\rm RS} + \widetilde
  C_1^{\rm RS} + 114.8 \, \bigg (1 + 0.14 \, \ln \left ( \frac{\mu_{\rm
        KK}}{3 \, {\rm TeV}} \right ) \! \!  \bigg ) \left( C_4^{\rm RS}
    + \frac{C_5^{\rm RS}}{3.1} \right) ,
\eeq
where the large scale-independent coefficient consists of a factor of
about 15 arising from the chiral enhancement of the hadronic matrix
elements \cite{Beall:1981ze, Gabbiani:1996hi}, and a factor of about
8 due to the RG evolution \cite{Bagger:1997gg, Ciuchini:1998ix,
  Buras:2000if} from $3 \, {\rm TeV}$ down to $2 \, {\rm GeV}$.
Assuming that all Wilson coefficients are of similar size, this
relation implies that $C_4^{\rm RS}$ gives the dominant contribution
to both $\Delta m_K$ and $|\Delta\epsilon_K|$ in the RS framework. In
the case of $B_{d,s}$--$\bar B_{d,s}$ and $D$--$\bar D$ mixing the
chiral enhancement of the matrix elements of the mixed-chirality
operators is much less pronounced and amounts to at most a factor of
2. Taking further into account that for bottom and charm mesons the
Wilson coefficients are only evolved down to $4.6 \, {\rm GeV}$ and
$2.8 \, {\rm GeV}$, respectively, the appropriate factors that replace
the numerical coefficients $114.8$ and $3.1$ in the expression
(\ref{eq:HeffKKapp}) read $7.4$ and $2.7$ ($13.3$ and $2.9$) in the
$B_{d,s}$ ($D$) sector. These moderate enhancement factors suggest
that in the case of the $B_{d,s}$--$\bar B_{d,s}$ and $D$--$\bar D$
mixing amplitudes, the contribution from the sum of $C_1^{\rm RS}$ and
$\widetilde{C}_1^{\rm RS}$ can compete with the one arising from
$C_4^{\rm RS}$ and $C_5^{\rm RS}$. Our numerical analysis will confirm
this model-independent conclusion.

\begin{figure}[!t]
  \begin{center}
    \includegraphics[height=6.5cm]{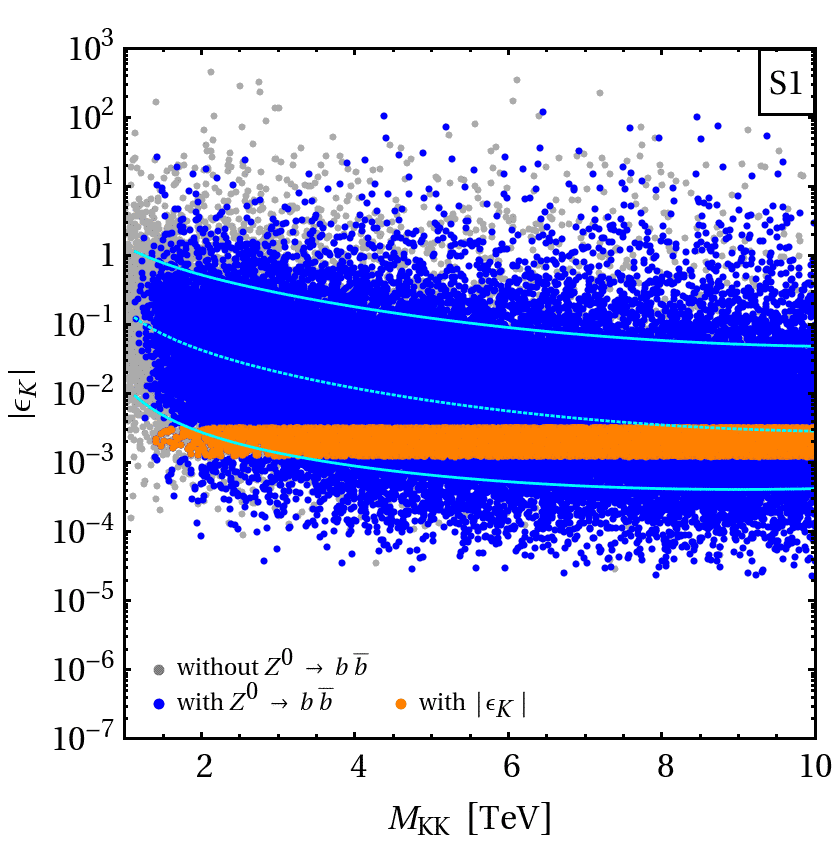}
    \hspace{0.75cm}
    \includegraphics[height=6.5cm]{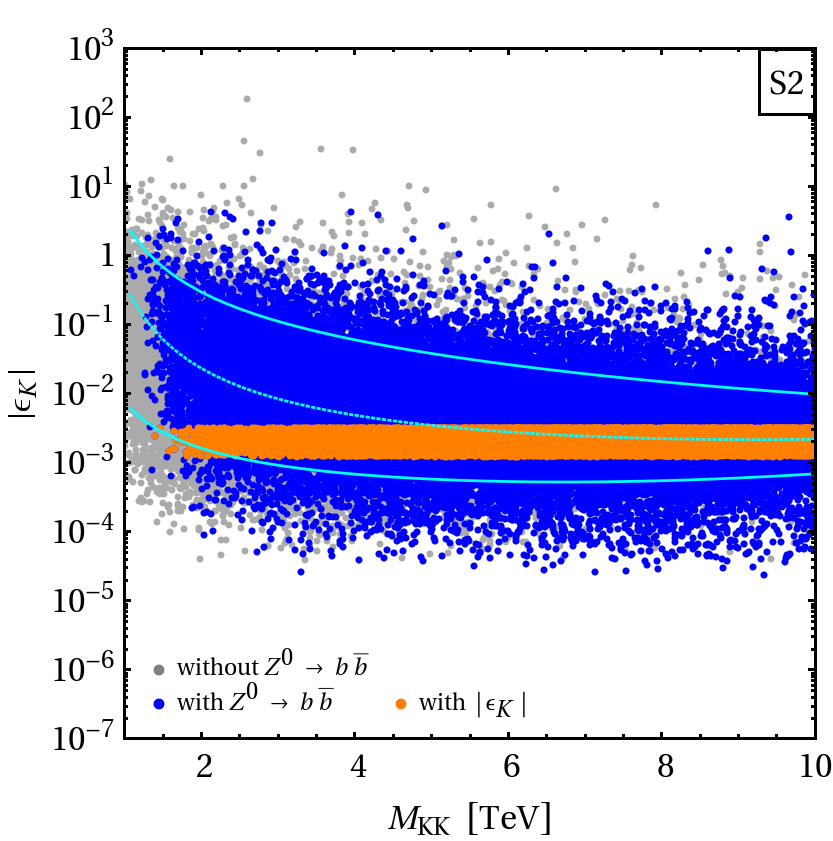}

    \vspace{4mm}

    \includegraphics[height=6.5cm]{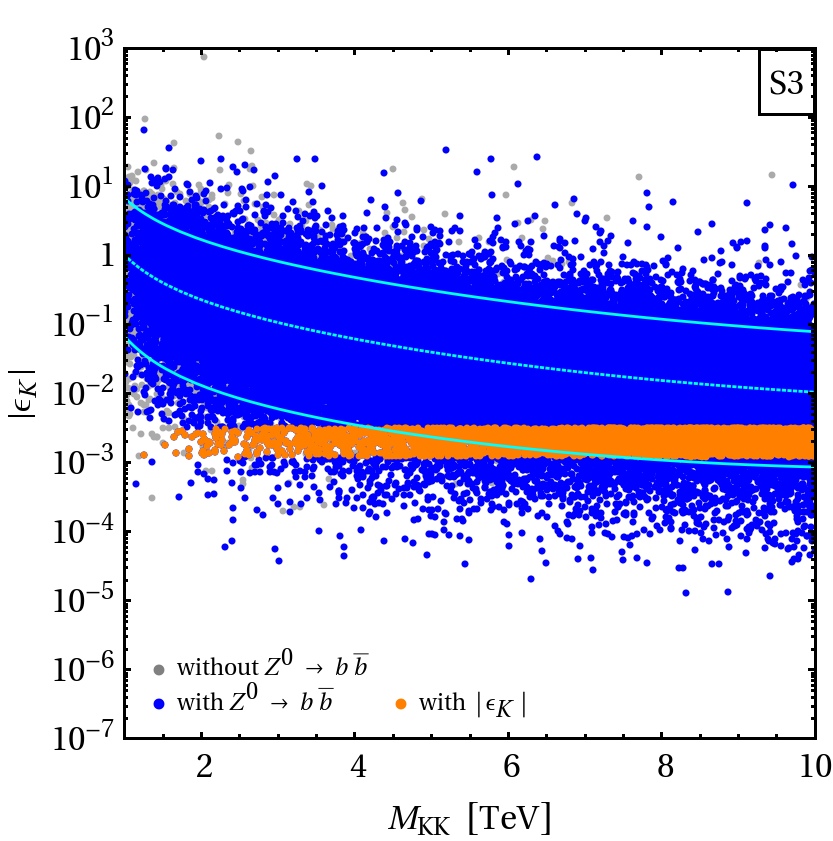}
    \hspace{0.75cm}
    \includegraphics[height=6.5cm]{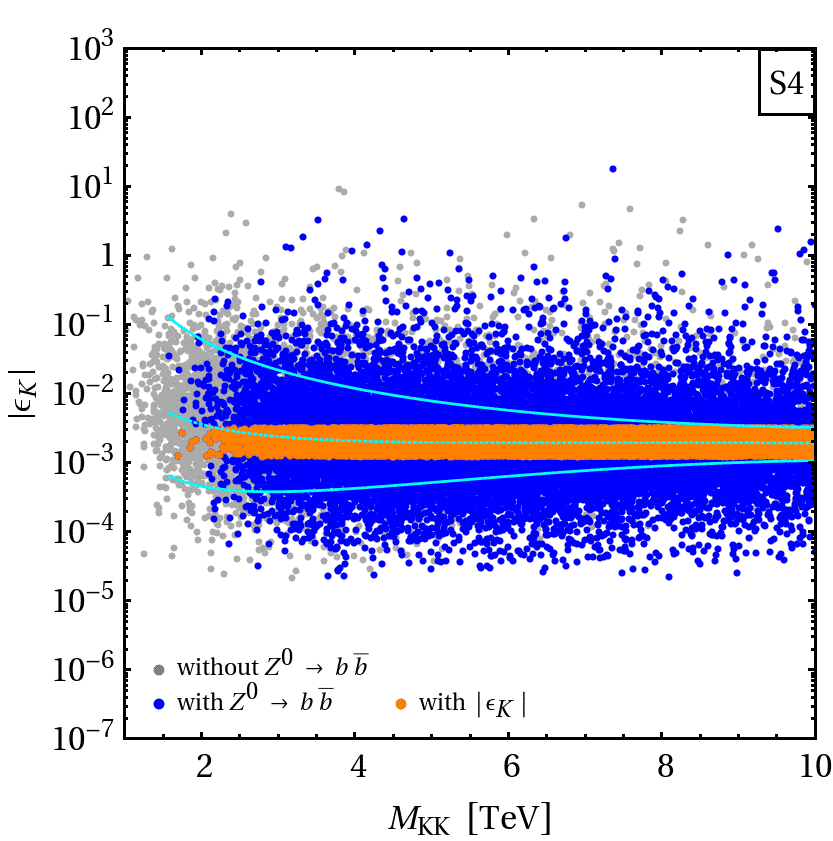}
  \end{center}
  \vspace{-8mm}
  \begin{center}
    \parbox{15.5cm}{\caption{\label{fig:epsK} Predictions for
        $|\epsilon_K|$ as a function of $\Mkk$ in the different
        benchmark scenarios. The blue (light gray) points are consistent
        (inconsistent) with $Z^0 \to b \bar b$ at the 99\% CL. The
        subset of points that is in agreement with both the $Z^0 b \bar
        b$ couplings and the experimental value of $|\epsilon_K|$ at
        95\% CL is colored orange. The cyan lines illustrate the
        decoupling behavior with $\Mkk$ obtained from a fit to the
        $5\%$ (lower solid), $50\%$ (middle dashed), and $95\%$ (upper
        solid lines) quantile. See text for details.}}
  \end{center}
\end{figure}

We are now ready to analyze the impact of new-physics effects on
$K$--$\bar K$ mixing in the RS framework. The predictions for
$|\epsilon_K|$ in the four different benchmark scenarios introduced in
Section \ref{sec:benchmarks} are displayed in
Figure~\ref{fig:epsK}. All shown points reproduce the correct quark
masses and CKM parameters within errors. Scatter points that are
consistent with $Z^0 \to b\bar b$ are colored blue, while the subset
of scatter points that in addition satisfies $|\epsilon_K| \in [1.2,
3.2] \cdot 10^{-3}$ is colored in orange. The latter constraint
guarantees that the theoretical prediction for $|\epsilon_K|$ obtained
by adding the SM and RS contributions is consistent with the
measurement at 95\% CL after combining theoretical and errors. The
decoupling behavior of the RS corrections to $|\epsilon_K|$ is
illustrated by the cyan dashed line which represents the fit to the
median value of $|\epsilon_K|$ for fixed $\Mkk$. While the average
value of $|\epsilon_K|$ becomes consistent with the measurement only
for $\Mkk \gtrsim 8 \,\text{TeV}$, the $5\%$ quantile (lower cyan
solid line) crosses the experimentally allowed range already at about
$2\,\text{TeV}$. For such low values of $\Mkk$ one can immediately see
from all three panels that the values of $|\epsilon_K|$ are
typically a factor of about 100 bigger than the SM prediction
\cite{Csaki:2008zd, Blanke:2008zb, Agashe:2008uz,
  Bauer:2008xb, Santiago:2008vq}. This is the generic case, for which
the Wilson coefficient $C_4^{\rm RS}$ gives the by far largest
contribution to $|\epsilon_K|$. The order-of-magnitude enhancement of
$|\epsilon_K|$ is explained by the following observations. First, it
turns out that even for low KK scales the magnitude of the RS
contribution to $\Delta m_K$ typically does not exceed the SM
contribution by an unacceptably large amount (given the huge SM
uncertainties). Second, the ratio of imaginary to real part of the
$K$--$\bar K$ mixing amplitude is strongly suppressed in the SM due to
the smallness of ${\rm Im} \, (V_{ts}^\ast V_{td})$. Numerically, one
finds $({\rm Im} \, M_{12}^{K \ast}/{\rm Re} \, M_{12}^{K \ast})_{\rm
  SM} \approx -2 A^4 \lambda^{10} (1 - \bar{\rho}) \hspace{0.2mm} \bar
\eta \, \eta_{tt} S_0 (m_t^2/m_W^2)/(\lambda^2 \, \eta_{cc} \,
m_c^2/m_W^2 ) \approx -1/150$, where $S_0(m_t^2/m_W^2) \approx 2.35$
originates from the one-loop box diagrams containing a top quark,
while $\eta_{tt} \approx 0.57$ \cite{Buras:1990fn} and $\eta_{cc}
\approx 1.50$ \cite{Herrlich:1993yv} summarize higher-order QCD
corrections in the top- and charm-quark sectors. Under the natural
assumption that the phase in $(M_{12}^{K \ast})_{\rm RS}$ is of
$\ord(1)$, it follows then that $|\epsilon_K|_{\rm
  RS}/|\epsilon_K|_{\rm SM} = |({\rm Im} \, M_{12}^{K \ast})_{\rm
  RS}/({\rm Im} \, M_{12}^{K \ast})_{\rm SM}| \sim |({\rm Re} \,
M_{12}^{K \ast})_{\rm RS}/({\rm Im} \, M_{12}^{K \ast})_{\rm SM}| \sim
|({\rm Re} \, M_{12}^{K \ast})_{\rm SM}/({\rm Im} \, M_{12}^{K
  \ast})_{\rm SM}| \sim 100$. Large deviations of $|\epsilon_K|_{\rm
  RS}$ from this generic value require a tuning of parameters, for
example an adjustment of the phase of the associated combination of
Yukawa couplings so that $({\rm Im} \, M_{12}^{K \ast})_{\rm RS}/({\rm
  Im} \, M_{12}^{K \ast})_{\rm SM}$ becomes of $\ord{(1)}$.

\begin{figure}[!t]
  \begin{center}
    \includegraphics[height=6.5cm]{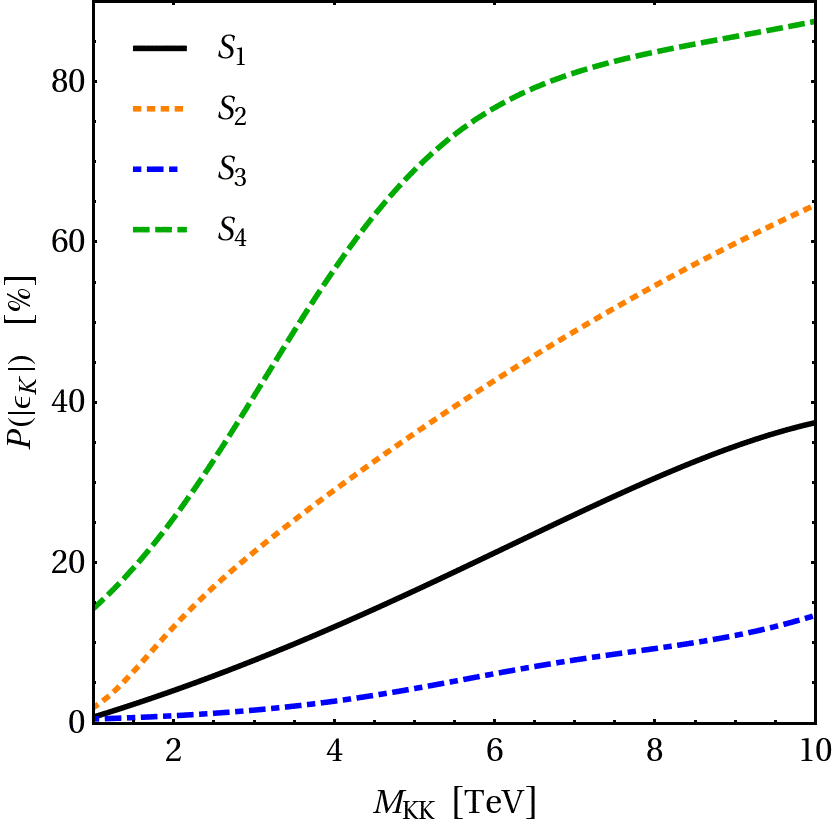}
    \hspace{0.75cm}
    \includegraphics[height=6.5cm]{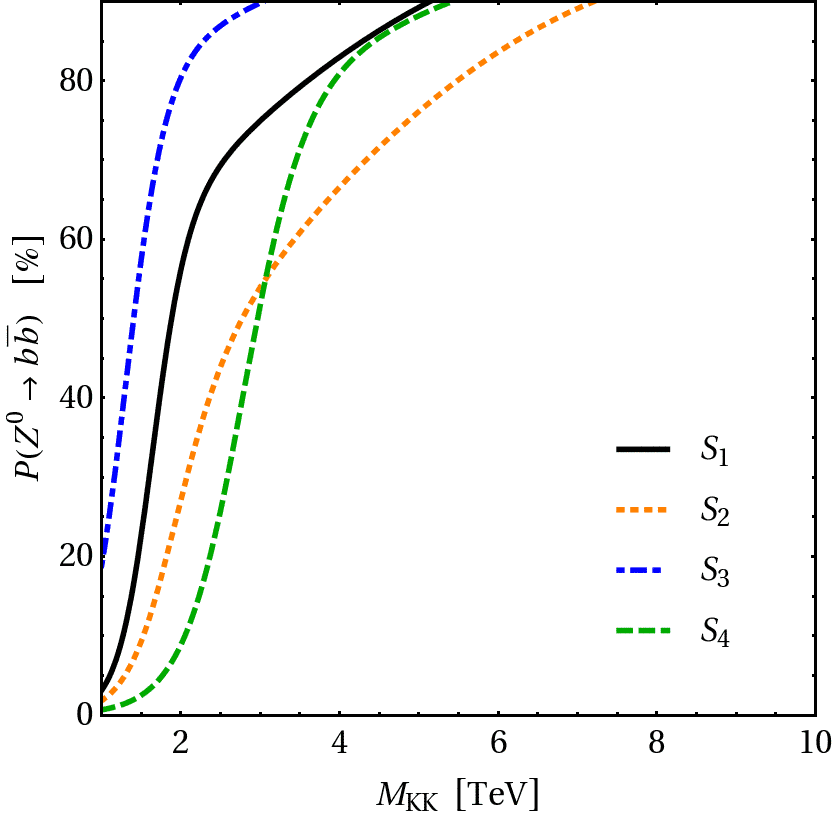}

    \vspace{4mm}

    \includegraphics[height=6.5cm]{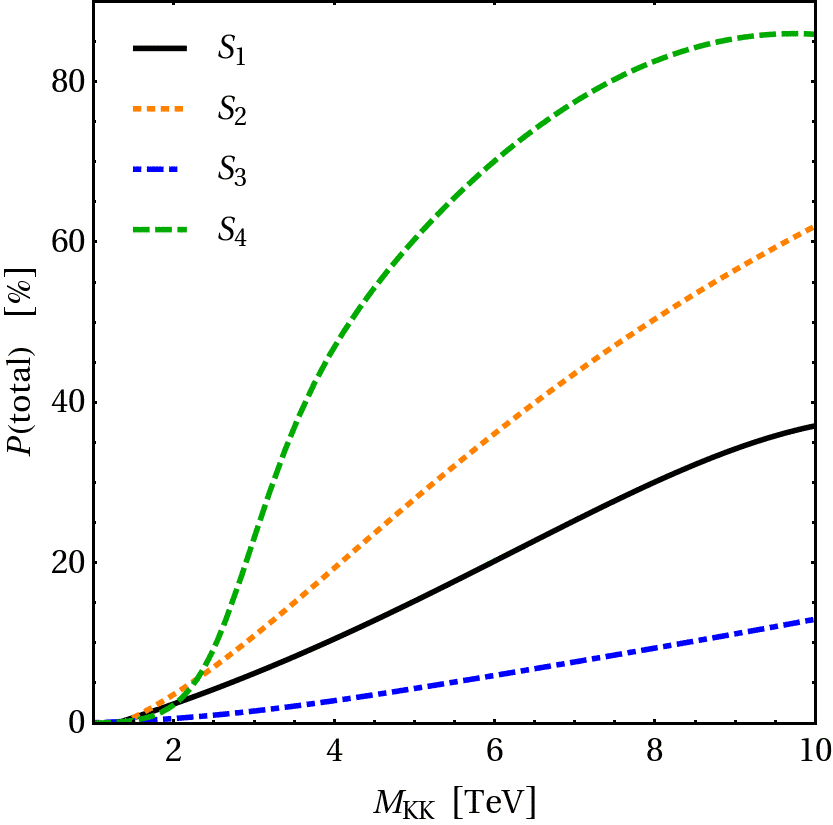}
  \end{center}
  \vspace{-8mm}
  \begin{center}
    \parbox{15.5cm}{\caption{\label{fig:tuning} Percentage of scatter
        points as a function of $\Mkk$ that are consistent with the
        experimental values of $|\epsilon_K|$ (upper left), $Z^0 \to b
        \bar b$ (upper right), and both constraints (lower panel). The
        shown lines correspond to our four different parameter
        scenarios. See text for details.}}
  \end{center}
\end{figure}

In order to achieve an acceptable amount of CP violation in the kaon
sector, the contribution arising from $C_4^{\rm RS}$ needs to be
suppressed. One possibility to protect $|\epsilon_K|$ from
excessive corrections is to arrange for common bulk mass parameters
$c_d \equiv c_{d_i}$ in the sector of the right-handed down-type
quarks \cite{Santiago:2008vq}. This subclass of models is considered
in our benchmark scenario S2. In the case of ``alignment''
of down-type quark bulk mass parameters, certain mixing matrices
differ from the ones present in the RS model with ``hierarchical''
$c_{d_i}$ parameters. The phenomenologically most important change
occurs in the case of $(\widetilde{\Delta}_D)_{mn} \otimes
(\widetilde{\Delta}_d)_{mn}$, which becomes quasi-diagonal. To derive
an expression for the off-diagonal elements one has to expand the
flavor matrices defined in (\ref{Deltaotimes}) up to $\ord
(v^2/\Mkk^2)$. Employing (I:70), we find that the relation
(\ref{eq:DDcomponent}) receives the correction
\beq \label{eq:ddcomponent}
\begin{split}
  (\widetilde{\delta}_D)_{mn} \otimes (\widetilde{\delta}_d)_{mn} & =
  \frac{m_{d_m} m_{d_n}}{\Mkk^2} \left [ \big( U_d^\dagger \big)_{mi}
    \hspace{0.5mm} \big( U_d \big)_{in} \hspace{0.5mm}
    (\widetilde{\delta}_D)_{ij} \hspace{0.5mm} \big( U_d^\dagger
    \big)_{m j} \hspace{0.5mm} \big( U_d \big)_{j n} \right. \\ &
  \left. \hspace{2.25cm} + \big( W_d^\dagger \big)_{mi} \hspace{0.5mm}
    \big( W_d \big)_{in} \hspace{0.5mm} (\widetilde{\delta}_d)_{ij}
    \hspace{0.5mm} \big( W_d^\dagger \big)_{m j} \hspace{0.5mm} \big(
    W_d \big)_{j n}\right ]
\end{split}
\eeq 
from the terms in (\ref{Deltaotimes}) involving both even
$\bm{C}^{(A)}_m (\phi)$ and odd $\bm{S}^{(A)}_m (\phi)$ fermion
profiles. Here a summation over the indices $i,j$ is
understood. Neglecting terms suppressed by $F^2(c_{Q_i})$ and
$F^2(c_{Q_i}) F^2(c_{Q_j})$, the elements of
$(\widetilde{\delta}_D)_{ij}$ take the form
\beq \label{eq:dDij}
  (\widetilde{\delta}_D)_{ij} = \frac{2 (3 + c_{Q_i} - c_{Q_j})}{(3 +
    2 c_{Q_i}) (3 - 2 c_{Q_j}) (2 + c_{Q_i} - c_{Q_j}) } \,
  \frac{F^2(c_{Q_i})}{F^2(c_{Q_j})} \,.
\eeq
An analogous expression holds in the case of
$(\widetilde{\delta}_d)_{ij}$ with $c_{Q_i}$ replaced by
$c_{d_i}$. Since all bulk mass parameters are close to $-1/2$ it is
also a good approximation to replace the rational function of
$c_{Q_i}$ and $c_{Q_j}$ in (\ref{eq:dDij}) by the numerical factor
3/8.

Using the scaling relations (I:96) and (\ref{eq:UqWq}), it is
straightforward to deduce from (\ref{eq:DDcomponent}) and
(\ref{eq:ddcomponent}) that to leading order in hierarchies one has
\begin{equation} \label{eq:DDDdscaling}
(\widetilde{\Delta}_D)_{mn} \otimes
(\widetilde{\Delta}_d)_{mn} \sim 
\begin{cases}
  \, F(c_{Q_m}) F(c_{Q_n}) F(c_{d_m}) F(c_{d_n})
  \,, & \text{``hierarchical''} \,, \\ \\
  \, F(c_{Q_m}) F(c_{Q_n}) F^2(c_d) \left ( \delta_{mn} +
    \displaystyle \frac{Y_d^2 v^2}{2 \Mkk^2} \right ) , &
  \text{``aligned''} \,.
\end{cases}
\end{equation}
For clarity we have not shown higher-order terms in $v/\Mkk$ in the
flavor-diagonal contribution. Notice that the $\ord (v^2/\Mkk^2)$
correction arising in the "aligned" case is solely due to the first
term in (\ref{eq:ddcomponent}) involving left-handed rotations
$\bm{U}_d$. Compared to the latter correction, the contribution from
the second term in (\ref{eq:ddcomponent}) is further suppressed, since
the universality of $(\widetilde{\delta}_d)_{ij} \sim 1$ in
combination with the unitarity of the $\bm{W}_d$ matrices renders it
negligibly small.

The relation (\ref{eq:DDDdscaling}) implies that there is a
suppression of the Wilson coefficient $C_4^{\rm RS}$ in the
``aligned'' relative to the ``hierarchical'' case. We find the
following scaling
\beq \label{eq:C4ratio}
  \frac{(C_4^{\rm RS})_{\text{``aligned''}}}{(C_4^{\rm RS})_{\text
      {``hierarchical''}}} \, \sim \, \frac{Y_d^2 v^2}{2 \Mkk^2}
  \frac{F(c_{d_1})}{F(c_{d_2})} \, \sim \, \frac{Y_d^2 v^2}{2 \Mkk^2}
  \frac{m_d}{m_s} \frac{1}{\lambda} \, \approx \, 8 \cdot 10^{-3} \,,
\eeq
where in the first step we used that $F(c_d) \approx F(c_{d_2})$, in
the second employed the scaling relation (I:107), and finally set
$\Mkk = 3 \, {\rm TeV}$ and $Y_d = 3$ to obtain the numerical
estimate. Our results (\ref{eq:DDDdscaling}) and (\ref{eq:C4ratio})
agree with the findings in \cite{Santiago:2008vq} derived by means of
the mass-insertion approximation. The equivalence of our exact
approach and the mass-insertion approximation follows from the ``IR
dominance'' of the overlap integrals (\ref{Deltaotimes}) and the
boundary conditions in (I:63), which connect the odd $\bm{S}^{(A)}_n
(\pi^-)$ with the even $\bm{C}^{(A)}_n (\pi)$ fermion profiles. While
the suppression factor of $\ord(100)$ appearing in (\ref{eq:C4ratio})
looks very promising in itself, only a rigorous numerical analysis
including all contributions can tell us to which extent RS
realizations with ``alignment'' allow one to relax the stringent
constraint from $|\epsilon_K|$. The results of such an analysis are
depicted in the upper right panel of Figure~\ref{fig:epsK}, which
shows a visible improvement relative to the benchmark scenario S1.

To quantify the improvement, we calculate the fraction
$P(|\epsilon_K|)$ of points fulfilling the $|\epsilon_K|$ constraint
by dividing the range $\Mkk \in [1, 10] \, {\rm TeV}$ into bins of
$100 \, {\rm GeV}$ and fitting the binned data using an appropriate
function. In the case at hand, we find that the combination of a Fermi
function and a second-order polynomial is perfectly suited to
reproduced the observed decoupling behavior.\footnote{It is not
  possible to attach a rigorous statistical meaning to the calculated
  fractions $P(|\epsilon_K|)$, $P(Z^0 \to b \bar b)$, and
  $P(\rm{total})$, since they depend on the way in which the results are
  binned and the fit is performed. The same statement applies to the
  fraction $P(\epseps)$ introduced below.} From the upper left panel in
Figure~\ref{fig:tuning} it is apparent that, relative to the standard
scenario S1, in scenario S2 corresponding to the ``aligned'' case it
is easier to obtain a consistent description of the experimental data
on $|\epsilon_K|$. Numerically we find that compared to our default
scenario S1, in the benchmark scenario S4 the fraction of points that
satisfy the latter constraint is enhanced from 19\% to 38\%, and that
the bound from requiring $P(|\epsilon_K|) > 10\%$ is lowered from
$\Mkk > 3.6 \, {\rm TeV}$ to $\Mkk > 1.9 \, {\rm TeV}$. We have also
studied the effect of a slight ``misalignment'' of the bulk mass
parameters $c_d$, which effectively simulates the presence of flavor
non-universal brane kinetic terms. In agreement with
\cite{Santiago:2008vq}, we find that the obtained results depend very
sensitively on the exact amount of non-universality and that already
small deviations from $c_{d_1} = c_{d_2} = c_{d_3}$ can easily spoil
the ${\cal O} (v^2/\Mkk^2)$ suppression in (\ref{eq:C4ratio}). One
also observes from the orange dotted line in the upper right panel of
Figure~\ref{fig:tuning} that the $Z^0 b \bar b$ constraint turns out
to be more stringent in the ``aligned'' RS framework. This feature is
easy to explain. In RS realizations with ``alignment'', the singlet
$b_R$ lives further away from the IR brane compared to the original
``hierarchical'' case. Typically one has $c_{b_R} = c_d \approx -0.60$
instead of $c_{b_R} \approx -0.58$. In order to obtain the correct
value of the bottom-quark mass, the doublet $(t _L, b_L)$ thus needs
to be localized more closely to the IR brane. This is not a favorable
option, since placing $(t_L, b_L)$ too close to the IR brane generates
an unacceptably large correction to the $Z^0 b_L \bar b_L$
coupling. Quantitatively, we find that demanding $P(Z^0 \to b \bar b)
> 25\%$ sets the lower limit $\Mkk > 2.0 \, {\rm TeV}$, what is
slightly higher than $\Mkk > 1.6 \, {\rm TeV}$ in the standard
scenario S1.  Notice that in warped extra-dimension models with
custodial protection of the $Z^0 b_L \bar b_L$ vertex this problem is
not present, because the corrections to the $Z^0 \to b \bar b$
couplings are small and largely independent of the bulk mass
parameters. Nevertheless, as demonstrated by the orange dotted curve
in the lower plot of Figure~\ref{fig:tuning}, universal bulk mass
parameters $c_d$ nevertheless help to lower the bound on the KK scale
following from a combination of constraints. In order for the fraction
$P(\text{total})$ of the points that fulfill both constraints to reach
10\%, one has to require $\Mkk > 3.9 \, {\rm TeV}$ in scenario S1,
whereas the weaker limit $\Mkk > 2.9 \, {\rm TeV}$ applies in the case
of benchmark scenario S4.

The presence of the ``volume factor'' $L$ in (\ref{eq:Cmix}) suggests
that another alternative cure for the ``flavor problem'' in the
$\Delta S = 2$ sector is to reduce the UV cutoff sufficiently below
the Planck scale. Indeed, the alleviation of the $|\epsilon_K|$
constraint has been mentioned as one of the attractive features of
``little'' RS scenarios \cite{Davoudiasl:2008hx}. However, a careful
analysis reveals that the naive conjecture of a reduction of terms
proportional to $L$ is flawed in the case of $|\epsilon_K|$
\cite{Bauer:2008xb}. This feature is illustrated by the lower left
and upper left plots Figures~\ref{fig:epsK} and \ref{fig:tuning}. We
see that in scenario S3 featuring $L = \ln (10^3) \approx 7$ instead
of $L = \ln (10^{16}) \approx 37$, the percentage of points that are
in agreement with the measured value of $|\epsilon_K|$ amounts to only
5\% and that, in particular $P(|\epsilon_K|)$ is strictly smaller than
in the standard scenario S1 for all values of $\Mkk$.
The origin of the enhancement of flavor-changing
$\Delta S = 2$ effects is the phenomenon of ``UV dominance'', which
arises whenever the bulk mass parameters determining the strange-quark
mass satisfy the critical condition $c_{Q_2} + c_{d_2} < -2$
\cite{Bauer:2008xb}. The relevant overlap integrals
(\ref{Deltaotimes}) are then dominated by the region near the UV
brane, thereby partially evading the RS-GIM mechanism. New physics
contributions to $|\epsilon_K|$ are then exponentially enhanced with
respect to the standard scenario S1 with $L = \ln (10^{16}) \approx
37$. While in ``little'' RS scenarios thus no improvement concerning
$|\epsilon_K|$ can be achieved, unless the UV cutoff is raised above
several $10^3$ TeV, this class of models still helps to relax the
constraint arising from $Z^0 \to b \bar b$. This can be seen from the
blue dashed-dotted curve in the upper right panel of
Figure~\ref{fig:tuning}, which reaches $P(Z^0 \to b \bar b) > 25\%$
once $\Mkk > 1.1 \, {\rm TeV}$ is satisfied. On the other hand, it is
clearly visible from the blue dashed-dotted curve in the lower plot of
Figure~\ref{fig:tuning} that if one requires the constraints from both
$|\epsilon_K|$ and $Z^0 \to b \bar b$ to hold, then the bounds on
$\Mkk$ are even stronger in scenario S3 than in the default scenario
S1.  Demanding as usual $P({\rm total}) > 10\%$ translates in the case
of the ``little'' RS model into the rather severe limit $\Mkk > 8.4 \,
{\rm TeV}$.  Further details on RS variants with ``volume-truncated''
background can be found in \cite{Bauer:2008xb}.

Finally, the most obvious way to reduce the magnitude of $C_4^{\rm
  RS}$ consists in increasing the overall size of the elements of the
down-type Yukawa matrix. This is due to the scaling relation $C_4^{\rm
  RS} \sim (L/\Mkk^2) \, (4 \pi \alpha_s/Y_d^2) \, (2 m_s m_d/v^2)$ as
first noted in \cite{Csaki:2008zd}. Here $Y_d$ represents a
combination of elements of the down-type Yukawa matrix. How effective
this suppression is, can be assessed by comparing the total number of
orange points in the upper left and right plot of
Figure~\ref{fig:epsK}. Numerically, we find that 75\% of the total
number of points satisfy the latter constraint and that even for $\Mkk
= 1 \, {\rm TeV}$ roughly $14\%$ of the points satisfy the
$|\epsilon_K|$ constraint, which constitutes a noticeable improvement
relative to the case of the standard scenario S1, in which this
fraction is less than $1\%$. This improvement however comes with a
price. This is illustrated by the green dashed curve in the upper
right plot in Figure~\ref{fig:tuning}, which shows the fraction $P(Z^0
\to b \bar b)$ of points that lead to an agreement with the measured
$Z^0 b \bar b$ couplings as a function of $\Mkk$. For $P(Z^0 \to b
\bar b) > 25\%$ we obtain $\Mkk > 2.5 \, {\rm TeV}$ in scenario S4,
which allows for larger Yukawa couplings. This feature is readily
understood from the scaling relation $(g_L^b)_{\rm RS} \sim
-(F^2(c_{b_L})/\Mkk^2) \, ( L m_Z^2 + Y_d^2 v^2 )$
\cite{Casagrande:2008hr} of the RS corrections to the $Z^0 b_L \bar
b_L$ vertex. The individual terms arise here from the deviation of the
$Z^0$-boson profile from a constant and the mixing of zero and KK
fermion modes. For larger Yukawa couplings, effects due to the mixing
of fermion zero-modes with their KK excitations get enhanced, and as a
consequence the $Z^0 \to b \bar b$ constraint tends to become more
stringent. Yet since the $|\epsilon_K|$ constraint is generically more
difficult to satisfy than the one originating from $Z^0 \to b \bar b$,
the overall picture still improves when one allows for larger Yukawa
couplings. The improvement can be read off from the green dashed curve
in the lower plot of Figure~\ref{fig:tuning}.  Numerically, we obtain
for $P(\text{total}) > 10\%$ the bound $\Mkk > 2.6 \, {\rm TeV}$,
which constitutes an improvement relative to the ``hierarchical''
case.

It should have become clear from the above discussion and explanations
that obtaining a natural solution to the ``flavor problem'' in the
$\Delta S = 2$ sector, with KK gauge-boson masses in the reach of the
LHC, is difficult in the framework of warped extra dimensions
\cite{Csaki:2008zd, Santiago:2008vq, Blanke:2008zb, Agashe:2008uz,
  Bauer:2008xb}. Our detailed numerical analysis of anarchic RS
realizations shows that requiring only a moderate amount of
fine-tuning, $P(|\epsilon_K|) = 10\%$, implies that the mass of the
first-level KK gluon mode has to fulfill
\beq
M_{g^{(1)}} \approx 2.45\,\Mkk \gtrsim 10 \, {\rm TeV} \,.
\eeq
However, consistency with $|\epsilon_K|$ can be achieved for masses of
the first KK excitation in the ballpark of 3 TeV either by
``alignment'' in the right-handed down-type quark sector, or by
allowing for larger Yukawa couplings. Both solutions have their
shortcomings.  In the first case the natural flavor suppression
(\ref{eq:C4ratio}) in $|\epsilon_K|$ can be impaired by the presence
of flavor non-universal brane kinetic terms \cite{Santiago:2008vq},
making this setup radiatively unstable, while in the second case the
constraint from $B \to X_s \gamma$ typically becomes more stringent
\cite{Agashe:2008uz}.

\boldmath \subsubsection{Numerical Analysis of $B_{d,s}$--$\bar
  B_{d,s}$ Mixing} \unboldmath
\label{sec:num_BBmix}

In the following we will study the impact of new-physics effects on
the $B_{d,s}$--$\bar B_{d,s}$ mixing amplitudes, taking into account
the stringent constraints on the parameter space imposed by
$|\epsilon_K|$. Since the new CP-odd phases appearing in the $s \to d$
and $b \to d, s$ transitions are highly uncorrelated, we will see that
large departures in both $S_{\psi \phi}$ and $A_{\rm SL}^{s}$ from
their SM values are possible, which are consistent with all remaining
constraints in the $\Delta F = 2$ sector.

\begin{figure}[!t]
  \begin{center}
    \includegraphics[height=6.5cm]{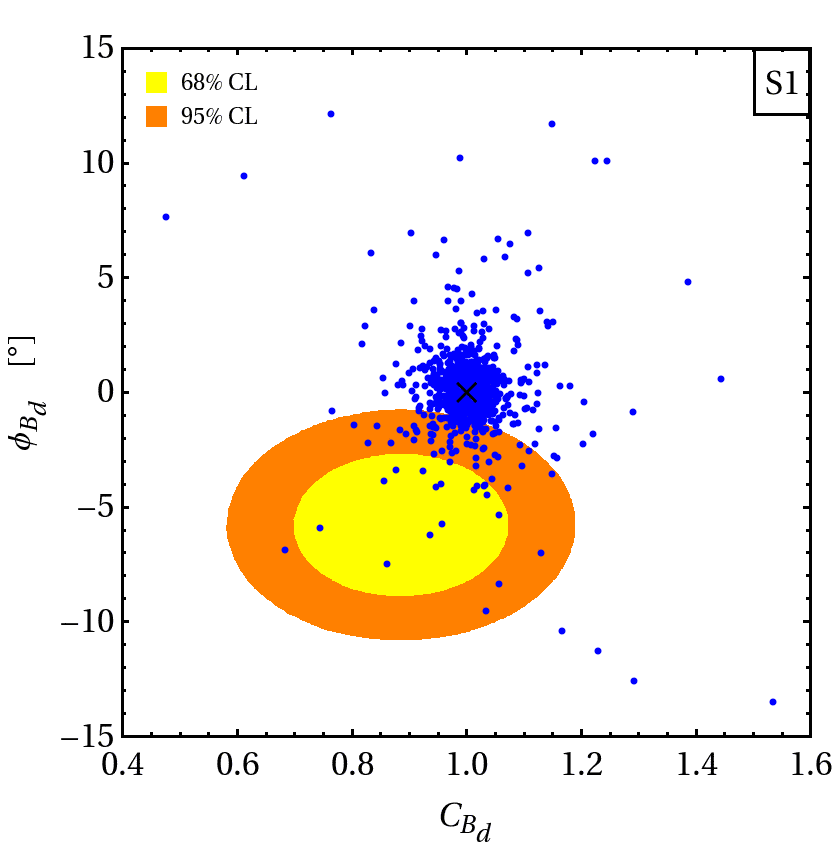}
    \hspace{0.75cm}
    \includegraphics[height=6.4cm]{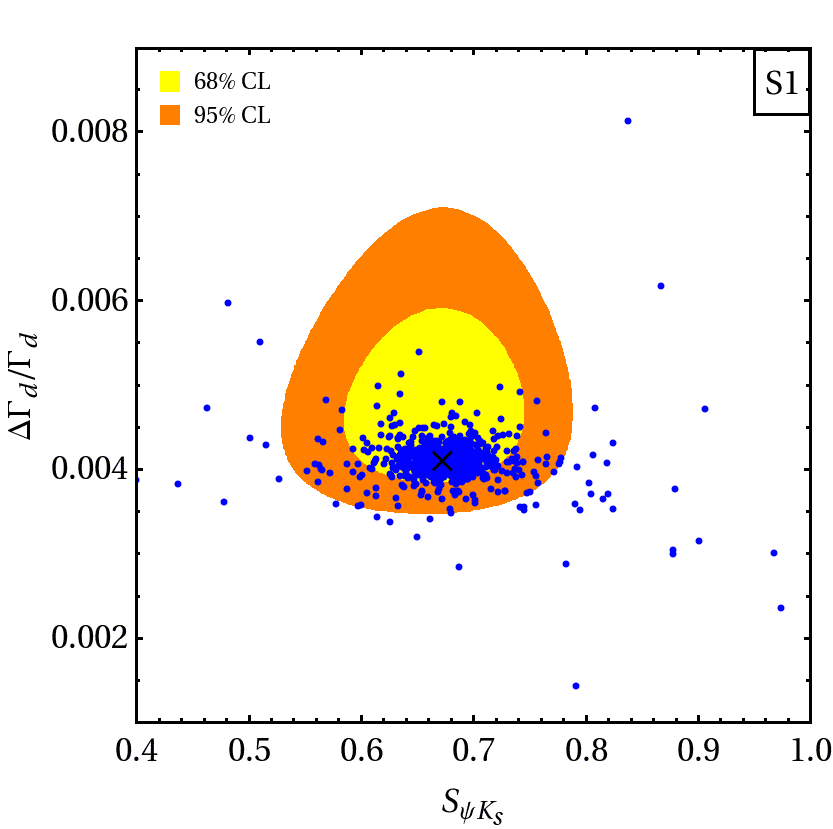}

    \vspace{4mm}

    \includegraphics[height=6.4cm]{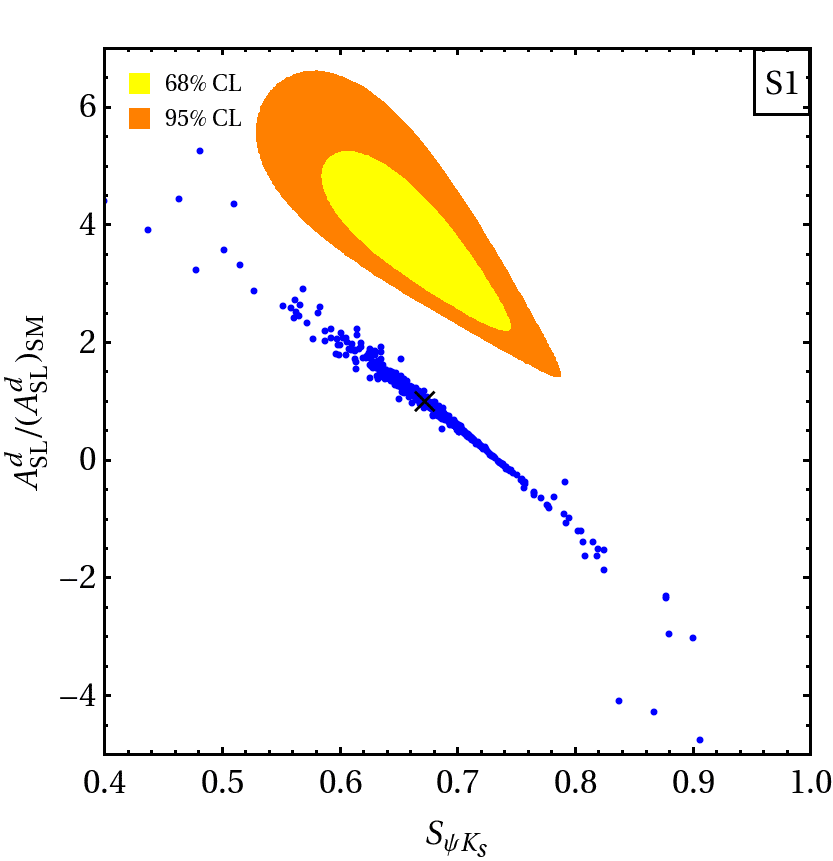}
  \end{center}
  \vspace{-8mm}
  \begin{center}
    \parbox{15.5cm}{\caption{\label{fig:Bdplot} Predictions for
        $\phi_{B_d}$ versus $C_{B_d}$ (upper left), as well as $\Delta
        \Gamma_d/\Gamma_d$ (upper right) and $A_{\rm SL}^{d}/(A_{\rm
          SL}^{d})_{\rm SM}$ (lower panel) versus $S_{\psi K_S}$. The
        blue points reproduce the measured values of $|\epsilon_K|$ and
        the $Z^0 b \bar b$ couplings at the 95\% and 99\% CL,
        respectively. The black crosses indicate the SM predictions and
        the yellow (orange) contours the experimentally favored regions
        of 68\% (95\%) probability. See text for details.}}
  \end{center}
\end{figure}

We start our survey in the $B_d$ system. In Figure~\ref{fig:Bdplot} we
display the predictions for $\phi_{B_d}$ versus $C_{B_d}$ as well as
$\Delta \Gamma_d/\Gamma_d$ and $A_{\rm SL}^d/(A_{\rm SL}^d)_{\rm SM}$
versus $S_{\psi K_S}$ obtained in scenario S1. Our findings are
however largely independent of the choice of the benchmark
scenario. In all three panels the shown blue points reproduce the
correct quark masses and mixings within errors and satisfy the
constraints from both $|\epsilon_K|$ and $Z^0 \to b \bar b$. The SM
prediction is indicated by a black cross, and the yellow (orange)
colored contour corresponds to the experimentally preferred 68\%
(95\%) CL region. Turning our attention to the predictions in the
$C_{B_d}$--$\phi_{B_d}$ plane, we observe that while shifts of more
than $\pm 0.3$ and $\pm 10^\circ$ are possible in $C_{B_d}$ and
$\phi_{B_d}$, the corrections are on average small and thus no cause
of concern. Notice that there is a slight tension \cite{Bona:2007vi,
  Buras:2008nn, Lenz:2006hd, Lunghi:2008aa} of $2.1 \sigma$ between the
SM expectation, $(\phi_{B_d})_{\rm SM} = 0^\circ$, and the value
$\phi_{B_d} = (-5.8 \pm 2.8)^\circ$ extracted from our global fit. The
coefficient $C_{B_d} = 0.89 \pm 0.17$, on the other hand, shows good
agreement with the SM value $(C_{B_d})_{\rm SM} = 1$. In our analysis
the bulk of the discrepancy in $\phi_{B_d}$ is due to the correlation
between the observable ${\cal B} (B \to \tau \nu_\tau)$ and the angle
$\beta$ \cite{Deschamps:2008de}. Obviously, the disagreement is not
large enough to exclude the possibility of a statistical origin of the
tension, but it is interesting to note that the RS model can naturally
accommodate a small phase $\phi_{B_d}$ of the order of a few degrees
with both signs. In the case of the observables $\Delta
\Gamma_d/\Gamma_d$, $A_{\rm SL}^d/(A_{\rm SL}^d)_{\rm SM}$, and
$S_{\psi K_S}$, we find that the RS predictions are in general
compatible with the current direct experimental determinations, which
read $(\Delta \Gamma_d/\Gamma_d)_{\rm exp} = 0.009 \pm 0.037$,
$(A_{\rm SL}^d)_{\rm exp} = -0.0047 \pm 0.0046$, and $S_{\psi K_S} =
0.672 \pm 0.023$ \cite{Barberio:2008fa}. Notice that the values
$\Delta \Gamma_d/\Gamma_d = 0.0049 \pm 0.0010$ and $A_{\rm SL}^d =
-0.0018 \pm 0.0007$ following from our global analysis are much more
precise than the corresponding direct extractions, and that the slight
disagreement in $\phi_{B_d}$ translates into a tension in $A_{\rm
  SL}^d/(A_{\rm SL}^d)_{\rm SM}$. We conclude that the RS-GIM mechanism
makes warped extra-dimension models naturally consistent with the
available data on $\Delta m_d$, while it does not exclude the presence
of a new-physics contribution to the mixing phase $\phi_{B_d}$ of the
right size to be directly tested with improved measurements of $\Delta
\Gamma_d/\Gamma_d$, $A_{\rm SL}^d$, and $S_{\psi K_S}$. Similar
conclusions have been drawn in \cite{Blanke:2008zb}.

\begin{figure}[!t]
  \begin{center}
    \includegraphics[height=6.4cm]{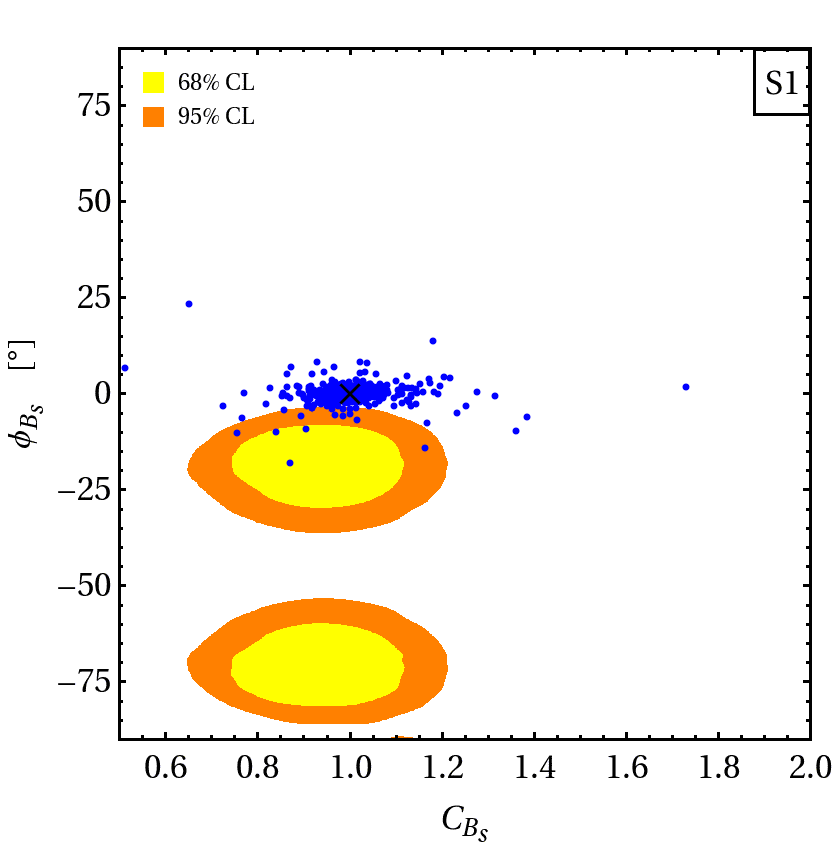}
    \hspace{0.75cm}
    \includegraphics[height=6.5cm]{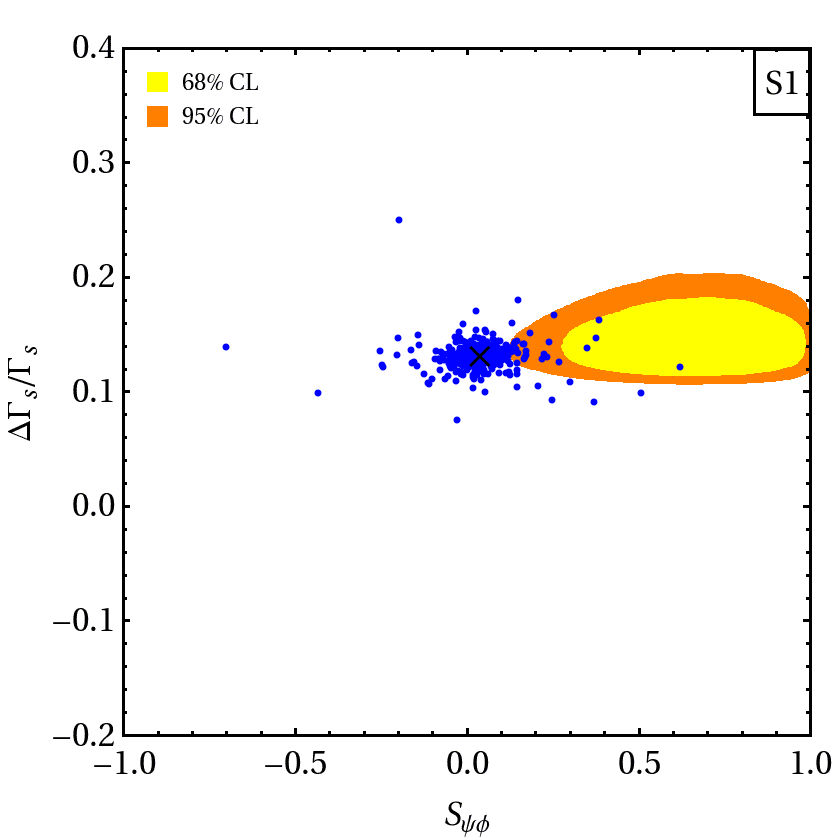}

    \vspace{4mm}

    \includegraphics[height=6.4cm]{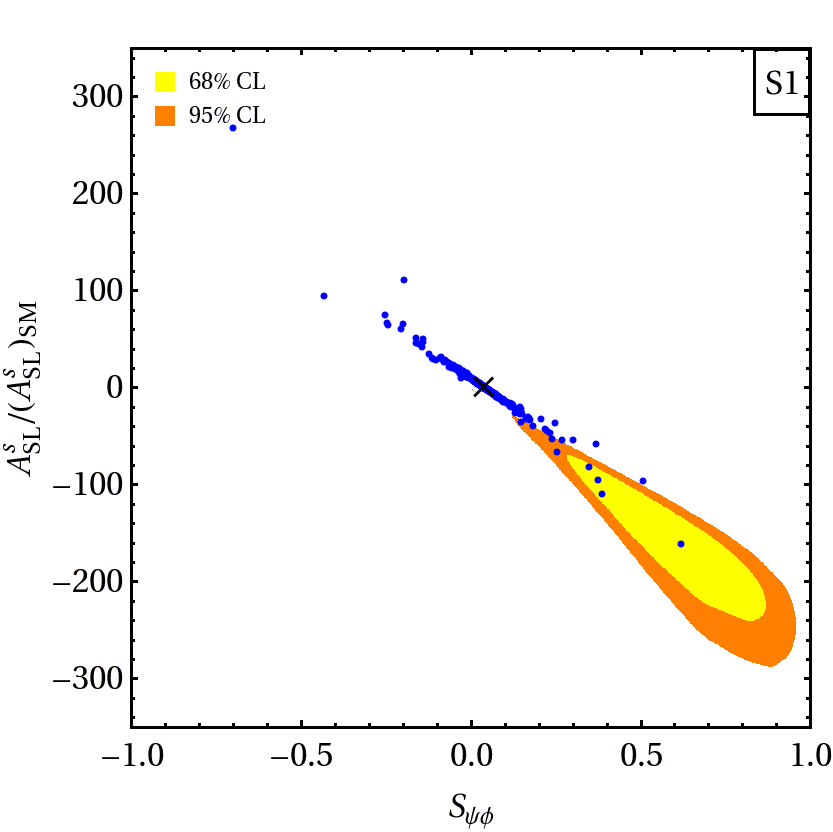}
  \end{center}
  \vspace{-8mm}
  \begin{center}
    \parbox{15.5cm}{\caption{\label{fig:Bsplot} Predictions for
        $\phi_{B_s}$ versus $C_{B_s}$ (upper left), as well as $\Delta
        \Gamma_s/\Gamma_s$ (upper right) and $A_{\rm SL}^{s}/(A_{\rm
          SL}^{s})_{\rm SM}$ (lower panel) versus $S_{\psi \phi}$. The
        blue points reproduce the measured values of $|\epsilon_K|$, the
        $Z^0 b \bar b$ couplings, and $B_d$--$\bar B_d$ mixing at the
        95\%, 99\%, and 95\% CL.  The black crosses indicate the SM
        predictions and the yellow (orange) contours the experimentally
        preferred regions of 68\% (95\%) probability. See text for
        details.}}
  \end{center}
\end{figure}

We now move onto the $B_s$ system. Our results for $\phi_{B_s}$ versus
$C_{B_s}$, as well as $\Delta \Gamma_s/\Gamma_s$ and $A_{\rm
  SL}^d/(A_{\rm SL}^s)_{\rm SM}$ versus $S_{\psi \phi}$ obtained in the
parameter scenario S1 are shown in Figure~\ref{fig:Bsplot}. In this
case we only include points in the plots that satisfy the constraints
from $|\epsilon_K|$, $Z^0 \to b \bar b$, and $B_d$--$\bar B_d$
mixing. The SM predictions are indicated by black crosses, while the
yellow (orange) colored contours resemble the experimentally favored
regions at 68\% (95\%) CL. Focusing first on the predictions in the
$C_{B_s}$--$\phi_{B_s}$ plane, we see that shifts of up to $\pm 0.4$
in $C_{B_s}$ and large corrections in $\phi_{B_s}$ are possible in the
RS model. For comparison, we show the results of a model-independent
analysis of new-physics contributions to $B_s$--$\bar B_s$ mixing
employing the parametrization (\ref{eq:BBmixpara}). We obtain two
solution for $\phi_{B_s}$, reflecting the twofold ambiguity inherent
in the measurement of time-dependent tagged angular analysis of $B_s
\to \psi \phi$ decays, {\it i.e.}, $\varphi_s \leftrightarrow 90^\circ
- \varphi_s$ and $\Delta \Gamma_s \leftrightarrow -\Delta
\Gamma_s$. The numerical results of the two solutions are $\phi_{B_s}
= (-19.0 \pm 10.8)^\circ$ and $\phi_{B_s} = (-69.9 \pm 10.1)^\circ$,
which implies a deviation of $\varphi_s = |\beta_s| - \phi_{B_s}$ of
more than $2.5 \sigma$ from its SM value $(\varphi_s)_{\rm SM} \approx
1^\circ$. For the magnitude of the $B_s$--$\bar B_s$ mixing amplitude
we find $C_{B_s} = 0.93 \pm 0.19$, in agreement with the SM
expectation. Our global fit is based on the combined CDF and {D\O}
measurement of $\Delta m_s$ \cite{Abulencia:2006ze, D0note1} and the
two-dimensional CL for $\phi_{B_s}$ and $\Delta \Gamma_s$ obtained
from the flavor-tagged analysis of mixing-induced CP violation in $B_s
\to \psi \phi$ \cite{Aaltonen:2007he, Abazov:2008fj}.
Concerning the remaining observables in the $B_s$ system, we observe
that compared to the SM value $(S_{\psi \phi})_{\rm SM} \approx 0.04$
the large range $[-0.5, 0.5]$ of $S_{\psi \phi}$ is attainable in the
RS framework, and that also the semileptonic CP asymmetry $A_{\rm
  SL}^s$ can be enhanced by more than two orders of magnitude relative
to its SM value $(A_{\rm SL}^s)_{\rm SM} \approx 2 \cdot 10^{-5}$. In
particular, the values $S_{\psi \phi} = 0.63 \pm 0.35$ and $A_{\rm
  SL}^s = -0.0032 \pm 0.0020$ favored by the existing data can be
obtained. On the other hand, the predicted corrections in $\Delta
\Gamma_s/\Gamma_s$ are typically small and compatible with both the
experimentally favored range $\Delta \Gamma_s/\Gamma_s = 0.15 \pm
0.03$ and the SM expectation $(\Delta \Gamma_s/\Gamma_s)_{\rm SM}
\approx 0.13$.

The plots in Figures~\ref{fig:Bdplot} and~\ref{fig:Bsplot} display
strong correlations between the various observables. In order to
better understand the pattern of deviations we express $\Delta
\Gamma_{d,s}/\Gamma_{d,s}$ and $A_{\rm SL}^{d,s}/(A_{\rm
  SL}^{d,s})_{\rm SM}$ through $S_{\psi K_S}$ and $S_{\psi \phi}$,
respectively. For the $B_d$ system, one finds (keeping terms to first
order in the deviation of $S_{\psi K_S}$ from its SM value
$\sin2\beta$)
\begin{equation} \label{eq:patternBd}
\begin{aligned}
  \frac{\Delta \Gamma_d}{\Gamma_d} & \approx - \left ( \frac{\Delta
      m_{d}}{\Gamma_{d}} \right )_{\rm exp} \left [ {\rm Re} \left (
      \Gamma_{12}^d/M_{12}^d \right )_{\rm SM} + \frac{{\rm Im} \left
        ( \Gamma_{12}^d/M_{12}^d \right )_{\rm SM}}{\cos2\beta} \, 
        \big ( \sin2\beta -
    S_{\psi K_S} \big ) \right ] C_{B_d}^{-1} \\
  & = \big ( 0.004 - 0.0005 \, S_{\psi K_S} \big ) \, C_{B_d}^{-1} \,,
   \\[2mm]
  \frac{A_{\rm SL}^d}{(A_{\rm SL}^d)_{\rm SM}} 
  & \approx \left [ 1 + \frac{{\rm Re} \left (
        \Gamma_{12}^d/M_{12}^d \right )_{\rm SM}}{{\rm Im} \left (
        \Gamma_{12}^d/M_{12}^d \right )_{\rm SM}}\, 
        \frac{1}{\cos2\beta}\, \big ( \sin2\beta
    - S_{\psi K_S} \big ) \right ] C_{B_d}^{-1} \\
  & = \big ( 11.7 - 15.3 \, S_{\psi K_S} \big ) \,
  C_{B_d}^{-1} \,,
\end{aligned}
\end{equation}
where in the last step we have inserted the central values of the SM
predictions quoted in (\ref{eq:ALUN}).

In the case of the $B_s$ system the corresponding expressions
simplify, since $\sin 2 |\beta_s| \approx 0$ and $\cos 2 |\beta_s|
\approx 1$. Keeping terms to first order in $S_{\psi \phi}$ but
neglecting small corrections due to $|\beta_s|\neq 0$, one obtains
\begin{equation} \label{eq:patternBs}
\begin{aligned}
  \frac{\Delta \Gamma_s}{\Gamma_s} & \approx - \left ( \frac{\Delta
      m_{s}}{\Gamma_{s}} \right )_{\rm exp} \Big [ {\rm Re} \left (
    \Gamma_{12}^s/M_{12}^s \right )_{\rm SM} + {\rm Im} \left (
    \Gamma_{12}^s/M_{12}^s \right )_{\rm SM} \, S_{\psi \phi} \Big ]
  \, C_{B_s}^{-1} \\
  & = \big ( 0.131 - 0.0006 \, S_{\psi \phi} \big ) \, C_{B_s}^{-1} \,,
  \\[2mm]
  \frac{A_{\rm SL}^s}{(A_{\rm SL}^s)_{\rm SM}} 
  & \approx \frac{{\rm Re} \left ( \Gamma_{12}^s/M_{12}^s
    \right )_{\rm SM}}{{\rm Im} \left ( \Gamma_{12}^s/M_{12}^s
    \right )_{\rm SM}} \, S_{\psi \phi} \, C_{B_s}^{-1} \\
  & = - 238.1 \, S_{\psi \phi} \, C_{B_s}^{-1} \,,
\end{aligned}
\end{equation}
where the final numerical values correspond to the SM predictions
given in (\ref{eq:ALUN}). The strong correlation between $A_{\rm
  SL}^s$ and $S_{\psi \phi}$ has been first pointed out in
\cite{Ligeti:2006pm}. That this correlation also exists in the RS
scenario has been emphasized in \cite{Blanke:2008zb}. Given the
model-independent character \cite{Blanke:2006ig} of the linear
relation between $A_{\rm SL}^{s}/(A_{\rm SL}^{s})_{\rm SM}$ and
$S_{\psi \phi}$, which can only be violated if there is sizable direct
CP-violation due to new physics in the decay, the observation of a
correlation that follows the pattern in (\ref{eq:patternBs}) is
however not a ``smoking gun'' signal unique to the RS model. The same
statement applies to the remaining correlations following from the
relations (\ref{eq:patternBd}) and (\ref{eq:patternBs}).

We close this section by adding some comments concerning the model
dependence of the obtained results. As already mentioned earlier, in
RS scenarios with custodial $SU(2)$ symmetry electroweak corrections
to $B_{d,s}$--$\bar B_{d,s}$ mixing can compete with the corrections
due to KK gluon exchange. In rough accordance with
\cite{Blanke:2008zb}, we find that purely electroweak effects in
$C_1^{\rm RS}$, $\tilde{C}_1^{\rm RS}$, and $C_5^{\rm RS}$ are
relative to (\ref{eq:Cmix}) modified by factors of about $2.5$,
$94.6$, and $-10.9$. We remark that in the custodial model a
cancellation between the strong and electroweak part in $C_5^{\rm RS}$
occurs since the individual contributions have approximately equal
size but opposite signs. The ratios of the full RS contributions in
the custodial relative to the original model are thus $1.3$, $1.9$,
and $-0.3$. Compared to Figures \ref{fig:Bdplot} and \ref{fig:Bsplot},
these changes, in combination with the relaxation of the $Z^0 \to b
\bar b$ constraint due to custodial protection \cite{Agashe:2006at},
allow for somewhat larger effects in the $B_{d,s}$--$\bar B_{d,s}$
mixing observables. The pattern of departures from the SM
expectations, however, remains unchanged. We leave a detailed analysis
of neutral-meson mixing in the RS model with custodial symmetry
$SU(2)_L \times SU(2)_R \times P_{LR}$ for future work.

\boldmath \subsubsection{Numerical Analysis of $D$--$\bar D$ Mixing}
\unboldmath \label{sec:num_DDmix}

Short-distance contributions from new physics can also affect the
dispersive part of the $D$--$\bar D$ mixing amplitude $M_{12}^D$ in a
significant way. Similarly to the case of the $K_L$--$K_S$ mass
difference, the calculation of $|M_{12}^D|$ is plagued by
long-distance contributions \cite{Petrov:2006nc}. In order to grasp
the impact of non-perturbative effects on the obtained results we
proceed in the following way. We write the full amplitude $M_{12}^D$
as the sum of the RS amplitude $(M_{12}^D)_{\rm RS} = |M_{12}^D|_{\rm
  RS} \, e^{-2 i \varphi_D}$ and the real SM amplitude,
$(M_{12}^D)_{\rm SM}$, containing both short- and long-distance
contributions. We then consider two diametrically opposed cases. In
the first case, the SM contribution to $(M_{12}^D)_{\rm SM}$ is set to
zero, and the constraint on $|M_{12}^D|$ and the phase $\varphi_D$ is
directly applied to the RS contribution. In the second case, we take
$(M_{12}^D)_{\rm SM}$ to be flatly distributed in the range $[-0.02,
0.02] \, {\rm ps}^{-1}$, so that the SM contribution alone can
saturate the experimental bound. We will see that even with the latter
conservative treatment of the theoretical uncertainties entering the
SM prediction, the available experimental data on $D$--$\bar D$ mixing
have a non-trivial impact on the allowed model parameters in the RS
framework.

\begin{figure}[!t]
  \begin{center}
    \includegraphics[height=6.5cm]{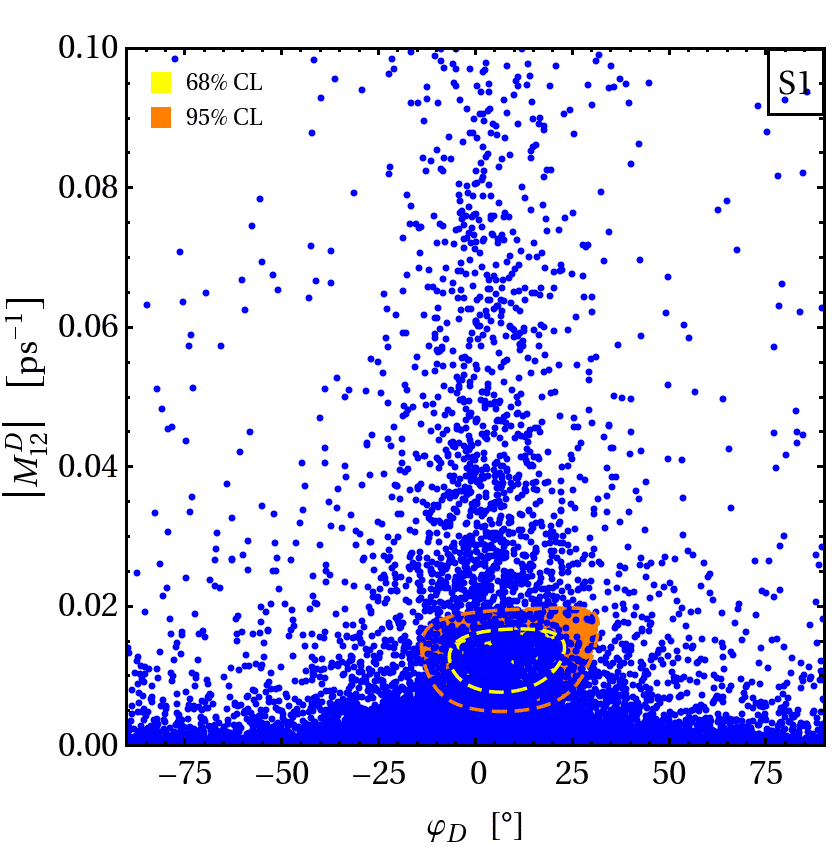}
    \hspace{0.75cm}
    \includegraphics[height=6.5cm]{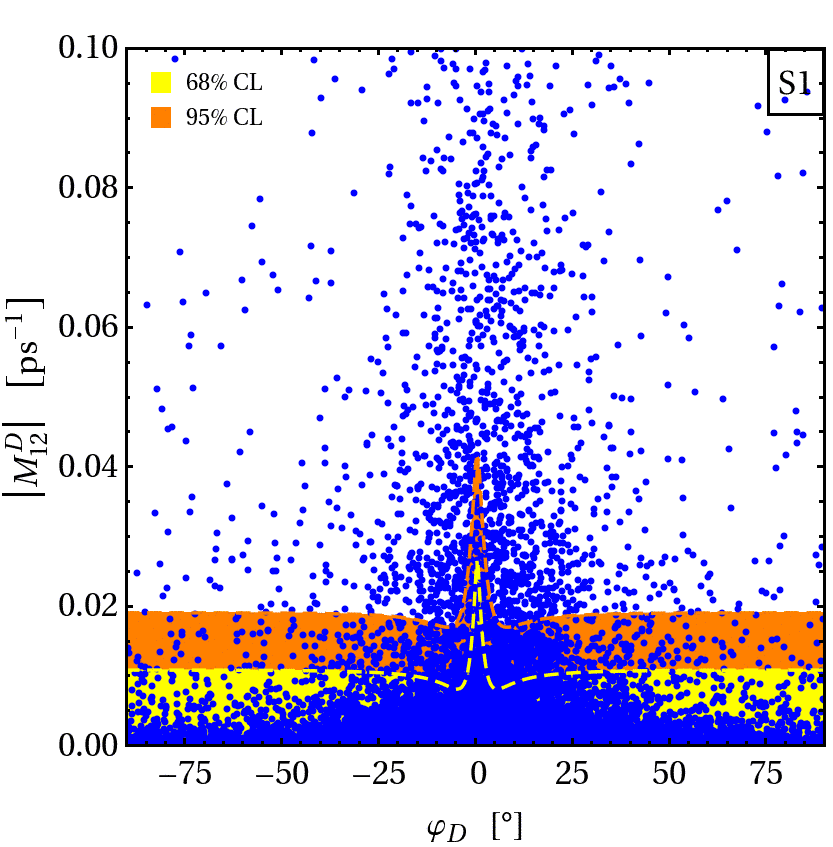}
  \end{center}
  \vspace{-8mm}
  \begin{center}
    \parbox{15.5cm}{\caption{\label{fig:DDMix} Predictions for
        $|M_{12}^D|_{\rm RS}$ versus $\varphi_D$ compared to the
        experimentally allowed 68\% (95\%) CL regions shown in yellow
        (orange), assuming $(M_{12}^D)_{\rm SM} = 0$ (left) and allowing
        for $(M_{12}^D)_{\rm SM} \in [-0.02, 0.02] \, {\rm ps}^{-1}$
        (right). The blue scatter points reproduce the measured values
        of $|\epsilon_K|$, the $Z^0 \to b \bar b$ couplings, and
        $B_d$--$\bar B_d$ mixing at 95\%, 99\%, and 95\% CL. See text
        for details.}}
  \end{center}
\end{figure}

In Figure~\ref{fig:DDMix} we show the predictions of the RS model in
the $\varphi_D$--$|M_{12}^D|_{\rm RS}$ plane obtained from a general
scan in the benchmark scenario S1. The blue points satisfy the
constraints from $|\epsilon_K|$, $Z^0 \to b \bar b$, and $B_d$--$\bar
B_d$ mixing. For comparison, the 68\% (95\%) CL regions favored by the
data are underlaid in yellow (orange). They have been obtained from a
simultaneous fit to the data from BaBar \cite{Aubert:2007wf}, Belle
\cite{Staric:2007dt, Abe:2007rd}, and CDF \cite{Aaltonen:2007uc},
employing the two different treatments of theoretical uncertainties
described above. Our fit is based on the experimental inputs $x_D =
0.0098^{+0.0024}_{-0.0026}$, $y_D = 0.0083 \pm 0.0016$, $(x_D^2 +
y_D^2)/2 \leq (1.3 \pm 2.7)\cdot 10^{-4}$, $|q/p|_D =
0.87^{+0.17}_{-0.15}$, and $\tau_D = (0.4101 \pm 0.0015) \, {\rm ps}$,
as well as the corresponding correlation matrix
\cite{Barberio:2008fa, Schwartz:2009jv}. These numbers imply for the
dispersive and absorptive parts of the off-diagonal mixing elements
$(M_{12}^D)_{\rm exp} \approx \pm 0.012 \,{\rm ps}^{-1}$ and
$(\Gamma_{12}^D)_{\rm exp} \approx \pm 0.020 \,{\rm ps}^{-1}$. The
distribution of scatter points shows that large corrections to both
the magnitude and phase of the $D$--$\bar D$ mixing amplitude can
occur in the RS framework. In particular, the entire range of
CP-violating phases $\varphi_D \in [-90^\circ, 90^\circ]$ can be
populated. These findings agree with \cite{Gedalia:2009kh}. From the
inspection of the two panels it is evident that the power of the
$D$--$\bar D$ constraint depends in a crucial way on the treatment of
the large theoretical uncertainties plaguing the calculation of
$(M_{12}^D)_{\rm SM}$. If the constraint on $|M_{12}^D|$ and the phase
$\varphi_D$ is applied to the RS contribution alone, neglecting the SM
contribution, then the impact of the $D$--$\bar D$ constraint is quite
stringent. On the other hand, it turns out to be much weaker, though
non-negligible, if one allows the SM contribution alone to saturate
the experimental bounds. This feature implies that a breakthrough in
the theoretical control over the quantity $(M_{12}^D)_{\rm SM}$ is
required to resolve whether the observed size of $\Delta M_D$ is
completely due to SM dynamics or contains a sizable or maybe even
leading contribution due to new physics.

\begin{figure}[!t]
  \begin{center}
    \includegraphics[height=6.5cm]{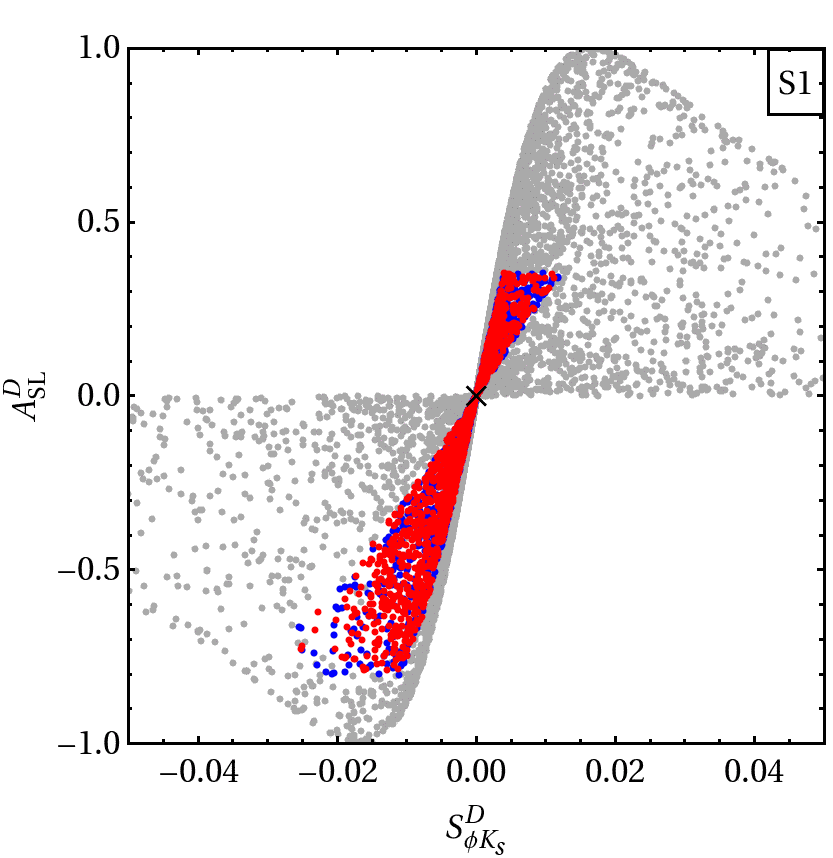}
    \hspace{0.75cm}
    \includegraphics[height=6.5cm]{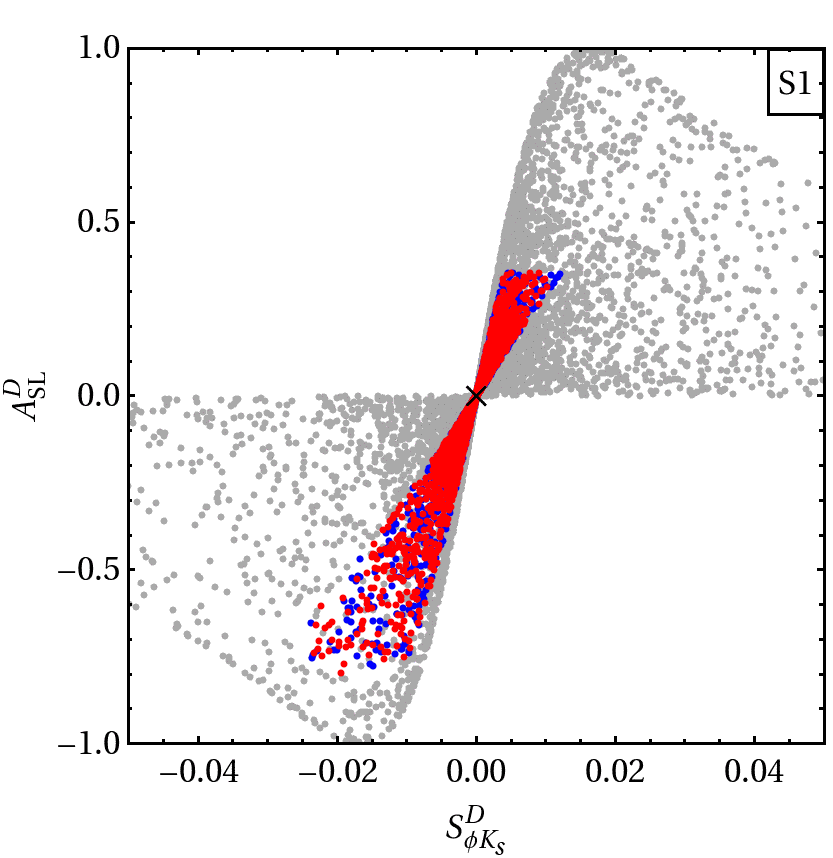}
  \end{center}
  \vspace{-8mm}
  \begin{center}
    \parbox{15.5cm}{\caption{\label{fig:ASLD} Predictions for $S_{\phi
          K_S}^D$ versus $A_{\rm SL}^D$ assuming $(M_{12}^D)_{\rm SM} =
        0$ (left) and allowing for $(M_{12}^D)_{\rm SM} \in [-0.02,
        0.02] \, {\rm ps}^{-1}$ (right). All the shown scatter points
        reproduce the measured values of $|\epsilon_K|$, the $Z^0 \to b
        \bar b$ couplings, and $B_d$--$\bar B_d$ mixing at 95\%, 99\%,
        and 95\% CL. Blue (red) points are obtained for $\Gamma_{12}^D =
        +0.020 \, {\rm ps}^{-1}$ ($\Gamma_{12}^D = -0.020 \, {\rm
          ps}^{-1}$). They satisfy the constraints arising from the
        $D$--$\bar D$ mixing measurements at 95\% CL, while the light
        gray points do not. See text for details.}}
  \end{center}
\end{figure}

Our above findings suggest that KK gluon exchange in the RS model can
potentially lead to sizable CP asymmetries in various $D$-meson decay
channels. The Cabibbo-favored channel with the best trade-off between
theoretical simplicity and experimental accessibility is $D \to \phi
K_S$ \cite{Bigi:2009df}. In Figure~\ref{fig:ASLD} we show the
correlation between $S_{\psi K_S}^D$ and $A_{\rm SL}^D$ for a
parameter scan in the benchmark scenario S1. The shown points are
consistent with the constraints arising from $|\epsilon_K|$, $Z^0 \to
b \bar b$, and $B_d$--$\bar B_d$ mixing. The black crosses indicate
the SM prediction. In the left panel we have set $(M_{12}^D)_{\rm SM}
= 0$, while in the right panel $(M_{12}^D)_{\rm SM}$ is allowed to
vary freely in the range $[-0.02, 0.02] \, {\rm ps}^{-1}$. In both
panels, blue (red) points indicate solutions with $\Gamma_{12}^D =+
0.020 \,{\rm ps}^{-1}$ ($\Gamma_{12}^D =-0.020 \,{\rm ps}^{-1}$) that
are in agreement with the measured values of $x_D$, $y_d$, and
$|q/p|_{D}$. The shown light gray points are not consistent with these
constraints. While {\it a priori} RS dynamics could generate values
for $S_{\phi K_S}^D$ that exceed $\pm 0.05$, the experimental
constraint on $|q/p|_D$ and consequently on $A_{\rm SL}^D$ implies an
allowed range of at most $S_{\phi K_s}^D \in [-0.02, 0.01]$ due to the
strong correlation between the two CP asymmetries. We observe that for
realistic values of $A_{\rm SL}^D$ there is a strict linear
correlation between these two CP asymmetries, with the slope
determined by the prefactor $y_D/(x^2_D+ y^2_D) \approx 50$ entering
(\ref{eq:ASLD}). Any violation of this correlation would signal the
presence of direct CP violation in the decay $D \to \phi K_S$. Like in
the case of $B_{d,s}$--$\bar B_{d,s}$ mixing, an observation of the
correlation between $A_{\rm SL}^D$ and $S_{\phi K_s}^D$ will however
not tell us whether it is due to KK exchange or some other kind of new
physics. For example, it seems difficult to disentangle the
corrections predicted in the RS framework from those arising in the
littlest-Higgs model with $T$-parity studied in
\cite{Bigi:2009df}. CP violation due to RS dynamics can also be
searched for in the non-leptonic channels $D \to K_S K^+ K^-$, $K^+
K^-$, $\pi^+ \pi^-$, and $K^+ \pi^-$. A comprehensive study of these
transitions is beyond the scope of this article.

Let us again comment on the model dependence of our results for
$D$--$\bar D$ mixing. Similar to $B_{d,s}$--$\bar B_{d,s}$
oscillations, also in this case electroweak corrections can compete
with the corrections due to KK gluon exchange in models with $SU(2)_L
\times SU(2)_R \times P_{LR}$ gauge symmetry in the bulk. Numerically,
we find that purely electroweak effects in $C_1^{\rm RS}$,
$\tilde{C}_1^{\rm RS}$, and $C_5^{\rm RS}$ are relative to
(\ref{eq:Cmix}) modified by factors of about $3.4$, $2.5$, and $-0.9$.
While this feature allows for somewhat larger effects, the pattern of
deviations from the SM expectations remains unchanged. In this
context, we also point out that the custodial protection mechanism
{\em simultaneously} suppresses corrections to the $Z^0 d_{i
  \hspace{0.25mm} L} \bar d_{j \hspace{0.25mm} L}$ as well as the $Z^0
u_{i \hspace{0.25mm} R} \bar u_{j \hspace{0.25mm} R}$ couplings if the
quark sector is implemented as in \cite{Blanke:2008zb, Blanke:2008yr,
  Carena:2007ua, Medina:2007hz}. The same conclusion has been drawn
independently in \cite{Buras:2009ka}. This feature implies, in
particular, that the chirality of the $Z^0 tc$ interactions in this
specific RS variant is predicted not to be right-handed
\cite{Agashe:2006wa} but left-handed. Of course, other choices of the
quantum numbers of the right-handed up-type quarks are possible, so
that the RS framework does not lead to a firm prediction of the
chirality of the $Z^0 tc$ interactions.

\boldmath \subsection{Rare Leptonic Decays of Kaons and $B$ Mesons}
\unboldmath

This section is devoted to detailed studies of non-standard effects in
rare decay modes of kaons and $B$ mesons arising from the tree-level
exchange of neutral gauge bosons, their KK excitations, and the Higgs
boson. In the former case, the special role of the $K \to \pi \nu \bar
\nu$ and $K_L \to \pi^0 l^+ l^-$ modes is emphasized, which due to
their theoretical cleanliness and their enhanced sensitivity to
non-standard flavor and CP violation are unique tools to discover or,
if no deviation is found, to set severe constraints on the parameter
space of RS models. In the latter case, we begin our discussion with
$B_q \to \mu^+ \mu^-$ and $B \to X_q \nu \bar \nu$ and stress the
power of these decay modes in probing the short-distance physics
related to $Z^0$-penguin diagrams. We conclude our explorations in the
$B$-meson sector by studying the numerical impact of the exchange of
the $Z^0$ boson and its KK modes on $B \to X_s l^+ l^-$ and $B \to
K^\ast l^+ l^-$. In this context we investigate whether the
predictions for the decay distributions lie within the experimentally
allowed bounds. A discussion on how the obtained results depend on the
choice of the bulk gauge group and the implementation of the fermionic
sector complement our phenomenological survey.

\boldmath \subsubsection{Numerical Analysis of $K \to \pi \nu \bar
  \nu$, $K_L \to \pi^0 l^+ l^-$, and $K_L \to \mu^+ \mu^-$}
\unboldmath \label{sec:numkaons}

\begin{figure}[!p]
  \begin{center}
    \includegraphics[height=6.5cm]{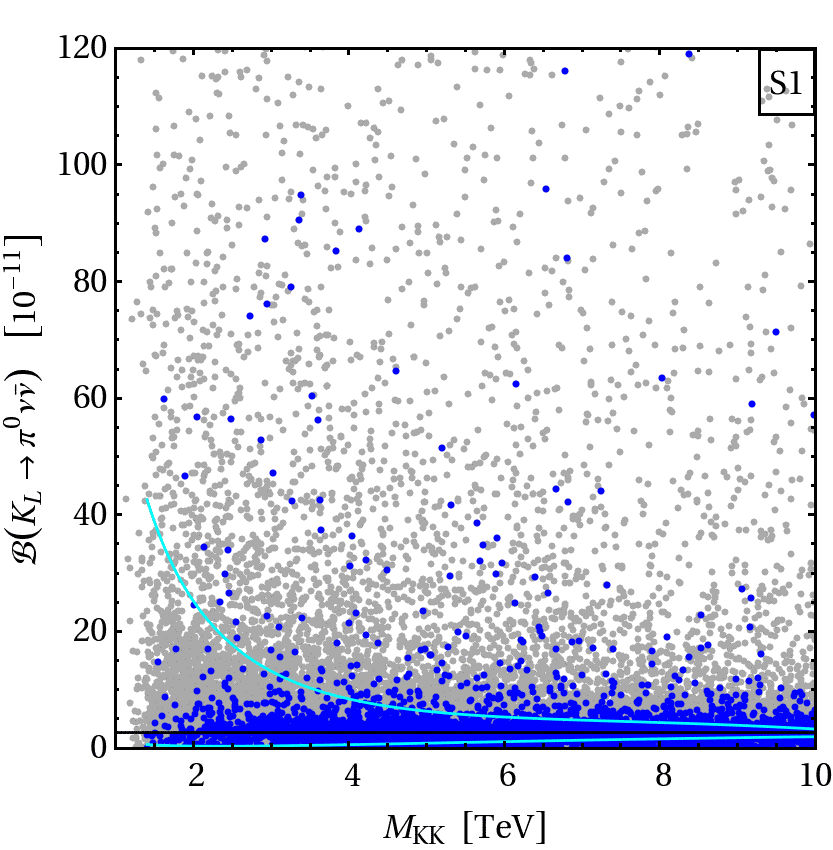} \\ \vspace{0.75cm}
    \includegraphics[height=6.5cm]{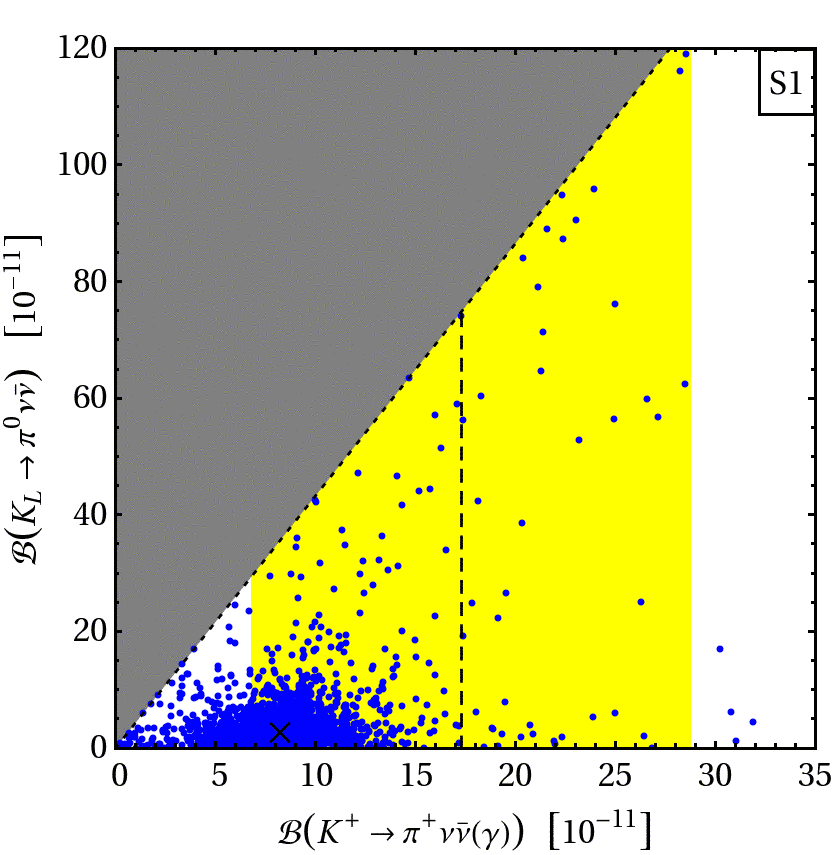} \hspace{0.75cm}
    \includegraphics[height=6.5cm]{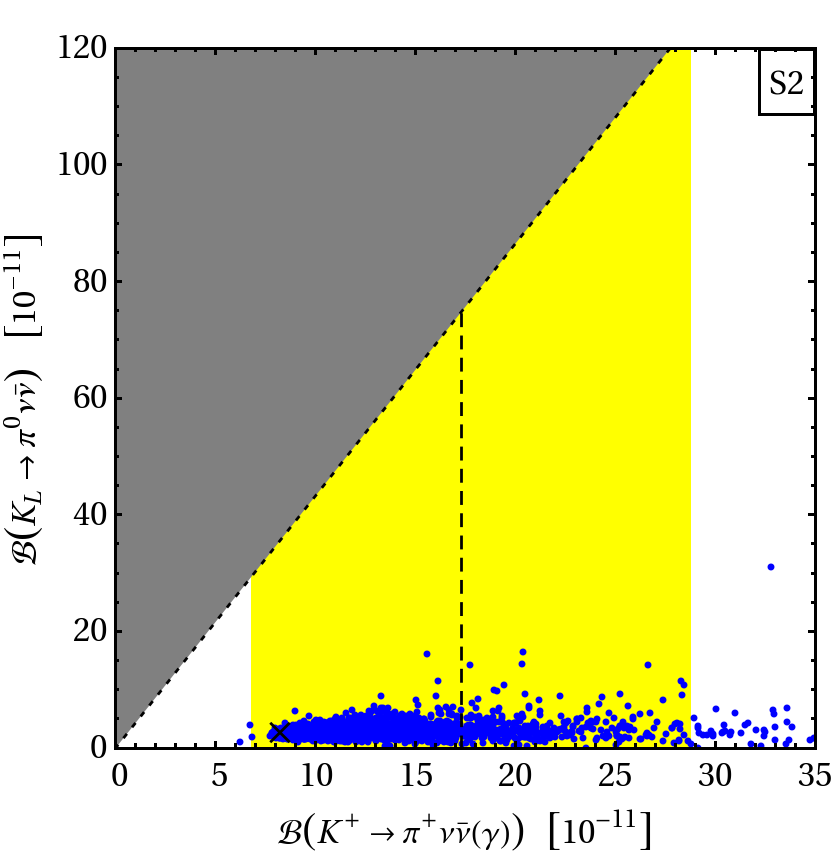}
  \end{center} \vspace{-8mm}
  \begin{center}
    \parbox{15.5cm}{\caption{\label{fig:KLvsKp} Predictions for ${\cal
          B} (K_L \to \pi^0 \nu \bar \nu)$ as a function of $\Mkk$
        (upper panel). The solid line represents the SM prediction. The
        lower left (right) panel shows the results for ${\cal B} (K^+
        \to \pi^+ \nu \bar \nu (\gamma))$ versus ${\cal B} (K_L \to
        \pi^0 \nu \bar \nu)$ in the benchmark scenario S1 (S2). In both
        panels the black cross indicates the SM point, the straight
        dotted black line and the light gray shaded area shows the GN
        bound, and the vertical dashed black line and the yellow band
        display the experimental central value and 68\% CL range for
        ${\cal B} (K^+ \to \pi^+ \nu \bar \nu (\gamma))$. The gray
        scatter points in the upper panel reproduce the measured values
        of the $Z^0\to b \bar b$ couplings at 99\% CL, while the blue
        points additionally reproduce the measured values of
        $|\epsilon_K|$ and $B_d$--$\bar B_d$ mixing parameters at 95\%
        CL. The cyan lines illustrate the decoupling behavior with
        $\Mkk$ obtained from a fit to the $5\%$ (lower line) and $95\%$
        (upper line) quantile of the points that are consistent with all
        three mentioned constraints. See text for details.}}
  \end{center}
\end{figure}

We begin our investigations by studying the rare decays $K_L \to \pi^0
\nu \bar \nu$ and $K^+ \to \pi^+ \nu \bar \nu$, which offer the
cleanest window into the sector of $s \rightarrow d$ transitions. In
Figure~\ref{fig:KLvsKp} we display the predictions for the branching
ratio of the neutral mode versus that of the charged one. The blue
points correspond to parameter values that reproduce the correct quark
masses and mixings and satisfy the constraints from $|\epsilon_K|$,
$Z^0 \to b \bar b$, and $B_d$--$\bar B_d$ mixing. For comparison, the
central value and the 68\% CL range of the experimental world average
${\cal B} (K^+ \to \pi^+ \nu \bar \nu (\gamma))_{\rm exp} = \left
  (1.73^{+1.15}_{-1.05} \right ) \cdot 10 ^{-10}$ based on seven
events \cite{Artamonov:2008qb} is indicated by the vertical dashed
black line and the yellow band. The experimental 90\% CL upper limit
${\cal B} (K_L \to \pi^0 \nu \bar \nu)_{\rm exp} < 2.6 \cdot 10^{-8}$
\cite{Ahn:2009gb} is not displayed in the figure. The central values
of the SM predictions (\ref{eq:KvvSM}) are indicated by the black
cross. As can be seen from the left panel showing the results of a
parameter scan in the benchmark scenario S1, the branching fractions
of both $K \to \pi \nu \bar \nu$ channels can be significantly
enhanced compared to the SM prediction. In particular, it is even
possible to saturate the model-independent Grossman-Nir (GN) bound
\cite{Grossman:1997sk}
\beq \label{eq:GNbound} 
  {\cal B} (K_L \to \pi^0 \nu \bar \nu) \hspace{0.25mm} \leq
  \hspace{0.25mm} \frac{\kappa_L}{\kappa_+ (1 + \Delta_{\rm EM}) } \,
  {\cal B} (K^+ \to \pi^+ \nu \bar \nu (\gamma)) \hspace{0.25mm} \approx
  \hspace{0.25mm} 4.3 \, {\cal B} (K^+ \to \pi^+ \nu \bar \nu (\gamma))
  \,,
\eeq 
which represents an enhancement of ${\cal B} (K_L \to \pi^0 \nu \bar
\nu)$ by almost a factor of 50 over the SM expectation for values of
${\cal B} (K^+ \to \pi^+ \nu \bar \nu (\gamma))$ at the upper end of
the experimentally favored range. The inequality (\ref{eq:GNbound}) is
indicated by the straight dotted black line in the figure, and the
inaccessible area is colored light gray. Because the weak phase
entering the $s \to d \nu \bar \nu$ transition can take essentially
any value in the standard scenario S1, the predictions for the charged
and neutral mode are highly uncorrelated, so that, similar to what is
observed in the case of the general minimal supersymmetric SM (MSSM)
\cite{Buras:2004qb}, every point in the ${\cal B} (K^+ \to \pi^+ \nu
\bar \nu (\gamma)) \hspace{0.25mm} $--$ \hspace{0.25mm} {\cal B} (K_L
\to \pi^0 \nu \bar \nu)$ plane that satisfies the GN bound can be
attained. Interestingly, due to the slow decoupling of the RS
corrections, enhancements of the rate of the neutral mode by almost an
order of magnitude are possible even for $M_{\rm KK} = 10 \, {\rm
  TeV}$. This feature, illustrated in the upper panel of
Figure~\ref{fig:KLvsKp}, implies that measurements of $K_L \to \pi^0
\nu \bar \nu$ can in principle probe scales far above those accessible
to direct searches at the LHC. As we will show in Section
\ref{sec:epsepsnum}, the possible enhancements in the $K \to \pi \nu
\bar \nu$ system are reduced noticeably if one takes into account the
constraint following from $\epseps$, because large shifts in the
neutrino modes typically appear together with an unacceptable amount
of direct CP violation in $K \to \pi \pi$.

While the pattern of deviations found in the benchmark scenario
S3 and S4 follows that observed in the standard case S1, a
strikingly different picture emerges in the scenario S2,
which features common bulk mass parameters $c_{d_i}$. This is
illustrated in the lower right panel of Figure~\ref{fig:KLvsKp}. To
understand the observed pattern, we first recall that in the benchmark
scenario S2 right-handed currents entering $|\epsilon_K|$ in
form of the Wilson coefficients $\tilde C_1^{\rm RS}$, $C_4^{\rm RS}$,
and $C_5^{\rm RS}$ as well as $K \to \pi \nu \bar \nu$ in form of
$\tilde C_\nu^{\rm RS}$ are parametrically suppressed by factors of
$v^2/\Mkk^2$ and thus numerically subleading compared to the
left-handed corrections $C_1^{\rm RS}$ and $C_\nu^{\rm RS}$, which to
leading order in $L$ are proportional to $(\Delta_D)_{12} \otimes
(\Delta_D)_{12}$ and $(\Delta_D)_{12}$. Furthermore, the fact that the
left-handed contribution $(\Delta_D)_{12} \otimes (\Delta_D)_{12}$
factorizes as $(\Delta_D)_{12} \otimes (\Delta_D)_{12} \approx \left
  ((\Delta_D)_{12} \right )^2 = \left |(\Delta_D)_{12} \right |^2 e^{2
  \hspace{0.25mm} i \hspace{0.25mm} \varphi_{12}}$ implies that
$|\epsilon_K|$ as well as $K \to \pi \nu \bar \nu$ are governed by the
same weak phase $\varphi_{12}$.  The requirement $|\epsilon_K|_{\rm
  RS} \propto {\rm Im} \hspace{0.mm} \left ((\Delta_D)_{12} \otimes
  (\Delta_D)_{12} \right ) \approx 0$ then forces $\varphi_{12}$ in
the standard CKM phase convention (\ref{eq:standardckm}) to satisfy
$\varphi_{12} \approx n \hspace{0.25mm} \pi/2$ with $n = 0,1,2,3$. One
can show in a model-independent fashion that in such a situation only
two branches of solutions in the ${\cal B} (K^+ \to \pi^+ \nu \bar \nu
(\gamma)) \hspace{0.25mm}$--$\hspace{0.25mm} {\cal B} (K_L \to \pi^0
\nu \bar \nu)$ plane are allowed \cite{Blanke:2009pq}. The first
branch features ${\cal B} (K_L \to \pi^0 \nu \bar \nu) \approx {\cal
  B} (K_L \to \pi^0 \nu \bar \nu)_{\rm SM}$, while the second one runs
through the SM point (\ref{eq:KvvSM}) with a slope approximately equal
to the one of the GN bound (\ref{eq:GNbound}).

In order to understand why the second branch of solutions is absent in
the case at hand requires further analysis. To leading order in
hierarchies, the $12$ elements of the charged-current matrix ${\bm
  V}_L$ and the mixing matrix ${\bm \Delta}_D$ take in the case of
``alignment'' the form\footnote{It turns out that the Froggatt-Nielsen
  analysis performed in \cite{Casagrande:2008hr} is still applicable
  in the ``aligned'' case once the expressions for down-quark masses
  $m_d, m_s, m_b$ and the rotations matrices ${\bm U}_d$ and ${\bm
    W}_d$ have been adjusted appropriately \cite{Sandro}. In
  particular, it is relatively easy to show that the relations
  (\ref{eq:alignedZMA}) simply follow from the ZMA results (I:A.1) and
  (I:A.2) by applying the replacements ${\bm Y}_d \to {\bm Y}_d {\bm
    Y}_d^\dagger$ and ${\bm M}_d \to {\bm M}_d {\bm M}_d^\dagger$.}
\begin{gather} 
  (V_L)_{12} = \left [ \frac{({\bm M}_d {\bm M}_d^\dagger)_{21}}{({\bm
        M}_d {\bm M}_d^\dagger)_{11}} - \frac{(M_u)_{21}}{(M_u)_{11}}
  \right ] \! \frac{F(c_{Q_1})}{F(c_{Q_2})} \,, \nonumber \\[-4mm]
  \label{eq:alignedZMA} \\[-4mm]
  (\Delta_D)_{12} = -\frac{1}{2} \left [ \frac{({\bm M}_d {\bm
        M}_d^\dagger)_{21}}{({\bm M}_d {\bm M}_d^\dagger)_{11}} +
    \frac{({\bm Y}_d {\bm Y}_d^\dagger)_{23}^\ast}{({\bm Y}_d {\bm
        Y}_d^\dagger)_{33}} \, \frac{({\bm M}_d {\bm
        M}_d^\dagger)_{31}}{({\bm M}_d {\bm M}_d^\dagger)_{11}} \right
  ] \! F(c_{Q_1}) \hspace{0.25mm} F(c_{Q_2}) \,. \nonumber
\end{gather}
Here ${\bm M}_d$ denotes the minors of ${\bm Y}_d$, \ie, the element
$({\bm M}_d)_{ij}$ gives the determinant of the matrix obtained by
deleting the $i^{\rm th}$ row and the $j^{\rm th}$ column of ${\bm
  Y}_d$. The same definitions apply in the case of $(M_u)_{ij}$.

It is clear that sizable corrections in $K \to \pi \nu \bar \nu$
correspond to large values of $|(\Delta_D)_{12}|$. Looking at
(\ref{eq:alignedZMA}) we see that $|(\Delta_D)_{12}|$ grows with
decreasing $({\bm Y}_d {\bm Y}_d^\dagger)_{33}$ and/or decreasing
$({\bm M}_d {\bm M}_d^\dagger)_{11}$. The first option requires
however a tuning of parameters, because $({\bm Y}_d {\bm
  Y}_d^\dagger)_{33} = \sum_i |(Y_d)_{3i}|^2$, which tells us that in
an anarchic setup the latter object has a natural size of ${\cal
  O}(Y_{\rm max}^2)$.  Barring accidental cancellations, we are thus
left with the second option. Yet in the limit $({\bm M}_d {\bm
  M}_d^\dagger)_{11} \to 0$, the expressions (\ref{eq:alignedZMA}) are
approximated by
\begin{gather} 
  (V_L)_{12} \approx \frac{({\bm M}_d {\bm M}_d^\dagger)_{21}}{({\bm
      M}_d {\bm M}_d^\dagger)_{11}} \, \frac{F(c_{Q_1})}{F(c_{Q_2})}
  \,, \nonumber \\[-4mm] \label{eq:limitZMA} \\[-4mm]
  (\Delta_D)_{12} \approx -\frac{1}{2} \left [ 1 + \frac{({\bm Y}_d
      {\bm Y}_d^\dagger)_{22}}{ ({\bm Y}_d {\bm Y}_d^\dagger)_{33}}
  \right ] \frac{({\bm M}_d {\bm M}_d^\dagger)_{21}}{({\bm M}_d {\bm
      M}_d^\dagger)_{11}} \, F(c_{Q_1}) \hspace{0.25mm} F(c_{Q_2}) \,,
  \nonumber
\end{gather}
so that in the standard CKM phase convention (\ref{eq:standardckm}),
\ie, after rotating away the phase of $(V_L)_{12}$, one ends up with
\beq \label{eq:DeltaDfinal} 
  (\Delta_D)_{12} \, e^{-i \, {\rm arg} \hspace{0.25mm} \left
      ((V_L)_{12} \right )} \approx -\frac{1}{2} \left [ 1 + \frac{({\bm
        Y}_d {\bm Y}_d^\dagger)_{22}}{ ({\bm Y}_d {\bm
        Y}_d^\dagger)_{33}} \right ] \frac{\big | ({\bm M}_d {\bm
      M}_d^\dagger)_{21} \big |}{({\bm M}_d {\bm M}_d^\dagger)_{11}} \,
  F(c_{Q_1}) \hspace{0.25mm} F(c_{Q_2}) \, < \, 0 \,.
\eeq 
Using (\ref{eq:Sigmas}) and (\ref{eq:Cnus}), we find that this
inequality implies that ${\rm Re} \, (X_{\rm RS}) < 0$ and ${\rm Im}
\, (X_{\rm RS}) \approx 0$ for the quantity $X_{\rm RS}$ defined in
(\ref{eq:XSMRS}).  Recalling that in the SM one has ${\rm Re} \,
(X_{\rm SM}) \approx -1.2$ and ${\rm Im} \, (X_{\rm SM}) \approx 0.3$,
we then deduce from (\ref{eq:BRKpivv}) that in the limit
(\ref{eq:limitZMA}) the ``aligned'' scenario predicts constructive
interference in the branching ratio of $K^+ \to \pi^+ \nu \bar \nu$,
while the $K_L \to \pi^0 \nu \bar \nu$ rate is expected to
approximately take its SM value. Invoking the $|\epsilon_K|$
constraint further drives the solutions toward (\ref{eq:DeltaDfinal}),
as it singles out solutions with $\varphi_{12} \approx \pi$, so that
${\rm Re} \, (X_{\rm RS}) < 0$ and ${\rm Im} \, (X_{\rm RS}) \approx
0$ turn out to hold in the ``aligned'' case even when
$|(\Delta_D)_{12}|$ is small. We conclude that in the benchmark
scenario S2 the allowed points in the ${\cal B} (K^+ \to \pi^+ \nu
\bar \nu (\gamma)) \hspace{0.25mm}$--$\hspace{0.25mm} {\cal B} (K_L
\to \pi^0 \nu \bar \nu)$ plane should show the striking feature that
they all lie on a horizontal line to the right of the SM point
(\ref{eq:KvvSM}). This feature is clearly exhibited in the right panel
of Figure~\ref{fig:KLvsKp}.

The above discussion highlights that the $K \to \pi \nu \bar \nu$
decays offer a unique tool to study the fermion geography in the
down-type quark sector of models with warped extra dimensions, since
they provide, in combination with $|\epsilon_K|$, a powerful way to
test the universality of new-physics contributions in $\Delta S = 1$
and $\Delta S = 2$ transitions. Precision measurements of the $K_L \to
\pi^0 \nu \bar \nu$ and $K^+ \to \pi^+ \nu \bar \nu$ branching
fractions feasible at high-intensity proton-beam facilities such as
NA62, J-PARC, and Project X should therefore be primary goals of the
future flavor-physics program and pursued with great vigor.

\begin{figure}[!t]
  \begin{center}
    \includegraphics[height=6.5cm]{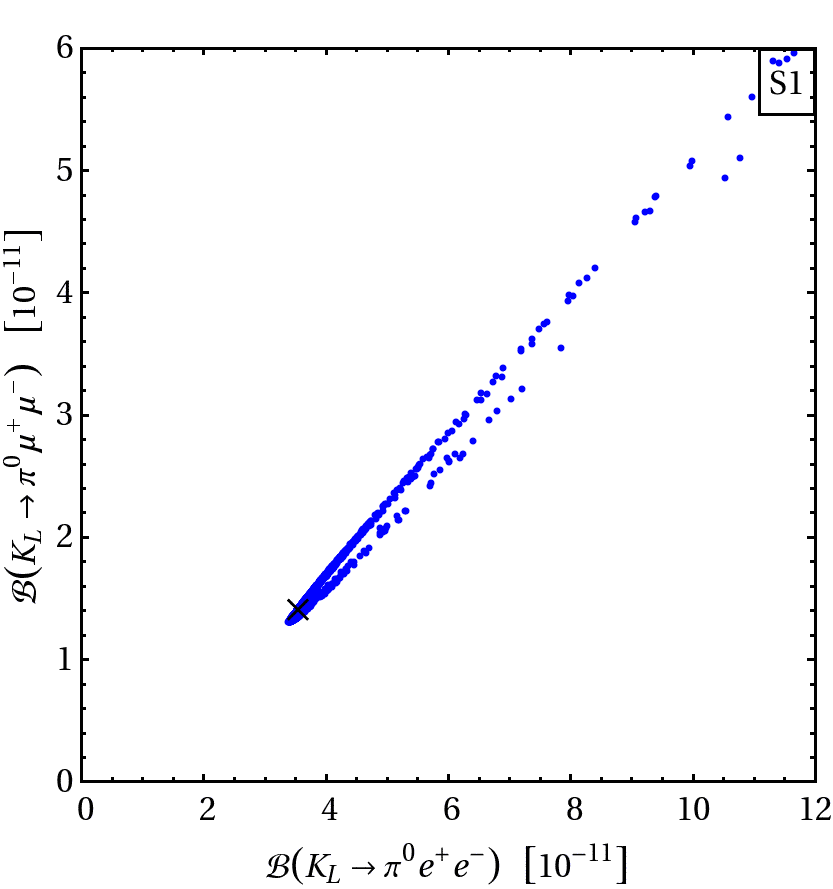}
    \hspace{0.75cm}
    \includegraphics[height=6.5cm]{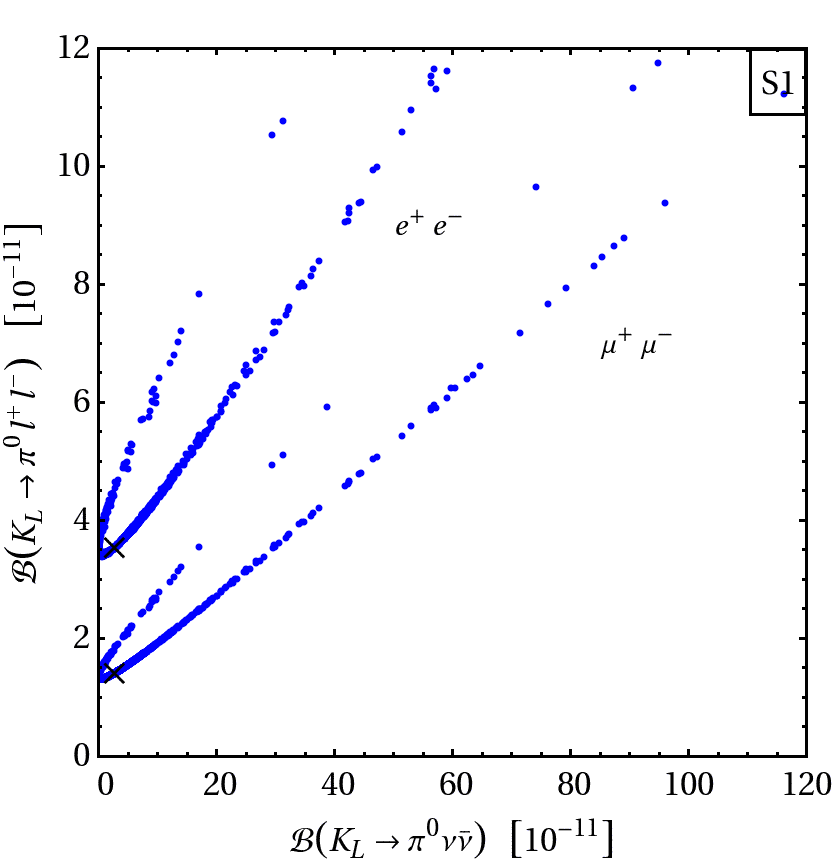}

    \vspace{4mm}

    \includegraphics[height=6.5cm]{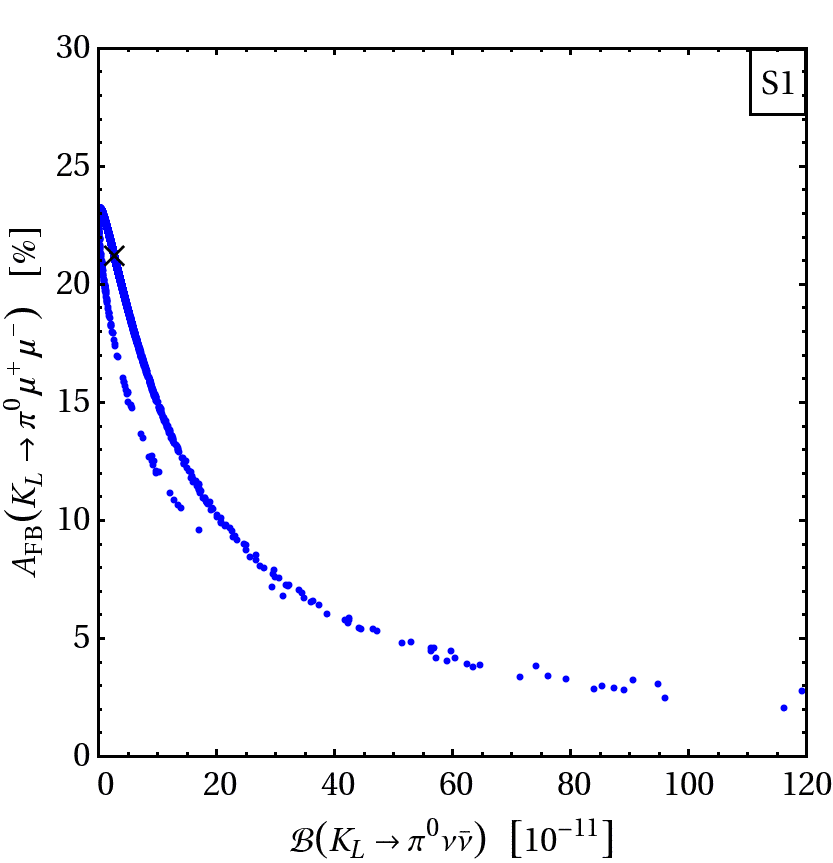}
  \end{center}
  \vspace{-8mm}
  \begin{center}
    \parbox{15.5cm}{\caption{\label{fig:Kllplot} Prediction for ${\cal
          B} (K_L \to \pi^0 e^+ e^-)$ versus ${\cal B} (K_L \to \pi^0
        \mu^+ \mu^-)$ (upper left), ${\cal B} (K_L \to \pi^0 \nu \bar
        \nu)$ versus ${\cal B} (K_L \to \pi^0 l^+ l^-)$ (upper right),
        and ${\cal B} (K_L \to \pi^0 \nu \bar \nu)$ versus $A_{\rm FB}
        (K_L \to \pi^0 \mu^+ \mu^-)$ (lower panel). The black crosses
        indicate the SM points while the blue scatter points reproduce
        the measured values of $|\epsilon_K|$, the $Z^0 b \bar b$
        couplings, and $B_d$--$\bar B_d$ mixing at 95\%, 99\%, and 95\%
        CL. See text for details.}}
  \end{center}
\end{figure}

While theoretically not as clean as $K \to \pi \nu \bar \nu$, the $K_L
\to \pi^0 l^+ l^-$ channels offer the unique opportunity to look for
and to constraint additional $\Delta S = 1$ effective operators that
are not accessible to the neutrino modes. In Figure~\ref{fig:Kllplot}
we show various predictions for $K_L \to \pi^0 l^+ l^-$ and their
correlation with the $K_L \to \pi^0 \nu \bar \nu$ mode. All plots
correspond to our default scenario S1. The predictions in the
remaining benchmark scenarios are essentially indistinguishable from
the ones displayed. The shown blue points reproduce the correct quark
masses and mixings and satisfy the constraints from $|\epsilon_K|$,
$Z^0 \to b \bar b$, and $B_d$--$\bar B_d$ mixing. The central values
of the SM predictions (\ref{eq:KvvSM}), (\ref{eq:BRKllSM}), and
(\ref{eq:AFBmmSM}) are indicated by black crosses. In the case of the
$K_L \to \pi^0 l^+ l^-$ observables we have assumed constructive
interference between the direct and indirect CP-violating
amplitudes. We observe that enhancements of the branching ratio of
both $K_L \to \pi^0 l^+ l^-$ modes by a factor of about 5 are possible
without violating any constraints. On the other hand, the RS
predictions for $A_{\rm FB} (K_L \to \pi^0 \mu^+ \mu^-)$ can be
smaller than the SM value by almost a factor of 10. Notice that
enhancements/suppressions in ${\cal B} (K_L \to \pi^0 \mu^+ \mu^-)$
and $A_{\rm FB} (K_L \to \pi^0 l^+ l^-)$ are anti-correlated, since
the branching ratio enters (\ref{eq:AFBKll}) in the denominator. The
pattern of the correlations seen in the upper left panel of
Figure~\ref{fig:Kllplot} arises because the coefficient of the
semileptonic vector operator, $Y_V$, is suppressed with respect to the
coefficient of the axial-vector operator, $Y_A$, by a factor of about
$(1-4\sws)\approx 0.08$, and scalar interactions play essentially no
role. This factor stems from the coupling of the $Z^0$ boson and its
KK excitations to the charged lepton pair. The ratio $Y_V/Y_A$
determines the angle between the two possible branches of ${\cal B}
(K_L \to \pi^0 \mu^+ \mu^-)$ as a function of ${\cal B} (K_L \to \pi^0
e^+ e^-)$. Since $Y_V/Y_A$ is generically small in the RS model, the
two branches are hardly visible in the latter figure. The correlations
observed in the upper right and lower panel of
Figure~\ref{fig:Kllplot} have a similar origin. In this case they are
a result of the interplay of the coupling of the $Z^0$ boson and its
KK excitations to a pair of charged and neutral leptons. While the
former is mostly axial-vector like, the latter is purely
left-handed. The observed correlations between $K_L \to \pi^0 \nu \bar
\nu$ and $K_L \to \pi^0 l^+ l^-$ should therefore be considered a
generic feature of models where the couplings of heavy neutral gauge
bosons to leptons are SM-like, rather than a specific characteristic
of the RS framework.

\begin{figure}[!t]
  \begin{center}
    \includegraphics[height=6.5cm]{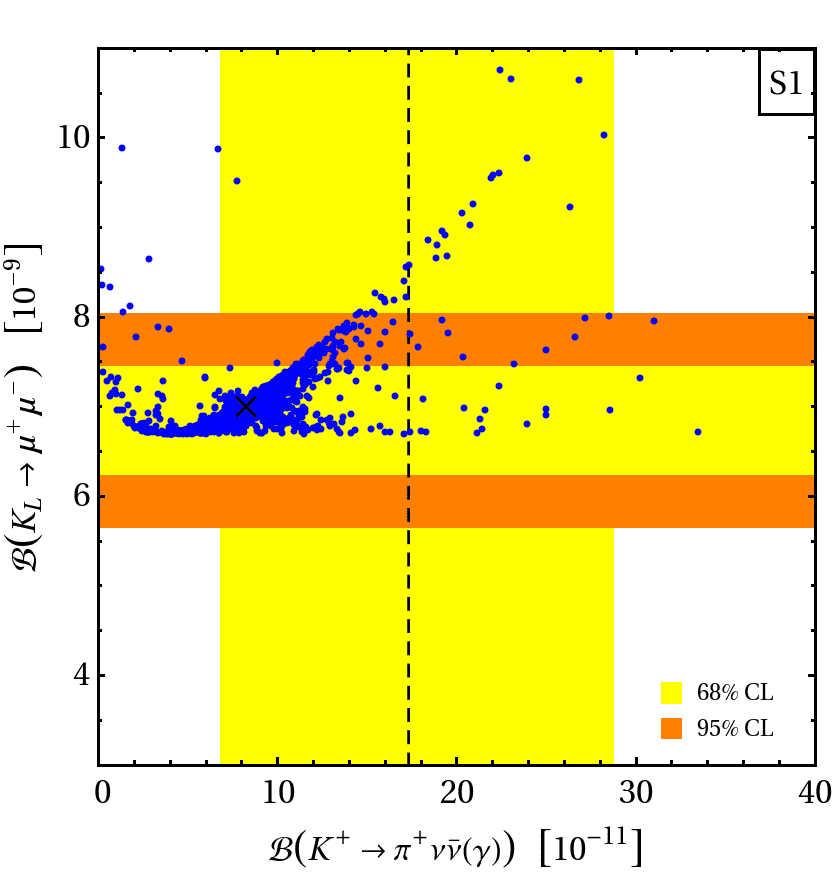}
    \hspace{0.75cm}
    \includegraphics[height=6.5cm]{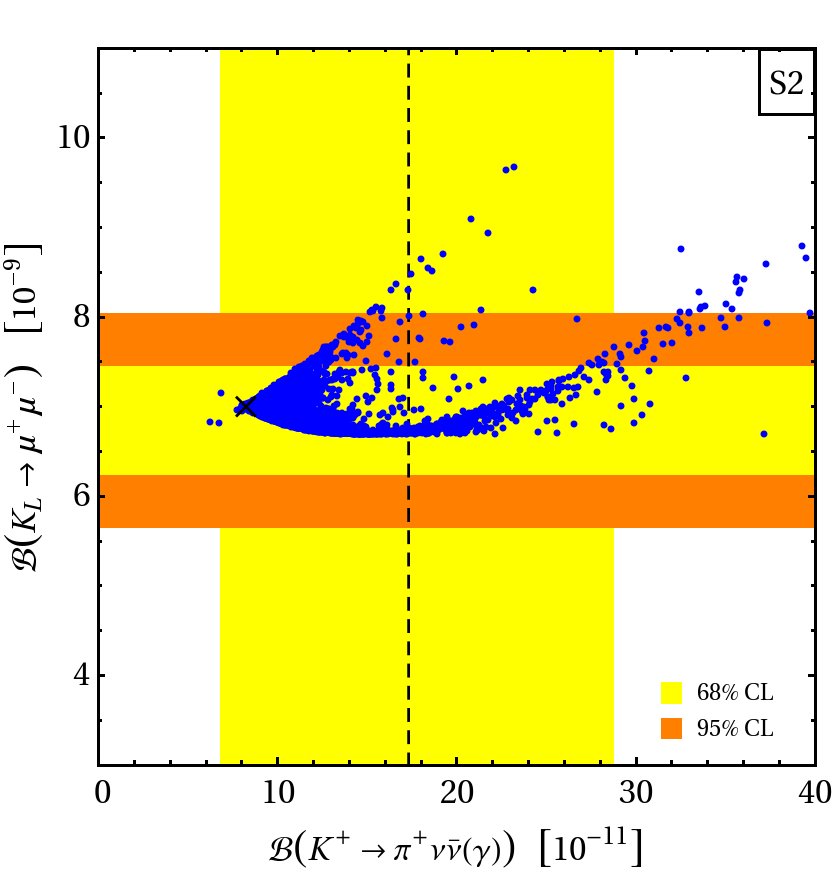}
  \end{center}
  \vspace{-8mm}
  \begin{center}
    \parbox{15.5cm}{\caption{\label{fig:Kmmplot} Prediction for ${\cal
          B} (K^+ \to \pi^+ \nu \bar \nu (\gamma))$ versus ${\cal B}
        (K_L \to \mu^+ \mu^-)$ in scenario S1 (left) and S2 (right). The
        black crosses indicate the SM point, while the blue scatter
        points reproduce the measured values of $|\epsilon_K|$, the
        $Z^0\to b \bar b$ couplings, and $B_d$--$\bar B_d$ mixing at
        95\%, 99\%, and 95\% CL. For comparison, the experimental
        central value (vertical dashed line) and 68\% CL range for
        ${\cal B} (K^+ \to \pi^+ \nu \bar \nu (\gamma))$ (yellow
        vertical band) as well as the 68\% (yellow horizontal band) and
        95\% (orange horizontal band) CL regions for ${\cal B} (K_L \to
        \mu^+ \mu^-)$ including both experimental and theoretical errors
        are shown. See text for details.}}
  \end{center}
\end{figure}

Interesting complementary information can be obtained from the $K_L
\to \mu^+ \mu^-$ decay mode, which has been measured precisely. The
predictions for ${\cal B} (K_L \to \mu^+ \mu^-)$ as a function of
${\cal B} (K^+ \to \pi^+ \nu \bar \nu (\gamma))$ in the parameter
scenarios S1 and S2 are displayed in the left and right
panels of Figure~\ref{fig:Kmmplot}, respectively. The blue scatter
points again reproduce the correct quark masses and the CKM elements
and satisfy the constraints from $|\epsilon_K|$, $Z^0 \to b \bar b$,
and $B_d$--$\bar B_d$ mixing. The central values of the SM predictions
(\ref{eq:KvvSM}) and (\ref{eq:KmmSM}) are indicated by black
crosses. In obtaining the figures we have assumed that the two-photon
amplitude in $K_L \to \mu^+ \mu^-$ has positive sign. The 68\% (95\%)
CL regions are underlaid in yellow (orange). In the case of the $K_L
\to \mu^+ \mu^-$ branching ratio the error band includes the theory
error, which in fact by far dominates over the experimental
uncertainty. We observe that the branching ratio of $K_L \to \mu^+
\mu^-$ can reach values up to $10^{-8}$, which represents an
enhancement of more than 40\% relative to both the SM prediction and
the measured value, ${\cal B} (K_L \to \mu^+ \mu^-)_{\rm exp} = (6.84
\pm 0.11) \cdot 10^{-9}$ \cite{Amsler:2008zz}. A positive linear
correlation between $K^+ \to \pi^+ \nu \bar \nu $ and $K_L \to \mu^+
\mu^-$ is also visible in the panels. This correlation originates from
the fact that $K^+ \to \pi^+ \nu \bar \nu$ measures the vector, while
$K_L \to \mu^+ \mu^-$ measures the axial-vector component of the $Z^0
d \bar s$ vertex. Since the SM flavor-changing $Z^0$ penguin is purely
left-handed and the RS contribution is dominated in our case by the
very same component, the SM and new-physics contributions enter both
decay modes with the same sign. Notice that the correlation is more
pronounced in scenario S2 than in S1, because the
right-handed contributions to the $Z^0 d \bar s$ vertex are further
suppressed in the former benchmark scenario. In a RS variant with
custodial protection, the correlation between $K^+ \to \pi^+ \nu \bar
\nu$ and $K_L \to \mu^+ \mu^-$ has been found to be an inverse one
\cite{Blanke:2008yr}. Precision measurements of $K^+ \to \pi^+ \nu
\bar \nu$ accompanied by major theoretical progress in the prediction
of $K_L \to \mu^+ \mu^-$ thus would allow one to identify the chiral
structure of the $Z^0 d \bar s$ vertex and in this way to select
between different models of non-standard interactions.

We close this section with a critical comparison of our results with
the findings of \cite{Blanke:2008yr}, which analyses the rare decays
$K \to \pi \nu \bar \nu$, $K_L \to \pi^0 l^+ l^-$, and $K_L \to \mu^+
\mu^-$ in the context of a warped extra-dimension scenario with
custodial protection of the $Z^0 d_{i \hspace{0.25mm} L} \bar d_{j
  \hspace{0.25mm} L}$ vertices. Our discussion will shed some light on
the model dependence of the obtained results and applies in a similar
fashion to the rare $B$-meson decays examined below. We start with a
comparison of the left- and right-handed couplings of the $Z^0$ boson
to down-type quarks. Focusing on the corrections arising from the
non-trivial overlap integrals of gauge-boson and fermion profiles and
employing the implementation of the quark sector as described in
\cite{Blanke:2008zb, Blanke:2008yr, Carena:2007ua, Medina:2007hz}, we
find that the ratios of the flavor-changing corrections to the
$Z^0$-boson couplings in the two models under consideration are given
by
\begin{align} \label{eq:downcouplings}
  \frac{(g_L^d)_{ij}^{\text{custodial}}}{(g_L^d)_{ij}^{\text{
        original}}} & = -\frac{(1/2 - \sws/3) \,
    (\Delta_D^\prime)_{ij}}{(1/2 - \sws/3) \, (L \, (\Delta_D)_{ij} -
    (\Delta_D^\prime)_{ij})} \approx -\frac{1}{L} \,, \nonumber \\[1mm]
  \frac{(g_R^d)_{ij}^{\text{custodial}}}
  {(g_R^d)_{ij}^{\text{original}}} & = -\frac{\cws \, L \,
    (\Delta_d)_{ij} - \sws/3 \, (\Delta_d^\prime)_{ij}}{\sws/3 \, (L
    \, (\Delta_d)_{ij} - (\Delta_d^\prime)_{ij})} \approx -\frac{3
    \cws}{\sws} \,,
\end{align}
where in the last step we have used that $(\Delta_{D,d})_{ij} \approx
(\Delta_{D,d}^\prime)_{ij}$ and neglected subleading terms in $L
\approx 37$. The $\bm{\Delta}_{D,d}^{(\prime)}$ matrices are defined
in (I:122). From the first relation it is obvious that only the
leading term in $L$ appearing in the left-handed couplings is
protected by the combination of the custodial and the left-right
exchange symmetry, while no such protection is active for the
subleading terms. This implies that the corrections to the $Z^0 d_{i
  \hspace{0.25mm} L}\bar d_{j \hspace{0.25mm} L}$ vertices arising
from the gauge sector are {\em parametrically} suppressed by a factor
of $L$ in the $SU(2)_L \times SU(2)_R \times P_{LR}$ model relative to
the original RS model, rather than by a purely numerical factor of
about 100 as claimed in \cite{Blanke:2008zb, Blanke:2008yr}. On the
other hand, we find that the $Z^0 d_{i \hspace{0.25mm} R} \bar d_{j
  \hspace{0.25mm} R}$ vertices in the model with custodial protection
are enhanced in magnitude by a factor of about 10 relative to the
original RS formulation, which is in accordance with
\cite{Blanke:2008zb, Blanke:2008yr}. We presented a more detailed
discussion of the implementation of the custodial protection
mechanism, treating the effects of a brane-localized Higgs in the
$SU(2)_L \times SU(2)_R \times P_{LR}$ model exactly, in another
publication \cite{Casagrande:2010si}.

\begin{figure}[!t]
  \begin{center}
    \includegraphics[height=6.45cm]{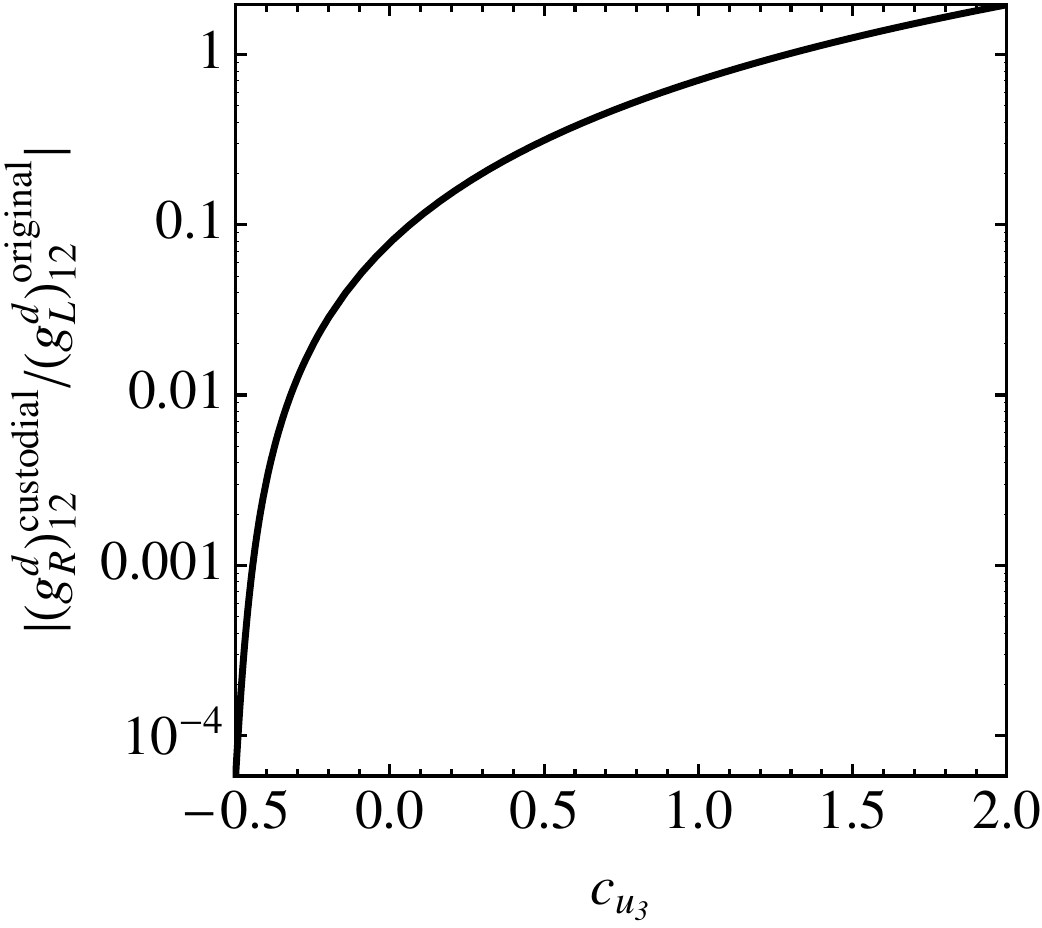}
    \hspace{0.5cm}
    \includegraphics[height=6.5cm]{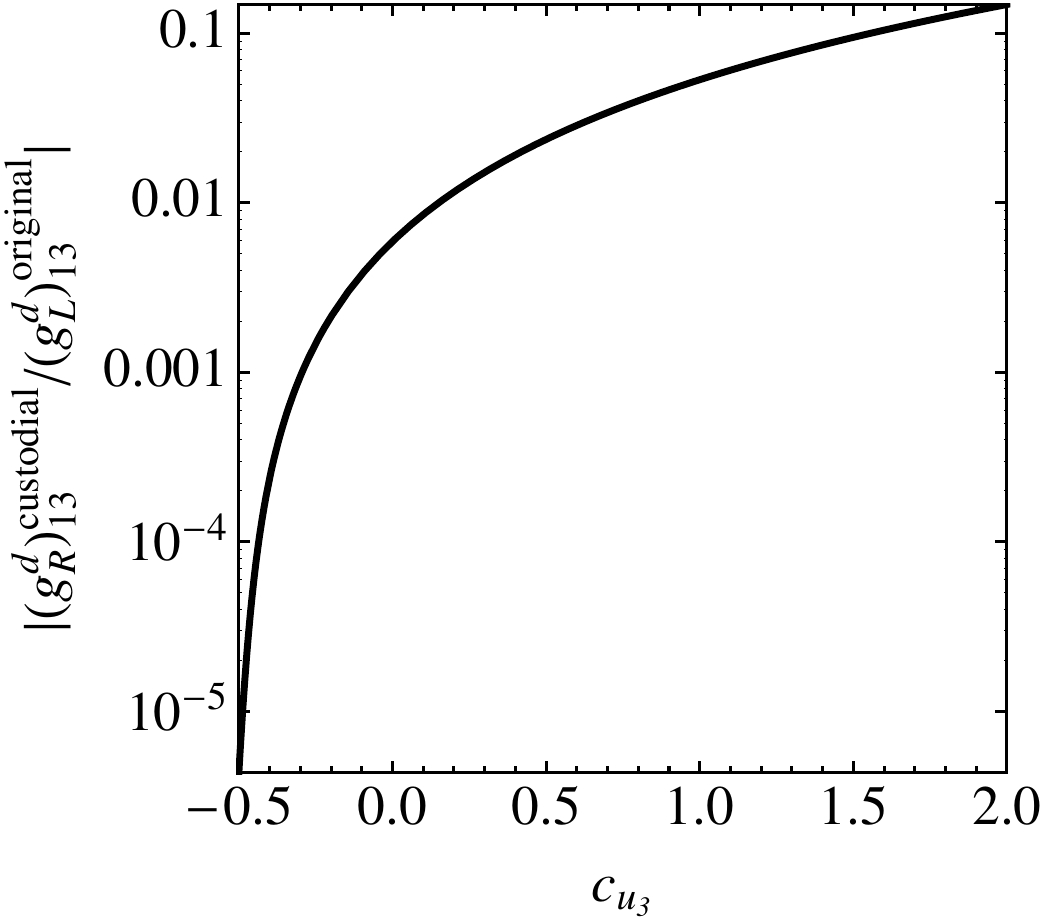}
  \end{center}
  \vspace{-8mm}
  \begin{center}
    \parbox{15.5cm}{\caption{\label{fig:cusvsorg} Predictions for $\big
        |(g_R^d)_{ij}^{\text{custodial}}/(g_L^d)_{ij}^{\text{original}}
        \big |$ as a function of the bulk mass parameter $c_{u_3}$ for
        the cases $ij=12$ (left) and $13$ (right). The shown curves
        correspond to the choice $Y_u = Y_d = 1$. See text for details.
      }}
  \end{center}
\end{figure} 

From (\ref{eq:downcouplings}) we observe that
$(g_R^d)_{ij}^{\text{custodial}}/(g_L^d)_{ij}^{\text{original}}
\approx -\cws/(1/2 - \sws/3) \, (\Delta_d)_{ij}/(\Delta_D)_{ij}
\approx -2 \, (\Delta_d)_{ij}/(\Delta_D)_{ij}$. The scaling relations
(I:128) then imply that $|(g_R^d)_{ij}^{\text{
    custodial}}/(g_L^d)_{ij}^{\text{original}}| \sim (F(c_{d_i})
F(c_{d_j}))/(F(c_{Q_i}) F(c_{Q_j}))$ up to a factor of $\ord(1)$,
which tells us that in order to have effects in rare kaon decays of
similar magnitude in the custodially protected and the original RS
model requires $F(c_{d_1}) F(c_{d_2}) \approx F(c_{Q_1})
F(c_{Q_2})$. Notice that to arrive at these relations, we used the
fact that the bulk masses do not depend on whether one considers the
custodially protected or the original RS model, since the $c_{A_i}$
parameters are determined by the quark masses and mixings. Yet the
locations of the zero-modes are not unique solutions
\cite{Huber:2003tu}, a feature that manifests itself in the
invariance under a set of reparametrization transformations
\cite{Casagrande:2008hr}. In particular, a simultaneous rescaling of
the fermion profiles for $SU(2)_L$ doublet and singlet fields by
opposite factors, while leaving the 5D Yukawa couplings invariant,
allows one to redistributes effects between the left- and right-handed
sectors. This freedom in combination with the scaling relations
(I:128) can be used to express $(F(c_{d_1}) F(c_{d_2}))/(F(c_{Q_1})
F(c_{Q_2}))$ through the quark masses, Wolfenstein parameters, and
$F(c_{u_3})$. To leading power in hierarchies we find
\beq \label{eq:cusvsorg} 
\left |\frac{(g_R^d)_{12}^{\text{custodial}}}
  {(g_L^d)_{12}^{\text{original}}} \right| \sim \frac{m_d m_s \, Y_u^4
  v^2}{2 A^4 \, \lambda^{10} \, m_t^4 \, Y_d^2 } \left[ F(c_{u_3})
\right]^4 = 0.08 \; \frac{Y_u^4}{Y_d^2} \left[ F(c_{u_3}) \right]^4 ,
\eeq 
which implies that the ratio of right- to left-handed $Z^0 d \bar s$
couplings can be enhanced by localizing the right-handed top quark
closer to the IR brane. Analogous formulas hold in the case of the
$Z^0 \to b \bar d \, (b\bar s)$ transitions with $(m_d m_s)/(A^4
\lambda^{10})$ replaced by $(m_d m_b)/(A^2 \lambda^6)$ $\big ( (m_s
  m_b)/(A^2 \lambda^4) \big)$. The corresponding numerical factors
replacing 0.08 in (\ref{eq:cusvsorg}) are 0.006 and 0.005,
respectively.

The above considerations make it clear that the degree of
compositeness of the top quark plays a crucial role in the context of
FCNC interactions as it determines the relative size of left- to
right-handed couplings. To make the latter statement more quantitative
we plot in Figure~\ref{fig:cusvsorg} the ratio $ \big
|(g_R^d)_{ij}^{\text{custodial}}/(g_L^d)_{ij}^{\text{original}} \big
|$ as a function of $c_{u_3}$ for the cases $ij=12$ and $13$. The case
$ij=23$ is not shown explicitly, since it basically resembles the one
of $ij=13$. The displayed curves have been obtained by assuming for
simplicity that the various elements of the up- and down-type Yukawa
matrices are all equal to 1. The strong dependence of the results on
$c_{u_3}$, arising from the factor $[F(c_{u_3})]^4 \approx
(1+2c_{u_3})^2$ in (\ref{eq:cusvsorg}), is clearly visible in both
panels. We furthermore see that requiring $\big
|(g_R^d)_{12}^{\text{custodial}}/(g_L^d)_{12}^{\text{original}} \big |
\gtrsim 1$ translates into the limit $c_{u_3} \gtrsim 1$ on the bulk
mass parameter of the right-handed top quark. In the case of the $Z^0
\to b \bar d \, (b\bar s)$ transitions the corresponding limit is
$c_{u_3} \gtrsim 6$. We conclude that in order to obtain $Z^0 d_{i
  \hspace{0.25mm} R} \bar d_{j \hspace{0.25mm} R}$ couplings in the
custodially protected model comparable in magnitude to the $Z^0 d_{i
  \hspace{0.25mm} L} \bar d_{j \hspace{0.25mm} L}$ couplings in the
original RS framework requires, barring a conspiracy of undetermined
${\cal O} (1)$ factors, parametrically large values of $c_{u_3}$.
Such a choice seems however unnatural, since $c_{u_3} > 1$ implies
that the corresponding bulk mass exceeds the curvature scale, in which
case the right-handed top quark should be treated as a brane-localized
and not a bulk fermion. The finding of \cite{Blanke:2008yr} that
${\cal O} (1)$ enhancements in $K \to \pi \nu \bar \nu$, $K_L \to
\pi^0 l^+ l^-$, and $K_L \to \mu^+ \mu^-$ due to right-handed
interactions are possible in the RS model with custodial protection
has thus to be taken with a grain of salt, because these solutions
typically feature bulk mass parameters $c_{u_3}$ that lie quite
significantly above $-1/2$. In this context it is also important to
realize that in the limit from the right $c_{u_3} \to -1/2^+$, which
is sufficient to generate the large mass of the top quark, the ratio
$\big |(g_R^d)_{12}^{\text{custodial}}/(g_L^d)_{12}^{\text{original}}
\big |$ scales as $1/L^2$ and consequently $\big
|(g_R^d)_{12}^{\text{custodial}}/(g_L^d)_{12}^{\text{custodial}} \big
|$ becomes proportional to $1/L$. This means that in this case the
right-handed couplings are not $L$-enhanced but rather $L$-suppressed
relative to the custodially protected left-handed couplings, and that
the branching fractions of $K \to \pi \nu \bar \nu$, $K_L \to \pi^0
l^+ l^-$, and $K_L \to \mu^+ \mu^-$ are predicted to be SM-like. A
detailed analysis of rare kaon decays in the $SU(2)_L \times SU(2)_R
\times P_{LR}$ model will be presented elsewhere.

\boldmath \subsubsection{Numerical Analysis of $B_q \to \mu^+ \mu^-$
  and $B \to X_q \nu \bar \nu$} \unboldmath

\begin{figure}[!t]
  \begin{center}
    \includegraphics[height=6.5cm]{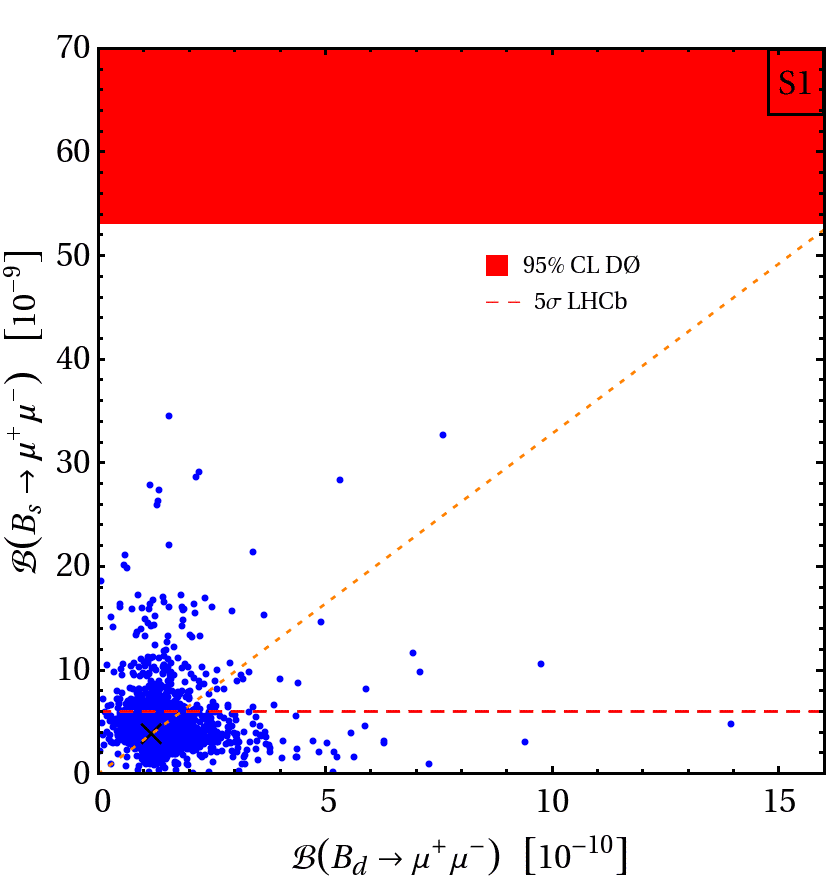}
    \hspace{0.75cm}
    \includegraphics[height=6.5cm]{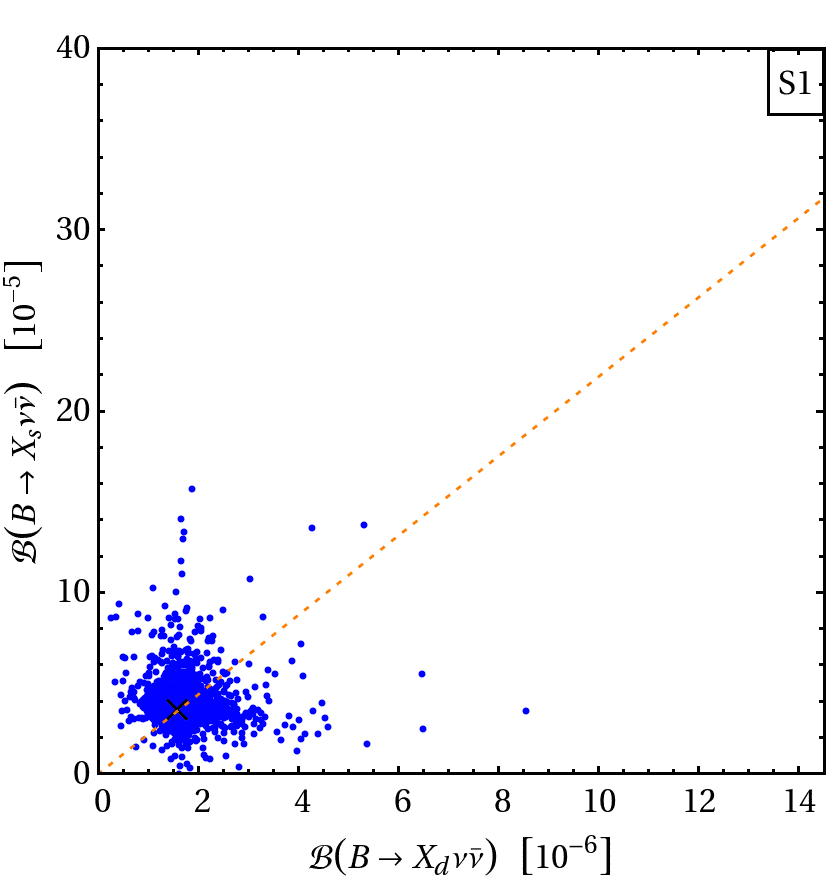}

    \vspace{4mm}

    \includegraphics[height=6.5cm]{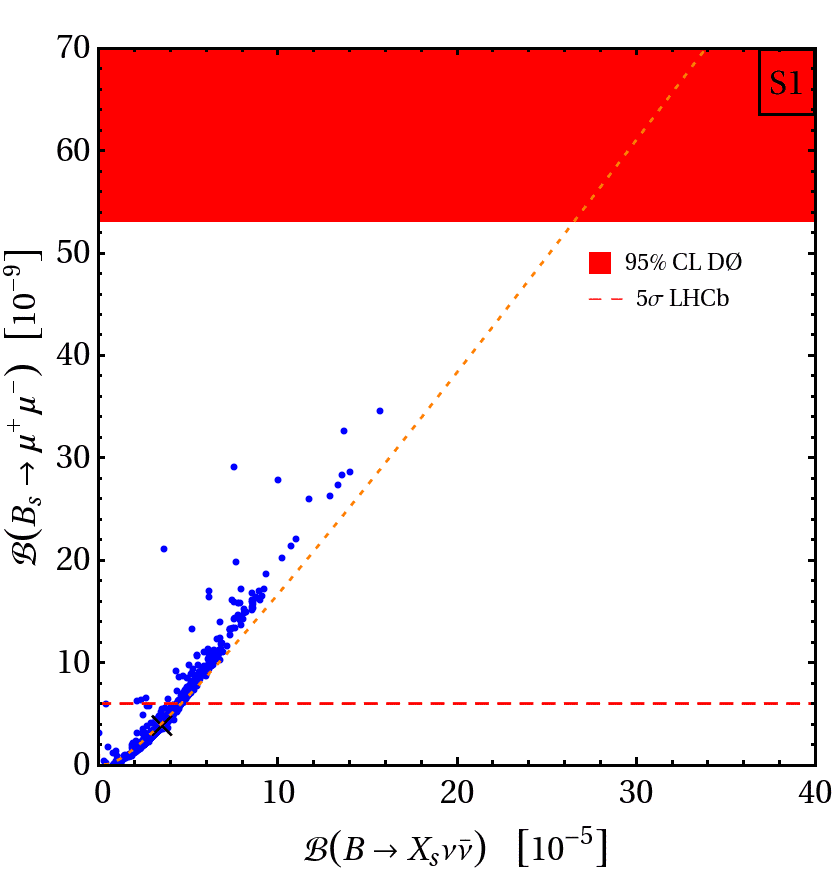}
  \end{center}
  \vspace{-8mm}
  \begin{center}
    \parbox{15.5cm}{\caption{\label{fig:rareB} Prediction for ${\cal B}
        (B_d \to \mu^+ \mu^-)$ versus ${\cal B} (B_s \to \mu^+ \mu^-)$
        (upper left), ${\cal B} (B \to X_d \nu \bar \nu)$ versus ${\cal
          B} (B \to X_s \nu \bar \nu)$ (upper right), and ${\cal B} (B
        \to X_s \nu \bar \nu)$ versus ${\cal B} (B_s \to \mu^+ \mu^-)$
        (lower panel). All panels show results obtained in benchmark
        scenario S1. The black crosses indicate the SM point, while the
        blue scatter points reproduce the measured values of
        $|\epsilon_K|$, the $Z^0 b \bar b$ couplings, and $B_d$--$\bar
        B_d$ mixing at 95\%, 99\%, and 95\% CL. In the upper left panel
        the current 95\% CL upper limit on ${\cal B} (B_s \to \mu^+
        \mu^-)$ from {D\O} and the minimum branching fraction allowing
        for a $5\sigma$ discovery at LHCb are indicated by the red band
        and dashed line, respectively. The orange dotted lines in the
        upper panels represent the CMFV correlation between the two
        purely leptonic/semileptonic modes, while the orange dotted
        curve in the lower panel indicates the model-independent
        prediction obtained under the assumption that only left-handed
        operators contribute to the branching fractions. See text for
        details.}}
  \end{center}
\end{figure}

In this section we perform numerical studies of the impact of virtual
KK exchange on the predictions of $B_q \to \mu^+ \mu^-$ and $B \to X_q
\nu \bar \nu$, where $q = d, s$. We begin with the purely semileptonic
modes. The predictions for ${\cal B} (B_s \to \mu^+ \mu^-)$ as a
function of ${\cal B} (B_d \to \mu^+ \mu^-)$ obtained from a parameter
scan in the scenario S1 are displayed in the upper left panel of
Figure \ref{fig:rareB}. This plot is also representative for the
results obtained in the other benchmark scenarios. The blue points are
in agreement with the observed quark masses and CKM elements and
satisfy the constraints from $|\epsilon_K|$, $Z^0 \to b \bar b$, and
$B_d$--$\bar B_d$ mixing. The central values of the SM predictions
(\ref{eq:BqmmSM}) are indicated by the black cross. The distribution
of points indicates that large uncorrelated enhancements by a factor
of 10 are possible in both the $B_d$ and $B_s$ decay modes without
violating existing constraints. We also see that large deviations from
the relation ${\cal B} (B_s \to \mu^+ \mu^-) = (f_{B_s}^2 m_{B_s}
\tau_{B_s} |V_{ts}|^2)/(f_{B_d}^2 m_{B_d} \tau_{B_d} |V_{td}|^2) \,
{\cal B} (B_d \to \mu^+ \mu^-) \approx 32.8 \, {\cal B} (B_d \to \mu^+
\mu^-)$ \cite{Buras:2003td}, which is valid in models with
constrained MFV (CMFV) \cite{Buras:2000dm}, can naturally appear in
the RS model. This feature is indicated by the orange dotted
line. Notice that the 95\% CL upper limit ${\cal B}(B_s \to \mu^+
\mu^-)_{\rm exp} < 5.3 \cdot 10^{-8}$ based on $5 \ {\rm fb}^{-1}$ of
data at {D\O} \cite{D0note2}, which is a factor of 14 above the SM
expectation,\footnote{An unofficial average of the CDF
  \cite{Aaltonen:2007kv} and {D\O} data yields an upper bound of $4.5
  \cdot 10^{-8}$ at 95\% CL \cite{Punzi}.} almost starts to constrain
the parameter space of the RS model. The experimentally disfavored
region is indicated by the red band in the left panel of
Figure~\ref{fig:rareB}. Both CDF and {D\O} should reach limits in the
ballpark of $2 \cdot 10^{-8}$ for $8 \ {\rm fb}^{-1}$ of data. For
comparison we have also included in the plot the minimum of ${\cal B}
(B_s \to \mu^+ \mu^-)$ allowing for a signal discovery with $5\sigma$
significance with $2 \ {\rm fb}^{-1}$ of integrated luminosity, which
is expected to be $6 \cdot 10^{-9}$ at LHCb \cite{LHCbnote}. This
sensitivity is indicated by the dashed red line in the panel. A
discovery at the SM level requires about $6 \ {\rm fb}^{-1}$ ($100 \
{\rm fb}^{-1}$) of data at LHCb (ATLAS or CMS) and will require
several years of LHC running \cite{LHCbnote, Aad:2009wy,
  Langenegger:2006jk}. The existing limit in the case of
$B_d\to\mu^+\mu^-$ is much weaker, ${\cal B}(B_d \to \mu^+ \mu^-)_{\rm
  exp} < 1.8 \cdot 10^{-8}$ at 95\% CL \cite{Aaltonen:2007kv}. It is
therefore not shown in the plot. While it should be possible to
improve the latter bound at LHCb by an order of magnitude with $10 \,
{\rm fb}^{-1}$, an observation/discovery of $B_d \to \mu^+ \mu^-$ at
the SM level is even challenging for a super flavor factory.

We now move onto the rare semileptonic modes. The predictions for
${\cal B} (B \to X_d\nu \bar \nu)$ versus ${\cal B} (B \to X_s \nu
\bar \nu)$ corresponding to the benchmark scenario S1 are shown in the
upper right panel of Figure~\ref{fig:rareB}. Similar results are
obtained in the remaining scenarios. Like in the case of $B_q \to
\mu^+ \mu^-$, we see that enhancements of an order of magnitude are
allowed in both $B \to X_q \nu \bar \nu$ decay modes and that the
deviations do not follow the linear relation ${\cal B} (B \to X_s \nu
\bar \nu) = |V_{ts}|^2/|V_{td}|^2 \, {\cal B} (B \to X_d \nu \bar \nu)
\approx 21.9 \ {\cal B} (B \to X_d \nu \bar \nu)$ predicted in CMFV
scenarios. The latter correlation is indicated by the orange dotted
line in the panel. While theoretically the cleanest and thus the most
attractive mode, the existing bound on the branching ratio of the
inclusive $b \to s \nu \bar \nu$ transition is very weak, ${\cal B} (B
\to X_s \nu \bar \nu)_{\rm exp} < 6.4 \cdot 10^{-4}$ at 90\% CL
\cite{Barate:2000rc}. Stronger bounds exist in the case of the
exclusive modes, ${\cal B} (B \to K \nu \bar \nu) < 1.4 \cdot 10^{-5}$
\cite{Chen:2007zk} and ${\cal B} (B \to K^\ast \nu \bar \nu) < 8.0
\cdot 10^{-5}$ \cite{Aubert:2008fr} both at 90\% CL. While fully
inclusive measurements will be notoriously difficult to perform even
with the highest available luminosity at a super flavor factory, an
observation of $B \to K^{\ast} \nu \bar \nu$ is expected to be
possible with $50 \ {\rm ab}^{-1}$ of integrated luminosity at such a
machine \cite{Renga}. Since the pattern of deviations observed in the
exclusive transitions follows the one in $B \to X_s \nu \bar \nu$, our
discussion of non-standard effects in $B \to X_s \nu \bar \nu$
essentially also apply to the modes $B \to K^{(\ast)} \nu \bar \nu$.

In the lower panel of Figure~\ref{fig:rareB}, we finally show the
correlation between ${\cal B} (B \to X_s \nu \bar \nu)$ and ${\cal B}
(B_s \to \mu^+ \mu^-)$. The plot clearly shows that for large values
of the branching ratios there is a linear relation between the rates
of $B \to X_s \nu \bar \nu$ and $B_s \to \mu^+ \mu^-$. This pattern
arises since both modes receive in effect only corrections from
tree-level exchange of the $Z^0$ boson and its KK excitations. Scalar
contributions to $B_s \to \mu^+ \mu^-$ are highly suppressed, and
corrections due to KK photons do not contribute to the purely leptonic
rate since they couple vectorial. Given that in the RS model without
custodial protection the contribution from right-handed operators are
suppressed with respect to the left-handed ones, it is easy to show
that the branching ratios in question satisfy approximately
\beq \label{eq:rareBcorr}
{\cal B} (B_s \to \mu^+ \mu^-) \approx \left ( 1.1 -3.3 \left [
    \frac{{\cal B} (B \to X_s \nu \bar \nu)}{10^{-5}} \right ]^{1/2} +
  2.6 \; \frac{{\cal B} (B \to X_s \nu \bar \nu)}{10^{-5}} \right )
\cdot 10^{-9} \,.
\eeq
This relation is indicated by the orange dotted curve in the lower
panel of Figure~\ref{fig:rareB}. Deviation from the behavior
(\ref{eq:rareBcorr}) measure the strength of the $Z^0 b_R \bar s_R$
relative to the $Z^0 b_L \bar s_L$ coupling. Notice that while ${\cal
  B} (B \to X_s \nu \bar \nu)$ depends quadratically on the $Z^0 b_R
\bar s_R$ coupling, ${\cal B} (B_s \to \mu^+ \mu^-)$ contains also
linear terms, so that small right-handed contributions will have a
bigger impact in the purely leptonic than in the semileptonic mode.
Furthermore, in the $B \to X_s \nu \bar \nu$ decay non-zero
right-handed couplings necessarily enhance the branching fraction,
whereas in the $B_s \to \mu^+ \mu^-$ mode both constructive and
destructive interference is possible. From (\ref{eq:Cl}),
(\ref{eq:BRBqmm}), and (\ref{eq:CACAp}) one can convince oneself that
in the minimal RS model right-handed operators typically add to the
$B_s \to \mu^+ \mu^-$ rate, which implies that the slope of the
correlation between ${\cal B} (B \to X_s \nu \bar \nu)$ and ${\cal B}
(B_s \to \mu^+ \mu^-)$ should be steeper than the one predicted in
(\ref{eq:rareBcorr}). This is indeed what is observed in the
plot. Notice that in the case where right-handed new-physics
contributions dominate over left-handed ones, $B \to X_s \nu \bar \nu$
and $B_s \to \mu^+ \mu^-$ are again correlated. In the absence of
additional contributions to the $Z^0 b_L \bar s_L$ vertex, the
predictions in the ${\cal B} (B \to X_s \nu \bar \nu)$--${\cal B} (B_s
\to \mu^+ \mu^-)$ plane form a parabola-like curve that is tilted to
the right. Depending on whether the right-handed contributions
interfere constructively or destructively in $B_s \to \mu^+ \mu^-$,
the predictions fall on the upper or lower branch of the curve.

\boldmath \subsubsection{Numerical Analysis of $B \to X_s l^+ l^-$ and
  $B \to K^{\ast} l^+ l^-$} \unboldmath

Our explorations in the sector of semileptonic $B$-meson decays are
rounded off by a study of the numerical impact of the exchange of the
$Z^0$ boson and its KK excitations on the decays $B \to X_s l^+ l^-$
and $B \to K^\ast l^+ l^-$. Before we present our findings, let us
stress again that effects of electromagnetic dipole operators entering
the effective Hamiltonian for $b \to s l^+ l^-$ first at the one-loop
level are not included in our analysis. The possibility that such
loop-suppressed effects could have a non-negligible impact on the
obtained results for the various $b \to s l^+ l^-$ observables should
be kept in mind. We leave a detailed analysis of this issue for future
work.

\begin{figure}[!t]
  \begin{center}
    \includegraphics[height=6.5cm]{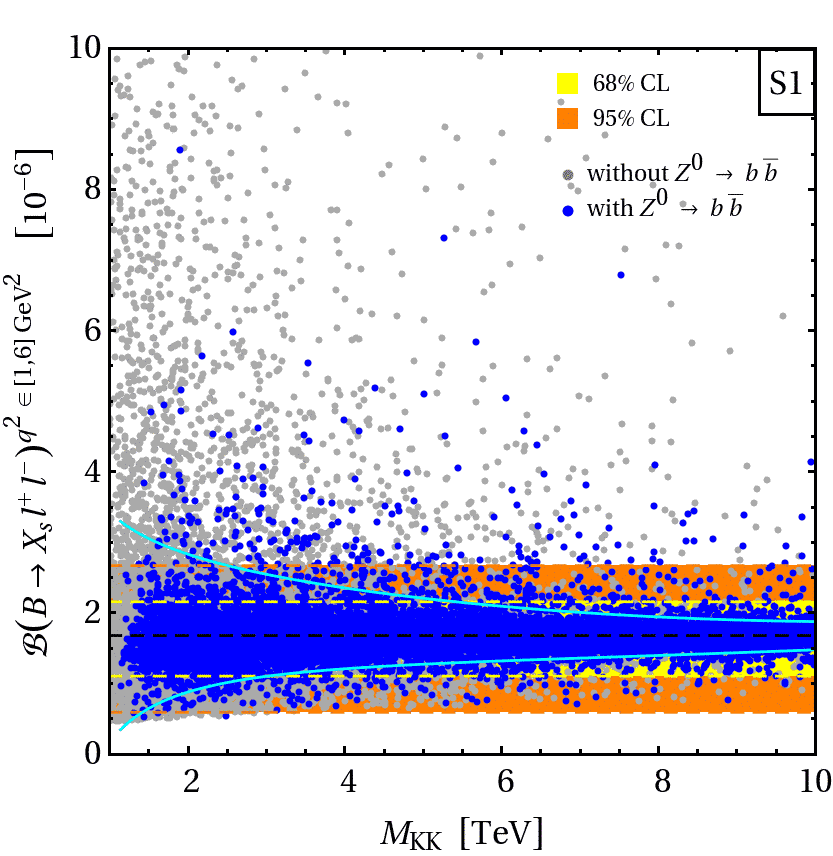}
    \hspace{0.75cm}
    \includegraphics[height=6.5cm]{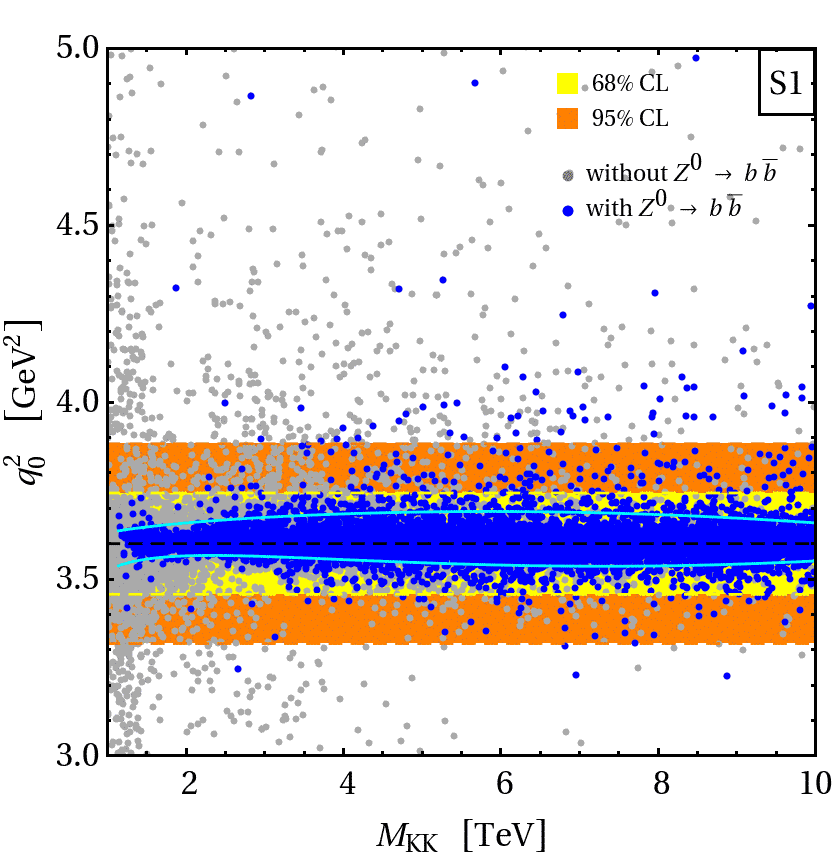}

    \vspace{4mm}

    \includegraphics[height=6.5cm]{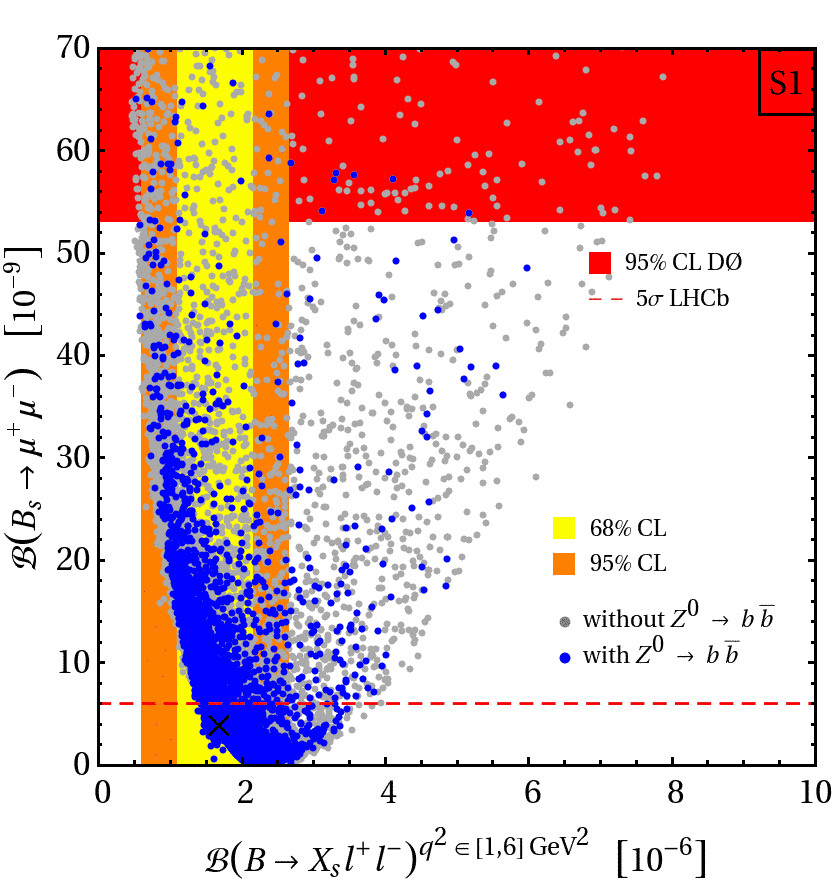}
  \end{center}
  \vspace{-8mm}
  \begin{center}
    \parbox{15.5cm}{\caption{\label{fig:BXsll} Prediction for ${\cal B}
        (B \to X_s l^+ l^-)$ and $q_0^2$ versus $M_{\rm KK}$ (upper
        panels) and ${\cal B} (B \to X_s \nu \bar \nu)$ versus ${\cal B}
        (B_s \to \mu^+ \mu^-)$ (lower panel). The shown results
        correspond to benchmark scenario S1. The blue (light gray)
        points are consistent (inconsistent) with the measured $Z^0 \to
        b \bar b$ couplings at the 99\% CL. The solid black lines and
        the black cross indicate the SM expectation. The cyan lines
        illustrate the decoupling behavior with $\Mkk$ obtained from a
        fit to the $1\%$ (lower lines) and $99\%$ (upper lines) quantile
        of the points that are consistent with $Z^0 \to b \bar b$. For
        comparison the regions of 68\% (yellow) and 95\% (orange) CL are
        also displayed. In the lower panel the 95\% CL exclusion of
        ${\cal B} (B_s \to \mu^+ \mu^-)$ and the minimum of the
        branching fraction allowing for a discovery with $5\sigma$ at
        LHCb are indicated by the red band and dashed line. See text for
        details.}}
  \end{center}
\end{figure}

The predictions for the inclusive $B \to X_s l^+ l^-$ channel are
shown in Figure~\ref{fig:BXsll}. All panels show results for the
benchmark scenario S1 that reproduce the correct hierarchies in the
quark sector.  Scatter points that fulfill (violate) the constraints
arising from the measurements of the $Z^0 \to b \bar b$ couplings are
colored blue (light gray). In the upper left panel we display ${\cal
  B} (B \to X_s l^+ l^-)^{q^2 \in [1, 6] \, {\rm GeV}^2}$ as a
function of $M_{\rm KK}$. The central value of the SM expectation
(\ref{eq:BRBXsllSM}) is indicated by the black line, while the yellow
(orange) band represents the 68\% (95\%) CL interval of the
experimental world average ${\cal B} (B \to X_s l^+ l^-)_{\rm
  exp}^{q^2 \in [1,6] \, {\rm GeV}^2} = (1.63 \pm 0.53) \cdot 10^{-6}$
\cite{Aubert:2004it, Iwasaki:2005sy}. It is evident from the figure
that, although enhancements of a factor of 2 are possible, the vast
majority of points that satisfy the $Z^0 \to b \bar b$ constraints
lead to values of the inclusive branching ratio that lie within the
experimentally allowed range. This feature is expected
\cite{Haisch:2007ia} in models in which the modification of the
flavor structure is closely connected to the third generation. We also
add that in warped setups with custodial protection of the $Z^0 b_L
\bar b_L$ vertex the correlation between $Z^0 \to b \bar b$ and $b \to
s l^+ l^-$ is in general different compared to the case discussed
here. However, at the end of Section~\ref{sec:numkaons} we have shown
that $\big |(g_R^d)_{ij}^{\rm custodial}/(g_L^d)_{ij}^{\rm original}
\big | \ll 1$ for the case $ij=23$ unless the bulk mass parameter
$c_{u_3}$ describing the localization of the right-handed top quark is
much bigger than 1. We thus expect the effects in $b \to s l^+ l^-$
due to exchange of the $Z^0$ boson and its KK modes to be even smaller
in the custodially protected RS model for natural choices of
$c_{u_3}$. A detailed study of this question is postponed to a
forthcoming publication.

An observable that is theoretically even cleaner than the $B \to X_s
l^+ l^-$ branching ratio, is the zero of the forward-backward
asymmetry, $q_0^2$. The results for $q_0^2$ as a function of the
new-physics scale $M_{\rm KK}$ are given in the upper right plot of
Figure~\ref{fig:BXsll}. The solid black line and the yellow (orange)
band indicate the central value and the 68\% (95\%) CL interval of the
theory prediction (\ref{eq:q02SM}). Like in the case of the branching
ratio, we see that points that satisfy the $Z^0 \to b \bar b$
constraint usually lead to deviations in $q_0^2$ that do not exceed
the level of 20\%. The moderate size of the corrections will make it
difficult, but certainly not impossible at a super flavor factory, to
distinguish the RS model from the SM on the basis of a measurement of
$q_0^2$. In the lower panel of Figure~\ref{fig:BXsll} we finally
present the correlation between ${\cal B} (B \to X_s l^+ l^-)^{q^2 \in
  [1, 6] \, {\rm GeV}^2}$ and ${\cal B} (B_s \to \mu^+ \mu^-)$. We
observe that without the $Z^0 \to b \bar b$ constraint the predictions
form a tilted parabola-shaped area, which is transformed into a
boomerang-shaped area by the $Z^0 \to b \bar b$ constraint, which
eliminates in particular points leading to a simultaneous enhancement
of both branching fractions. Enhancements in the purely leptonic mode
thus occur more frequently for values of ${\cal B} (B \to X_s l^+
l^-)^{q^2 \in [1, 6] \, {\rm GeV}^2}$ below the SM expectation
(\ref{eq:BRBXsllSM}). This feature is easily understood. Both decay
modes receive the dominant contribution from the new-physics
contributions $C_{10}^{l \, \rm RS}$ and $C_A^{\rm RS}$ to the
axial-vector couplings $C_{10}^{l} \approx -4.4 + C_{10}^{l \, \rm
  RS}$ and $C_A \approx 0.96 + C_A^{\rm RS}$. Since the coefficients
$C_{10}^{l \, \rm RS}$ and $C_A^{\rm RS}$ are aligned in flavor space
and the SM contribution has opposite sign, constructive interference
in $B \to X_s l^+ l^-$ typically implies destructive interference in
$B_s \to \mu^+ \mu^-$ and {\it vice versa}.

\begin{figure}[!t]
  \begin{center}
    \includegraphics[height=6.5cm]{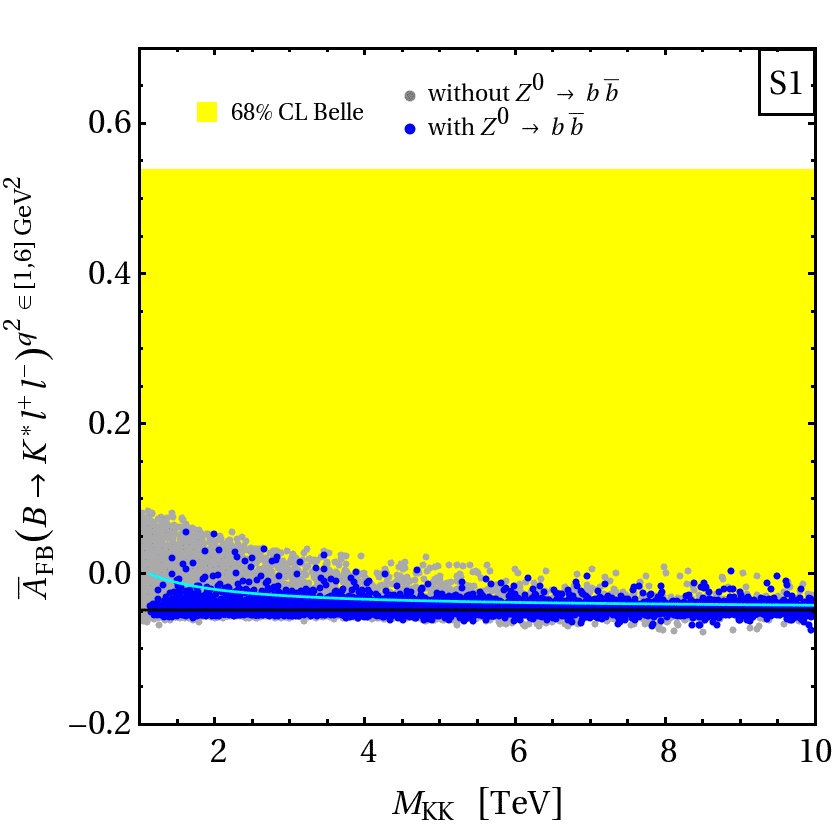}
    \hspace{0.75cm}
    \includegraphics[height=6.65cm]{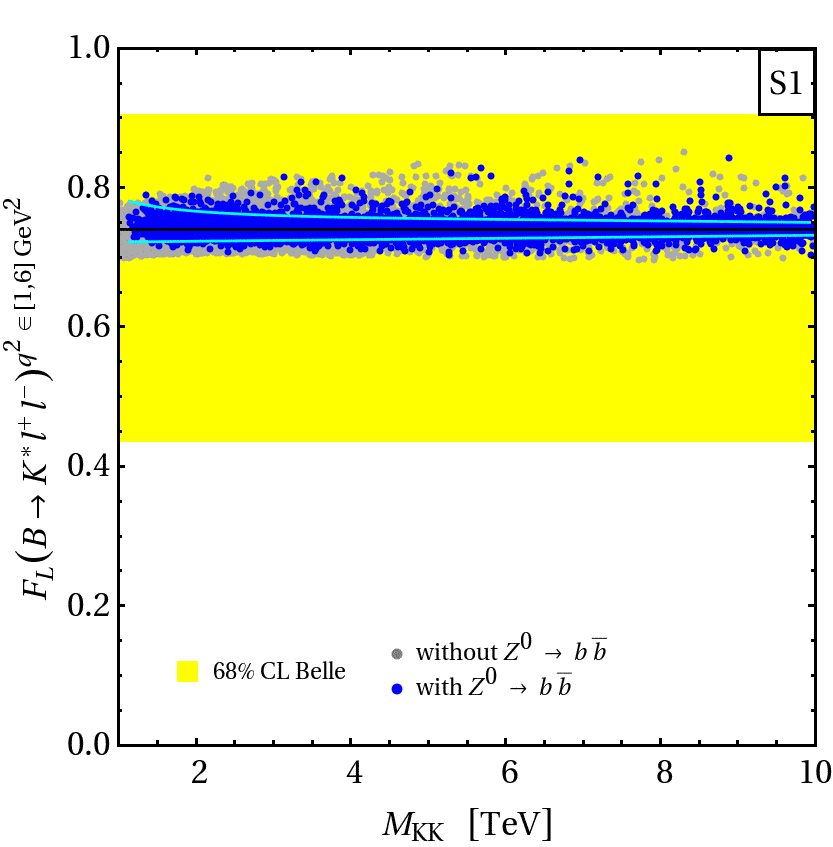}

    \vspace{4mm}

    \includegraphics[height=6.5cm]{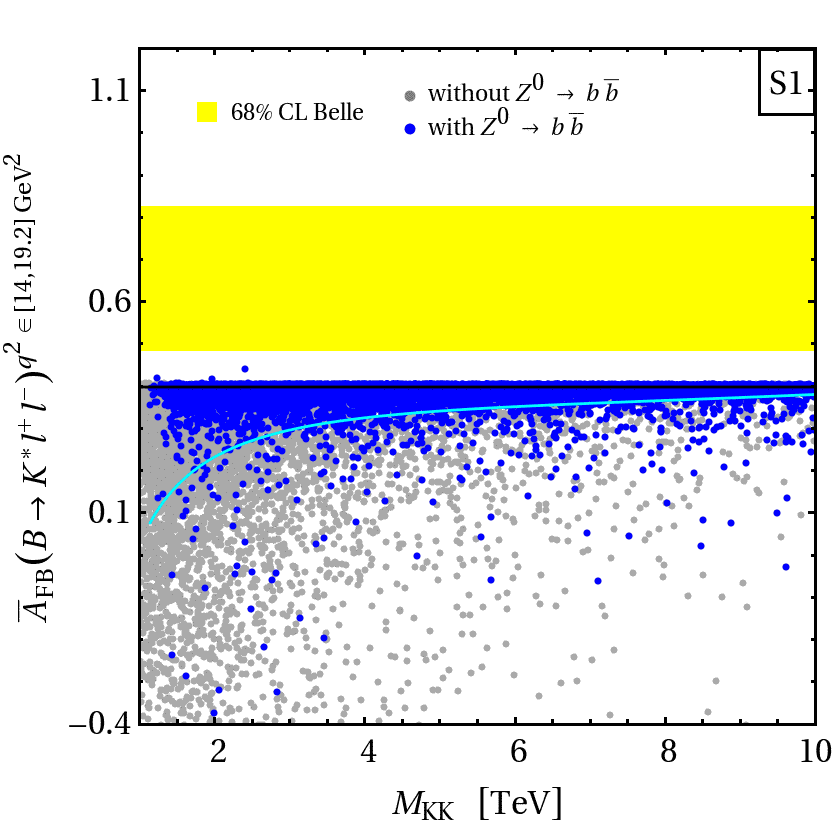}
    \hspace{0.75cm}
    \includegraphics[height=6.65cm]{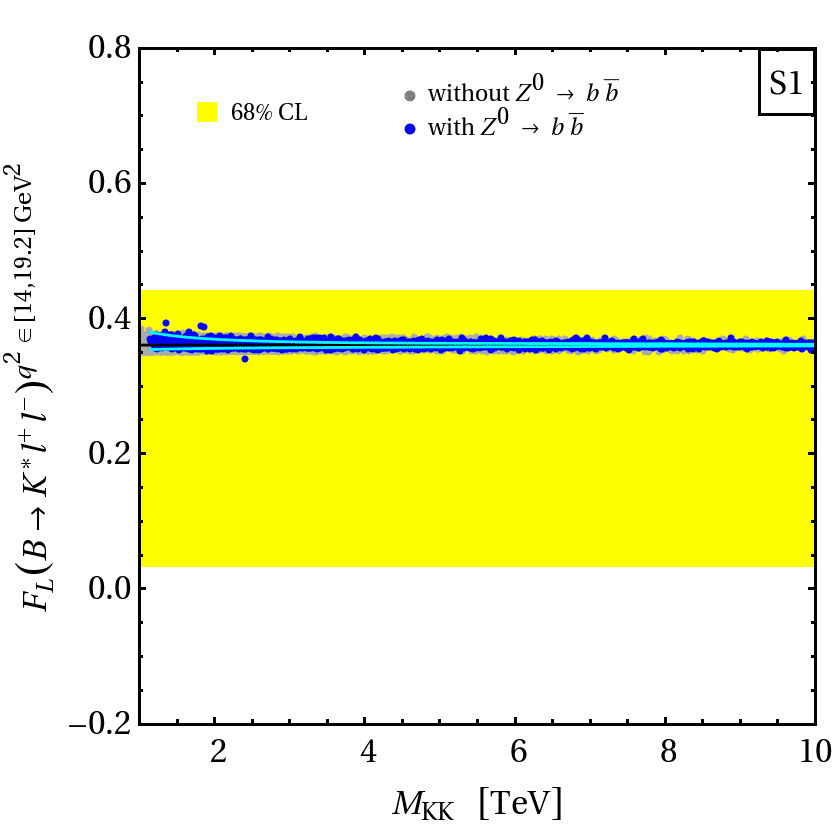}
  \end{center}
  \vspace{-8mm}
  \begin{center}
    \parbox{15.5cm}{\caption{\label{fig:bsllexcl} Predictions for $\bar
        A_{\rm FB} (B \to K^\ast l^+ l^-)^{q^2 \in [q_1^2, q_2^2]}$ and
        $F_{\rm L} (B \to K^\ast l^+ l^-)^{q^2 \in [q_1^2, q_2^2]}$ as a
        function of $M_{\rm KK}$ with $[q_1^2, q_2^2] = [1, 6] \, {\rm
          GeV}^2$ ($[q_1^2, q_2^2] = [14, 19.2] \, {\rm GeV}^2$) shown
        in the upper (lower) row. All displayed results have been
        obtained in benchmark scenario S1. The blue (light gray) scatter
        points are consistent (inconsistent) with the measured $Z^0 \to
        b \bar b$ couplings at the 99\% CL. The solid black lines
        indicate the central values of the SM expectations. The cyan
        lines illustrate the decoupling behavior with $\Mkk$ obtained
        from a fit to the $1\%$ (lower lines) and $99\%$ (upper lines)
        quantile of the points that are consistent with $Z^0 \to b \bar
        b$. In the case of the forward-backward asymmetry we show only
        the higher and lower quantile for the low- and high-$q^2$ range,
        respectively. For comparison the regions of 68\% (yellow) CL
        favored by experiment are also displayed. See text for
        details.}}
  \end{center}
\end{figure}

We now leave behind the sector of inclusive $b \to s l^+ l^-$ decay
distributions and continue our survey in the area of exclusive
decays. We emphasize that our study will be exploratory and will focus
on $B \to K^\ast l^+ l^-$ observables that can already be accessed at
BaBar \cite{Aubert:2006vb, Aubert:2008ju} and Belle
\cite{Ishikawa:2006fh, Wei:2009zv}. An extension of the discussion to
other decay channels \cite{Bobeth:2007dw}, angular distributions
\cite{Kruger:2005ep, Lunghi:2006hc, Egede:2008uy,
  Altmannshofer:2008dz}, and CP asymmetries
\cite{Altmannshofer:2008dz, Bobeth:2008ij} is straightforward, but
reserved for the future. The panels in Figure \ref{fig:bsllexcl} show
the results for the normalized forward-backward asymmetry $\bar A_{\rm
  FB} (B \to K^\ast l^+ l^-)$ and the longitudinal $K^\ast$
polarization $F_{\rm L} (B \to K^\ast l^+ l^-)$ integrated over the
low- and high-$q^2$ regions for points reproducing the measured
hierarchies in the quark sector. Scatter points that fulfill (violate)
the constraints arising from the measurements of the $Z^0 \to b \bar
b$ couplings are colored blue (light gray). The central values of the
SM expectations are indicated by the solid black lines. In the upper
left panel the measurement $\bar A_{\rm FB} (B \to K^\ast l^+
l^-)^{q^2 \in [1, 6] \, {\rm GeV}^2}_{\rm exp} =
0.26{^{+0.27}_{-0.30}}_{\rm stat} \pm 0.07_{\rm syst}$
\cite{Wei:2009zv} is underlaid in form of a yellow band. The quoted
Belle result agrees well with the latest BaBar measurement
\cite{Aubert:2008ju}. It has slightly smaller errors but includes
events with dilepton invariant masses down to $0.1 \, {\rm GeV}^2$,
which is theoretically problematic due to the presence of very light
$q \bar q$ resonances. Compared to the experimental central value,
which is shifted toward the positive side from the SM expectation
(\ref{eq:bsllexcllowSM}), the new-physics corrections in the
normalized forward-backward asymmetry are quite small but positive. We
see that only values in the range $[-6\%, 5\%]$ ($[-6\%, 10\%$]) can
be achieved after (before) imposing the $Z^0 \to b \bar b$
constraint. The upper right panel of Figure~\ref{fig:bsllexcl}
demonstrates that the size of the possible corrections in $F_{\rm L}
(B \to K^\ast l^+ l^-)^{q^2 \in [1, 6] \, {\rm GeV}^2}$ due to the
exchange of the $Z^0$-boson and its KK modes is also limited. Values
of the integrated longitudinal $K^\ast$ polarization in the range of
$[0.70, 0.81]$, corresponding to relative shifts of $[-5\%, 9\%]$
compared to the SM prediction (\ref{eq:bsllexcllowSM}), can be
realized. The related experimental result $F_{\rm L} (B \to K^\ast l^+
l^-)^{q^2 \in [1, 6] \, {\rm GeV}^2}_{\rm exp} = 0.67\pm 0.23_{\rm
  stat} \pm 0.05_{\rm syst}$ \cite{Wei:2009zv} is indicated by the
yellow band. The central value of the most recent BaBar result
\cite{Aubert:2008ju} is smaller by almost a factor of 2 but agrees
with the cited number from Belle within errors. The smallness of the
observed deviations is a result of the weak dependence of the
observables on the numerically dominant semileptonic Wilson
coefficient $C_{10}^{l \, {\rm RS}}$. Since the sensitivity of $\bar
A_{\rm FB} (B \to K^\ast l^+ l^-)$ and $F_{\rm L} (B \to K^\ast l^+
l^-)$ to the electromagnetic dipole coefficients $C_{7, {\rm
    RS}}^{\gamma}$ and $\tilde C_{7, {\rm RS}}^{\gamma}$ is much more
pronounced, larger effects might occur if loop-suppressed corrections
are included in the analysis. To determine the exact impact of effects
due to $Q_7^{\gamma}$ and $\tilde Q_7^{\gamma}$, however, would
require a dedicated study.

The panels in the lower row of Figure~\ref{fig:bsllexcl} show the
high-$q^2$ predictions for the observables in question. In the case of
the normalized forward-backward asymmetry we find that the RS
tree-level corrections tend to reduce the SM value
(\ref{eq:bsllexclhighSM}) and allow for values in the range $[-39\%,
41\%$] if one does require the $Z^0 \to b \bar b$ constraint to be
fulfilled. The yellow shaded band in the same panel represents a naive
combination of the Belle measurements $\bar A_{\rm FB} (B \to K^\ast
l^+ l^-)^{q^2 \in [14.18, 16] \, {\rm GeV}^2}_{\rm exp} =
0.70{^{+0.16}_{-0.22}}_{\rm stat} \pm 0.07_{\rm syst}$ and $\bar
A_{\rm FB} (B \to K^\ast l^+ l^-)^{q^2 > 16 \, {\rm GeV}^2}_{\rm exp}
= 0.66{^{+0.11}_{-0.16}}_{\rm stat} \pm 0.04_{\rm syst}$
\cite{Wei:2009zv}, which are compatible with the BaBar result in the
high-$q^2$ bin \cite{Aubert:2008ju} but have better
precision. Concerning the observed high values of $\bar A_{\rm FB} (B
\to K^\ast l^+ l^-)$, we would like to point out that arbitrary
complex left-handed coefficients $C_{9,10}^l$ can enhance the
forward-backward asymmetry only by a couple of percent relative to the
SM value. Larger values of $\bar A_{\rm FB} (B \to K^\ast l^+ l^-)$ up
to about $55\%$ are possible if one allows for arbitrary complex
right-handed coefficients $\tilde C_{9,10}^l$. To achieve values in
the ballpark of $70\%$ requires the inclusion of the electromagnetic
dipole coefficients $C_7^\gamma$ and $\tilde C_7^\gamma$. In
particular, $C_7^\gamma$ needs to be complex and almost anti-parallel
to the SM contribution. In the lower right panel of
Figure~\ref{fig:bsllexcl} we finally show the predictions for $F_{\rm
  L} (B \to K^\ast l^+ l^-)^{q^2 \in [14, 19.2] \, {\rm GeV}^2}$. For
comparison, the naive average of the BaBar result $F_L (B \to K^\ast
l^+ l^-)^{q^2 > 10.24 \, {\rm GeV}^2}_{\rm exp} =
0.71{^{+0.20}_{-0.22}}_{\rm stat} \pm 0.04_{\rm syst}$
\cite{Aubert:2008ju} and the Belle measurements $F_{\rm L} (B \to
K^\ast l^+ l^-)^{q^2 \in [14.18, 16] \, {\rm GeV}^2}_{\rm exp} =
-0.15{^{+0.27}_{-0.23}}_{\rm stat} \pm 0.07_{\rm syst}$ and $F_{\rm L}
(B \to K^\ast l^+ l^-)^{q^2 > 16 \, {\rm GeV}^2}_{\rm exp} =
0.12{^{+0.15}_{-0.13}}_{\rm stat} \pm 0.02_{\rm syst}$
\cite{Wei:2009zv} is indicated by the yellow band. We see that the
predictions for the longitudinal $K^\ast$ polarization are all
essentially SM-like. This feature is easy to understand if one
realizes that as a consequence of (\ref{eq:Aihigh}) the dependence on
the dominant axial-vector coefficient $C_{10}^{l \, {\rm RS}}$ almost
drops out in the ratio (\ref{eq:AFBFLlow}) determining $F_{\rm L} (B
\to K^\ast l^+ l^-)$. We mention that values of $F_{\rm L} (B \to
K^\ast l^+ l^-)$ integrated over the high-$q^2$ region significantly
below the SM expectation (\ref{eq:bsllexclhighSM}) would hint towards
the presence of right-handed currents.

It has been argued in \cite{Aubert:2006vb, Aubert:2008ju,
  Ishikawa:2006fh, Wei:2009zv} that the available experimental data
shows a slight preference for new-physics scenarios where the sign of
$C_7^\gamma$ is opposite with respect to the SM expectation $\sgn \!
\left (C_{7, {\rm SM}}^\gamma \right ) = -1$. While much larger data
sets are certainly needed to draw a definite conclusion, it is
important to emphasize that models with $\sgn \left (C_{7}^\gamma
\right ) = +1$ and only small non-standard contributions to the
semileptonic coefficients $C_{9,10}^l$ are disfavored at the $3\sigma$
level by the combination of the ${\cal B} (B \to X_s \gamma)$ and
${\cal B} (B \to X_s l^+ l^-)$ measurements
\cite{Gambino:2004mv}. Hypothetical new-physics scenarios with
flipped-sign $C_7^\gamma$ only are thus at variance with the available
data on the inclusive $b \to s \gamma, l^+ l^-$ transitions. This
observation elucidates the fact that to bound the values of the
various Wilson coefficients one should exploit all available
experimental information in the $b \to s \gamma$ and $b \to s l^+ l^-$
sector, combining both inclusive and exclusive channels.

\boldmath \subsection{Rare Non-Leptonic Decays of Kaons and $B$
  Mesons} \unboldmath

We finally turn our attention to the vast array of non-leptonic kaon
and $B$-meson decays, focusing on a few of the most interesting
observables. We start out with a detailed numerical analysis of the
CP-violating ratio $\epseps$ and investigate its correlation with the
rare $K \to \pi \nu \bar \nu$ and $K_L \to \pi^0 l^+ l^-$
decays. Subsequently, we will consider the RS contributions to a
variety of exclusive $B$-meson decays into final states containing two
pseudoscalar mesons or a pseudoscalar and a vector meson. One
important question that we will address in this context is whether
possible discrepancies in the $b \to s$ sector, as suggested by
experiment, can be explained in the context of warped extra-dimension
models.

\boldmath \subsubsection{Numerical Analysis of $\epseps$}
\unboldmath \label{sec:epsepsnum}

Within the SM the smallness of direct CP violation in $K \to \pi \pi$
is the result of a destructive interference between the positive
contribution due to QCD penguins and the negative contribution arising
from electroweak penguin diagrams. In new-physics models in which the
$\Delta I = 1/2$ and $\Delta I = 3/2$ contributions to $s \to d q \bar
q$ processes are affected in a different way, this partial
cancellation is usually much less pronounced or even absent. A
qualitative understanding of the situation in the minimal RS model can
be obtained from the approximate relation
\beq \label{eq:epsepscrude}
  \left (\frac{\epsilon^\prime_K}{\epsilon_K} \right )_{\rm RS} 
  \propto 1 + [-0.1, 0.1]  \hspace{0.25mm}
  B_6^{(1/2)} -12 \hspace{0.25mm} B_8^{(3/2)} ,
\eeq 
which has been obtained from the second expression in (\ref{eq:feps})
by inserting typical values of the Wilson coefficients $C_{3-10}^{\rm
  RS}$ and $\tilde C_{3-10}^{\rm RS}$. The smallness (largeness) of
the coefficient multiplying the hadronic parameter $B_6^{(1/2)}$
($B_8^{(3/2)}$) is easily explained by the structure of $\epseps$
within the SM. While both the QCD and electroweak penguin
contributions are strongly enhanced by RG effects, the former
correction results mainly from the mixing of $Q_6$ with $Q_{1,2}$ and
is thus essentially unaffected by new physics. On the other hand,
mixing with the current-current operators plays only a very minor role
in the case of the electroweak penguins, so that any new-physics
contribution to the initial conditions in this sector directly feeds
through into $\epseps$. This implies that electroweak penguin
operators give the dominant correction to direct CP violation in the
kaon sector. Our numerical analysis confirms this model-independent
conclusion. Notice that the sign of $(\epseps)_{\rm RS}$ in
(\ref{eq:epsepscrude}) is not fixed, so that the prediction for the
total ratio $\epseps$ can be positive or negative. An enhancement
(depletion) of $\epseps$ occurs if the imaginary part of the enhanced
electroweak penguin contribution has opposite (identical) sign with
respect to the SM $Z^0$-penguin amplitude.

\begin{figure}[!t]
  \begin{center}
    \includegraphics[height=6.5cm]{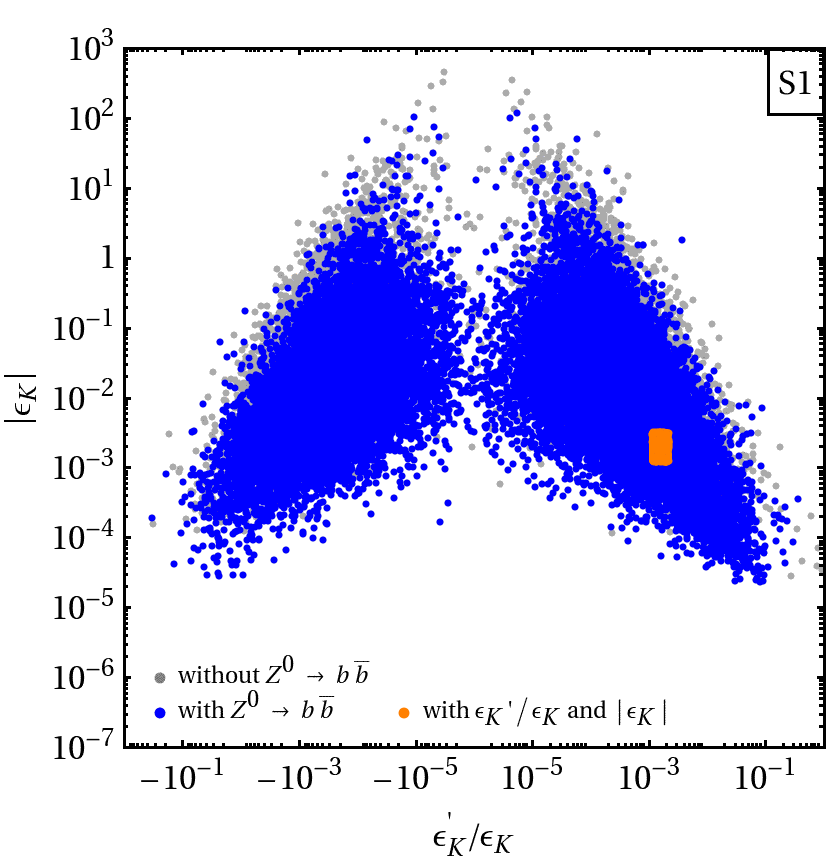}
    \hspace{0.75cm}
    \includegraphics[height=6.5cm]{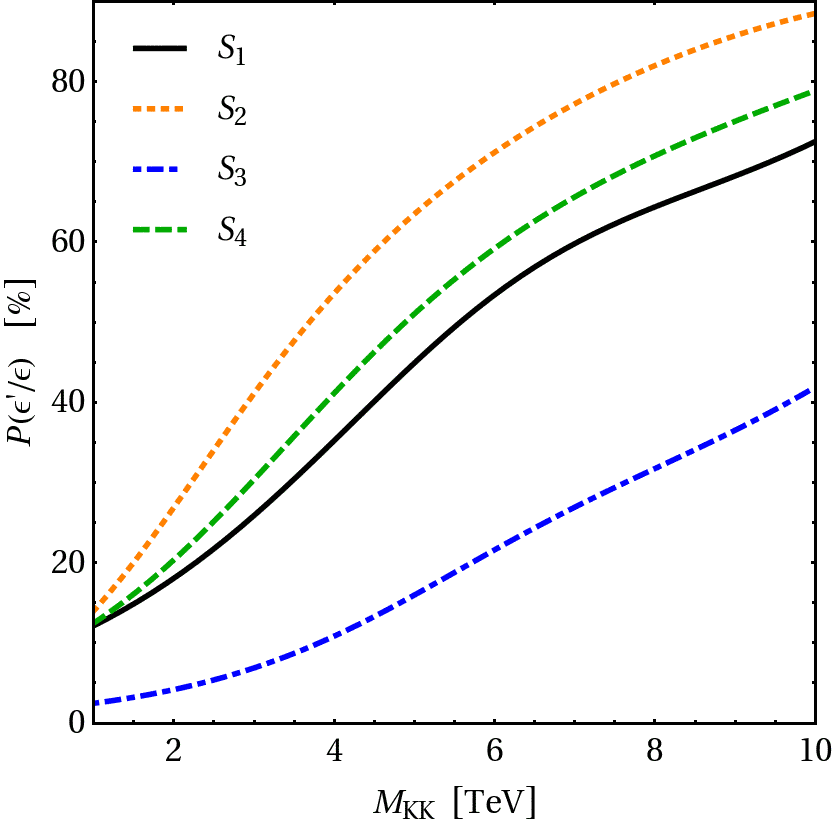}
  \end{center}
  \vspace{-8mm}
  \begin{center}
    \parbox{15.5cm}{\caption{\label{fig:epseps1} Left panel: Correlation
        between $\epseps$ and $|\epsilon_K|$ in the benchmark scenario
        S1. The blue (light gray) points are consistent (inconsistent)
        with $Z^0 \to b \bar b$ at the 99\% CL. The points that are in
        agreement with the experimental values of both $\epseps$ and
        $|\epsilon_K|$ at 95\% CL are marked in orange. Right panel:
        Percentage of scatter points as a function of $\Mkk$ that are
        consistent with $\epseps$. The shown lines correspond to our
        four different parameter scenarios. See text for details.}}
  \end{center}
\end{figure}

It is also important to bear in mind that the observables $\epseps$
and $|\epsilon_K|$ are not independent from each other if the
experimental constraint on $|\epsilon_K|$ is ignored and $\epseps$ is
calculated fully in theory. Since the central value of the theory
prediction (\ref{eq:epsepsSM}) is below the world average for
$\epseps$, models are disfavored in which there are new positive
contributions to $|\epsilon_K|$ and negative contributions to
$\epsilon'_K$. The correlation between $\epseps$ and $|\epsilon_K|$ in
the benchmark scenario S1 is shown in the left panel of
Figure~\ref{fig:epseps1}. All shown points reproduce the correct quark
masses and mixings. Scatter points that are consistent with $Z^0 \to
b\bar b$ are colored blue, while those points that lie inside the 95\%
CL ranges $\epseps \in [11.3, 21.7] \cdot 10^{-4}$ and $|\epsilon_K|
\in [1.2, 3.2] \cdot 10^{-3}$ are marked in orange. The plot has been
obtained by setting the non-perturbative parameters $R_{6,8}$ entering
(\ref{eq:feps}) equal to 1. It is evident from the distribution of
points that solutions featuring unacceptably large values of
$|\epsilon_K|$ are typically also in conflict with $\epseps$, since
they predict too low values for the latter observable. For points in
line with $|\epsilon_K|$, it is however still possible to obtain
values for $\epseps$ in the range $[-1, 1] \cdot 10^{-1}$. The wide
spread of viable results suggests that the data on $\epseps$ impose
non-trivial constraints on the allowed model parameters even if the
$|\epsilon_K|$ constraint is satisfied.

The right plot in Figure~\ref{fig:epseps1} illustrates how severe the
$\epseps$ constraint is in each of the four benchmark scenarios. In
our numerical analysis a point is considered compatible with
$\epseps$, if it is possible to bring the theoretical prediction into
agreement with the measured value at the 95\% CL by varying the
hadronic parameters and light quark masses in the ranges given in
(\ref{eq:bagranges}) and (\ref{eq:bagranges}), requiring in addition
$B_6^{(1/2)} > B_8^{(3/2)}$. Although calculations of the hadronic
matrix elements $\langle (\pi \pi)_0 | Q_6 | K \rangle$ and $\langle
(\pi \pi)_2 | Q_8 | K \rangle$ that are free from systematic
uncertainties do not exist at present, we believe that this is a very
conservative treatment of theoretical errors. The shown $P(\epseps)$
curves correspond again to the best fits to the data after binning in
$M_{\rm KK}$ with a bin size of 100 GeV and using the functional form
of a Fermi function times a quadratic polynomial. We find that $47\%$,
$62\%$, $20\%$, and $61\%$ of the total number of generated points are
consistent with $\epseps$ in the scenario S1, S2, S3, and S4
respectively. One can make sense out this percentages if one recalls
the pattern of enhancements and depletions of $\Delta F = 2$ and
$\Delta F = 1$ contributions in the different benchmark
scenarios. Compared to the default scenario S1, corrections to
$|\epsilon_K|$ are typically less pronounced in S2 and S4, because
these scenarios are specifically designed to suppress $|\epsilon_K|$
by either by eliminating flavor mixing that arises from the
non-universality of the right-handed down-type quark profiles or
allowing for larger Yukawa couplings. On the other hand the
corrections to $s \to d Z^0$ entering $\epseps$ are only indirectly
affected. They tend to be smaller as well, since the parameter
scenarios S2 and S4 show a preference for a stronger localization of
the first two SM-like doublets towards the UV to account for the
correct values of the quark mass. In the benchmark scenario S3 the
situation is reversed. Numerically, we find that demanding $P(\epseps)
> 25\%$ sets lower limits of $M_{\rm KK} > 2.9$, $1.9$, $6.7$, and
$2.5 \, {\rm TeV}$ in the scenario S1, S2, S3, and S4.\footnote{If we
  normalize $\epseps$ to the experimental rather than the theoretical
  value of $|\epsilon_K|$, as done in \cite{Gedalia:2009ws}, we find
  that requiring $P(\epseps) = 25\%$ sets the limits $4.8$, $3.2$,
  $3.1$, and $6.5 \, {\rm TeV}$ on the mass of the first KK excitation
  in scenario S1, S2, and S3. Notice that in this case the bound is
  solely due to the electroweak penguin contribution.} In terms of the
mass of the first KK mode this means $7.1$, $4.6$, $16.3$, and $6.1 \,
{\rm TeV}$.  These numbers imply that with our conservative treatment
of errors the $\epseps$ constraint is typically weaker than the
constraint from $|\epsilon_K|$ but stronger than the one arising from
$Z^0 \to b \bar b$.

\begin{figure}[!t]
  \begin{center}
    \includegraphics[height=6.5cm]{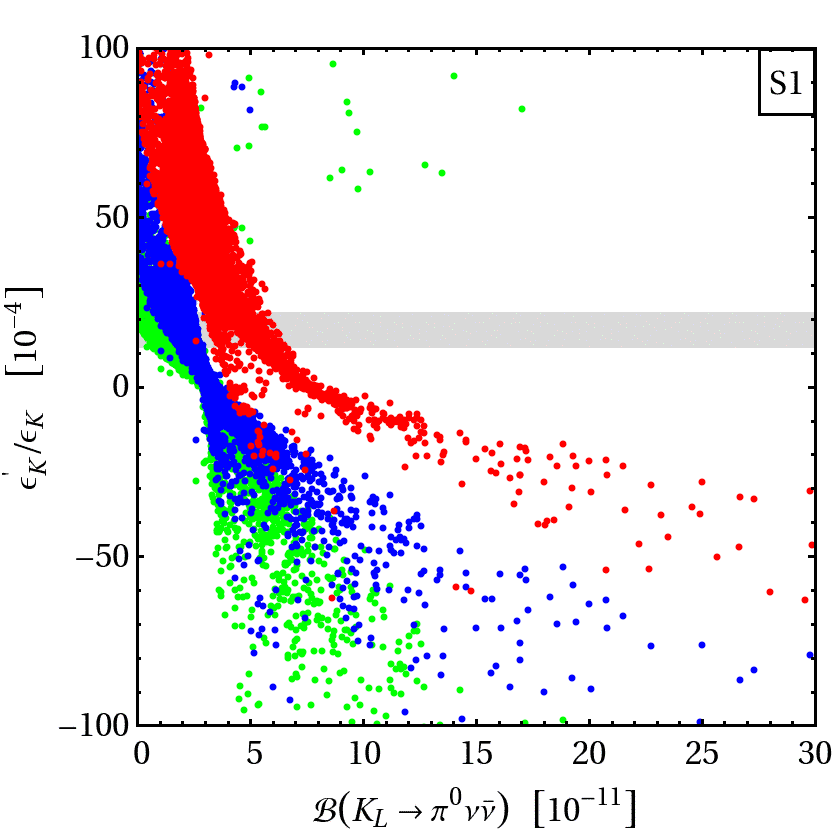}
    \hspace{0.75cm}
    \includegraphics[height=6.5cm]{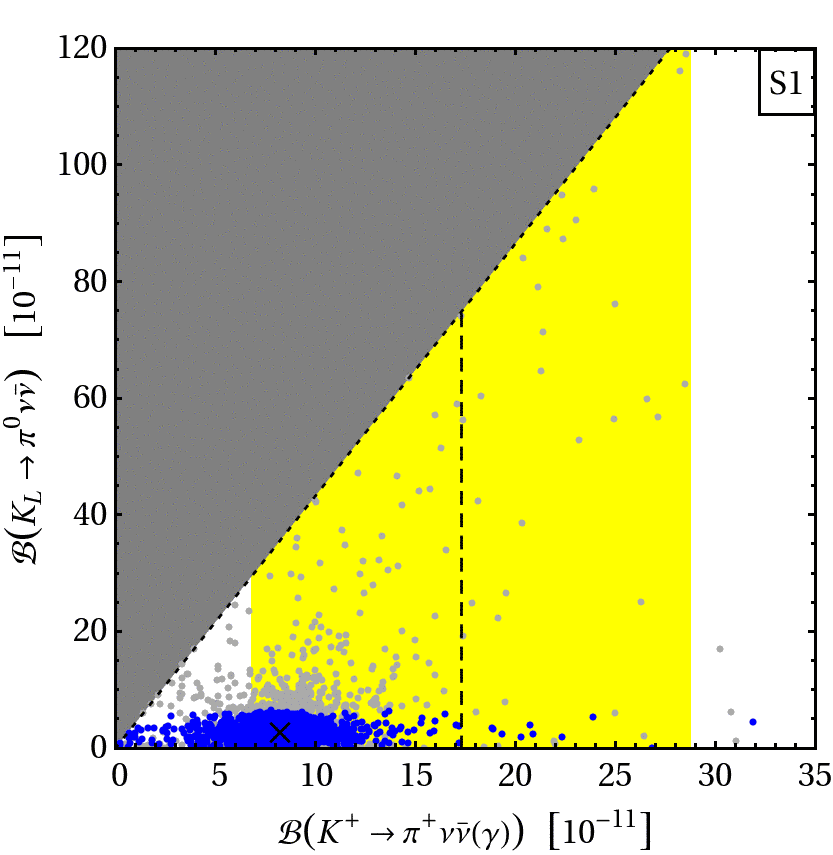}
  \end{center}
  \vspace{-8mm}
  \begin{center}
    \parbox{15.5cm}{\caption{\label{fig:epseps2} Left panel: Correlation
        between the $K_L \to \pi^0 \nu \bar \nu$ branching fraction and
        $\epsilon_K^\prime/\epsilon_K$ in benchmark scenario S1. All
        points reproduce the quark hierarchies and the measured values
        of $|\epsilon_K|$, $Z^0 \to b \bar b$, and $B_d$--$\bar B_d$
        mixing at 95\%, 99\%, and 95\% CL. The blue points correspond to
        the central value for $\epsilon_K^\prime/\epsilon_K$ obtained
        for $R_{6,8} = 1$, while the red (green) points illustrate the
        maximal (minimal) achievable values of
        $\epsilon_K^\prime/\epsilon_K$ for the same set of points
        obtained by varying the hadronic parameters. For comparison the
        experimental 95\% CL (light gray) is also displayed.  Right
        panel: Impact of $\epsilon_K^\prime/\epsilon_K$ on the
        predictions in the ${\cal B} (K^+ \to \pi^+ \nu \bar \nu
        (\gamma)) \hspace{0.25mm}$--${\cal B} (K_L \to \pi^0 \nu \bar
        \nu) \hspace{0.25mm}$ plane in parameter scenario S1. The blue
        (light gray) scatter points are consistent (inconsistent) with
        the measured value of $\epsilon_K^\prime/\epsilon_K$. See text
        for details.}}
  \end{center}
\end{figure}
 
We also mention that potentially large corrections to
$\epsilon'_K/\epsilon_K$ can arise from the presence of the
chromomagnetic dipole operator $Q_8^g$ and its chirality-flipped
partner $\tilde{Q}_8^g$. In the context of scenarios with hierarchical
fermion profiles, this issue has been analyzed in a model-independent
fashion in \cite{Davidson:2007si}. Interestingly, the contributions
to $\epsilon'_K/\epsilon_K$ from $C_8^g$ and $\tilde{C}_8^g$ and to
$\epsilon_K$ from $C_4^{\rm RS}$ depend in an opposite way on the
Yukawa couplings. This makes it difficult to decouple flavor-violating
effects using these parameters and will lead to a ``tension'' between
tree- and loop-level effects, similar to what happens in the case of
$B \to X_s \gamma$ and $\epsilon_K$ \cite{Agashe:2008uz}. A study of
this anti-correlation including the first KK level of the quarks and
the zero-mode of the Higgs field has been very recently presented in
\cite{Gedalia:2009ws}. There it has been pointed out that the
constraint stemming from the chromomagnetic dipole-operator
contributions to $\epsilon^\prime_K/\epsilon_K$ yields a lower bound
of $5.5 \, {\rm TeV}$ on the mass of the first KK excitation if the
Higgs boson is maximally spread into the bulk. This bound is raised to
$7.5 \, {\rm TeV}$ if one considers the case of the two-site model
\cite{Agashe:2008uz}.  It is important to realize that the quoted
bounds are obtained for values of the down-type quark Yukawa couplings
($Y_d = 4.6$ and $Y_d = 5.9$) that are larger than the value $Y_{\rm
  max} = 3$ allowed by perturbativity in three of our benchmark
scenarios. We emphasize that in a set-up with a brane-localized Higgs
sector and applying our more conservative treatment of theoretical
errors, the constraint stemming from the chromomagnetic
dipole-operator contributions to $\epseps$ leads to a limit of only
$M_{\rm KK} > 2 \, {\rm TeV}$. This bound is weaker than the one
arising from the electroweak penguin diagrams discussed above.

Our explicit calculation of the chromomagnetic dipole contribution to
$\epseps$, utilizing the formulas presented in \cite{Gedalia:2009ws},
confirms this result. Applying a Froggatt-Nielsen analysis to leading
order in the Cabibbo angle, we find that the additional contribution
to $\epseps$ arising from the operators $Q_8^g$ and $\tilde Q_8^g$ can
be expressed as $F_{\rm RS}^{g} \approx 2.5 \cdot 10^{-3} \,
B_{g}^{(1/2)} \, f_g \, \big(1 \, \text{TeV}/{\Mkk}\big)^2$, where
$B_{g}^{(1/2)} \approx 1$ denotes the hadronic parameter related to
the matrix element of $Q_8^g$, while $f_g$ is a function of ratios of
Yukawa couplings and thus naturally of ${\cal O} (1)$. The correction
$F_{\rm RS}^g$ has to be compared to the largest contribution to
$F_{\rm RS}$ (see (\ref{eq:feps})) which arises from the Wilson
coefficient $C_7^{\rm RS}$.  Using the central values for
$c_{Q_{1,2}}$ quoted in Table \ref{tab:bulkmasses}, we obtain $F_{\rm
  RS} \approx 1.0 \cdot 10^{-2} \, B_{8}^{(3/2)} \, f_8 \, \big(1 \,
\text{TeV}/{\Mkk}\big)^2$, where $f_8$ is another function of ratios
of Yukawa couplings, naturally expected to be of ${\cal O}(1)$.  It
follows that after requiring the correct quark hierarchies the
corrections due to the chromomagnetic dipole operators (encoded in
$F_{\rm RS}^g$) are typically an order of magnitude smaller than those
arising from the electroweak penguin sector (that dominate $F_{\rm
  RS}$). In the further discussion, we will thus simply ignore
effective couplings in $\epsilon_K'/\epsilon_K$ generated first at the
one-loop level. We have explicitly verified, using the full
expressions for the Wilson coefficients $C_8^g$ and $\tilde C_8^g$,
that this omission does not change any conclusion drawn hereafter
qualitatively.

The marked sensitivity of $\epseps$ to modifications of the
electroweak penguin sector leads to stringent correlations between
$\epseps$ and the $s \to d \nu \bar \nu$, $s\to d l^+ l^-$
observables. This feature is illustrated in Figure~\ref{fig:epseps2},
where the left panel shows the predictions for $\epseps$ as a function
of the branching fraction of $K_L \to \pi^0 \nu \bar \nu$ in scenario
S1. All scatter points reproduce the hierarchies in the quark sector
as well as the measured values of $|\epsilon_K|$, $Z^0 \to b \bar b$,
and $B_d$--$\bar B_d$ mixing. The blue points correspond to the values
for $\epsilon_K^\prime/\epsilon_K$ for fixed $R_{6,8} = 1$, while the
red (green) points are obtained by scanning the hadronic parameters
over the ranges (\ref{eq:bagranges}) and (\ref{eq:massranges}),
requiring in addition $B_6^{(1/2)} > B_8^{(3/2)}$, and determining the
maximal (minimal) values of $\epsilon_K^\prime/\epsilon_K$. The light
gray band represents the experimental 95\% CL. We observe that in our
default scenario S1 enhancements (suppressions) of ${\cal B} (K_L \to
\pi^0 \nu \bar \nu)$ tend to be accompanied by suppressions
(enhancements) of $\epsilon_K^\prime/\epsilon_K$. The large
enhancements of ${\cal B} (K_L \to \pi^0 \nu \bar \nu)$ found earlier
in Section~\ref{sec:numkaons} are hence disfavored by the measured
amount of direct CP violation in $K \to \pi \pi$. This is explicitly
demonstrated for our default scenario S1 in the right panel of
Figure~\ref{fig:epseps2}, which shows the predictions for ${\cal B}
(K^+ \to \pi^+ \nu \bar \nu (\gamma))$ versus ${\cal B} (K_L \to \pi^0
\nu \bar \nu)$ before (gray points) and after (blue points) imposing
the $\epseps$ constraint. We see that even with a conservative
treatment of errors, {\it i.e.}, scanning over hadronic parameters,
$\epseps$ has a non-negligible impact on the possible new-physics
effects in rare $K$ decays within the RS model. Notice that values of
$B_8^{(3/2)}$ near the upper end of the range shown in
(\ref{eq:bagranges}) are particularly problematic in this respect,
since they amplify the electroweak penguin contributions to
$\epseps$. As enhancements of the branching fractions $K_L \to \pi^0
\nu \bar \nu$ and $K_L \to \pi^0 l^+ l^-$ are strongly correlated with
each other, similar statements apply to the rare kaon decay modes with
charged leptons in the final state. The same strong anti-correlation
between ${\cal B} (K_L \to \pi^0 \nu \bar \nu)$ and
$\epsilon_K^\prime/\epsilon_K$ is also present in the parameter
scenarios S3 and S4. It arises because both observables receive the
dominant correction from the imaginary part of the left-handed $s \to
d Z^0$ amplitude, which relative to the SM contribution enters
linearly and with opposite sign in $\epsilon_K^\prime/\epsilon_K$,
whereas it appears quadratically and with the same sign in ${\cal B}
(K_L \to \pi^0 \nu \bar \nu)$. Interestingly, in new-physics models in
which the $s \to d$ transitions are dominated by right-handed
operators the correlation between $\epsilon_K^\prime/\epsilon_K$ and
${\cal B} (K_L \to \pi^0 \nu \bar \nu)$ is predicted to be a negative
one too. This model-independent conclusion follows from the fact that
in both $K_L \to \pi^0 \nu \bar \nu$ and
$\epsilon_K^\prime/\epsilon_K$ the left- and right-handed $Z^0$-boson
contribution interfere destructively, see (\ref{eq:Ki}) and
(\ref{eq:Cnus}). Studies of the correlation between ${\cal B} (K_L \to
\pi^0 \nu \bar \nu)$ and $\epsilon_K^\prime/\epsilon_K$ therefore will
not allow to test the chiral nature of the electroweak penguin
operators in the $s \to d$ sector. The correlation between ${\cal B}
(K^+ \to \pi^+ \nu \bar \nu)$ and $\epseps$ is weaker compared to the
one present in the neutral sector. As a result, larger departures from
the SM expectation in the charged mode are possible even after
enforcing the $\epseps$ constraint.

\boldmath \subsubsection{Numerical Analysis of Non-Leptonic $B$
  Decays} \unboldmath

\begin{figure}[!t]
  \begin{center}
    \includegraphics[height=6.5cm]{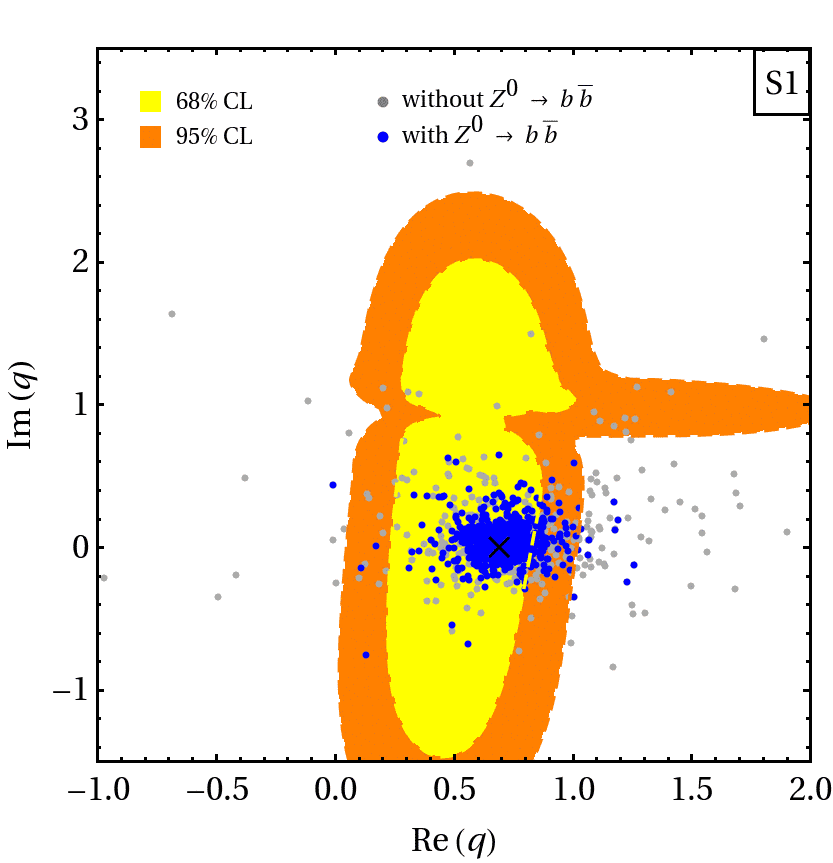}
    \hspace{0.75cm}
    \includegraphics[height=6.5cm]{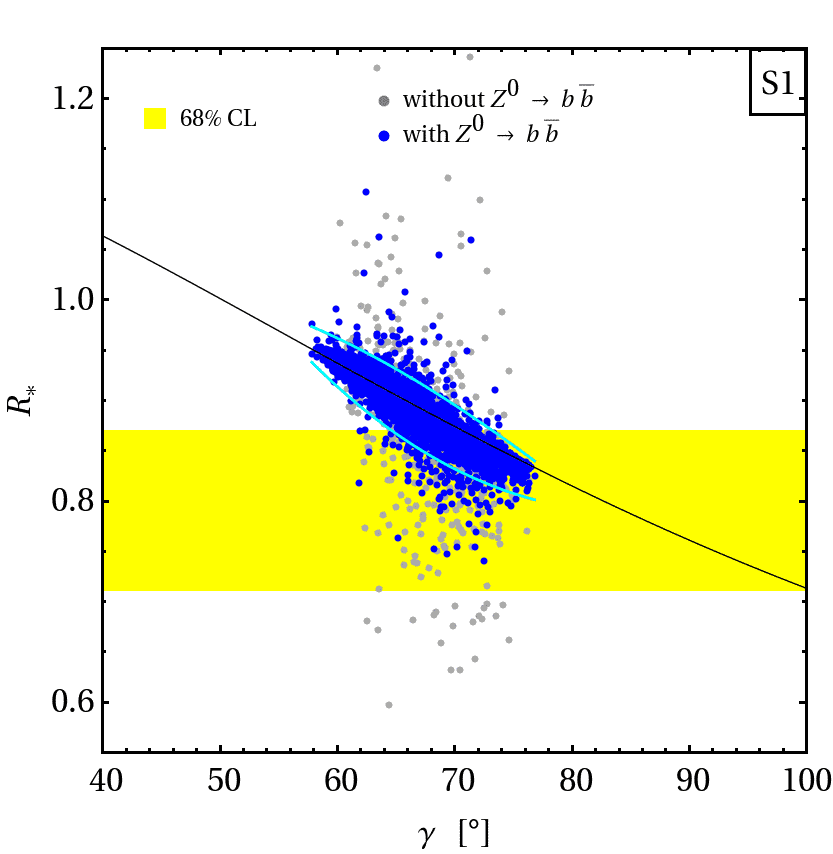}
  \end{center}
  \vspace{-8mm}
  \begin{center}
    \parbox{15.5cm}{\caption{\label{fig:nonlep1} Left panel: Predictions
        in the ${\rm Re} (q) \hspace{0.25mm}$--$\hspace{0.25mm} {\rm Im}
        (q)$ plane. The black cross indicates the SM expectation. For
        comparison the regions of 68\% (yellow) and 95\% CL (orange)
        following from a fit to the current $B \to K \pi, \pi \pi$ data
        are also displayed.  Right panel: Predictions for $R_\ast$ as a
        function of $\gamma$. The black line indicates the SM prediction
        while the yellow band corresponds to the experimental 68\%
        CL. The blue (light gray) points shown in both panels are
        consistent (inconsistent) with the measured $Z^0 \to b \bar b$
        couplings at the 99\% CL. The cyan lines illustrate the
        functional behavior with $\gamma$ of the region containing the
        typical effect obtained from a fit to the $1\%$ (lower lines)
        and $99\%$ (upper lines) quantile of the points that are
        consistent with $Z^0 \to b \bar b$. See text for details.}}
  \end{center}
\end{figure}

We conclude our comprehensive phenomenological survey with a numerical
analysis of the most interesting observables in the vast field of
exclusive hadronic $B$-meson decays. As the considered observables
turn out to be largely independent of the specific choice for the
parameter scenario, we will in the following only show results
corresponding to our default benchmark scenario S1. Moreover only
scatter points that reproduce the mass and mixing hierarchies in the
quark sector and the constraints arising from $|\epsilon_K|$ and
$B_d$--$\bar B_d$ mixing will be included in the plots. To illustrate
the impact of the $Z^0 \to b \bar b$ constraint, we discriminate
between points that are consistent (inconsistent) with the constraint
by coloring them in blue (light gray).

In the left panel of Figure~\ref{fig:nonlep1} we display our
predictions in the electroweak penguin parameter $q$ given in
(\ref{eq:qnaive}). The SM expectation is indicated by a black cross,
while the yellow (orange) colored contours show the regions of 68\%
(95\%) CL obtained in \cite{Fleischer:2008wb} from a $\chi^2$ fit to
the $B \to K \pi$ and $B \to \pi \pi$ data. The latter data has a
strong impact in the fit, yielding two almost degenerate minima with
$|q|_{\rm exp} = 0.8^{+0.2}_{-0.3}$, $\phi_{\rm exp} = \left (
  45^{+18}_{-28} \right)^\circ$ and $|q|_{\rm exp} = 1.3 \pm 0.4$,
$\phi_{\rm exp} = \left ( 63^{+10}_{-9} \right)^\circ$
\cite{Fleischer:2008wb}. We see that while most of the scatter points
fall into the range $|q|_{\rm SM} = 0.69 \pm 0.18$, $\phi_{\rm SM} =
(0.2 \pm 2.1)^\circ$ some solutions differ visibly from the SM
expectation both in magnitude and/or phase. These points all feature
an enhanced electroweak penguin sector. Notice that larger deviations
would be possible if the stringent constraint arising from $Z^0 \to b
\bar b$ would be relaxed. Even in such a case, however, corrections
that would accommodate the second minimum in the $\chi^2$
fit\footnote{Such corrections typically feature negative values of
  ${\rm Im} \hspace{0.25mm} K_9$ of a few $10^{-4}$ and positive
  values of ${\rm Re} \hspace{0.25mm} K_9$ that are smaller by an
  order of magnitude.} are disfavored by the existing $b \to s l^+
l^-$ data. This is a model-independent conclusion that holds in all
new-physics models where the couplings between neutral gauge bosons
and leptons are not suppressed relative to the couplings to light
quarks (like it happens to be the case in RS models), and where
electromagnetic dipole operators do not receive sizable corrections
that (partially) cancel the effects from the semileptonic
operators. Removing the $\epsilon_K$ constraint, on the other hand,
does not change the obtained results qualitatively, since the
new-physics contributions to the $s \to d$ and $b \to s$ sector are
highly uncorrelated in all RS models with flavor anarchy.

\begin{figure}[!t]
  \begin{center}
    \includegraphics[height=6.5cm]{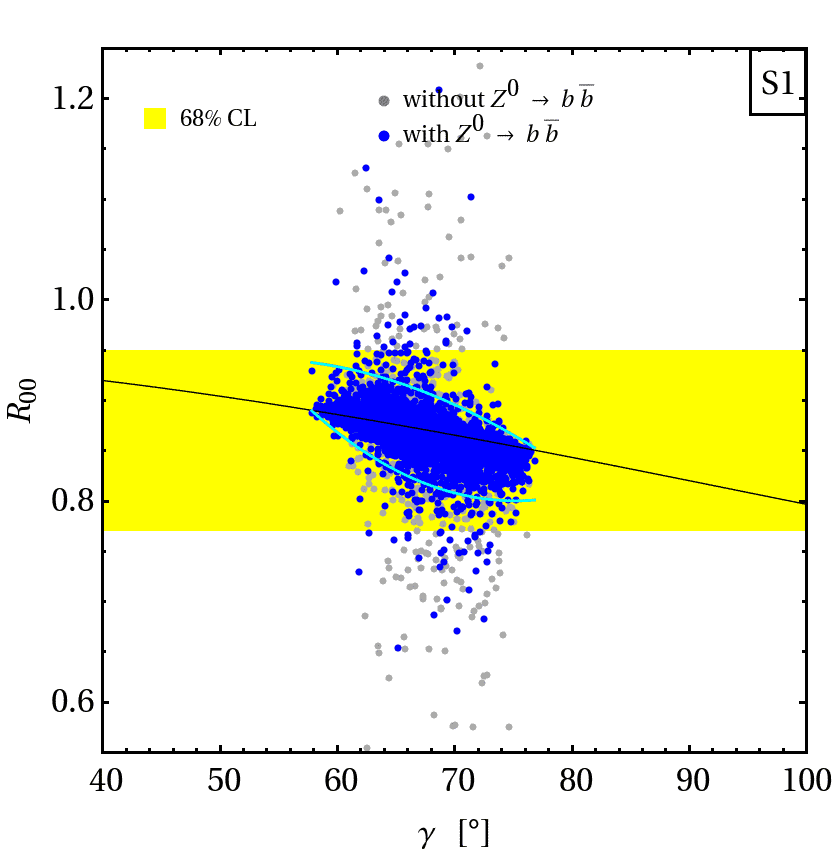}
    \hspace{0.75cm}
    \includegraphics[height=6.5cm]{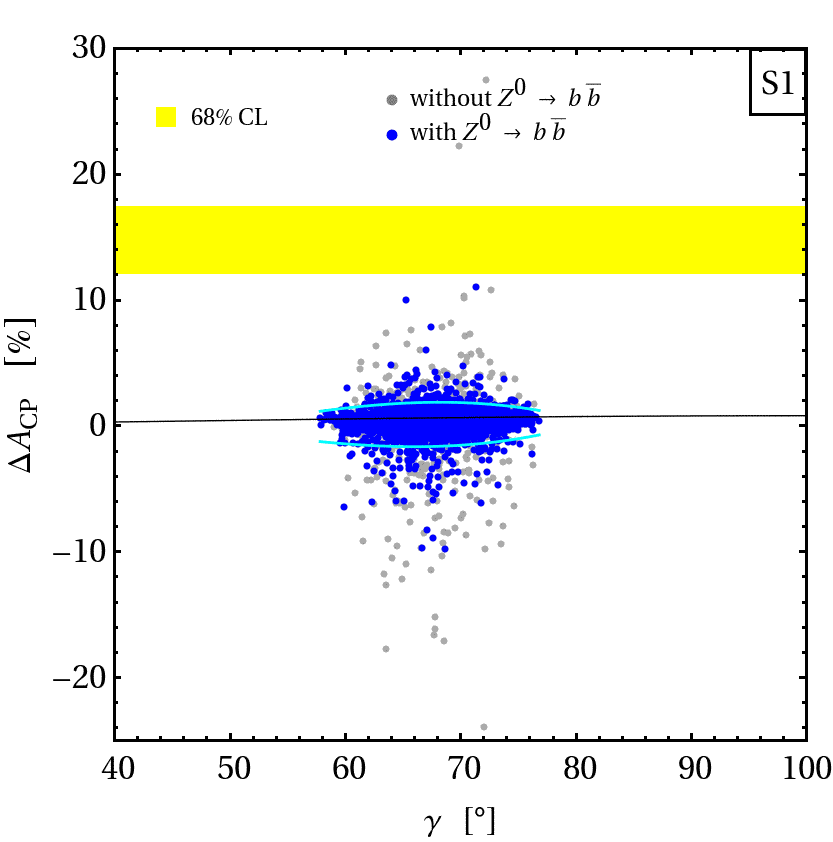}
  \end{center}
  \vspace{-8mm}
  \begin{center}
    \parbox{15.5cm}{\caption{\label{fig:nonlep2} Predictions for
        $R_{00}$ and $\Delta A_{\rm CP}$ as a function of $\gamma$. The
        black lines display the central value of the SM expectation
        while the yellow bands illustrate the corresponding experimental
        68\% CL. The shown blue (light gray) scatter points reproduce
        the values of the $Z^0 b \bar b$ couplings at 99\% CL. The cyan
        lines illustrate the functional behavior with $\gamma$ of the
        region containing the typical effect obtained from a fit to the
        $1\%$ (lower lines) and $99\%$ (upper lines) quantile of the
        points that are consistent with $Z^0 \to b \bar b$. See text for
        details.}}
  \end{center}
\end{figure}

In the right and left panel of Figure~\ref{fig:nonlep1} and
\ref{fig:nonlep2} we present the predictions for $R_\ast$ and $R_{00}$
as a function of the angle $\gamma$ of the unitarity triangle. The
black lines represent the central values of the SM expectations, and
the yellow bands are the experimentally preferred ranges at 68\%
probability. Like in the case of $q$, we observe that the RS
contributions to the ratios $R_\ast$ and $R_{00}$ typically do not
exceed the uncertainties of the measurements, which are similar in
size to the theoretical errors. We remark that an enhancement
(suppression) of $R_\ast$ and $R_{00}$ relative to the SM corresponds
to destructive (constructive) interference between the new-physics and
the SM electroweak penguin contribution. It is also interesting to
note that the RS contribution to $R_\ast$ ($R_{00}$) is a factor of
$7$ ($11$) more sensitive to the real than the imaginary part of
$C_9$. The weak dependence of $R_{00}$ on $\gamma$ in combination with
the fact that $R_{00} < 1$ is a practically model-independent SM
prediction (which can be understood in terms of a negative
interference between QCD and electroweak penguins) implies that
improved measurements of the latter ratio could turn out to be
valuable in probing the nature of the electroweak penguin sector of
warped extra-dimension models.

We now turn our attention to the so-called $\bar B \to \pi \bar K$
puzzle. Our predictions for $\Delta A_{\rm CP}$ are shown in the right
panel of Figure~\ref{fig:nonlep2}. The black line represents the
central value of the SM theory prediction as a function of
$\gamma$. For comparison we also show the 68\% CL range of $\Delta
A_{\rm CP}$ favored by the updated BaBar and Belle measurements of the
CP asymmetries in $\bar B^0 \to \pi^+ K^-$ and $B^- \to \pi^0 K^-$
decays (yellow band). Obviously the RS corrections in $\Delta A_{\rm
  CP}$ are typically at the level of a few percent only, \ie, too
small to explain the $3.5 \sigma$ discrepancy between experiment and
SM expectation. We remark that in order to achieve agreement in
$\Delta A_{\rm CP}$ within $1 \sigma$ of the combined experimental and
theoretical error, the coefficient $K_9$ should satisfy\footnote{The
  given inequality is only a crude approximation that should allow one
  to gain an understanding of how low the KK mass scale would have to
  be in order to reach values for $\Delta A_{\rm CP}$ in the ballpark
  of 15\%.}
\begin{equation} \label{eq:DeltaACPcond} 
  \left (0.5 + 45 \, {\rm Re}
    \hspace{0.5mm} K_9 \right ) \cdot 10^{-3} \, \lesssim \, {\rm Im}
  \hspace{0.5mm} K_9 \, \lesssim \, \left (0.9 - 75 \, {\rm Re} 
    \hspace{0.5mm} K_9 \right ) \cdot 10^{-3} \,.
\end{equation}
This formula has been obtained by setting $C_7^{\rm RS} = -\sws/\cws
\, C_9^{\rm RS} \approx -1/3 \, C_9^{\rm RS}$ and neglecting the
matching corrections to the QCD penguins, $C_{3-6}$, as well as the
chirality-flipped operators, $\tilde C_{3-10}$ (which are all good
approximations in the considered setup). The relatively weak
dependence on ${\rm Re} \hspace{0.5mm} K_9$ implies that $\Delta
A_{\rm CP}$ essentially measures the imaginary part of $K_9$, which
has to be positive in order to enhance $\Delta A_{\rm CP}$. Assuming
now that $K_9$ has the correct phase to fulfill
(\ref{eq:DeltaACPcond}), the relations (\ref{eq:DeltaZMA}),
(\ref{eq:Cpenguin}), and (\ref{eq:Ki}) can be used to derive the rough
bound
\begin{equation} \label{eq:ACPMKK}
  \Mkk \, \lesssim \, \sqrt{ \, 5 \hspace{0.5mm} L 
    \hspace{0.25mm} F(c_{Q_2}) \hspace{0.5mm} F(c_{Q_3})} \; {\rm  TeV} 
  \approx 1.7 \, {\rm TeV}
\end{equation}
on the KK mass scale. Here only the $L$-enhanced corrections in
(\ref{eq:Cpenguin}) have been included, and the quoted numerical
result corresponds to the central values for $c_{Q_{2,3}}$ in our
default benchmark scenario S1 as given in Table~\ref{tab:bulkmasses}.
Recalling that the $Z^0 \to b \bar b$ constraint typically forces the
KK scale to lie above 2 TeV in the original RS model, it is clear that
there is a generic tension between large effects in $\Delta A_{\rm
  CP}$ due to a modified electroweak penguin sector and the precision
measurements involving bottom quarks at the $Z^0$-pole. We also
emphasize that in RS variants with custodial protection of the $Z^0
d_{i \hspace{0.25mm} L} \bar d_{j \hspace{0.25mm} L}$ couplings the
effects in $\Delta A_{\rm CP}$ are even smaller than in the minimal RS
scenario, because the right-handed quark profile functions are
naturally more UV-localized than their left-handed counterparts. A
resolution of the $\bar B \to \pi \bar K$ puzzle therefore seems to be
difficult in the RS framework in general.

\begin{figure}[!t]
  \begin{center}
    \includegraphics[height=6.5cm]{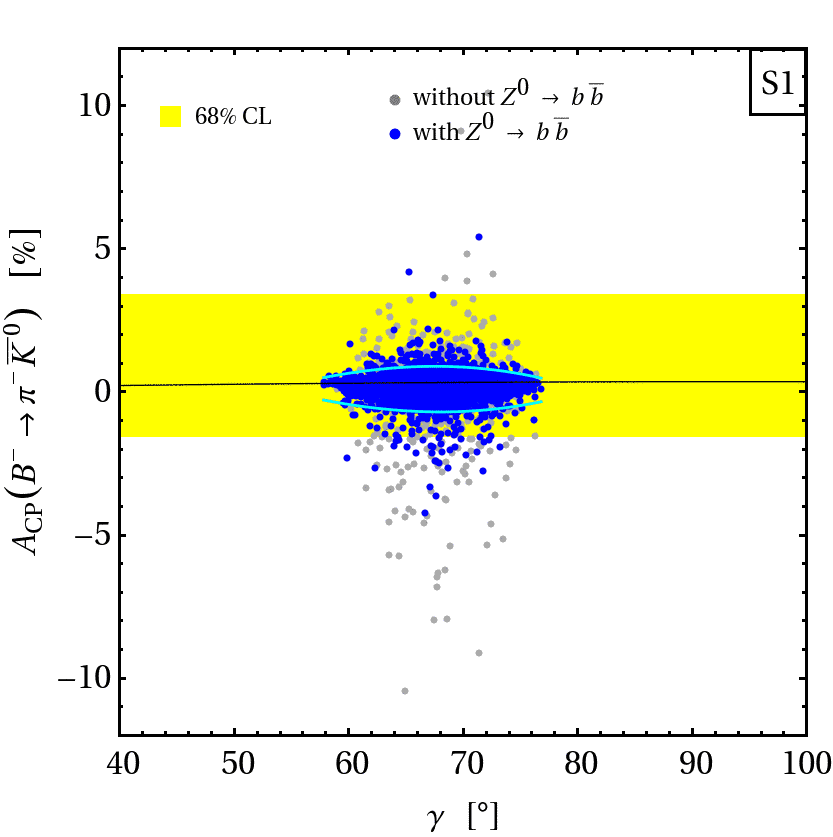}
    \hspace{0.75cm}
    \includegraphics[height=6.5cm]{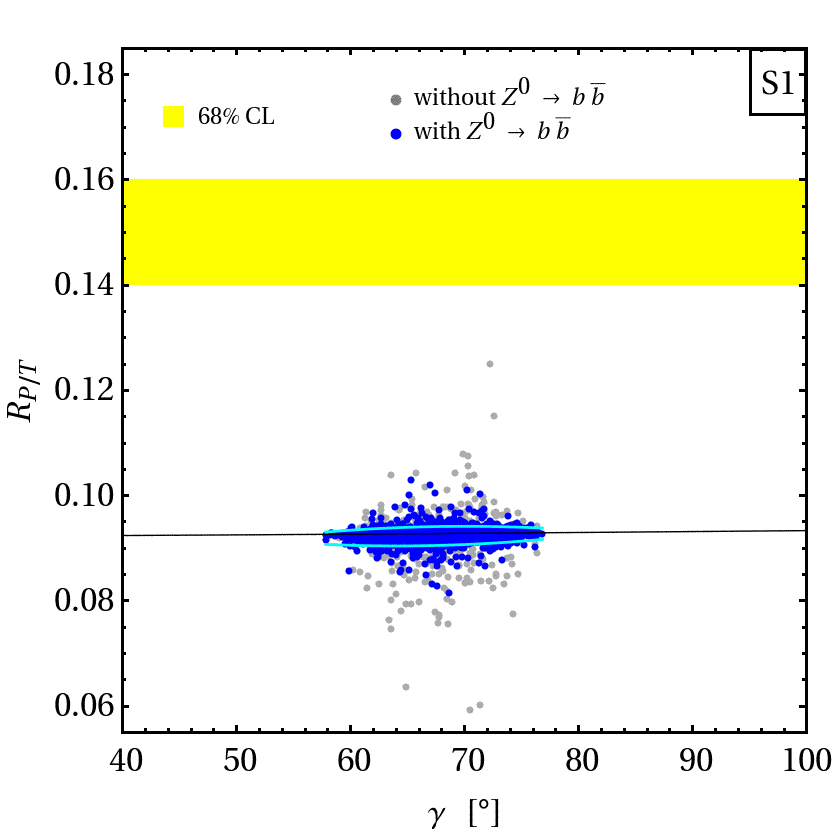}
  \end{center}
  \vspace{-8mm}
  \begin{center}
    \parbox{15.5cm}{\caption{\label{fig:nonlep3} Predictions for $A_{\rm
          CP} (B^- \to \pi^- \bar K^0)$ and $R_{P/T}$ as a function of
        $\gamma$. The black lines and yellow bands represent the central
        value of the SM expectation and the corresponding experimental
        68\% CL, respectively. Blue (light gray) points are consistent
        (inconsistent) with $Z^0 \to b \bar b$ at 99\% CL. The cyan
        lines illustrate the functional behavior with $\gamma$ of the
        region containing the typical effect obtained from a fit to the
        $1\%$ (lower lines) and $99\%$ (upper lines) quantile of the
        points that are consistent with $Z^0 \to b \bar b$. See text
        for details.}}
  \end{center}
\end{figure}

Our predictions for $A_{\rm CP} (B^- \to \pi^- \bar K^0)$ and
$R_{P/T}$ are shown in the left and right panels of
Figure~\ref{fig:nonlep3}. The black lines and yellow bands depict the
central values of the SM prediction and the experimental 68\% CL
ranges.  While in the case of $A_{\rm CP} (B^- \to \pi^- \bar K^0)$ we
see that it is possible to saturate the experimental limits,
corresponding to shifts in the CP asymmetry of a couple of percent,
the typical deviations in the penguin-to-tree ratio $R_{P/T}$ are much
smaller than the difference between the central value of the theory
prediction and the experimental determination. Although the
discrepancy between $(R_{P/T})_{\rm exp}$ and $(R_{P/T})_{\rm SM}$ is
in view of the sizable SM error not significant, it is still
interesting to ask how an electroweak penguin sector that would
improve the agreement between experiment and theory may look
like. Neglecting again subleading contributions to $R_{P/T}$ from QCD
penguins and operators with opposite chirality, and setting $C_7^{\rm
  RS} = -\sws/\cws \, C_9^{\rm RS}$, it is easy show that points that
lead to the same value for $R_{P/T}$ lie on circles in the ${\rm Re}
\hspace{0.25mm} K_9 \hspace{0.5mm}$--$\hspace{0.5mm} {\rm Im}
\hspace{0.25mm} K_9$ plane. For example, to bring $R_{P/T}$ within $1
\sigma$ of the experimental central value, the following condition has
to be fulfilled
\begin{equation}
  5.0 \cdot 10^{-6} \, \lesssim \, \left ( {\rm Re} \hspace{0.25mm} K_9 + 
    0.0014 \right )^2 + \left ( {\rm Im} \hspace{0.25mm} K_9 - 
    0.0030 \right )^2 \, \lesssim \, 6.1 \cdot 10^{-6} \,.
\end{equation}
This inequality implies that, like in the case of $\Delta A_{\rm CP}$,
an enhancement of $R_{P/T}$ relative to its SM value requires the
imaginary part of $K_9$ to be positive and of order $7 \cdot
10^{-4}$. Such values of ${\rm Im} \hspace{0.25mm} K_9$ again call for
$\Mkk \lesssim 2$ TeV, at variance with the parameter region preferred
by the $Z^0 b \bar b$ couplings.

\begin{figure}[!t]
  \begin{center}
    \includegraphics[height=6.5cm]{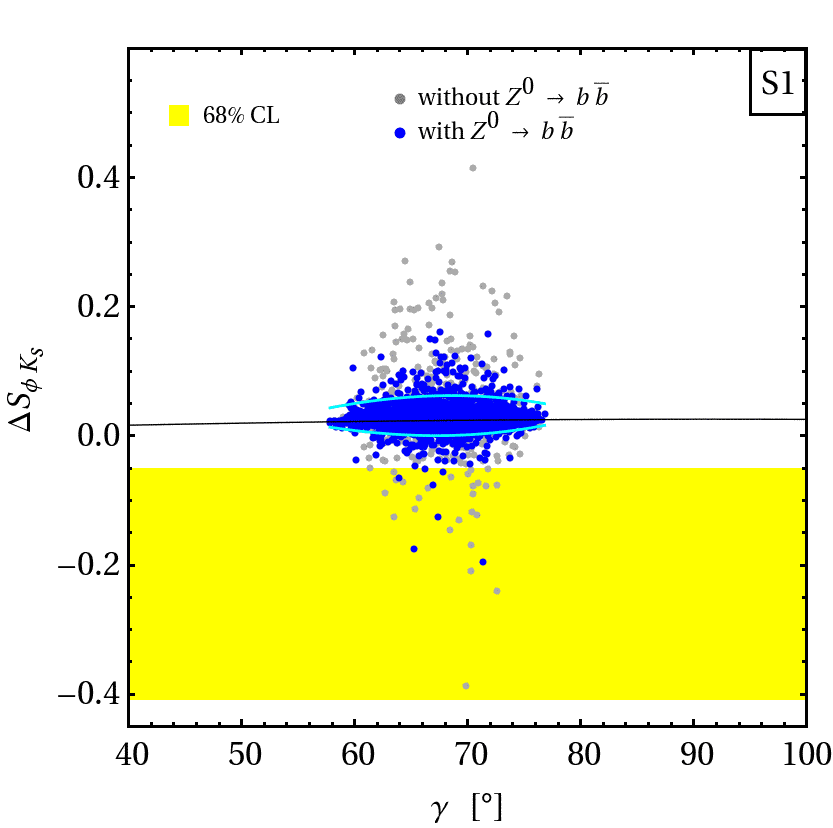}
    \hspace{0.75cm}
    \includegraphics[height=6.5cm]{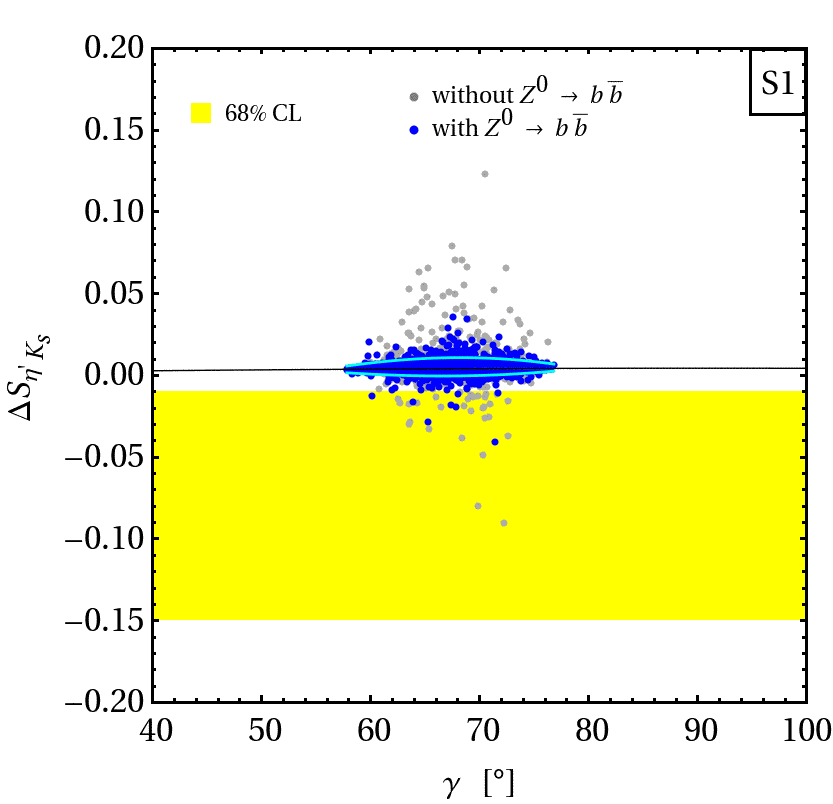}

    \vspace{4mm}

    \includegraphics[height=6.5cm]{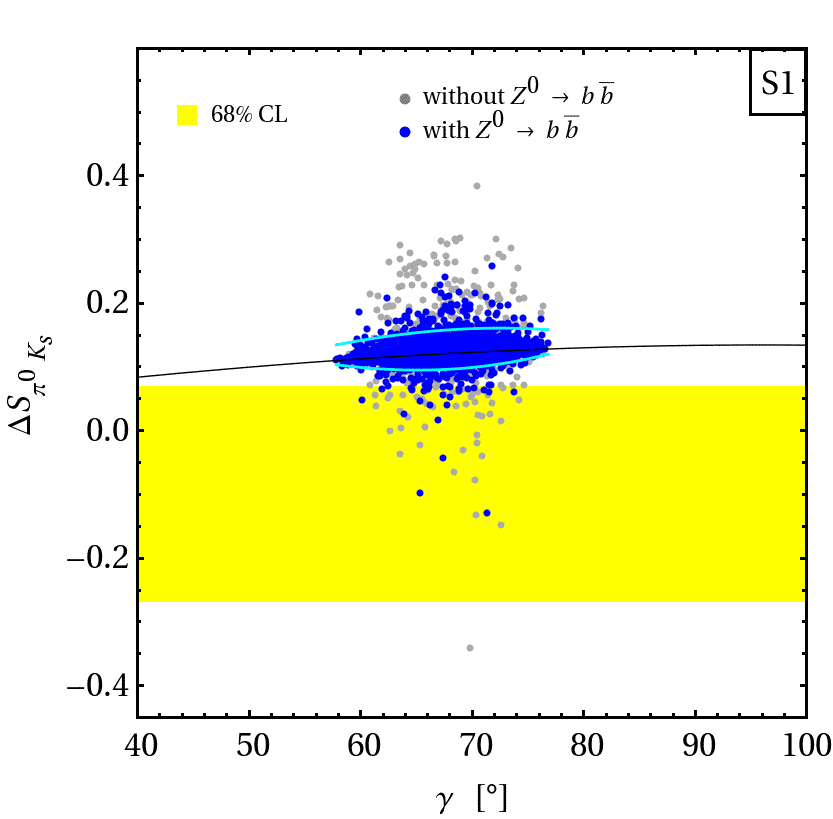}
  \end{center}
  \vspace{-8mm}
  \begin{center}
    \parbox{15.5cm}{\caption{\label{fig:nonlep4} Predictions for $\Delta
        S_{\phi K_S}$, $\Delta S_{\eta^\prime K_S}$, and $\Delta
        S_{\pi^0 K_S}$ as a function of $\gamma$. The black lines and
        yellow bands correspond to the central values expected in the SM
        and the associated experimental 68\% CL regions. Scatter points
        colored blue (light gray) satisfy (fail to satisfy) the $Z^0 \to
        b \bar b$ constraint at 99\% CL. The cyan lines illustrate the
        functional behavior with $\gamma$ of the region containing the
        typical effect obtained from a fit to the $1\%$ (lower lines)
        and $99\%$ (upper lines) quantile of the points that are
        consistent with $Z^0 \to b \bar b$. See text for details.}}
  \end{center}
\end{figure}

In Figure~\ref{fig:nonlep4}, we finally show the predictions for the
shifts $\Delta S_{\phi K_S}$, $\Delta S_{\eta^\prime K_S}$, and
$\Delta S_{\pi^0 K_S}$ as a function of $\gamma$. The central values
of the SM predictions (black lines) and the experimental 68\% CL
ranges (yellow bands) are displayed for comparison. It is evident that
in all three cases it is possible to improve the agreement between
theory and experiment. The maximal deviations in $\Delta S_{\phi K_S}$
and $\Delta S_{\pi^0 K_S}$ ($\Delta S_{\eta^\prime K_S}$) are at the
level of $\pm 0.2$ ($\pm 0.02$), corresponding to corrections in
$K_{7-10}$ of order $10^{-4}$. In order to explain this pattern, we
first note that $\Delta S_{\phi K_S}$, $\Delta S_{\eta^\prime K_S}$,
and $\Delta S_{\pi^0 K_S}$ depend only very weakly on the real parts
of the electroweak penguin coefficients $K_{7-10}$. Setting again
$C_7^{\rm RS} = -\sws/\cws \, C_9^{\rm RS}$ and ignoring all
corrections from QCD penguin and chirality-flipped operators, we find
that the dependence of $\Delta S_{\phi K_S}$ and $\Delta S_{\pi^0
  K_S}$ on ${\rm Im} \hspace{0.25mm} K_9$ is more pronounced by a
factor of roughly 10 than the one of $\Delta S_{\eta^\prime
  K_S}$. Furthermore, positive values of ${\rm Im} \hspace{0.25mm}
K_9$ lead to negative shifts $\Delta S_{\phi K_S}$ and $\Delta
S_{\pi^0 K_S}$ and {\it vice versa}, while in the case of $\Delta
S_{\eta^\prime K_S}$ the correlation is a positive one.  A
simultaneous improvement in all three observables is therefore in
general not possible in the RS framework.

\section{Conclusions and Outlook}
\label{sec:concl}

We have presented a detailed and comprehensive study of
flavor-violating effects in the original RS scenario, which features
an $SU(2)_L \times U(1)_Y$ bulk gauge symmetry and a minimal,
brane-localized Higgs sector. The complete analytic expressions up to
${\cal O} (v^2/\Mkk^2)$ for all relevant effective four-fermion
interactions induced by the exchange of gauge or Higgs bosons have
been derived. In contrast to previous studies of quark flavor physics
in RS models, the given expressions encode the summation over the
entire KK tower and not only one (or a few) KK excitations. Employing
the methods and results presented in the first part of this sequel, it
would be easy to include higher-order contributions in $v^2/\Mkk^2$ in
closed form as well.

Next we have applied these general findings to weak decay processes,
focusing on observables in the quark sector. We first have obtained
the complete tree-level expressions for the relevant effective $\Delta
F = 2$ Hamiltonians, giving rise to the mixing of neutral $K$,
$B_{d,s}$, and $D$ mesons with their antiparticles. Next followed a
derivation of the effective $\Delta F = 1$ Hamiltonians describing the
$b \to s q \bar q$, $s \to d \nu \bar \nu$, $b \to s l^+ l^-$, and $b
\to u l \bar \nu$ transitions arising from the tree-level exchange of
KK gluons and photons, the $Z^0$ and $W^\pm$ bosons and their KK
excitations, as well as the Higgs boson. We then collected the
formulas needed for the phenomenological analysis of neutral-meson
mixing, as well as rare non-leptonic and leptonic decays of kaons and
$B$ mesons. Besides quantities that have already received much
attention in the literature, such as $K$--$\bar K$ and $B$--$\bar B$
mixing, the discussed observables include many that were not often
considered before, ranging from $D$--$\bar D$ mixing to the
(semi-)leptonic processes $B \to \tau \nu_\tau$, $B\to\mu^+\mu^-$, $B
\to X_s (K^\ast) \, l^+ l^-$ to $\epseps$ and non-leptonic decays such
as $\bar B \to \pi \bar K$. Throughout our discussion we tried to be
as general as possible, so that the given formulas can be applied to
any SM extension with flavor-violating contributions to both left- and
right-handed dimension-six operators.

After this preparatory work, we have presented dedicated numerical
analyses of the various flavor observables. In order to investigate
how sensitively the predictions depend on the most relevant parameters
of the minimal RS model, we have studied four different parameter
scenarios in detail, investigating the numerical and perturbative
stability of each parameter point using the Barbieri and Giudice fine-tuning
measure. Besides the default scenario (S1, ``standard''),
we investigated the consequences of common right-handed down-type bulk
masses (S2, ``aligned''), the impact of a reduction of the
``volume'' of the extra dimension (S3, ``little''), and of larger
Yukawa couplings (S2, ``large''). The latter three benchmark scenarios
are designed specifically to suppress harmful contributions to
electroweak precision and quark flavor observables, and we have
quantified the improvement that can be achieved in the $Z^0 \to b \bar
b$ and $\epsilon_K$ observables in each case. While the average amount
of tuning required in order to satisfy the stringent constraints from
the bottom-quark $Z^0$-pole observables and indirect CP violation in
the kaon sector depend on the particular scenario, we found that most
other flavor observables are largely insensitive to such a choice.

We began our phenomenological survey in the charged-current sector,
emphasizing that the exchange of the entire KK tower of the $W^\pm$
bosons generically induces unitarity violations in the columns and
rows of the quark-mixing matrices. While the non-standard effects in
the elements of the generalized CKM matrix involving only light quarks
are small after correct normalization to the Fermi constant $G_F$, the
corrections in the heavy-quark sector can be of the order of a few
percent. This leads to a non-closure of the unitary triangle. While
this effect is severely constrained by the $Z^0\to b\bar{b}$ precision
observables in the minimal RS scenario, it can still be at the level
of the current experimental uncertainties. We have furthermore pointed
out that the prospects for a detection of a unitarity violation of the
left-handed quark-mixing matrix or deviations in the branching ratios
of semileptonic and purely leptonic $B$ decays are better in RS
realizations with extended bulk gauge symmetry, because the custodial
symmetry cannot simultaneously protect the $Z^0 b_L \bar b_L$ and the
$W u_{i \hspace{0.25mm} L} d_{j \hspace{0.25mm} L}$ vertices.

A first key observation gleaned from the analyses of $\Delta F = 2$
observables is that the four-quark operators induced by KK gluon
exchange give the by far dominant (leading) contributions to the
effective Hamiltonians describing $K$--$\bar K$ ($B_{d,s}$--$\bar
B_{d,s}$ and $D$--$\bar D$) mixing. This implies that mixing phenomena
mainly probe the extra-dimensional aspects of the strong interactions,
but are to first approximation insensitive to the precise embedding of
the electroweak gauge symmetry in the higher-dimensional geometry. A
second important finding is that, in spite of the RS-GIM mechanism, a
residual ``little CP problem'' in the $\Delta F = 2$ sector is found
in all considered benchmark scenarios in the form of excessive
contributions to $\epsilon_K$ arising from dimension-six operators
involving quarks of different chiralities. Our detailed numerical
analysis showed that fulfilling the latter constraint with only a
moderate amount of tuning implies that the mass of the first-level KK
gluon mode has to lie above 10 TeV in the parameter scenario S1, and
even higher in scenario S3.  An acceptable amount of indirect CP
violation in the kaon sector can be achieved for masses of the first
KK excitation below 5 TeV by ``alignment'' in the right-handed down-type
quark sector (S2), or by allowing for larger down-type Yukawa
couplings (S4). Yet both solutions have their limitations.
The former scenario has potential problems with loop-induced misalignment
and additional flavor violation from brane kinetic terms, while in the
second case loop contributions to dipole operators are parametrically
enhanced, making it impossible to fully decouple flavor-violating
effects. Since the new CP-odd phases appearing in the $s \to d$,
$b \to s$, and $c \to u$ transitions are highly uncorrelated, we found
that even after enforcing the $\epsilon_K$ constraint it is possible
to obtain CP-violating effects in $B_s$--$\bar B_s$ and $D$--$\bar D$
mixing that exceed by far the SM expectations. While both $A_{\rm
  SL}^s$ and $S_{\psi \phi}$ as well as $A_{\rm SL}^D$ and $S_{\psi
  K_s}^D$ turn out to be strongly correlated, we have stressed that
the same linear correlations arise in a wide class of scenarios of new
physics, so that they cannot be regarded as unmistakable signals of
warped extra-dimension models.

Subsequently we have argued that, in contrast to the $\Delta F = 2$
sector, the predictions for $\Delta F = 1$ observables depend strongly
on the exact realization of both the gauge and fermionic sectors,
because they receive the dominant contributions from tree-level
exchange of the $Z^0$ boson and its KK excitations. While these
effects are enhanced by the logarithm of the warp factor $L$ in models
with $SU(2)_L \times U(1)_Y$ gauge symmetry, they are absent in the
$Z^0 d_{i \hspace{0.25mm} L} \bar d_{j \hspace{0.25mm} L}$ couplings
if the bulk gauge group is extended to $SU(2)_L \times SU(2)_R\times
U(1)_X \times P_{LR}$ and an appropriate embedding of the left-handed
down-type quarks is chosen. If the right-handed down-type quarks are
embedded into triplet representations, which is necessary to arrive at
$U(1)_X$ invariant Yukawa couplings, then the $Z^0 d_{i
  \hspace{0.25mm} R} \bar d_{j \hspace{0.25mm} R}$ couplings are
enhanced by one order of magnitude relative to the minimal RS
model. Despite this enhancement, we found that right-handed currents
in the $b \to d, s$ sector remain small in the custodial RS model,
since the involved right-handed quark profiles are naturally more
UV-localized than their left-handed counterparts. Larger effects are
possible in the $s \to d$ sector, but this would require the bulk mass
parameter of the right-handed top quark to be larger than about 1,
which is rather far away from the critical value $-1/2$ required for
IR localization.  We concluded that, while the pattern of new-physics
effects in processes such as $B_{d, s} \to \mu^+ \mu^-$, $B \to
X_{d,s} \nu \bar \nu$, $K_L \to \mu^+ \mu^-$, $K \to \pi \nu \bar
\nu$, and $K_L \to \pi^0 l^+ l^-$ is model dependent,
order-of-magnitude enhancements of the branching fractions of rare
$B$- and $K$-meson decays seem only possible in the minimal RS
scenario, after satisfying the $Z^0 \to b \bar b$ constraint by
tuning. We emphasize that while sizable effects in $K^+ \to \pi^+ \nu
\bar \nu$ and $B_{d, s} \to \mu^+ \mu^-$ are {\it not} the generic
prediction of the RS framework, they can and do occur (also
simultaneously) even after imposing the $\epsilon_K$ constraint, since
the flavor-violating contributions due to virtual KK gluon and
$Z^0$-boson exchange are largely uncorrelated in the considered model.

We have also extended earlier studies of $\Delta F = 1$ quark
processes by examining observables which have not (or only partly)
been considered in the literature. Our work covers $B \to \tau
\nu_\tau$, $B \to X_s (K^\ast) \, l^+ l^-$, $\epseps$, and $\bar B \to
\pi \bar K$ as well as other observables in the field of exclusive
hadronic $B$-meson decays. The key question that we have addressed in
the context of the decays $B \to \tau \nu_\tau$ and $\bar B \to \pi
\bar K$ is whether possible discrepancies, as suggested by experiment,
may be explained in scenarios with warped extra dimensions.  We found
that it is not possible to explain the tension in $B \to \tau
\nu_\tau$, since the corrections to the $b \to u l \bar \nu$
transition that are consistent with $Z^0 \to b \bar b$ do not exceed
the level of 1\%. The new-physics effects in $\Delta A_{\rm CP}$
usually also turn out to be too small to resolve the $\bar B \to \pi
\bar K$ puzzle. From general considerations we have deduced that a
notable improvement of the overall consistency between experiment and
theory in exclusive hadronic $B$-meson decays would call for masses of
the first KK excitations below 5 TeV, which in the minimal RS
framework is problematic in view of the stringent $Z^0 \to b \bar b$
constraint. In the case of the CP-violating parameter $\epseps$, we
have pointed out that the marked sensitivity of this ratio to the
electroweak penguin sector leads to a strong correlation between
$\epseps$ and the $s \to d \nu \bar \nu$ and $s \to d l^+ l^-$
observables. We have explicitly shown that even if hadronic
uncertainties are treated conservatively, the $\epseps$ constraint has
a non-negligible impact on possible new-physics effects in rare $K$
decays. In particular, large enhancements in $K_L \to \pi^0 \nu \bar
\nu$ decay have been shown to be disfavored by the measured amount of
direct CP violation in $K \to \pi \pi$. As enhancements of the
branching fractions for $K_L \to \pi^0 \nu \bar \nu$ and $K_L \to
\pi^0 l^+ l^-$ are strongly correlated with each other, similar
statements apply to the rare kaon decay modes with charged leptons in
the final state. The correlation between $K^+ \to \pi^+ \nu \bar \nu$
and $\epseps$ is weaker, and consequently departures from the SM
expectation in the charged mode are possible even after enforcing the
$\epseps$ constraint.

The present work, which is a sequel to \cite{Casagrande:2008hr} and
\cite{Bauer:2008xb}, concludes our detailed and rather comprehensive
survey of quark flavor physics in the minimal RS model. We hope that
it will provide a useful reference for many experimental analyses
searching for new-physics signals at present and future flavor
factories such as LHCb and a super flavor factory. Extensions of the
minimal model with an enlarged electroweak gauge group in the bulk,
and with a more complicated matter sector, will be explored in a
future publication.

\subsubsection*{Acknowledgments}

We are grateful to L.~Gr\"under for collaboration at early stages of
this work. It is also a pleasure to thank A.~Azatov, C.~Bobeth,
J.~Brod, F.~Goertz, M.~Gorbahn, S.~J\"ager, A.~Lenz, T.~Pfoh,
M.~Toharia, G.~Zanderighi, L.~Zhu, and J.~Zupan for useful discussions
and helpful correspondence. The research of SC is supported by the DFG
cluster of excellence ``Origin and Structure of the Universe''.

\begin{appendix}

\section{RG Evolution of Penguin Coefficients}
\label{app:RGE}

\renewcommand{\theequation}{A\arabic{equation}}
\setcounter{equation}{0}

In this appendix we present analytic and numerical formulas relating
the weak-scale Wilson coefficients $C_{3-10} (m_W)$ of the QCD and
electroweak penguins with their initial conditions $C_{3-10} (\Mkk)$
evaluated at $\Mkk\gg m_t$. If not stated otherwise, the results are
obtained at leading-logarithmic order in QCD and take into account
that in the RS model one has $C_4 (\Mkk) = -3 \, C_5 (\Mkk) = C_6
(\Mkk)$ and $C_8 (\Mkk) = C_{10} (\Mkk) = 0$ at tree level. Analog
formulas for $\tilde C_{3-10} (m_W)$ are not explicitly shown. Since
QCD is chirality-blind they can simply be obtained by the replacing
the original Wilson coefficients by the chirality-flipped coefficients
in the expressions presented below.

For new-physics scales $\Mkk \in [1, 10] \, {\rm TeV}$, we find to
excellent approximation 
\begin{align}
  C_3 (m_W) & = \left ( 0.416 \, \eta^{2} - 0.962 \, \eta + 1.570
  \right ) C_3 (\Mkk) - \left ( 0.002 \, \eta^{2} -
    0.242 \, \eta + 0.263 \right ) C_6 (\Mkk) \,, \nonumber \\
  C_4 (m_W) & = -\left ( 0.505 \, \eta^{2} - 1.465 \, \eta + 1.017
  \right ) C_3 (\Mkk) - \left ( 0.416 \, \eta^{2} - 1.190 \, \eta
    - 0.189 \right ) C_6 (\Mkk) \,, \nonumber \\
  C_5 (m_W) & = - \left ( 0.034 \, \eta^{2} - 0.036 \, \eta - 0.001
  \right ) C_3 (\Mkk) + \left ( 0.192 \, \eta^{2} - 0.592 \, \eta
    + 0.090 \right ) C_6 (\Mkk) \,, \nonumber \\
  C_6 (m_W) & = \left ( 0.002 \, \eta^{2} + 0.093 \, \eta - 0.102
  \right ) C_3 (\Mkk) + \left ( 1.376 \, \eta^{2} - 3.005 \, \eta +
    2.703 \right ) C_6 (\Mkk) \,,\nonumber \\
  C_{7} (m_W) & = 0.987 \, \eta ^{0.143} \, C_7 (\Mkk) \,, \nonumber \\
  C_{8} (m_W) & = \left( 0.371 \, \eta^{-1.143} -
    0.329 \, \eta^{0.143} \right) C_7 (\Mkk) \,, \\
  C_{9} (m_W) & = \left( 0.528 \, \eta^{-0.571} +
    0.487 \, \eta^{0.286} \right) C_9 (\Mkk) \,, \nonumber \\
  C_{10} (m_W) & = \left (-0.528 \, \eta^{-0.571} + 0.487 \,
    \eta^{0.286} \right ) C_9 (\Mkk) \,, \nonumber
\end{align}
where $\eta \equiv \alpha_s(\Mkk)/\alpha_s (m_t)$ is the relevant
ratio of strong coupling constants. The numerically suppressed
contributions $C_{7, 9} (\Mkk)$ entering the QCD penguin coefficients
$C_{3-6} (m_W)$ through mixing have been suppressed here for
simplicity.

Inserting $\eta \approx 0.73$, which corresponds to $\Mkk = 3 \, {\rm
  TeV}$, into the above expressions and restoring the dependence on
the coefficient $C_9 (3 \, {\rm TeV})$ in the QCD penguins, we obtain
numerically \\
\begin{equation} 
\begin{split}
  C_3 (m_W) & = 1.09 \, C_3 (3 \, {\rm TeV}) - 0.09 \, C_6 (3 \, {\rm
    TeV}) - 0.01 \, C_9 (3 \, {\rm TeV}) \,, \mkern+140mu \\
  C_4 (m_W) & = - 0.22 \, C_3 (3 \, {\rm TeV}) + 0.84 \, C_6 (3 \,
  {\rm TeV}) + 0.02 \, C_9 (3 \, {\rm TeV})  \,, \\
  C_5 (m_W) & = 0.01 \, C_3 (3 \, {\rm TeV}) - 0.24 \, C_6 (3 \,
  {\rm TeV})  - 0.01 \, C_9 (3 \, {\rm TeV})  \,, \\
  C_6 (m_W) & = - 0.03 \, C_3 (3 \, {\rm TeV}) + 1.26 \, C_6 (3 \,
  {\rm TeV}) + 0.03 \, C_9 (3 \, {\rm TeV}) \,, \\
  C_7 (m_W) & = 0.94 \, C_7 (3 \, {\rm TeV}) \,, \\
  C_8 (m_W) & = 0.21 \, C_7 (3 \, {\rm TeV}) \,, \\
  C_9 (m_W) & = 1.08 \, C_9 (3 \, {\rm TeV}) \,, \\
  C_{10} (m_W) & = -0.18 \, C_9 (3 \, {\rm TeV}) \,.
\end{split}
\end{equation}
Contributions proportional to $C_7 (3 \, {\rm TeV})$ in $C_{3-6}
(m_W)$ receive coefficients smaller than 0.01 and have been discarded.

We also present numerical formulas for the Wilson coefficients
$C_{3-10} (m_b)$ and $C_{3-10} (m_c)$ evaluated at the bottom- and
charm-quark mass scale, respectively. Performing the evolution below
$m_W$ at next-to-leading logarithmic accuracy in QCD and QED
\cite{Buras:1991jm, Buras:1992tc, Buras:1992zv, Ciuchini:1992tj,
  Ciuchini:1993vr}, we arrive at
\begin{equation} 
\begin{split}
  C_3 (m_b) & = 1.25 \, C_3 (3 \, {\rm TeV}) - 0.17 \, C_6 (3 \, {\rm
    TeV}) + 0.01 \, C_7 (3 \, {\rm TeV}) - 0.03 \, C_9 (3 \, {\rm
    TeV}) \,, \\
  C_4 (m_b) & = - 0.53 \, C_3 (3 \, {\rm TeV}) + 0.62 \, C_6 (3 \,
  {\rm TeV}) - 0.01 \, C_7 (3 \, {\rm TeV}) + 0.06 \, C_9 (3 \, {\rm
    TeV}) \,, \\
  C_5 (m_b) & = 0.03 \, C_3 (3 \, {\rm TeV}) - 0.06 \, C_6 (3 \,
  {\rm TeV}) - 0.01 \, C_9 (3 \, {\rm TeV})  \,, \\
  C_6 (m_b) & = - 0.08 \, C_3 (3 \, {\rm TeV}) + 1.97 \, C_6 (3 \,
  {\rm TeV}) - 0.01 \, C_7 (3 \, {\rm TeV}) + 0.09 \, C_9 (3 \, {\rm
    TeV}) \,, \\
  C_7 (m_b) & = 0.88 \, C_7 (3 \, {\rm TeV}) - 0.01 \, C_9 (3 \, {\rm
    TeV}) \,, \\
  C_8 (m_b) & = -0.01 \, C_6 (3 \, {\rm TeV}) + 0.78 \, C_7 (3 \,
  {\rm TeV}) \,, \\
  C_9 (m_b) & = -0.01 \, C_7 (3 \, {\rm TeV}) + 1.24 \, C_9 (3 \,
  {\rm TeV}) \,, \\
  C_{10} (m_b) & = -0.48 \, C_9 (3 \, {\rm TeV}) \,,
\end{split}
\end{equation}
and 
\begin{equation} 
\begin{split} \label{pip}
  C_3 (m_c) & = 1.43 \, C_3 (3 \, {\rm TeV}) - 0.17 \, C_6 (3 \, {\rm
    TeV}) + 0.02 \, C_7 (3 \, {\rm TeV}) - 0.07 \, C_9 (3 \, {\rm
    TeV}) \,, \\
  C_4 (m_c) & = - 0.81 \, C_3 (3 \, {\rm TeV}) + 0.41 \, C_6 (3 \,
  {\rm TeV}) - 0.03 \, C_7 (3 \, {\rm TeV}) + 0.12 \, C_9 (3 \, {\rm
    TeV}) \,,  \\
  C_5 (m_c) & = 0.05 \, C_3 (3 \, {\rm TeV}) + 0.19 \, C_6 (3 \,
  {\rm TeV}) - 0.01 \, C_7 (3 \, {\rm TeV})  \,, \\
  C_6 (m_c) & = - 0.12 \, C_3 (3 \, {\rm TeV}) + 3.19 \, C_6 (3 \,
  {\rm TeV}) - 0.06 \, C_7 (3 \, {\rm TeV}) + 0.20 \, C_9 (3 \, {\rm
    TeV}) \,,  \\
  C_7 (m_c) & = 0.90 \, C_7 (3 \, {\rm TeV}) - 0.01 \, C_9 (3 \, {\rm
    TeV}) \,,   \\
  C_8 (m_c) & = -0.03 \, C_6 (3 \, {\rm TeV}) + 1.72 \, C_7 (3 \,
  {\rm TeV}) -0.01 \, C_9 (3 \, {\rm TeV})  \,,  \\
  C_9 (m_c) & = 0.01 \, C_3 (3 \, {\rm TeV}) -0.02 \, C_7 (3 \, {\rm
    TeV}) + 1.43 \, C_9 (3 \, {\rm TeV}) \,,   \\
  C_{10} (m_c) & = 0.01 \, C_6 (3 \, {\rm TeV}) - 0.76 \, C_9 (3 \,
  {\rm TeV}) \,.
\end{split}
\end{equation}
The above relations imply that neglecting the running between $\Mkk$
and $m_W$ would lead to results for $C_{6,8,10} (m_b)$ that are
smaller than the correct ones by about $25\%$ to $60\%$. In the case
of $C_{6,8,10} (m_c)$ the associated reductions range between $30\%$
and $45\%$. For the remaining Wilson coefficients the effects are less
pronounced. We conclude that the inclusion of the RG evolution between
the KK scale and the weak scale is important in order to obtain
accurate results.

\section{Input Parameters}
\label{app:input}

\renewcommand{\theequation}{B\arabic{equation}}
\renewcommand{\thetable}{B\arabic{table}}
\setcounter{equation}{0}
\setcounter{table}{0}

Here we collect the values of the experimental and theoretical
parameters used in our numerical analysis. They are thematically
ordered in Tables~\ref{tab:input1} to \ref{tab:LCSRfit}.  In the case
of the exclusive hadronic $B$-meson decay observables, parameters
related to $\eta^{(\prime)}$ mesons, such as their decay constants,
form factors, and the $\eta$--$\eta^\prime$ mixing angles in the
quark-flavor basis have not been spelled out. They can be found in
\cite{Beneke:2002jn, Beneke:2003zv}.

\begin{table}[!b]
\begin{center}
\begin{tabular}{|c|c|c|c|c|}
  \hline
  Parameter & Value $\pm$ Error & Reference \\[0.5mm] 
  \hline 
  $G_F$ & $1.16637 \cdot 10^{-5} \, {\rm GeV}^{-2}$ & 
  \cite{Amsler:2008zz} \\
  $\sws$ & $0.2312$ & \cite{LEPEWWG:2005ema} \\
  $m_W$ & $80.398 \, {\rm GeV}$ & \cite{LEPEWWG:2005ema} \\
  $m_Z$ & $91.1875 \, {\rm GeV}$ & \cite{LEPEWWG:2005ema} \\
  $\alpha(m_Z)$ & $1/127.9$ & \cite{LEPEWWG:2005ema} \\
  $\alpha_s(m_Z)$ & $0.118 \pm 0.003$ & \cite{Amsler:2008zz} \\
  $m_t$ & $(144 \pm 5) \, {\rm GeV}$ & 
  \cite{Group:2008nq} \\
  $m_b$ & $(2.3 \pm 0.1) \, {\rm GeV}$ & 
  \cite{Amsler:2008zz} \\
  $m_c$ & $(560 \pm 40) \, {\rm MeV}$ & 
  \cite{Amsler:2008zz} \\
  $m_s$ & $(50 \pm 15) \, {\rm MeV}$ & 
  \cite{Amsler:2008zz} \\
  $m_d$ & $(3.0 \pm 2.0) \, {\rm MeV}$ & 
  \cite{Amsler:2008zz} \\
  $m_u$ & $(1.5 \pm 1.0) \, {\rm MeV}$ & 
  \cite{Amsler:2008zz} \\
  $\lambda$ & $0.2265 \pm 0.0008$ & \cite{Charles:2004jd} \\
  $A$ & $0.807 \pm 0.018$ & \cite{Charles:2004jd} \\
  $\bar \rho$ & $0.147^{+0.029}_{-0.017}$ & \cite{Charles:2004jd} \\
  $\bar \eta$ & $0.343 \pm 0.016$ & \cite{Charles:2004jd} \\
  \hline 
\end{tabular}
\end{center}
\vspace{-7mm}
\begin{center}
  \parbox{15.5cm}{\caption{\label{tab:input1} Parameters used in the
      SM predictions and the generation of RS parameter points. The
      quoted values of the quark masses correspond to $\overline{\rm
        MS}$ masses evaluated at a scale of $1 \, {\rm TeV}$.}}
\end{center}
\end{table}

\begin{table}
\begin{center}
\begin{tabular}{|c|c|c|c|c|}
  \hline
  Parameter & Value $\pm$ Error & Reference \\[0.5mm] 
  \hline 
  $m_K$ & $497.6 \, {\rm MeV}$ & \cite{Amsler:2008zz} \\
  $f_K$ & $156.1 \, {\rm MeV}$ & \cite{Amsler:2008zz} \\
  $B_1^{sd}$ & $0.527 \pm 0.022$ & \cite{Aubin:2009jh}\\
  $B_2^{sd}$ & $0.7 \pm 0.2$ & \cite{Lubicz:2008am} \\
  $B_3^{sd}$ & $1.0 \pm 0.4$ & \cite{Lubicz:2008am} \\
  $B_4^{sd}$ & $0.9 \pm 0.2$ & \cite{Lubicz:2008am} \\
  $B_5^{sd}$ & $0.6 \pm 0.1$ & \cite{Lubicz:2008am} \\
  $(m_s + m_d)$ & $(135 \pm 18) \, {\rm MeV}$ & \cite{Lubicz:2008am} \\
  $\phi_\epsilon$ & $(43.51 \pm 0.05)^\circ$ & \cite{Amsler:2008zz} \\
  $\kappa_{\epsilon}$ & $0.92 \pm 0.02$ & \cite{Buras:2008nn} \\
  $(\Delta m_K)_{\rm exp}$ & $3.4833 \cdot 10^{-15} \, {\rm GeV}$ &
  \cite{Amsler:2008zz} \\
  $m_t (m_t)$ & $(163.8 \pm 1.3) \, {\rm GeV}$ & \cite{Group:2008nq} \\
  $m_c (m_c)$ & $\left (1.27^{+0.07}_{-0.11} \right )\, {\rm GeV}$ & 
  \cite{Amsler:2008zz} \\
  $\eta_{tt}$ & $0.57 \pm 0.01$ & \cite{Buras:1990fn} \\
  $\eta_{cc}$ & $1.50 \pm 0.37$ & \cite{Herrlich:1993yv} \\
  $\eta_{ct}$ & $0.47 \pm 0.05$  & \cite{Herrlich:1995hh, Herrlich:1996vf} \\
  \hline 
\end{tabular}
\end{center}
\vspace{-7mm}
\begin{center}
  \parbox{15.5cm}{\caption{\label{tab:input2} Parameters used in the
      $K$--$\bar K$ mixing observables. The given values for the
      $B_i^{sd}$ parameters and the sum $(m_s + m_d)$ correspond to
      the RI-MOM scheme taken at the scale $2 \, {\rm GeV}$.}}
\end{center}
\end{table}

\begin{table}
\begin{center}
\begin{tabular}{|c|c|c|c|c|}
  \hline
  Parameter & Value $\pm$ Error & Reference \\[0.5mm] 
  \hline 
  $m_{B_d}$ & $5.2795 \, {\rm GeV}$ & \cite{Amsler:2008zz} \\
  $f_{B_d}$ & $(200 \pm 20) \, {\rm MeV}$ & \cite{Lubicz:2008am} \\
  $B_1^{bd}$ & $0.81 \pm 0.08$ & \cite{Lubicz:2008am} \\
  $B_2^{bd}$ & $0.84 \pm 0.10$ & \cite{Lubicz:2008am} \\
  $B_3^{bd}$ & $0.89 \pm 0.13$ & \cite{Lubicz:2008am} \\
  $B_4^{bd}$ & $1.14 \pm 0.13$ & \cite{Lubicz:2008am} \\
  $B_5^{bd}$ & $1.72 \pm 0.19$ & \cite{Lubicz:2008am} \\
  $(m_b + m_d)$ & $(4.22 \pm 0.08) \, {\rm GeV}$ & \cite{Lubicz:2008am} \\
  $\eta_{B}$ & $0.55 \pm 0.01$ & \cite{Buras:1990fn} \\
  $\sin 2 \beta$ & $0.672 \pm 0.023$ & \cite{Barberio:2008fa} \\
  $(\Delta m_d/\Gamma_d)_{\rm exp}$ & $0.78 \pm 0.01$ &  
  \cite{Barberio:2008fa} \\
  ${\rm Re} \, ( \Gamma_{12}^{d}/M_{12}^{d})$ & 
  $\left (-5.26^{+1.15}_{-1.28} \right )\cdot 10^{-3}$ & \cite{Lenz:2006hd} \\ 
  ${\rm Im} \, ( \Gamma_{12}^{d}/M_{12}^{d} )$ & 
  $\left (-4.8^{+1.0}_{-1.2} \right )\cdot 10^{-4}$ & \cite{Lenz:2006hd} \\ 
  $m_{B_s}$ & $5.3661 \, {\rm GeV}$ & \cite{Amsler:2008zz} \\
  $f_{B_s}$ & $(245 \pm 25) \, {\rm MeV}$ & \cite{Lubicz:2008am} \\
  $B_1^{bs}$ & $0.80 \pm 0.08$ & \cite{Lubicz:2008am} \\
  $B_2^{bs}$ & $0.85 \pm 0.10$ & \cite{Lubicz:2008am} \\
  $B_3^{bs}$ & $0.90 \pm 0.13$ & \cite{Lubicz:2008am} \\
  $B_4^{bs}$ & $1.15 \pm 0.13$ & \cite{Lubicz:2008am} \\
  $B_5^{bs}$ & $1.74 \pm 0.19$ & \cite{Lubicz:2008am} \\
  $(m_b + m_s)$ & $(4.30 \pm 0.08) \, {\rm GeV}$ & \cite{Lubicz:2008am} \\
  $\sin 2 \beta_s$ & $\left ( -0.0368^{+0.0018}_{-0.0017} \right )$ & 
  \cite{Charles:2004jd} \\
  $(\Delta m_s/\Gamma_s)_{\rm exp}$ & $26.3 \pm 0.6$ &  
  \cite{Barberio:2008fa} \\
  ${\rm Re} \, ( \Gamma_{12}^{s}/M_{12}^{s} )$ & 
  $(-4.97\pm 0.94)\cdot 10^{-3}$ &  \cite{Lenz:2006hd} \\
  ${\rm Im} \, ( \Gamma_{12}^{s}/M_{12}^{s} )$ & 
  $(2.06\pm 0.57)\cdot 10^{-5}$ &  \cite{Lenz:2006hd} \\
  \hline 
\end{tabular}
\end{center}
\vspace{-7mm}
\begin{center}
  \parbox{15.5cm}{\caption{\label{tab:input3} Parameters used in the
      $B_{d,s}$--$\bar B_{d,s}$ mixing observables. The given values
      for the $B_i^{bd}$ and $B_i^{bs}$ parameters and the sums $(m_b
      + m_{d,s})$ are $\overline{\rm MS}$ quantities normalized at the
      scale $4.2 \, {\rm GeV}$.}}
\end{center}
\end{table}

\begin{table}
\begin{center}
\begin{tabular}{|c|c|c|c|c|}
  \hline
  Parameter & Value $\pm$ Error & Reference \\[0.5mm] 
  \hline 
  $m_D$ & $1.8645 \, {\rm MeV}$ & \cite{Amsler:2008zz} \\
  $f_D$ & $(212 \pm 14) \, {\rm MeV}$ & \cite{Lubicz:2008am} \\
  $B_1^{cu}$ & $0.85 \pm 0.09$ & \cite{Lubicz:2008am} \\
  $B_2^{cu}$ & $0.82 \pm 0.09$ & \cite{Lubicz:2008am} \\
  $B_3^{cu}$ & $1.07 \pm 0.12$ & \cite{Lubicz:2008am} \\
  $B_4^{cu}$ & $1.10 \pm 0.11$ & \cite{Lubicz:2008am} \\
  $B_5^{cu}$ & $1.37 \pm 0.14$ & \cite{Lubicz:2008am} \\
  $(m_c + m_u)$ & $(1.17 \pm 0.12) \, {\rm GeV}$ & 
  \cite{Amsler:2008zz, Chetyrkin:1999pq} \\
  $x_D$ & $0.0098^{+0.0024}_{-0.0026}$ & \cite{Barberio:2008fa, 
    Schwartz:2009jv} \\
  $y_D$ & $0.0083 \pm 0.0016$ & \cite{Barberio:2008fa, 
    Schwartz:2009jv} \\
  $(x_D^2 + y_D^2)/2$ & $\leq (1.3 \pm 2.7)\cdot 10^{-4}$ &
  \cite{Barberio:2008fa, Schwartz:2009jv} \\
  $|q/p|_D$ & $0.87^{+0.17}_{-0.15}$ & \cite{Barberio:2008fa, 
    Schwartz:2009jv} \\
  $\tau_D$ & $(0.4101 \pm 0.0015) \, {\rm ps}$ & 
  \cite{Barberio:2008fa} \\
  \hline 
\end{tabular}
\end{center}
\vspace{-7mm}
\begin{center}
  \parbox{15.5cm}{\caption{\label{tab:input4} Parameters used in the
      $D$--$\bar D$ mixing observables. The quoted values for the
      $B_i^{cu}$ parameters and the sum of masses $(m_c + m_u)$ is
      obtained in the RI-MOM scheme at the scale $2.8 \, {\rm GeV}$.}}
\end{center}
\end{table}

\begin{table}
\begin{center}
\begin{tabular}{|c|c|c|c|c|}
  \hline
  Parameter & Value $\pm$ Error & Reference \\[0.5mm] 
  \hline 
  $B_6^{(1/2)}$ & $[0.8, 2.0]$ & \cite{Buras:2003zz} \\
  $B_8^{(3/2)}$ & $[0.8, 1.2]$ & \cite{Buras:2003zz} \\ 
  $m_s(m_c)$ & $(115 \pm 20) \, {\rm MeV}$ & \cite{Amsler:2008zz} \\
  $m_d(m_c)$ & $(6 \pm 2) \, {\rm MeV}$ & \cite{Amsler:2008zz} \\
  $|V_{cb}|$ & $0.041 \pm 0.002$ & \cite{Beneke:2003zv} \\
  $|V_{ub}/V_{cb}|$ & $0.08 \pm 0.02$ &  \cite{Beneke:2003zv} \\
  $\gamma$ & $(70 \pm 20)^\circ$ & \cite{Beneke:2003zv}  \\
  $\tau_{B_u}$ & $1.67 \, {\rm ps}$ & \cite{Beneke:2003zv}  \\ 
  $\tau_{B_d}$ & $1.54 \, {\rm ps}$ & \cite{Beneke:2003zv}  \\ 
  $f_\pi/f_K$ & $0.82$ & \cite{Beneke:2003zv}  \\
  $f_\phi$ & $221 \, {\rm MeV}$ &  \cite{Beneke:2003zv} \\
  $f_\phi^\perp$ & $(175 \pm 25) \, {\rm MeV}$ &  \cite{Beneke:2003zv} \\
  $F_0^{B \to \pi} (0)$ & $0.25 \pm 0.05$ &  \cite{Beneke:2003zv} \\
  $F_0^{B \to K} (0)$ & $0.31 \pm 0.05$ &  \cite{Beneke:2003zv} \\
  $m_s$ & $(80 \pm 20)\, {\rm MeV}$ &  \cite{Beneke:2003zv} \\
  $\lambda_B$ & $(200 \pm 150) \, {\rm MeV}$ & \cite{Beneke:2003zv}  \\
  $\alpha_2^\pi$ & $0.3 \pm 0.3$ &  \cite{Beneke:2003zv} \\
  $\alpha_1^{\bar K}$ & $0.2 \pm 0.2$ &  \cite{Beneke:2003zv} \\
  $\alpha_2^{\bar K}$ & $0.1 \pm 0.3$ &  \cite{Beneke:2003zv} \\
  $\alpha_{2}^\phi, \alpha_{2, \perp}^\phi$ & $0 \pm 0.3$ &  
  \cite{Beneke:2003zv} \\
  $\rho_H$ & $1$ &  y\cite{Beneke:2003zv}\\
  $\rho_A$ & $1$ &  \cite{Beneke:2003zv} \\
  $\varphi_A^{PP}$ & $-55^\circ$ &  \cite{Beneke:2003zv} \\
  $\varphi_A^{PV}$ & $-20^\circ$ &  \cite{Beneke:2003zv} \\
  \hline 
\end{tabular}
\end{center}
\vspace{-7mm}
\begin{center}
  \parbox{15.5cm}{\caption{\label{tab:input5} Parameters used to
      predict the ratio $\epseps$ and the various non-leptonic
      $B$-meson decays. All scale-dependent quantities refer to $2 \,
      {\rm GeV}$ unless indicated otherwise. See text for details.}}
\end{center}
\end{table}

\begin{table}
\begin{center}
\begin{tabular}{|c|c|c|c|c|}
  \hline
  Parameter & Value $\pm$ Error & Reference \\[0.5mm] 
  \hline 
  $\kappa_L$ & $(2.353 \pm 0.014) \cdot 10^{-10}$ &  
  \cite{Mescia:2007kn} \\
  $\kappa_+$ & $(0.5455 \pm 0.0026) \cdot 10^{-10}$ &  
  \cite{Mescia:2007kn} \\
  $\Delta_{\rm EM}$ & $-0.003$ & \cite{Mescia:2007kn} \\
  $X_t$ & $1.464 \pm 0.041$ & \cite{Misiak:1999yg, Buchalla:1998ba} \\
  $P_{c,u}$ & $0.43 \pm 0.04$ & \cite{Isidori:2005xm, Buras:2005gr,
    Buras:2006gb, Brod:2008ss} \\
  $y_c$ & $(-0.20 \pm 0.03)$ & \cite{Gorbahn:2006bm} \\
  $y_{\gamma \gamma}$ & $0.4 \pm 0.5$ & \cite{Isidori:2003ts} \\
  $y_A$ & $(-0.68\pm 0.03)$ & \cite{Buras:1994qa} \\
  $|a_S|$ & $1.20 \pm 0.20$ & \cite{Batley:2003mu,Batley:2004wg} \\
  $y_V$ & $0.73 \pm 0.04$ & \cite{Buras:1994qa} \\
  ${\cal B} (B \to X_c l \bar \nu)_{\rm exp}$ & $0.1064 \pm 0.0011$ &
  \cite{Barberio:2008fa} \\
  $C$ & $0.546^{+0.023}_{-0.033}$ & \cite{Gambino:2008fj} \\
  $\tau_{B_d}$ & $(1.525 \pm 0.009 ) \, {\rm ps}$ & \cite{Barberio:2008fa} \\
  $\tau_{B_s}$ & $\left ( 1.472^{+0.024}_{-0.026} \right ) {\rm ps}$ & 
  \cite{Barberio:2008fa} \\
  $c_A$ & $0.96 \pm 0.02$ & \cite{Misiak:1999yg, Buchalla:1998ba} \\
  $m_\mu$ &  $105.66 \, \, {\rm MeV}$ & \cite{Amsler:2008zz} \\
  $m_\tau$ & $1.777 \, {\rm GeV}$ & \cite{Amsler:2008zz} \\
  \hline 
\end{tabular}
\end{center}
\vspace{-7mm}
\begin{center}
  \parbox{15.5cm}{\caption{\label{tab:input6} Parameters entering the
      predictions for the rare $K$- and $B$-meson decays.}}
\end{center}
\end{table}

\clearpage

Our predictions for the $B \to K^\ast l^+ l^-$ observables depend on
the $B \to K^\ast$ matrix elements, which can be parametrized in terms
of seven $q^2$-dependent QCD form factors $V$, $A_{0,1,2}$, and
$T_{1,2,3}$ as 
\begin{eqnarray}
\begin{aligned}
  & \left \langle K^\ast(p_B-q) | \bar{s} \gamma_\mu (1-\gamma_5) b
    |B(p_B) \right \rangle = - 2 \hspace{0.25mm}
  \epsilon_{\mu\nu\alpha\beta} \, \varepsilon^{\ast \, \nu} p_B^\alpha
  q^\beta \, \frac{V }{m_B + m_{K^\ast}} - i \varepsilon_\mu^\ast
  \left (m_B + m_{K^\ast} \right ) A_1   \\
  & \phantom{xx} + i \left (2 p_B - q \right )_\mu
  (\varepsilon^\ast\cdot q) \, \frac{A_2 }{m_B + m_{K^\ast}} + i q_\mu
  \left (\varepsilon^\ast\cdot q \right ) \, \frac{2
    m_{K^\ast}}{q^2} \, \left [ A_3  - A_0  \right ] \,, \\[2mm]
  & \left \langle K^\ast (p_B-q) | \bar{s} \sigma_{\mu\nu} q^\nu
    (1+\gamma_5) b | B(p_B) \right \rangle = -2 i\,
  \epsilon_{\mu\nu\alpha\beta} \, \varepsilon^{\ast \, \nu} \,
  p_B^\alpha q^\beta\, T_1
  \\
  & \phantom{xx} + \left [\varepsilon_\mu^\ast \left (m_B^2 -
      m_{K^\ast}^2 \right ) - (\varepsilon^\ast\cdot q) (2 p_B -
    q)_\mu \right ] T_2 + (\varepsilon^\ast\cdot q) \bigg[q_\mu -
  \frac{q^2}{m_B^2 - m_{K^\ast}^2} \, (2 p_B - q)_\mu \bigg]\, T_3 \,,
\end{aligned}
\end{eqnarray}
where
\begin{equation}
  A_3 \equiv \frac{m_B + m_{K^\ast}}{2 m_{K^\ast}} \, A_1  
  - \frac{m_B - m_{K^\ast}}{2 m_{K^\ast}} \, A_2  \,.
\end{equation}
Here $\varepsilon^{\ast \, \mu}$ denotes the polarization vector of
the $K^\ast$ and $p_B^\mu$ the four momentum of the $B$ meson.

In the large recoil limit, the QCD form factors obey symmetry
relations and can be expressed at leading order in the $1/E$ expansion
in terms of two universal form factors $\xi_{\perp,\parallel}$
\cite{Charles:1998dr}. Symmetry-breaking corrections at order
$\alpha_s$ have been calculated using QCD factorization
\cite{Beneke:2000wa} and are included in our numerical
analysis. Within the QCD factorization approach, we employ a
factorization scheme where the universal form factors
$\xi_{\perp, \parallel}$ are related to $V$ and $A_{1,2}$ via
\cite{Beneke:2004dp}
\begin{equation} \label{eq:xidef} 
  \xi_\perp = \frac{m_B}{m_B +
    m_{K^\ast}} \, V \,, \qquad \xi_\parallel = \frac{m_B +
    m_{K^\ast}}{2 E} \, A_1 - \frac{m_B - m_{K^\ast}}{m_B} \, A_2 \,.
\end{equation}
Here $E$ denotes the energy of the emitted $K^\ast$ in the $B$-meson
rest frame.

In order to describe the $q^2$-dependence of the universal form
factors $\xi_{\perp, \parallel}$, we follow the light-cone sum rule
calculation \cite{Ball:2004rg}, which utilizes the parametrizations
\beq
\begin{split}
  V(q^2) & = \frac{r_1}{1 - q^2/m_{R}^2} + \frac{r_2}{1 -
    q^2/m_{\rm fit}^2} \, ,  \\
  A_1(q^2) & = \frac{r_2}{1 - q^2/m_{\rm fit}^2} \, ,  \\
  A_2(q^2) & = \frac{r_1}{1 - q^2/m_{\rm fit}^2} + \frac{r_2}{(1 -
    q^2/m_{\rm fit}^2)^2} \, .
\end{split}
\eeq
In Table~\ref{tab:LCSRfit} we collect the values of the fit parameters
$r_{1,2}$, $m^2_{R}$, and $m^2_{\rm fit}$ for $V$ and $A_{1,2}$
derived in \cite{Ball:2004rg}. The central values and errors of the
Gegenbauer moments $a_{1, K^\ast}^{\perp, \parallel}$ and $a_{2,
  K^\ast}^{\perp, \parallel}$ are also taken from the latter
reference. The relative uncertainty of the form factors $V$, $A_1$,
and $A_{2}$ at zero-momentum transfer amounts to $11\%$, $12\%$, and
$14\%$, adding individual sources of uncertainty in quadrature. In our
numerical analysis we use the total relative uncertainty at maximal
recoil to estimate the form-factor uncertainties for $q^2 > 0$. The
quoted errors translate into uncertainties of 11\% and 14\% for
$\xi_\perp$ and $\xi_\parallel$, respectively.

\begin{table}
\begin{center}
\begin{tabular}{|c|c|c|c|c|}
  \hline
  $ $ & $r_1$ & $r_2$ & $m_R^2 \left [{\rm GeV}^2 \right ]$ & 
  $m_{\rm fit}^2\,\left [{\rm GeV}^2 \right ]$ \\
  \hline
  $V$   & $\phantom{-}0.923$  & $-0.511$ & $28.30$ & $49.40$ 
  \\
  $A_1$ & \text{--} & $\phantom{-}0.290$ & \text{--} & $40.38$  
  \\
  $A_2$ & $-0.084$ &  $\phantom{-}0.342$ & \text{--}  & $52.00$  
  \\
  $T_1$   & $\phantom{-}0.823$  & $-0.491$ & $28.30$ & $46.31$ 
  \\
  $T_2$ &  \text{--} & $\phantom{-}0.333$ & \text{--} & $41.41$  
  \\
  $\tilde T_3$ & $-0.036$ &  $\phantom{-}0.368$ & \text{--} & $48.10$  
  \\
  \hline 
\end{tabular}
\end{center}
\vspace{-7mm}
\begin{center}
  \parbox{15.5cm}{\caption{\label{tab:LCSRfit}
      Parameters describing the $q^2$-dependence of the
      form factors $V$, $A_{1,2}$, $T_{1,2}$, and $\tilde T_3$. All
      results are taken from \cite{Ball:2004rg}. See text for
      details.}}
\end{center}
\end{table}

The $q^2$-dependent form factors $f$, $g$, $a_+$, $g_\pm$, and $h$
relevant for $B \to K^\ast l^+ l^-$ decay in the low-recoil limit depend on
$V$, $A_{1,2}$, and $T_{1,2,3}$ via \cite{Grinstein:2004vb}
\begin{gather}
  f = \left (m_B + m_{K^\ast} \right ) A_1 \,, \qquad g =
  -\frac{V}{m_B + m_{K^\ast}} \,, \qquad
  a_+ = -\frac{A_2}{m_B + m_{K^\ast}} \,, \nonumber \\[-3.5mm]
 \\[-3.5mm]
  g_+ = T_1 \,, \qquad g_- = \frac{m_B^2 - m_{K^\ast}^2}{q^2} \, \left
    (T_2 - T_1 \right ) \,, \qquad h = \frac{T_1 - T_2}{q^2} -
  \frac{T_3}{m_B^2 - m_{K^\ast}^2} \,, \nonumber 
\end{gather}
with  
\begin{equation}
  T_3 \equiv \frac{m_B^2 - m_{K^\ast}^2}{q^2} \, \big ( 
  \tilde T_3 - T_2 \big ) .
\end{equation}
The $q^2$-dependence of the form factors $T_{1,2}$ and $\tilde T_3$ is
modeled by \cite{Ball:2004rg}
\beq
\begin{split}
  T_1(q^2) & = \frac{r_1}{1 - q^2/m_{R}^2} + \frac{r_2}{1 -
    q^2/m_{\rm fit}^2} \, ,  \\
  T_2(q^2) & = \frac{r_2}{1 - q^2/m_{\rm fit}^2} \, ,  \\
  \tilde T_3(q^2) & = \frac{r_1}{1 - q^2/m_{\rm fit}^2} +
  \frac{r_2}{(1 - q^2/m_{\rm fit}^2)^2} \, ,
\end{split}
\eeq
and the corresponding fit parameters $r_{1,2}$, $m^2_{R}$, and
$m^2_{\rm fit}$ can be found in Table~\ref{tab:LCSRfit}. The QCD form
factors $f$, $g$, $a_+$, $g_\pm$, and $h$ all have an uncertainty at
the level of 10\%.

\end{appendix}

\end{document}